\documentclass[aps,prd,twocolumn,longbibliography,nofootinbib,superscriptaddress,floatfix]{revtex4-2}
\usepackage{color}
\newcommand{\llmfig}{ Generated with LLM assistance per Appendix~\ref{app:llm}.}
\ifdefined\DeclareUnicodeCharacter
\DeclareUnicodeCharacter{2500}{--}
\fi

\usepackage{hyperref}
\usepackage{amsmath}
\usepackage{amssymb}
\usepackage{graphicx}
\usepackage{booktabs}
\usepackage{bm}
\usepackage{xr}
\makeatletter
\ifdefined\XR@@nocite
\makeatletter
\@namedef{r@sup-FirstPage}{{}{1}{}{}{}}
\@namedef{r@sup-sup:population}{{SI}{1}{}{}{}}
\@namedef{r@sup-sup:rates}{{SII}{1}{}{}{}}
\@namedef{r@sup-sec:return:implications}{{SIII}{1}{}{}{}}
\@namedef{r@sup-tab:rate-normalization}{{SI}{1}{}{}{}}
\@namedef{r@sup-sup:isolated}{{SIII\,A}{1}{}{}{}}
\@namedef{r@sup-sup:torque-tests}{{SIII\,B}{1}{}{}{}}
\@namedef{r@sup-fig:chaotic}{{S1}{2}{}{}{}}
\@namedef{r@sup-sec:return:clock}{{SIII\,C}{2}{}{}{}}
\@namedef{r@sup-eq:recurrence}{{S1}{2}{}{}{}}
\@namedef{r@sup-eq:shell-response}{{S2}{2}{}{}{}}
\@namedef{r@sup-fig:radius}{{S2}{3}{}{}{}}
\@namedef{r@sup-fig:returnmap}{{S3}{3}{}{}{}}
\@namedef{r@sup-sec:return:mach}{{SIII\,D}{3}{}{}{}}
\@namedef{r@sup-eq:mach-weight}{{S3}{3}{}{}{}}
\@namedef{r@sup-fig:tidal-scaling-test}{{S4}{4}{}{}{}}
\@namedef{r@sup-tab:return-clock}{{SII}{4}{}{}{}}
\@namedef{r@sup-sup:validity}{{SIV}{4}{}{}{}}
\@namedef{r@sup-tab:return-audit}{{SIII}{5}{}{}{}}
\@namedef{r@sup-fig:clock}{{S5}{5}{}{}{}}
\@namedef{r@sup-tab:drag-law}{{SIV}{5}{}{}{}}
\@namedef{r@sup-tab:mach-audit}{{SV}{5}{}{}{}}
\@namedef{r@sup-tab:validity}{{SVI}{5}{}{}{}}
\@namedef{r@sup-sec:hill-numerics}{{SV}{6}{}{}{}}
\@namedef{r@sup-eq:hill-kernel-estimator}{{S4}{6}{}{}{}}
\@namedef{r@sup-eq:hill-kernel-literature-rate}{{S5}{6}{}{}{}}
\@namedef{r@sup-eq:hill-kernel-jacobi}{{S6}{6}{}{}{}}
\@namedef{r@sup-fig:hill-estimator-diagram}{{S6}{7}{}{}{}}
\@namedef{r@sup-eq:hill-kernel-cancellation}{{S7}{7}{}{}{}}
\@namedef{r@sup-eq:hill-kernel-free-limit}{{S8}{7}{}{}{}}
\@namedef{r@sup-fig:hill-numerical-controls}{{S7}{8}{}{}{}}
\@namedef{r@sup-app:llm}{{SVI}{9}{}{}{}}
\@namedef{r@sup-LastBibItem}{{8}{9}{}{}{}}
\@namedef{r@sup-LastPage}{{}{9}{}{}{}}
\makeatother
\else
\makeatletter
\@namedef{r@sup-FirstPage}{{}{1}{}{}{}}
\@namedef{r@sup-sup:population}{{SI}{1}{}{}{}}
\@namedef{r@sup-sup:rates}{{SII}{1}{}{}{}}
\@namedef{r@sup-sec:return:implications}{{SIII}{1}{}{}{}}
\@namedef{r@sup-tab:rate-normalization}{{SI}{1}{}{}{}}
\@namedef{r@sup-sup:isolated}{{SIII\,A}{1}{}{}{}}
\@namedef{r@sup-sup:torque-tests}{{SIII\,B}{1}{}{}{}}
\@namedef{r@sup-fig:chaotic}{{S1}{2}{}{}{}}
\@namedef{r@sup-sec:return:clock}{{SIII\,C}{2}{}{}{}}
\@namedef{r@sup-eq:recurrence}{{S1}{2}{}{}{}}
\@namedef{r@sup-eq:shell-response}{{S2}{2}{}{}{}}
\@namedef{r@sup-fig:radius}{{S2}{3}{}{}{}}
\@namedef{r@sup-fig:returnmap}{{S3}{3}{}{}{}}
\@namedef{r@sup-sec:return:mach}{{SIII\,D}{3}{}{}{}}
\@namedef{r@sup-eq:mach-weight}{{S3}{3}{}{}{}}
\@namedef{r@sup-fig:tidal-scaling-test}{{S4}{4}{}{}{}}
\@namedef{r@sup-tab:return-clock}{{SII}{4}{}{}{}}
\@namedef{r@sup-sup:validity}{{SIV}{4}{}{}{}}
\@namedef{r@sup-tab:return-audit}{{SIII}{5}{}{}{}}
\@namedef{r@sup-fig:clock}{{S5}{5}{}{}{}}
\@namedef{r@sup-tab:drag-law}{{SIV}{5}{}{}{}}
\@namedef{r@sup-tab:mach-audit}{{SV}{5}{}{}{}}
\@namedef{r@sup-tab:validity}{{SVI}{5}{}{}{}}
\@namedef{r@sup-sec:hill-numerics}{{SV}{6}{}{}{}}
\@namedef{r@sup-eq:hill-kernel-estimator}{{S4}{6}{}{}{}}
\@namedef{r@sup-eq:hill-kernel-literature-rate}{{S5}{6}{}{}{}}
\@namedef{r@sup-eq:hill-kernel-jacobi}{{S6}{6}{}{}{}}
\@namedef{r@sup-fig:hill-estimator-diagram}{{S6}{7}{}{}{}}
\@namedef{r@sup-eq:hill-kernel-cancellation}{{S7}{7}{}{}{}}
\@namedef{r@sup-eq:hill-kernel-free-limit}{{S8}{7}{}{}{}}
\@namedef{r@sup-fig:hill-numerical-controls}{{S7}{8}{}{}{}}
\@namedef{r@sup-app:llm}{{SVI}{9}{}{}{}}
\@namedef{r@sup-LastBibItem}{{8}{9}{}{}{}}
\@namedef{r@sup-LastPage}{{}{9}{}{}{}}
\makeatother
\fi
\makeatother

\providecommand\prd{Phys.~Rev.~D}
\providecommand\mnras{MNRAS}\providecommand\apjl{ApJL}

\providecommand\aap{A\&A}\providecommand\apj{ApJ}\providecommand\araa{ARA\&A}
\providecommand\nat{Nature}

\newcommand\unit[1]{{\rm #1}}
\newcommand\Msun{M_\odot}
\newcommand\Rsun{R_\odot}
\newcommand\Mstar{M_\star}
\newcommand\Rstar{R_\star}
\newcommand\Mbh{M_\bullet}
\newcommand\rp{r_{\rm p}}
\newcommand\rH{r_{\rm H}}
\newcommand\vH{v_{\rm H}}
\newcommand\vinf{v_\infty}
\newcommand\Mach{\mathcal{M}}
\newcommand\Msmbh{M_{\rm SMBH}}
\newcommand\Sigmag{\Sigma_g}
\newcommand\mt{m_{\rm t}}
\newcommand\te{t_e}
\newcommand\Sturb{S_{\rm turb}}
\newcommand\Sstir{S_{\rm stir}}
\newcommand\Sdrag{S_{\rm drag,e}}
\newcommand\Cturb{C_{\rm turb}}

\providecommand\result[1]{%
  \ifcsname #1\endcsname\csname #1\endcsname\else\textbf{??#1??}\fi}

\expandafter\providecommand\csname BhMassMax\endcsname{}\expandafter\renewcommand\csname BhMassMax\endcsname{\ensuremath{40}}
\expandafter\providecommand\csname BhMassMin\endcsname{}\expandafter\renewcommand\csname BhMassMin\endcsname{\ensuremath{5}}
\expandafter\providecommand\csname BhMeanMass\endcsname{}\expandafter\renewcommand\csname BhMeanMass\endcsname{\ensuremath{11.9}}
\expandafter\providecommand\csname BinHighChaoticCollisions\endcsname{}\expandafter\renewcommand\csname BinHighChaoticCollisions\endcsname{\ensuremath{87.5}}
\expandafter\providecommand\csname BinHighCollisionRate\endcsname{}\expandafter\renewcommand\csname BinHighCollisionRate\endcsname{\ensuremath{1013}}
\expandafter\providecommand\csname BinHighCollisionRatio\endcsname{}\expandafter\renewcommand\csname BinHighCollisionRatio\endcsname{\ensuremath{2.56}}
\expandafter\providecommand\csname BinHighRate\endcsname{}\expandafter\renewcommand\csname BinHighRate\endcsname{\ensuremath{4.395\times10^{4}}}
\expandafter\providecommand\csname BinHighRateRatio\endcsname{}\expandafter\renewcommand\csname BinHighRateRatio\endcsname{\ensuremath{1.11}}
\expandafter\providecommand\csname BinLowChaoticCollisions\endcsname{}\expandafter\renewcommand\csname BinLowChaoticCollisions\endcsname{\ensuremath{38.1}}
\expandafter\providecommand\csname BinLowCollisionRate\endcsname{}\expandafter\renewcommand\csname BinLowCollisionRate\endcsname{\ensuremath{204.6}}
\expandafter\providecommand\csname BinLowCollisionRatio\endcsname{}\expandafter\renewcommand\csname BinLowCollisionRatio\endcsname{\ensuremath{0.517}}
\expandafter\providecommand\csname BinLowRate\endcsname{}\expandafter\renewcommand\csname BinLowRate\endcsname{\ensuremath{3.82\times10^{4}}}
\expandafter\providecommand\csname BinLowRateRatio\endcsname{}\expandafter\renewcommand\csname BinLowRateRatio\endcsname{\ensuremath{0.966}}
\expandafter\providecommand\csname BinaryAMin\endcsname{}\expandafter\renewcommand\csname BinaryAMin\endcsname{\ensuremath{0.05AU}}
\expandafter\providecommand\csname BinaryBinaryCollisionFraction\endcsname{}\expandafter\renewcommand\csname BinaryBinaryCollisionFraction\endcsname{\ensuremath{14}}
\expandafter\providecommand\csname BinaryBinaryCollisionRate\endcsname{}\expandafter\renewcommand\csname BinaryBinaryCollisionRate\endcsname{\ensuremath{13.12}}
\expandafter\providecommand\csname BinaryBinaryMedianQ\endcsname{}\expandafter\renewcommand\csname BinaryBinaryMedianQ\endcsname{\ensuremath{0.0689}}
\expandafter\providecommand\csname BinaryBinaryMedianU\endcsname{}\expandafter\renewcommand\csname BinaryBinaryMedianU\endcsname{\ensuremath{0}}
\expandafter\providecommand\csname BinaryBinaryMedianX\endcsname{}\expandafter\renewcommand\csname BinaryBinaryMedianX\endcsname{\ensuremath{5.15}}
\expandafter\providecommand\csname BinaryBinaryMicroTdeRate\endcsname{}\expandafter\renewcommand\csname BinaryBinaryMicroTdeRate\endcsname{\ensuremath{6.847}}
\expandafter\providecommand\csname BinaryBinaryMultiplicity\endcsname{}\expandafter\renewcommand\csname BinaryBinaryMultiplicity\endcsname{\ensuremath{2}}
\expandafter\providecommand\csname BinaryBinaryQsmallXsmall\endcsname{}\expandafter\renewcommand\csname BinaryBinaryQsmallXsmall\endcsname{\ensuremath{12.7}}
\expandafter\providecommand\csname BinaryBinaryQsmallXsmallRate\endcsname{}\expandafter\renewcommand\csname BinaryBinaryQsmallXsmallRate\endcsname{\ensuremath{11.96}}
\expandafter\providecommand\csname BinaryBinaryRate\endcsname{}\expandafter\renewcommand\csname BinaryBinaryRate\endcsname{\ensuremath{93.87}}
\expandafter\providecommand\csname BinaryBinaryVolumetricThousand\endcsname{}\expandafter\renewcommand\csname BinaryBinaryVolumetricThousand\endcsname{\ensuremath{93.87}}
\expandafter\providecommand\csname BinaryFractionBh\endcsname{}\expandafter\renewcommand\csname BinaryFractionBh\endcsname{\ensuremath{0.10}}
\expandafter\providecommand\csname BinaryFractionStarBh\endcsname{}\expandafter\renewcommand\csname BinaryFractionStarBh\endcsname{\ensuremath{0.050}}
\expandafter\providecommand\csname BinaryMedianSemimajor\endcsname{}\expandafter\renewcommand\csname BinaryMedianSemimajor\endcsname{\ensuremath{0.397}}
\expandafter\providecommand\csname BinaryMedianVoverVc\endcsname{}\expandafter\renewcommand\csname BinaryMedianVoverVc\endcsname{\ensuremath{0.587}}
\expandafter\providecommand\csname BinarySingleCollisionFraction\endcsname{}\expandafter\renewcommand\csname BinarySingleCollisionFraction\endcsname{\ensuremath{14.1}}
\expandafter\providecommand\csname BinarySingleCollisionRate\endcsname{}\expandafter\renewcommand\csname BinarySingleCollisionRate\endcsname{\ensuremath{256}}
\expandafter\providecommand\csname BinarySingleMedianQ\endcsname{}\expandafter\renewcommand\csname BinarySingleMedianQ\endcsname{\ensuremath{0.069}}
\expandafter\providecommand\csname BinarySingleMedianU\endcsname{}\expandafter\renewcommand\csname BinarySingleMedianU\endcsname{\ensuremath{0}}
\expandafter\providecommand\csname BinarySingleMedianX\endcsname{}\expandafter\renewcommand\csname BinarySingleMedianX\endcsname{\ensuremath{5.12}}
\expandafter\providecommand\csname BinarySingleMicroTdeRate\endcsname{}\expandafter\renewcommand\csname BinarySingleMicroTdeRate\endcsname{\ensuremath{132.1}}
\expandafter\providecommand\csname BinarySingleQsmallXsmall\endcsname{}\expandafter\renewcommand\csname BinarySingleQsmallXsmall\endcsname{\ensuremath{12.8}}
\expandafter\providecommand\csname BinarySingleQsmallXsmallRate\endcsname{}\expandafter\renewcommand\csname BinarySingleQsmallXsmallRate\endcsname{\ensuremath{233.3}}
\expandafter\providecommand\csname BinarySingleRate\endcsname{}\expandafter\renewcommand\csname BinarySingleRate\endcsname{\ensuremath{1820}}
\expandafter\providecommand\csname BinarySingleVolumetricThousand\endcsname{}\expandafter\renewcommand\csname BinarySingleVolumetricThousand\endcsname{\ensuremath{1820}}
\expandafter\providecommand\csname BranchDestroying\endcsname{}\expandafter\renewcommand\csname BranchDestroying\endcsname{\ensuremath{32.4}}
\expandafter\providecommand\csname BranchExchange\endcsname{}\expandafter\renewcommand\csname BranchExchange\endcsname{\ensuremath{31.4}}
\expandafter\providecommand\csname BranchFlyby\endcsname{}\expandafter\renewcommand\csname BranchFlyby\endcsname{\ensuremath{67.6}}
\expandafter\providecommand\csname BranchIonization\endcsname{}\expandafter\renewcommand\csname BranchIonization\endcsname{\ensuremath{0.996}}
\expandafter\providecommand\csname BranchIonizationCold\endcsname{}\expandafter\renewcommand\csname BranchIonizationCold\endcsname{\ensuremath{0.0678}}
\expandafter\providecommand\csname BranchIonizationSonic\endcsname{}\expandafter\renewcommand\csname BranchIonizationSonic\endcsname{\ensuremath{3.16}}
\expandafter\providecommand\csname BranchIonizationStarSingle\endcsname{}\expandafter\renewcommand\csname BranchIonizationStarSingle\endcsname{\ensuremath{2.01}}
\expandafter\providecommand\csname BranchIonizationStarSingleTwoBody\endcsname{}\expandafter\renewcommand\csname BranchIonizationStarSingleTwoBody\endcsname{\ensuremath{0.124}}
\expandafter\providecommand\csname BranchIonizationTwoBody\endcsname{}\expandafter\renewcommand\csname BranchIonizationTwoBody\endcsname{\ensuremath{0.0611}}
\expandafter\providecommand\csname CalibrationARef\endcsname{}\expandafter\renewcommand\csname CalibrationARef\endcsname{\ensuremath{2.83}}
\expandafter\providecommand\csname CalibrationEtaMax\endcsname{}\expandafter\renewcommand\csname CalibrationEtaMax\endcsname{\ensuremath{3}}
\expandafter\providecommand\csname CalibrationEtaMin\endcsname{}\expandafter\renewcommand\csname CalibrationEtaMin\endcsname{\ensuremath{0.003}}
\expandafter\providecommand\csname CalibrationMassMax\endcsname{}\expandafter\renewcommand\csname CalibrationMassMax\endcsname{\ensuremath{50}}
\expandafter\providecommand\csname CalibrationMassMin\endcsname{}\expandafter\renewcommand\csname CalibrationMassMin\endcsname{\ensuremath{5}}
\expandafter\providecommand\csname CalibrationTrials\endcsname{}\expandafter\renewcommand\csname CalibrationTrials\endcsname{\ensuremath{1\times10^{6}}}
\expandafter\providecommand\csname ChaoticFractionAll\endcsname{}\expandafter\renewcommand\csname ChaoticFractionAll\endcsname{\ensuremath{4.84}}
\expandafter\providecommand\csname ChaoticFractionCollisions\endcsname{}\expandafter\renewcommand\csname ChaoticFractionCollisions\endcsname{\ensuremath{68}}
\expandafter\providecommand\csname ChaoticFractionMicroTde\endcsname{}\expandafter\renewcommand\csname ChaoticFractionMicroTde\endcsname{\ensuremath{71.4}}
\expandafter\providecommand\csname ContactUNinetyFive\endcsname{}\expandafter\renewcommand\csname ContactUNinetyFive\endcsname{\ensuremath{0.085}}
\expandafter\providecommand\csname ContactUNinetyNine\endcsname{}\expandafter\renewcommand\csname ContactUNinetyNine\endcsname{\ensuremath{0.12}}
\expandafter\providecommand\csname DiscOuterPc\endcsname{}\expandafter\renewcommand\csname DiscOuterPc\endcsname{\ensuremath{0.239}}
\expandafter\providecommand\csname DiscOuterRg\endcsname{}\expandafter\renewcommand\csname DiscOuterRg\endcsname{\ensuremath{5}}
\expandafter\providecommand\csname EdgeEncounterTime\endcsname{}\expandafter\renewcommand\csname EdgeEncounterTime\endcsname{\ensuremath{2.60}}
\expandafter\providecommand\csname EdgeHillRadius\endcsname{}\expandafter\renewcommand\csname EdgeHillRadius\endcsname{\ensuremath{359}}
\expandafter\providecommand\csname EdgeHillVelocity\endcsname{}\expandafter\renewcommand\csname EdgeHillVelocity\endcsname{\ensuremath{9.7}}
\expandafter\providecommand\csname EdgeRadiusAu\endcsname{}\expandafter\renewcommand\csname EdgeRadiusAu\endcsname{\ensuremath{49353}}
\expandafter\providecommand\csname EncountersPerStarPerMyr\endcsname{}\expandafter\renewcommand\csname EncountersPerStarPerMyr\endcsname{\ensuremath{39.6}}
\expandafter\providecommand\csname GasChaoticCollisions\endcsname{}\expandafter\renewcommand\csname GasChaoticCollisions\endcsname{\ensuremath{57.1}}
\expandafter\providecommand\csname GasCollisionRate\endcsname{}\expandafter\renewcommand\csname GasCollisionRate\endcsname{\ensuremath{471.5}}
\expandafter\providecommand\csname GasCollisionRatio\endcsname{}\expandafter\renewcommand\csname GasCollisionRatio\endcsname{\ensuremath{1.19}}
\expandafter\providecommand\csname GasRate\endcsname{}\expandafter\renewcommand\csname GasRate\endcsname{\ensuremath{3.956\times10^{4}}}
\expandafter\providecommand\csname GasRateRatio\endcsname{}\expandafter\renewcommand\csname GasRateRatio\endcsname{\ensuremath{1}}
\expandafter\providecommand\csname GravRadiusAu\endcsname{}\expandafter\renewcommand\csname GravRadiusAu\endcsname{\ensuremath{0.987}}
\expandafter\providecommand\csname ImmChaoticCollisions\endcsname{}\expandafter\renewcommand\csname ImmChaoticCollisions\endcsname{\ensuremath{78}}
\expandafter\providecommand\csname ImmCollisionRate\endcsname{}\expandafter\renewcommand\csname ImmCollisionRate\endcsname{\ensuremath{178.5}}
\expandafter\providecommand\csname ImmMicroTdeRate\endcsname{}\expandafter\renewcommand\csname ImmMicroTdeRate\endcsname{\ensuremath{87}}
\expandafter\providecommand\csname ImmOversubscription\endcsname{}\expandafter\renewcommand\csname ImmOversubscription\endcsname{\ensuremath{1.8}}
\expandafter\providecommand\csname KinColdChaoticCollisions\endcsname{}\expandafter\renewcommand\csname KinColdChaoticCollisions\endcsname{\ensuremath{7.08}}
\expandafter\providecommand\csname KinColdCollisionRate\endcsname{}\expandafter\renewcommand\csname KinColdCollisionRate\endcsname{\ensuremath{3974}}
\expandafter\providecommand\csname KinColdCollisionRatio\endcsname{}\expandafter\renewcommand\csname KinColdCollisionRatio\endcsname{\ensuremath{10}}
\expandafter\providecommand\csname KinColdRate\endcsname{}\expandafter\renewcommand\csname KinColdRate\endcsname{\ensuremath{4.226\times10^{4}}}
\expandafter\providecommand\csname KinColdRateRatio\endcsname{}\expandafter\renewcommand\csname KinColdRateRatio\endcsname{\ensuremath{1.07}}
\expandafter\providecommand\csname KinGammaHighChaoticCollisions\endcsname{}\expandafter\renewcommand\csname KinGammaHighChaoticCollisions\endcsname{\ensuremath{86}}
\expandafter\providecommand\csname KinGammaHighCollisionRate\endcsname{}\expandafter\renewcommand\csname KinGammaHighCollisionRate\endcsname{\ensuremath{288.7}}
\expandafter\providecommand\csname KinGammaHighCollisionRatio\endcsname{}\expandafter\renewcommand\csname KinGammaHighCollisionRatio\endcsname{\ensuremath{0.729}}
\expandafter\providecommand\csname KinGammaHighRate\endcsname{}\expandafter\renewcommand\csname KinGammaHighRate\endcsname{\ensuremath{3.306\times10^{4}}}
\expandafter\providecommand\csname KinGammaHighRateRatio\endcsname{}\expandafter\renewcommand\csname KinGammaHighRateRatio\endcsname{\ensuremath{0.836}}
\expandafter\providecommand\csname KinGammaZeroChaoticCollisions\endcsname{}\expandafter\renewcommand\csname KinGammaZeroChaoticCollisions\endcsname{\ensuremath{62.4}}
\expandafter\providecommand\csname KinGammaZeroCollisionRate\endcsname{}\expandafter\renewcommand\csname KinGammaZeroCollisionRate\endcsname{\ensuremath{434.1}}
\expandafter\providecommand\csname KinGammaZeroCollisionRatio\endcsname{}\expandafter\renewcommand\csname KinGammaZeroCollisionRatio\endcsname{\ensuremath{1.1}}
\expandafter\providecommand\csname KinGammaZeroRate\endcsname{}\expandafter\renewcommand\csname KinGammaZeroRate\endcsname{\ensuremath{4.036\times10^{4}}}
\expandafter\providecommand\csname KinGammaZeroRateRatio\endcsname{}\expandafter\renewcommand\csname KinGammaZeroRateRatio\endcsname{\ensuremath{1.02}}
\expandafter\providecommand\csname KinIsoChaoticCollisions\endcsname{}\expandafter\renewcommand\csname KinIsoChaoticCollisions\endcsname{\ensuremath{83.4}}
\expandafter\providecommand\csname KinIsoCollisionRate\endcsname{}\expandafter\renewcommand\csname KinIsoCollisionRate\endcsname{\ensuremath{315.7}}
\expandafter\providecommand\csname KinIsoCollisionRatio\endcsname{}\expandafter\renewcommand\csname KinIsoCollisionRatio\endcsname{\ensuremath{0.798}}
\expandafter\providecommand\csname KinIsoRate\endcsname{}\expandafter\renewcommand\csname KinIsoRate\endcsname{\ensuremath{2.761\times10^{4}}}
\expandafter\providecommand\csname KinIsoRateRatio\endcsname{}\expandafter\renewcommand\csname KinIsoRateRatio\endcsname{\ensuremath{0.698}}
\expandafter\providecommand\csname KinSonicChaoticCollisions\endcsname{}\expandafter\renewcommand\csname KinSonicChaoticCollisions\endcsname{\ensuremath{89.5}}
\expandafter\providecommand\csname KinSonicCollisionRate\endcsname{}\expandafter\renewcommand\csname KinSonicCollisionRate\endcsname{\ensuremath{272.9}}
\expandafter\providecommand\csname KinSonicCollisionRatio\endcsname{}\expandafter\renewcommand\csname KinSonicCollisionRatio\endcsname{\ensuremath{0.69}}
\expandafter\providecommand\csname KinSonicRate\endcsname{}\expandafter\renewcommand\csname KinSonicRate\endcsname{\ensuremath{3.212\times10^{4}}}
\expandafter\providecommand\csname KinSonicRateRatio\endcsname{}\expandafter\renewcommand\csname KinSonicRateRatio\endcsname{\ensuremath{0.812}}
\expandafter\providecommand\csname KinStirHighChaoticCollisions\endcsname{}\expandafter\renewcommand\csname KinStirHighChaoticCollisions\endcsname{\ensuremath{76.8}}
\expandafter\providecommand\csname KinStirHighCollisionRate\endcsname{}\expandafter\renewcommand\csname KinStirHighCollisionRate\endcsname{\ensuremath{343.7}}
\expandafter\providecommand\csname KinStirHighCollisionRatio\endcsname{}\expandafter\renewcommand\csname KinStirHighCollisionRatio\endcsname{\ensuremath{0.868}}
\expandafter\providecommand\csname KinStirHighRate\endcsname{}\expandafter\renewcommand\csname KinStirHighRate\endcsname{\ensuremath{3.76\times10^{4}}}
\expandafter\providecommand\csname KinStirHighRateRatio\endcsname{}\expandafter\renewcommand\csname KinStirHighRateRatio\endcsname{\ensuremath{0.95}}
\expandafter\providecommand\csname KinStirLowChaoticCollisions\endcsname{}\expandafter\renewcommand\csname KinStirLowChaoticCollisions\endcsname{\ensuremath{58.6}}
\expandafter\providecommand\csname KinStirLowCollisionRate\endcsname{}\expandafter\renewcommand\csname KinStirLowCollisionRate\endcsname{\ensuremath{466.1}}
\expandafter\providecommand\csname KinStirLowCollisionRatio\endcsname{}\expandafter\renewcommand\csname KinStirLowCollisionRatio\endcsname{\ensuremath{1.18}}
\expandafter\providecommand\csname KinStirLowRate\endcsname{}\expandafter\renewcommand\csname KinStirLowRate\endcsname{\ensuremath{4.087\times10^{4}}}
\expandafter\providecommand\csname KinStirLowRateRatio\endcsname{}\expandafter\renewcommand\csname KinStirLowRateRatio\endcsname{\ensuremath{1.03}}
\expandafter\providecommand\csname LiteratureBelowContactUNinetyFive\endcsname{}\expandafter\renewcommand\csname LiteratureBelowContactUNinetyFive\endcsname{\ensuremath{20}}
\expandafter\providecommand\csname LiteratureBelowContactUNinetyNine\endcsname{}\expandafter\renewcommand\csname LiteratureBelowContactUNinetyNine\endcsname{\ensuremath{20}}
\expandafter\providecommand\csname LiteratureInsideU\endcsname{}\expandafter\renewcommand\csname LiteratureInsideU\endcsname{\ensuremath{38}}
\expandafter\providecommand\csname LiteraturePoints\endcsname{}\expandafter\renewcommand\csname LiteraturePoints\endcsname{\ensuremath{48}}
\expandafter\providecommand\csname MassStarMedianAll\endcsname{}\expandafter\renewcommand\csname MassStarMedianAll\endcsname{\ensuremath{159}}
\expandafter\providecommand\csname MassStarMedianContact\endcsname{}\expandafter\renewcommand\csname MassStarMedianContact\endcsname{\ensuremath{176}}
\expandafter\providecommand\csname MassStarMedianShearAll\endcsname{}\expandafter\renewcommand\csname MassStarMedianShearAll\endcsname{\ensuremath{159}}
\expandafter\providecommand\csname MassStarMedianShearContact\endcsname{}\expandafter\renewcommand\csname MassStarMedianShearContact\endcsname{\ensuremath{194}}
\expandafter\providecommand\csname ModalContactQ\endcsname{}\expandafter\renewcommand\csname ModalContactQ\endcsname{\ensuremath{0.0657}}
\expandafter\providecommand\csname ModalContactU\endcsname{}\expandafter\renewcommand\csname ModalContactU\endcsname{\ensuremath{0.0424}}
\expandafter\providecommand\csname ModalContactX\endcsname{}\expandafter\renewcommand\csname ModalContactX\endcsname{\ensuremath{0.859}}
\expandafter\providecommand\csname ModalQ\endcsname{}\expandafter\renewcommand\csname ModalQ\endcsname{\ensuremath{0.0657}}
\expandafter\providecommand\csname ModalU\endcsname{}\expandafter\renewcommand\csname ModalU\endcsname{\ensuremath{0.0424}}
\expandafter\providecommand\csname ModalX\endcsname{}\expandafter\renewcommand\csname ModalX\endcsname{\ensuremath{385}}
\expandafter\providecommand\csname NAgnRowanPessah\endcsname{}\expandafter\renewcommand\csname NAgnRowanPessah\endcsname{\ensuremath{1}}
\expandafter\providecommand\csname NAgnRuleOfThumb\endcsname{}\expandafter\renewcommand\csname NAgnRuleOfThumb\endcsname{\ensuremath{10}}
\expandafter\providecommand\csname NAgnSeyfert\endcsname{}\expandafter\renewcommand\csname NAgnSeyfert\endcsname{\ensuremath{180}}
\expandafter\providecommand\csname NAgnSpreadDecades\endcsname{}\expandafter\renewcommand\csname NAgnSpreadDecades\endcsname{\ensuremath{2}}
\expandafter\providecommand\csname NBlackHole\endcsname{}\expandafter\renewcommand\csname NBlackHole\endcsname{\ensuremath{200}}
\expandafter\providecommand\csname NStar\endcsname{}\expandafter\renewcommand\csname NStar\endcsname{\ensuremath{1000}}
\expandafter\providecommand\csname NoBBChaoticCollisions\endcsname{}\expandafter\renewcommand\csname NoBBChaoticCollisions\endcsname{\ensuremath{66.9}}
\expandafter\providecommand\csname NoBBCollisionRate\endcsname{}\expandafter\renewcommand\csname NoBBCollisionRate\endcsname{\ensuremath{382.6}}
\expandafter\providecommand\csname NoBBCollisionRatio\endcsname{}\expandafter\renewcommand\csname NoBBCollisionRatio\endcsname{\ensuremath{0.967}}
\expandafter\providecommand\csname NoBBRate\endcsname{}\expandafter\renewcommand\csname NoBBRate\endcsname{\ensuremath{3.947\times10^{4}}}
\expandafter\providecommand\csname NoBBRateRatio\endcsname{}\expandafter\renewcommand\csname NoBBRateRatio\endcsname{\ensuremath{0.998}}
\expandafter\providecommand\csname NoPostMsChaoticCollisions\endcsname{}\expandafter\renewcommand\csname NoPostMsChaoticCollisions\endcsname{\ensuremath{69.1}}
\expandafter\providecommand\csname NoPostMsCollisionRate\endcsname{}\expandafter\renewcommand\csname NoPostMsCollisionRate\endcsname{\ensuremath{369.9}}
\expandafter\providecommand\csname NoPostMsCollisionRatio\endcsname{}\expandafter\renewcommand\csname NoPostMsCollisionRatio\endcsname{\ensuremath{0.935}}
\expandafter\providecommand\csname NoPostMsRate\endcsname{}\expandafter\renewcommand\csname NoPostMsRate\endcsname{\ensuremath{3.958\times10^{4}}}
\expandafter\providecommand\csname NoPostMsRateRatio\endcsname{}\expandafter\renewcommand\csname NoPostMsRateRatio\endcsname{\ensuremath{1}}
\expandafter\providecommand\csname NumberedBelowContactUNinetyNine\endcsname{}\expandafter\renewcommand\csname NumberedBelowContactUNinetyNine\endcsname{\ensuremath{1}}
\expandafter\providecommand\csname NumberedOutsideU\endcsname{}\expandafter\renewcommand\csname NumberedOutsideU\endcsname{\ensuremath{0}}
\expandafter\providecommand\csname NumberedTargets\endcsname{}\expandafter\renewcommand\csname NumberedTargets\endcsname{\ensuremath{8}}
\expandafter\providecommand\csname PostMsExtChaoticCollisions\endcsname{}\expandafter\renewcommand\csname PostMsExtChaoticCollisions\endcsname{\ensuremath{62}}
\expandafter\providecommand\csname PostMsExtCollisionRate\endcsname{}\expandafter\renewcommand\csname PostMsExtCollisionRate\endcsname{\ensuremath{463.1}}
\expandafter\providecommand\csname PostMsExtCollisionRatio\endcsname{}\expandafter\renewcommand\csname PostMsExtCollisionRatio\endcsname{\ensuremath{1.17}}
\expandafter\providecommand\csname PostMsExtOversubscription\endcsname{}\expandafter\renewcommand\csname PostMsExtOversubscription\endcsname{\ensuremath{4.63}}
\expandafter\providecommand\csname PostMsFourChaoticCollisions\endcsname{}\expandafter\renewcommand\csname PostMsFourChaoticCollisions\endcsname{\ensuremath{68.4}}
\expandafter\providecommand\csname PostMsFourCollisionRate\endcsname{}\expandafter\renewcommand\csname PostMsFourCollisionRate\endcsname{\ensuremath{363.7}}
\expandafter\providecommand\csname PostMsFourCollisionRatio\endcsname{}\expandafter\renewcommand\csname PostMsFourCollisionRatio\endcsname{\ensuremath{0.919}}
\expandafter\providecommand\csname PostMsFourOversubscription\endcsname{}\expandafter\renewcommand\csname PostMsFourOversubscription\endcsname{\ensuremath{3.64}}
\expandafter\providecommand\csname PostMsOldChaoticCollisions\endcsname{}\expandafter\renewcommand\csname PostMsOldChaoticCollisions\endcsname{\ensuremath{58.1}}
\expandafter\providecommand\csname PostMsOldCollisionRate\endcsname{}\expandafter\renewcommand\csname PostMsOldCollisionRate\endcsname{\ensuremath{468.9}}
\expandafter\providecommand\csname PostMsOldCollisionRatio\endcsname{}\expandafter\renewcommand\csname PostMsOldCollisionRatio\endcsname{\ensuremath{1.18}}
\expandafter\providecommand\csname PostMsOldOversubscription\endcsname{}\expandafter\renewcommand\csname PostMsOldOversubscription\endcsname{\ensuremath{4.69}}
\expandafter\providecommand\csname RadiusCompactChaoticCollisions\endcsname{}\expandafter\renewcommand\csname RadiusCompactChaoticCollisions\endcsname{\ensuremath{78}}
\expandafter\providecommand\csname RadiusCompactCollisionRate\endcsname{}\expandafter\renewcommand\csname RadiusCompactCollisionRate\endcsname{\ensuremath{178.5}}
\expandafter\providecommand\csname RadiusCompactCollisionRatio\endcsname{}\expandafter\renewcommand\csname RadiusCompactCollisionRatio\endcsname{\ensuremath{0.451}}
\expandafter\providecommand\csname RadiusCompactRate\endcsname{}\expandafter\renewcommand\csname RadiusCompactRate\endcsname{\ensuremath{3.956\times10^{4}}}
\expandafter\providecommand\csname RadiusCompactRateRatio\endcsname{}\expandafter\renewcommand\csname RadiusCompactRateRatio\endcsname{\ensuremath{1}}
\expandafter\providecommand\csname ResPooledChaoticCollisions\endcsname{}\expandafter\renewcommand\csname ResPooledChaoticCollisions\endcsname{\ensuremath{89.6}}
\expandafter\providecommand\csname ResPooledCollisionRate\endcsname{}\expandafter\renewcommand\csname ResPooledCollisionRate\endcsname{\ensuremath{1212}}
\expandafter\providecommand\csname ResPooledCollisionRatio\endcsname{}\expandafter\renewcommand\csname ResPooledCollisionRatio\endcsname{\ensuremath{3.06}}
\expandafter\providecommand\csname ResPooledRate\endcsname{}\expandafter\renewcommand\csname ResPooledRate\endcsname{\ensuremath{3.953\times10^{4}}}
\expandafter\providecommand\csname ResPooledRateRatio\endcsname{}\expandafter\renewcommand\csname ResPooledRateRatio\endcsname{\ensuremath{0.999}}
\expandafter\providecommand\csname ResonanceXres\endcsname{}\expandafter\renewcommand\csname ResonanceXres\endcsname{\ensuremath{96}}
\expandafter\providecommand\csname RhHundredChebFactor\endcsname{}\expandafter\renewcommand\csname RhHundredChebFactor\endcsname{\ensuremath{1.7}}
\expandafter\providecommand\csname RhHundredChebFrac\endcsname{}\expandafter\renewcommand\csname RhHundredChebFrac\endcsname{\ensuremath{8.6}}
\expandafter\providecommand\csname RhHundredExtChebFactor\endcsname{}\expandafter\renewcommand\csname RhHundredExtChebFactor\endcsname{\ensuremath{6.1}}
\expandafter\providecommand\csname RhHundredExtChebFrac\endcsname{}\expandafter\renewcommand\csname RhHundredExtChebFrac\endcsname{\ensuremath{8.4}}
\expandafter\providecommand\csname RhHundredExtHgFactor\endcsname{}\expandafter\renewcommand\csname RhHundredExtHgFactor\endcsname{\ensuremath{29}}
\expandafter\providecommand\csname RhHundredExtHgFrac\endcsname{}\expandafter\renewcommand\csname RhHundredExtHgFrac\endcsname{\ensuremath{0.35}}
\expandafter\providecommand\csname RhHundredExtRmax\endcsname{}\expandafter\renewcommand\csname RhHundredExtRmax\endcsname{\ensuremath{791}}
\expandafter\providecommand\csname RhHundredFiftyNewChebFactor\endcsname{}\expandafter\renewcommand\csname RhHundredFiftyNewChebFactor\endcsname{\ensuremath{0.097}}
\expandafter\providecommand\csname RhHundredFiftyNewChebFrac\endcsname{}\expandafter\renewcommand\csname RhHundredFiftyNewChebFrac\endcsname{\ensuremath{11}}
\expandafter\providecommand\csname RhHundredFiftyNewHgFactor\endcsname{}\expandafter\renewcommand\csname RhHundredFiftyNewHgFactor\endcsname{\ensuremath{0.16}}
\expandafter\providecommand\csname RhHundredFiftyNewHgFrac\endcsname{}\expandafter\renewcommand\csname RhHundredFiftyNewHgFrac\endcsname{\ensuremath{0.35}}
\expandafter\providecommand\csname RhHundredHgFactor\endcsname{}\expandafter\renewcommand\csname RhHundredHgFactor\endcsname{\ensuremath{5}}
\expandafter\providecommand\csname RhHundredHgFrac\endcsname{}\expandafter\renewcommand\csname RhHundredHgFrac\endcsname{\ensuremath{0.34}}
\expandafter\providecommand\csname RhHundredOldChebFactor\endcsname{}\expandafter\renewcommand\csname RhHundredOldChebFactor\endcsname{\ensuremath{33}}
\expandafter\providecommand\csname RhHundredOldChebFrac\endcsname{}\expandafter\renewcommand\csname RhHundredOldChebFrac\endcsname{\ensuremath{2.1}}
\expandafter\providecommand\csname RhHundredOldHgFactor\endcsname{}\expandafter\renewcommand\csname RhHundredOldHgFactor\endcsname{\ensuremath{29}}
\expandafter\providecommand\csname RhHundredOldHgFrac\endcsname{}\expandafter\renewcommand\csname RhHundredOldHgFrac\endcsname{\ensuremath{0.38}}
\expandafter\providecommand\csname RhHundredRmax\endcsname{}\expandafter\renewcommand\csname RhHundredRmax\endcsname{\ensuremath{215}}
\expandafter\providecommand\csname RhSixtyNewChebFactor\endcsname{}\expandafter\renewcommand\csname RhSixtyNewChebFactor\endcsname{\ensuremath{22}}
\expandafter\providecommand\csname RhSixtyNewChebFrac\endcsname{}\expandafter\renewcommand\csname RhSixtyNewChebFrac\endcsname{\ensuremath{7}}
\expandafter\providecommand\csname RhSixtyNewHgFactor\endcsname{}\expandafter\renewcommand\csname RhSixtyNewHgFactor\endcsname{\ensuremath{33}}
\expandafter\providecommand\csname RhSixtyNewHgFrac\endcsname{}\expandafter\renewcommand\csname RhSixtyNewHgFrac\endcsname{\ensuremath{0.34}}
\expandafter\providecommand\csname RhTwentyChebFactor\endcsname{}\expandafter\renewcommand\csname RhTwentyChebFactor\endcsname{\ensuremath{204}}
\expandafter\providecommand\csname RhTwentyChebFrac\endcsname{}\expandafter\renewcommand\csname RhTwentyChebFrac\endcsname{\ensuremath{5.3}}
\expandafter\providecommand\csname RhTwentyHgFactor\endcsname{}\expandafter\renewcommand\csname RhTwentyHgFactor\endcsname{\ensuremath{99}}
\expandafter\providecommand\csname RhTwentyHgFrac\endcsname{}\expandafter\renewcommand\csname RhTwentyHgFrac\endcsname{\ensuremath{0.36}}
\expandafter\providecommand\csname RhTwentyNewChebFactor\endcsname{}\expandafter\renewcommand\csname RhTwentyNewChebFactor\endcsname{\ensuremath{42}}
\expandafter\providecommand\csname RhTwentyNewChebFrac\endcsname{}\expandafter\renewcommand\csname RhTwentyNewChebFrac\endcsname{\ensuremath{5.2}}
\expandafter\providecommand\csname RhTwentyNewHgFactor\endcsname{}\expandafter\renewcommand\csname RhTwentyNewHgFactor\endcsname{\ensuremath{24}}
\expandafter\providecommand\csname RhTwentyNewHgFrac\endcsname{}\expandafter\renewcommand\csname RhTwentyNewHgFrac\endcsname{\ensuremath{0.34}}
\expandafter\providecommand\csname RhTwentyNewUntunedChebFactor\endcsname{}\expandafter\renewcommand\csname RhTwentyNewUntunedChebFactor\endcsname{\ensuremath{206}}
\expandafter\providecommand\csname RhTwentyNewUntunedChebFrac\endcsname{}\expandafter\renewcommand\csname RhTwentyNewUntunedChebFrac\endcsname{\ensuremath{5.2}}
\expandafter\providecommand\csname RhTwentyNewUntunedHgFactor\endcsname{}\expandafter\renewcommand\csname RhTwentyNewUntunedHgFactor\endcsname{\ensuremath{92}}
\expandafter\providecommand\csname RhTwentyNewUntunedHgFrac\endcsname{}\expandafter\renewcommand\csname RhTwentyNewUntunedHgFrac\endcsname{\ensuremath{0.34}}
\expandafter\providecommand\csname RowanDestroyed\endcsname{}\expandafter\renewcommand\csname RowanDestroyed\endcsname{\ensuremath{80}}
\expandafter\providecommand\csname RowanNbh\endcsname{}\expandafter\renewcommand\csname RowanNbh\endcsname{\ensuremath{20}}
\expandafter\providecommand\csname RowanRateHigh\endcsname{}\expandafter\renewcommand\csname RowanRateHigh\endcsname{\ensuremath{12}}
\expandafter\providecommand\csname RowanRateLow\endcsname{}\expandafter\renewcommand\csname RowanRateLow\endcsname{\ensuremath{2}}
\expandafter\providecommand\csname RowanScatterHigh\endcsname{}\expandafter\renewcommand\csname RowanScatterHigh\endcsname{\ensuremath{50}}
\expandafter\providecommand\csname RowanScatterLow\endcsname{}\expandafter\renewcommand\csname RowanScatterLow\endcsname{\ensuremath{5}}
\expandafter\providecommand\csname RowanSmbhMass\endcsname{}\expandafter\renewcommand\csname RowanSmbhMass\endcsname{\ensuremath{4}}
\expandafter\providecommand\csname ShearTwoBodyChaoticCollisions\endcsname{}\expandafter\renewcommand\csname ShearTwoBodyChaoticCollisions\endcsname{\ensuremath{20.7}}
\expandafter\providecommand\csname ShearTwoBodyCollisionRate\endcsname{}\expandafter\renewcommand\csname ShearTwoBodyCollisionRate\endcsname{\ensuremath{1345}}
\expandafter\providecommand\csname ShearTwoBodyCollisionRatio\endcsname{}\expandafter\renewcommand\csname ShearTwoBodyCollisionRatio\endcsname{\ensuremath{3.4}}
\expandafter\providecommand\csname ShearTwoBodyRate\endcsname{}\expandafter\renewcommand\csname ShearTwoBodyRate\endcsname{\ensuremath{1.883\times10^{4}}}
\expandafter\providecommand\csname ShearTwoBodyRateRatio\endcsname{}\expandafter\renewcommand\csname ShearTwoBodyRateRatio\endcsname{\ensuremath{0.476}}
\expandafter\providecommand\csname SingleSingleCollisionFraction\endcsname{}\expandafter\renewcommand\csname SingleSingleCollisionFraction\endcsname{\ensuremath{0.336}}
\expandafter\providecommand\csname SingleSingleCollisionRate\endcsname{}\expandafter\renewcommand\csname SingleSingleCollisionRate\endcsname{\ensuremath{126.6}}
\expandafter\providecommand\csname SingleSingleMedianQ\endcsname{}\expandafter\renewcommand\csname SingleSingleMedianQ\endcsname{\ensuremath{0.0673}}
\expandafter\providecommand\csname SingleSingleMedianU\endcsname{}\expandafter\renewcommand\csname SingleSingleMedianU\endcsname{\ensuremath{0.02266}}
\expandafter\providecommand\csname SingleSingleMedianX\endcsname{}\expandafter\renewcommand\csname SingleSingleMedianX\endcsname{\ensuremath{271}}
\expandafter\providecommand\csname SingleSingleMicroTdeRate\endcsname{}\expandafter\renewcommand\csname SingleSingleMicroTdeRate\endcsname{\ensuremath{55.58}}
\expandafter\providecommand\csname SingleSingleQsmallXsmall\endcsname{}\expandafter\renewcommand\csname SingleSingleQsmallXsmall\endcsname{\ensuremath{0.305}}
\expandafter\providecommand\csname SingleSingleQsmallXsmallRate\endcsname{}\expandafter\renewcommand\csname SingleSingleQsmallXsmallRate\endcsname{\ensuremath{114.8}}
\expandafter\providecommand\csname SingleSingleRate\endcsname{}\expandafter\renewcommand\csname SingleSingleRate\endcsname{\ensuremath{3.765\times10^{4}}}
\expandafter\providecommand\csname SingleSingleVolumetricThousand\endcsname{}\expandafter\renewcommand\csname SingleSingleVolumetricThousand\endcsname{\ensuremath{3.765\times10^{4}}}
\expandafter\providecommand\csname SmbhMass\endcsname{}\expandafter\renewcommand\csname SmbhMass\endcsname{\ensuremath{8}}
\expandafter\providecommand\csname StackCollisionFraction\endcsname{}\expandafter\renewcommand\csname StackCollisionFraction\endcsname{\ensuremath{1}}
\expandafter\providecommand\csname StackCollisionRate\endcsname{}\expandafter\renewcommand\csname StackCollisionRate\endcsname{\ensuremath{395.8}}
\expandafter\providecommand\csname StackMicroTdeRate\endcsname{}\expandafter\renewcommand\csname StackMicroTdeRate\endcsname{\ensuremath{194.5}}
\expandafter\providecommand\csname StackQsmall\endcsname{}\expandafter\renewcommand\csname StackQsmall\endcsname{\ensuremath{89.9}}
\expandafter\providecommand\csname StackQsmallXsmall\endcsname{}\expandafter\renewcommand\csname StackQsmallXsmall\endcsname{\ensuremath{0.91}}
\expandafter\providecommand\csname StackQsmallXsmallRate\endcsname{}\expandafter\renewcommand\csname StackQsmallXsmallRate\endcsname{\ensuremath{360.1}}
\expandafter\providecommand\csname StackRate\endcsname{}\expandafter\renewcommand\csname StackRate\endcsname{\ensuremath{3.956\times10^{4}}}
\expandafter\providecommand\csname StackUMedian\endcsname{}\expandafter\renewcommand\csname StackUMedian\endcsname{\ensuremath{0.0227}}
\expandafter\providecommand\csname StackUpFive\endcsname{}\expandafter\renewcommand\csname StackUpFive\endcsname{\ensuremath{0.119}}
\expandafter\providecommand\csname StackXMedian\endcsname{}\expandafter\renewcommand\csname StackXMedian\endcsname{\ensuremath{271}}
\expandafter\providecommand\csname StarBandInner\endcsname{}\expandafter\renewcommand\csname StarBandInner\endcsname{\ensuremath{2}}
\expandafter\providecommand\csname StarBandOuter\endcsname{}\expandafter\renewcommand\csname StarBandOuter\endcsname{\ensuremath{2}}
\expandafter\providecommand\csname StarInnerEncounterTime\endcsname{}\expandafter\renewcommand\csname StarInnerEncounterTime\endcsname{\ensuremath{12.99}}
\expandafter\providecommand\csname StarInnerHillRadius\endcsname{}\expandafter\renewcommand\csname StarInnerHillRadius\endcsname{\ensuremath{14.3}}
\expandafter\providecommand\csname StarInnerHillVelocity\endcsname{}\expandafter\renewcommand\csname StarInnerHillVelocity\endcsname{\ensuremath{48.7}}
\expandafter\providecommand\csname StarInnerRadiusAu\endcsname{}\expandafter\renewcommand\csname StarInnerRadiusAu\endcsname{\ensuremath{1974}}
\expandafter\providecommand\csname StarMassMax\endcsname{}\expandafter\renewcommand\csname StarMassMax\endcsname{\ensuremath{300}}
\expandafter\providecommand\csname StarMassMin\endcsname{}\expandafter\renewcommand\csname StarMassMin\endcsname{\ensuremath{50}}
\expandafter\providecommand\csname StarMeanMass\endcsname{}\expandafter\renewcommand\csname StarMeanMass\endcsname{\ensuremath{140}}
\expandafter\providecommand\csname StarMidEncounterTime\endcsname{}\expandafter\renewcommand\csname StarMidEncounterTime\endcsname{\ensuremath{5.81}}
\expandafter\providecommand\csname StarMidHillRadius\endcsname{}\expandafter\renewcommand\csname StarMidHillRadius\endcsname{\ensuremath{71.7}}
\expandafter\providecommand\csname StarMidHillVelocity\endcsname{}\expandafter\renewcommand\csname StarMidHillVelocity\endcsname{\ensuremath{21.8}}
\expandafter\providecommand\csname StarMidRadiusAu\endcsname{}\expandafter\renewcommand\csname StarMidRadiusAu\endcsname{\ensuremath{9871}}
\expandafter\providecommand\csname StarStarChaoticCollisions\endcsname{}\expandafter\renewcommand\csname StarStarChaoticCollisions\endcsname{\ensuremath{83}}
\expandafter\providecommand\csname StarStarCollisionRate\endcsname{}\expandafter\renewcommand\csname StarStarCollisionRate\endcsname{\ensuremath{742.6}}
\expandafter\providecommand\csname StarStarCollisionRatio\endcsname{}\expandafter\renewcommand\csname StarStarCollisionRatio\endcsname{\ensuremath{1.88}}
\expandafter\providecommand\csname StarStarRate\endcsname{}\expandafter\renewcommand\csname StarStarRate\endcsname{\ensuremath{4.022\times10^{4}}}
\expandafter\providecommand\csname StarStarRateRatio\endcsname{}\expandafter\renewcommand\csname StarStarRateRatio\endcsname{\ensuremath{1.02}}
\expandafter\providecommand\csname StarsFewChaoticCollisions\endcsname{}\expandafter\renewcommand\csname StarsFewChaoticCollisions\endcsname{\ensuremath{57.7}}
\expandafter\providecommand\csname StarsFewCollisionRate\endcsname{}\expandafter\renewcommand\csname StarsFewCollisionRate\endcsname{\ensuremath{142.2}}
\expandafter\providecommand\csname StarsFewCollisionRatio\endcsname{}\expandafter\renewcommand\csname StarsFewCollisionRatio\endcsname{\ensuremath{0.359}}
\expandafter\providecommand\csname StarsFewRate\endcsname{}\expandafter\renewcommand\csname StarsFewRate\endcsname{\ensuremath{1.229\times10^{4}}}
\expandafter\providecommand\csname StarsFewRateRatio\endcsname{}\expandafter\renewcommand\csname StarsFewRateRatio\endcsname{\ensuremath{0.311}}
\expandafter\providecommand\csname StarsInnerChaoticCollisions\endcsname{}\expandafter\renewcommand\csname StarsInnerChaoticCollisions\endcsname{\ensuremath{65.3}}
\expandafter\providecommand\csname StarsInnerCollisionRate\endcsname{}\expandafter\renewcommand\csname StarsInnerCollisionRate\endcsname{\ensuremath{346.8}}
\expandafter\providecommand\csname StarsInnerCollisionRatio\endcsname{}\expandafter\renewcommand\csname StarsInnerCollisionRatio\endcsname{\ensuremath{0.876}}
\expandafter\providecommand\csname StarsInnerRate\endcsname{}\expandafter\renewcommand\csname StarsInnerRate\endcsname{\ensuremath{3.126\times10^{4}}}
\expandafter\providecommand\csname StarsInnerRateRatio\endcsname{}\expandafter\renewcommand\csname StarsInnerRateRatio\endcsname{\ensuremath{0.79}}
\expandafter\providecommand\csname StarsManyChaoticCollisions\endcsname{}\expandafter\renewcommand\csname StarsManyChaoticCollisions\endcsname{\ensuremath{76.8}}
\expandafter\providecommand\csname StarsManyCollisionRate\endcsname{}\expandafter\renewcommand\csname StarsManyCollisionRate\endcsname{\ensuremath{1031}}
\expandafter\providecommand\csname StarsManyCollisionRatio\endcsname{}\expandafter\renewcommand\csname StarsManyCollisionRatio\endcsname{\ensuremath{2.61}}
\expandafter\providecommand\csname StarsManyRate\endcsname{}\expandafter\renewcommand\csname StarsManyRate\endcsname{\ensuremath{1.128\times10^{5}}}
\expandafter\providecommand\csname StarsManyRateRatio\endcsname{}\expandafter\renewcommand\csname StarsManyRateRatio\endcsname{\ensuremath{2.85}}
\expandafter\providecommand\csname SupplyLimitedRate\endcsname{}\expandafter\renewcommand\csname SupplyLimitedRate\endcsname{\ensuremath{100}}
\expandafter\providecommand\csname SupplyOversubscription\endcsname{}\expandafter\renewcommand\csname SupplyOversubscription\endcsname{\ensuremath{3.96}}
\expandafter\providecommand\csname SupportUmax\endcsname{}\expandafter\renewcommand\csname SupportUmax\endcsname{\ensuremath{1.358}}
\expandafter\providecommand\csname TAgn\endcsname{}\expandafter\renewcommand\csname TAgn\endcsname{\ensuremath{7}}
\expandafter\providecommand\csname TargetEightQ\endcsname{}\expandafter\renewcommand\csname TargetEightQ\endcsname{\ensuremath{0.1}}
\expandafter\providecommand\csname TargetEightRank\endcsname{}\expandafter\renewcommand\csname TargetEightRank\endcsname{\ensuremath{6}}
\expandafter\providecommand\csname TargetEightU\endcsname{}\expandafter\renewcommand\csname TargetEightU\endcsname{\ensuremath{1.049}}
\expandafter\providecommand\csname TargetEightWeightQx\endcsname{}\expandafter\renewcommand\csname TargetEightWeightQx\endcsname{\ensuremath{0.0477}}
\expandafter\providecommand\csname TargetEightX\endcsname{}\expandafter\renewcommand\csname TargetEightX\endcsname{\ensuremath{0.2247}}
\expandafter\providecommand\csname TargetFiveQ\endcsname{}\expandafter\renewcommand\csname TargetFiveQ\endcsname{\ensuremath{0.1429}}
\expandafter\providecommand\csname TargetFiveRank\endcsname{}\expandafter\renewcommand\csname TargetFiveRank\endcsname{\ensuremath{7}}
\expandafter\providecommand\csname TargetFiveU\endcsname{}\expandafter\renewcommand\csname TargetFiveU\endcsname{\ensuremath{0.35}}
\expandafter\providecommand\csname TargetFiveWeightQx\endcsname{}\expandafter\renewcommand\csname TargetFiveWeightQx\endcsname{\ensuremath{0.0224}}
\expandafter\providecommand\csname TargetFiveX\endcsname{}\expandafter\renewcommand\csname TargetFiveX\endcsname{\ensuremath{0.2}}
\expandafter\providecommand\csname TargetFourQ\endcsname{}\expandafter\renewcommand\csname TargetFourQ\endcsname{\ensuremath{0.05567}}
\expandafter\providecommand\csname TargetFourRank\endcsname{}\expandafter\renewcommand\csname TargetFourRank\endcsname{\ensuremath{4}}
\expandafter\providecommand\csname TargetFourU\endcsname{}\expandafter\renewcommand\csname TargetFourU\endcsname{\ensuremath{0.02475}}
\expandafter\providecommand\csname TargetFourWeightQx\endcsname{}\expandafter\renewcommand\csname TargetFourWeightQx\endcsname{\ensuremath{0.0671}}
\expandafter\providecommand\csname TargetFourX\endcsname{}\expandafter\renewcommand\csname TargetFourX\endcsname{\ensuremath{0.8914}}
\expandafter\providecommand\csname TargetOneQ\endcsname{}\expandafter\renewcommand\csname TargetOneQ\endcsname{\ensuremath{0.02857}}
\expandafter\providecommand\csname TargetOneRank\endcsname{}\expandafter\renewcommand\csname TargetOneRank\endcsname{\ensuremath{1}}
\expandafter\providecommand\csname TargetOneU\endcsname{}\expandafter\renewcommand\csname TargetOneU\endcsname{\ensuremath{0.6}}
\expandafter\providecommand\csname TargetOneWeightQx\endcsname{}\expandafter\renewcommand\csname TargetOneWeightQx\endcsname{\ensuremath{0.118}}
\expandafter\providecommand\csname TargetOneX\endcsname{}\expandafter\renewcommand\csname TargetOneX\endcsname{\ensuremath{0.2}}
\expandafter\providecommand\csname TargetSevenQ\endcsname{}\expandafter\renewcommand\csname TargetSevenQ\endcsname{\ensuremath{0.1}}
\expandafter\providecommand\csname TargetSevenRank\endcsname{}\expandafter\renewcommand\csname TargetSevenRank\endcsname{\ensuremath{5}}
\expandafter\providecommand\csname TargetSevenU\endcsname{}\expandafter\renewcommand\csname TargetSevenU\endcsname{\ensuremath{1.049}}
\expandafter\providecommand\csname TargetSevenWeightQx\endcsname{}\expandafter\renewcommand\csname TargetSevenWeightQx\endcsname{\ensuremath{0.0477}}
\expandafter\providecommand\csname TargetSevenX\endcsname{}\expandafter\renewcommand\csname TargetSevenX\endcsname{\ensuremath{0.2247}}
\expandafter\providecommand\csname TargetSixQ\endcsname{}\expandafter\renewcommand\csname TargetSixQ\endcsname{\ensuremath{0.1429}}
\expandafter\providecommand\csname TargetSixRank\endcsname{}\expandafter\renewcommand\csname TargetSixRank\endcsname{\ensuremath{8}}
\expandafter\providecommand\csname TargetSixU\endcsname{}\expandafter\renewcommand\csname TargetSixU\endcsname{\ensuremath{0.2}}
\expandafter\providecommand\csname TargetSixWeightQx\endcsname{}\expandafter\renewcommand\csname TargetSixWeightQx\endcsname{\ensuremath{0.0224}}
\expandafter\providecommand\csname TargetSixX\endcsname{}\expandafter\renewcommand\csname TargetSixX\endcsname{\ensuremath{0.25}}
\expandafter\providecommand\csname TargetThreeQ\endcsname{}\expandafter\renewcommand\csname TargetThreeQ\endcsname{\ensuremath{0.02857}}
\expandafter\providecommand\csname TargetThreeRank\endcsname{}\expandafter\renewcommand\csname TargetThreeRank\endcsname{\ensuremath{3}}
\expandafter\providecommand\csname TargetThreeU\endcsname{}\expandafter\renewcommand\csname TargetThreeU\endcsname{\ensuremath{0.6}}
\expandafter\providecommand\csname TargetThreeWeightQx\endcsname{}\expandafter\renewcommand\csname TargetThreeWeightQx\endcsname{\ensuremath{0.118}}
\expandafter\providecommand\csname TargetThreeX\endcsname{}\expandafter\renewcommand\csname TargetThreeX\endcsname{\ensuremath{0.2}}
\expandafter\providecommand\csname TargetTwoQ\endcsname{}\expandafter\renewcommand\csname TargetTwoQ\endcsname{\ensuremath{0.02857}}
\expandafter\providecommand\csname TargetTwoRank\endcsname{}\expandafter\renewcommand\csname TargetTwoRank\endcsname{\ensuremath{2}}
\expandafter\providecommand\csname TargetTwoU\endcsname{}\expandafter\renewcommand\csname TargetTwoU\endcsname{\ensuremath{0.6}}
\expandafter\providecommand\csname TargetTwoWeightQx\endcsname{}\expandafter\renewcommand\csname TargetTwoWeightQx\endcsname{\ensuremath{0.118}}
\expandafter\providecommand\csname TargetTwoX\endcsname{}\expandafter\renewcommand\csname TargetTwoX\endcsname{\ensuremath{0.2}}
\expandafter\providecommand\csname TimeToFirstCollision\endcsname{}\expandafter\renewcommand\csname TimeToFirstCollision\endcsname{\ensuremath{25.27}}
\expandafter\providecommand\csname TrapBoostChaoticCollisions\endcsname{}\expandafter\renewcommand\csname TrapBoostChaoticCollisions\endcsname{\ensuremath{61.9}}
\expandafter\providecommand\csname TrapBoostCollisionRate\endcsname{}\expandafter\renewcommand\csname TrapBoostCollisionRate\endcsname{\ensuremath{814.3}}
\expandafter\providecommand\csname TrapBoostCollisionRatio\endcsname{}\expandafter\renewcommand\csname TrapBoostCollisionRatio\endcsname{\ensuremath{2.06}}
\expandafter\providecommand\csname TrapBoostRate\endcsname{}\expandafter\renewcommand\csname TrapBoostRate\endcsname{\ensuremath{6.944\times10^{4}}}
\expandafter\providecommand\csname TrapBoostRateRatio\endcsname{}\expandafter\renewcommand\csname TrapBoostRateRatio\endcsname{\ensuremath{1.76}}
\expandafter\providecommand\csname TrapEncounterTime\endcsname{}\expandafter\renewcommand\csname TrapEncounterTime\endcsname{\ensuremath{21.95}}
\expandafter\providecommand\csname TrapHillRadius\endcsname{}\expandafter\renewcommand\csname TrapHillRadius\endcsname{\ensuremath{5.02}}
\expandafter\providecommand\csname TrapHillVelocity\endcsname{}\expandafter\renewcommand\csname TrapHillVelocity\endcsname{\ensuremath{82.3}}
\expandafter\providecommand\csname TrapRadiusAu\endcsname{}\expandafter\renewcommand\csname TrapRadiusAu\endcsname{\ensuremath{691}}
\expandafter\providecommand\csname ValidityAnalyticIon\endcsname{}\expandafter\renewcommand\csname ValidityAnalyticIon\endcsname{\ensuremath{0}}
\expandafter\providecommand\csname ValidityClamped\endcsname{}\expandafter\renewcommand\csname ValidityClamped\endcsname{\ensuremath{0}}
\expandafter\providecommand\csname ValidityExtrapMass\endcsname{}\expandafter\renewcommand\csname ValidityExtrapMass\endcsname{\ensuremath{0}}
\expandafter\providecommand\csname ValidityFallback\endcsname{}\expandafter\renewcommand\csname ValidityFallback\endcsname{\ensuremath{0.00771}}
\expandafter\providecommand\csname ValidityFiniteRadius\endcsname{}\expandafter\renewcommand\csname ValidityFiniteRadius\endcsname{\ensuremath{0.966}}
\expandafter\providecommand\csname ValidityFromTable\endcsname{}\expandafter\renewcommand\csname ValidityFromTable\endcsname{\ensuremath{4.59}}
\expandafter\providecommand\csname ValidityPairIdentity\endcsname{}\expandafter\renewcommand\csname ValidityPairIdentity\endcsname{\ensuremath{0}}
\expandafter\providecommand\csname ValidityParabolic\endcsname{}\expandafter\renewcommand\csname ValidityParabolic\endcsname{\ensuremath{4.79}}
\expandafter\providecommand\csname ValidityPhaseTrue\endcsname{}\expandafter\renewcommand\csname ValidityPhaseTrue\endcsname{\ensuremath{4.6}}
\expandafter\providecommand\csname ValidityScaled\endcsname{}\expandafter\renewcommand\csname ValidityScaled\endcsname{\ensuremath{0.237}}
\expandafter\providecommand\csname ValidityShear\endcsname{}\expandafter\renewcommand\csname ValidityShear\endcsname{\ensuremath{100}}
\expandafter\providecommand\csname VolAllRowanPessah\endcsname{}\expandafter\renewcommand\csname VolAllRowanPessah\endcsname{\ensuremath{3956}}
\expandafter\providecommand\csname VolAllRuleOfThumb\endcsname{}\expandafter\renewcommand\csname VolAllRuleOfThumb\endcsname{\ensuremath{3.956\times10^{4}}}
\expandafter\providecommand\csname VolAllSeyfert\endcsname{}\expandafter\renewcommand\csname VolAllSeyfert\endcsname{\ensuremath{7.121\times10^{5}}}
\expandafter\providecommand\csname VolCollisionsRowanPessah\endcsname{}\expandafter\renewcommand\csname VolCollisionsRowanPessah\endcsname{\ensuremath{39.58}}
\expandafter\providecommand\csname VolCollisionsRuleOfThumb\endcsname{}\expandafter\renewcommand\csname VolCollisionsRuleOfThumb\endcsname{\ensuremath{395.8}}
\expandafter\providecommand\csname VolCollisionsSeyfert\endcsname{}\expandafter\renewcommand\csname VolCollisionsSeyfert\endcsname{\ensuremath{7124}}
\expandafter\providecommand\csname VolMicroTdeRowanPessah\endcsname{}\expandafter\renewcommand\csname VolMicroTdeRowanPessah\endcsname{\ensuremath{19.45}}
\expandafter\providecommand\csname VolMicroTdeRuleOfThumb\endcsname{}\expandafter\renewcommand\csname VolMicroTdeRuleOfThumb\endcsname{\ensuremath{194.5}}
\expandafter\providecommand\csname VolMicroTdeSeyfert\endcsname{}\expandafter\renewcommand\csname VolMicroTdeSeyfert\endcsname{\ensuremath{3502}}
\expandafter\providecommand\csname VolSupplyRowanPessah\endcsname{}\expandafter\renewcommand\csname VolSupplyRowanPessah\endcsname{\ensuremath{10}}
\expandafter\providecommand\csname VolSupplyRuleOfThumb\endcsname{}\expandafter\renewcommand\csname VolSupplyRuleOfThumb\endcsname{\ensuremath{100}}
\expandafter\providecommand\csname VolSupplySeyfert\endcsname{}\expandafter\renewcommand\csname VolSupplySeyfert\endcsname{\ensuremath{1800}}
\expandafter\providecommand\csname YangAlignments\endcsname{}\expandafter\renewcommand\csname YangAlignments\endcsname{\ensuremath{5}}
\expandafter\providecommand\csname YangMicroTdeRate\endcsname{}\expandafter\renewcommand\csname YangMicroTdeRate\endcsname{\ensuremath{170}}
\expandafter\providecommand\csname YangSmbhMass\endcsname{}\expandafter\renewcommand\csname YangSmbhMass\endcsname{\ensuremath{6}}
\expandafter\providecommand\csname YangStarStarRate\endcsname{}\expandafter\renewcommand\csname YangStarStarRate\endcsname{\ensuremath{20}}

\providecommand\result[1]{%
  \ifcsname #1\endcsname\csname #1\endcsname\else\textbf{??#1??}\fi}

\expandafter\providecommand\csname ACount\endcsname{}\expandafter\renewcommand\csname ACount\endcsname{\ensuremath{3}}
\expandafter\providecommand\csname AHeat\endcsname{}\expandafter\renewcommand\csname AHeat\endcsname{\ensuremath{0}}
\expandafter\providecommand\csname ARadius\endcsname{}\expandafter\renewcommand\csname ARadius\endcsname{\ensuremath{1.0000}}
\expandafter\providecommand\csname AgnDeepRigidCount\endcsname{}\expandafter\renewcommand\csname AgnDeepRigidCount\endcsname{\ensuremath{3}}
\expandafter\providecommand\csname AgnDeepRigidFirstRecurrence\endcsname{}\expandafter\renewcommand\csname AgnDeepRigidFirstRecurrence\endcsname{\ensuremath{991.7}}
\expandafter\providecommand\csname AgnDeepRigidLastRecurrence\endcsname{}\expandafter\renewcommand\csname AgnDeepRigidLastRecurrence\endcsname{\ensuremath{2.837}}
\expandafter\providecommand\csname AgnDeepRigidRadius\endcsname{}\expandafter\renewcommand\csname AgnDeepRigidRadius\endcsname{\ensuremath{1.000}}
\expandafter\providecommand\csname AgnDeepRigidTime\endcsname{}\expandafter\renewcommand\csname AgnDeepRigidTime\endcsname{\ensuremath{998.8}}
\expandafter\providecommand\csname AgnDeepSoftCount\endcsname{}\expandafter\renewcommand\csname AgnDeepSoftCount\endcsname{\ensuremath{2}}
\expandafter\providecommand\csname AgnDeepSoftFirstRecurrence\endcsname{}\expandafter\renewcommand\csname AgnDeepSoftFirstRecurrence\endcsname{\ensuremath{991.7}}
\expandafter\providecommand\csname AgnDeepSoftLastRecurrence\endcsname{}\expandafter\renewcommand\csname AgnDeepSoftLastRecurrence\endcsname{\ensuremath{991.7}}
\expandafter\providecommand\csname AgnDeepSoftRadius\endcsname{}\expandafter\renewcommand\csname AgnDeepSoftRadius\endcsname{\ensuremath{1.025}}
\expandafter\providecommand\csname AgnDeepSoftTime\endcsname{}\expandafter\renewcommand\csname AgnDeepSoftTime\endcsname{\ensuremath{995.4}}
\expandafter\providecommand\csname AgnMiddleRigidCount\endcsname{}\expandafter\renewcommand\csname AgnMiddleRigidCount\endcsname{\ensuremath{2}}
\expandafter\providecommand\csname AgnMiddleRigidFirstRecurrence\endcsname{}\expandafter\renewcommand\csname AgnMiddleRigidFirstRecurrence\endcsname{\ensuremath{4.94}}
\expandafter\providecommand\csname AgnMiddleRigidLastRecurrence\endcsname{}\expandafter\renewcommand\csname AgnMiddleRigidLastRecurrence\endcsname{\ensuremath{4.94}}
\expandafter\providecommand\csname AgnMiddleRigidRadius\endcsname{}\expandafter\renewcommand\csname AgnMiddleRigidRadius\endcsname{\ensuremath{1.000}}
\expandafter\providecommand\csname AgnMiddleRigidTime\endcsname{}\expandafter\renewcommand\csname AgnMiddleRigidTime\endcsname{\ensuremath{9.968}}
\expandafter\providecommand\csname AgnMiddleSoftCount\endcsname{}\expandafter\renewcommand\csname AgnMiddleSoftCount\endcsname{\ensuremath{3}}
\expandafter\providecommand\csname AgnMiddleSoftFirstRecurrence\endcsname{}\expandafter\renewcommand\csname AgnMiddleSoftFirstRecurrence\endcsname{\ensuremath{4.93}}
\expandafter\providecommand\csname AgnMiddleSoftLastRecurrence\endcsname{}\expandafter\renewcommand\csname AgnMiddleSoftLastRecurrence\endcsname{\ensuremath{0.5135}}
\expandafter\providecommand\csname AgnMiddleSoftRadius\endcsname{}\expandafter\renewcommand\csname AgnMiddleSoftRadius\endcsname{\ensuremath{1.040}}
\expandafter\providecommand\csname AgnMiddleSoftTime\endcsname{}\expandafter\renewcommand\csname AgnMiddleSoftTime\endcsname{\ensuremath{10.41}}
\expandafter\providecommand\csname AgnShallowRigidCount\endcsname{}\expandafter\renewcommand\csname AgnShallowRigidCount\endcsname{\ensuremath{3}}
\expandafter\providecommand\csname AgnShallowRigidFirstRecurrence\endcsname{}\expandafter\renewcommand\csname AgnShallowRigidFirstRecurrence\endcsname{\ensuremath{8.42}}
\expandafter\providecommand\csname AgnShallowRigidLastRecurrence\endcsname{}\expandafter\renewcommand\csname AgnShallowRigidLastRecurrence\endcsname{\ensuremath{1.68}}
\expandafter\providecommand\csname AgnShallowRigidRadius\endcsname{}\expandafter\renewcommand\csname AgnShallowRigidRadius\endcsname{\ensuremath{1.000}}
\expandafter\providecommand\csname AgnShallowRigidTime\endcsname{}\expandafter\renewcommand\csname AgnShallowRigidTime\endcsname{\ensuremath{15.87}}
\expandafter\providecommand\csname AgnShallowSoftCount\endcsname{}\expandafter\renewcommand\csname AgnShallowSoftCount\endcsname{\ensuremath{3}}
\expandafter\providecommand\csname AgnShallowSoftFirstRecurrence\endcsname{}\expandafter\renewcommand\csname AgnShallowSoftFirstRecurrence\endcsname{\ensuremath{8.408}}
\expandafter\providecommand\csname AgnShallowSoftLastRecurrence\endcsname{}\expandafter\renewcommand\csname AgnShallowSoftLastRecurrence\endcsname{\ensuremath{2.023}}
\expandafter\providecommand\csname AgnShallowSoftRadius\endcsname{}\expandafter\renewcommand\csname AgnShallowSoftRadius\endcsname{\ensuremath{1.040}}
\expandafter\providecommand\csname AgnShallowSoftTime\endcsname{}\expandafter\renewcommand\csname AgnShallowSoftTime\endcsname{\ensuremath{17.2}}
\expandafter\providecommand\csname AngularError\endcsname{}\expandafter\renewcommand\csname AngularError\endcsname{\ensuremath{0.017}}
\expandafter\providecommand\csname BCount\endcsname{}\expandafter\renewcommand\csname BCount\endcsname{\ensuremath{3}}
\expandafter\providecommand\csname BHeat\endcsname{}\expandafter\renewcommand\csname BHeat\endcsname{\ensuremath{1}}
\expandafter\providecommand\csname BRadius\endcsname{}\expandafter\renewcommand\csname BRadius\endcsname{\ensuremath{1.0337}}
\expandafter\providecommand\csname CCount\endcsname{}\expandafter\renewcommand\csname CCount\endcsname{\ensuremath{3}}
\expandafter\providecommand\csname CHeat\endcsname{}\expandafter\renewcommand\csname CHeat\endcsname{\ensuremath{0}}
\expandafter\providecommand\csname CRadius\endcsname{}\expandafter\renewcommand\csname CRadius\endcsname{\ensuremath{1.0000}}
\expandafter\providecommand\csname DCount\endcsname{}\expandafter\renewcommand\csname DCount\endcsname{\ensuremath{3}}
\expandafter\providecommand\csname DHeat\endcsname{}\expandafter\renewcommand\csname DHeat\endcsname{\ensuremath{1}}
\expandafter\providecommand\csname DRadius\endcsname{}\expandafter\renewcommand\csname DRadius\endcsname{\ensuremath{1.0337}}
\expandafter\providecommand\csname DeepInputDepth\endcsname{}\expandafter\renewcommand\csname DeepInputDepth\endcsname{\ensuremath{0.064333}}
\expandafter\providecommand\csname ECount\endcsname{}\expandafter\renewcommand\csname ECount\endcsname{\ensuremath{12}}
\expandafter\providecommand\csname EHeat\endcsname{}\expandafter\renewcommand\csname EHeat\endcsname{\ensuremath{0}}
\expandafter\providecommand\csname ERadius\endcsname{}\expandafter\renewcommand\csname ERadius\endcsname{\ensuremath{1.0000}}
\expandafter\providecommand\csname EnergyError\endcsname{}\expandafter\renewcommand\csname EnergyError\endcsname{\ensuremath{0.952}}
\expandafter\providecommand\csname FCount\endcsname{}\expandafter\renewcommand\csname FCount\endcsname{\ensuremath{11}}
\expandafter\providecommand\csname FHeat\endcsname{}\expandafter\renewcommand\csname FHeat\endcsname{\ensuremath{1}}
\expandafter\providecommand\csname FRadius\endcsname{}\expandafter\renewcommand\csname FRadius\endcsname{\ensuremath{1.0237}}
\expandafter\providecommand\csname GCount\endcsname{}\expandafter\renewcommand\csname GCount\endcsname{\ensuremath{12}}
\expandafter\providecommand\csname GHeat\endcsname{}\expandafter\renewcommand\csname GHeat\endcsname{\ensuremath{0}}
\expandafter\providecommand\csname GRadius\endcsname{}\expandafter\renewcommand\csname GRadius\endcsname{\ensuremath{1.0000}}
\expandafter\providecommand\csname HCount\endcsname{}\expandafter\renewcommand\csname HCount\endcsname{\ensuremath{11}}
\expandafter\providecommand\csname HHeat\endcsname{}\expandafter\renewcommand\csname HHeat\endcsname{\ensuremath{1}}
\expandafter\providecommand\csname HRadius\endcsname{}\expandafter\renewcommand\csname HRadius\endcsname{\ensuremath{1.0236}}
\expandafter\providecommand\csname HundredDeepRigidCount\endcsname{}\expandafter\renewcommand\csname HundredDeepRigidCount\endcsname{\ensuremath{3}}
\expandafter\providecommand\csname HundredDeepRigidFirstRecurrence\endcsname{}\expandafter\renewcommand\csname HundredDeepRigidFirstRecurrence\endcsname{\ensuremath{1326}}
\expandafter\providecommand\csname HundredDeepRigidLastRecurrence\endcsname{}\expandafter\renewcommand\csname HundredDeepRigidLastRecurrence\endcsname{\ensuremath{8.128}}
\expandafter\providecommand\csname HundredDeepRigidRadius\endcsname{}\expandafter\renewcommand\csname HundredDeepRigidRadius\endcsname{\ensuremath{1.000}}
\expandafter\providecommand\csname HundredDeepRigidTime\endcsname{}\expandafter\renewcommand\csname HundredDeepRigidTime\endcsname{\ensuremath{1347}}
\expandafter\providecommand\csname HundredDeepSoftCount\endcsname{}\expandafter\renewcommand\csname HundredDeepSoftCount\endcsname{\ensuremath{2}}
\expandafter\providecommand\csname HundredDeepSoftFirstRecurrence\endcsname{}\expandafter\renewcommand\csname HundredDeepSoftFirstRecurrence\endcsname{\ensuremath{1326}}
\expandafter\providecommand\csname HundredDeepSoftLastRecurrence\endcsname{}\expandafter\renewcommand\csname HundredDeepSoftLastRecurrence\endcsname{\ensuremath{1326}}
\expandafter\providecommand\csname HundredDeepSoftRadius\endcsname{}\expandafter\renewcommand\csname HundredDeepSoftRadius\endcsname{\ensuremath{1.026}}
\expandafter\providecommand\csname HundredDeepSoftTime\endcsname{}\expandafter\renewcommand\csname HundredDeepSoftTime\endcsname{\ensuremath{1337}}
\expandafter\providecommand\csname HundredMiddleRigidCount\endcsname{}\expandafter\renewcommand\csname HundredMiddleRigidCount\endcsname{\ensuremath{2}}
\expandafter\providecommand\csname HundredMiddleRigidFirstRecurrence\endcsname{}\expandafter\renewcommand\csname HundredMiddleRigidFirstRecurrence\endcsname{\ensuremath{14.48}}
\expandafter\providecommand\csname HundredMiddleRigidLastRecurrence\endcsname{}\expandafter\renewcommand\csname HundredMiddleRigidLastRecurrence\endcsname{\ensuremath{14.48}}
\expandafter\providecommand\csname HundredMiddleRigidRadius\endcsname{}\expandafter\renewcommand\csname HundredMiddleRigidRadius\endcsname{\ensuremath{1.000}}
\expandafter\providecommand\csname HundredMiddleRigidTime\endcsname{}\expandafter\renewcommand\csname HundredMiddleRigidTime\endcsname{\ensuremath{29.48}}
\expandafter\providecommand\csname HundredMiddleSoftCount\endcsname{}\expandafter\renewcommand\csname HundredMiddleSoftCount\endcsname{\ensuremath{2}}
\expandafter\providecommand\csname HundredMiddleSoftFirstRecurrence\endcsname{}\expandafter\renewcommand\csname HundredMiddleSoftFirstRecurrence\endcsname{\ensuremath{14.45}}
\expandafter\providecommand\csname HundredMiddleSoftLastRecurrence\endcsname{}\expandafter\renewcommand\csname HundredMiddleSoftLastRecurrence\endcsname{\ensuremath{14.45}}
\expandafter\providecommand\csname HundredMiddleSoftRadius\endcsname{}\expandafter\renewcommand\csname HundredMiddleSoftRadius\endcsname{\ensuremath{1.017}}
\expandafter\providecommand\csname HundredMiddleSoftTime\endcsname{}\expandafter\renewcommand\csname HundredMiddleSoftTime\endcsname{\ensuremath{20.95}}
\expandafter\providecommand\csname HundredShallowRigidCount\endcsname{}\expandafter\renewcommand\csname HundredShallowRigidCount\endcsname{\ensuremath{3}}
\expandafter\providecommand\csname HundredShallowRigidFirstRecurrence\endcsname{}\expandafter\renewcommand\csname HundredShallowRigidFirstRecurrence\endcsname{\ensuremath{25.19}}
\expandafter\providecommand\csname HundredShallowRigidLastRecurrence\endcsname{}\expandafter\renewcommand\csname HundredShallowRigidLastRecurrence\endcsname{\ensuremath{4.931}}
\expandafter\providecommand\csname HundredShallowRigidRadius\endcsname{}\expandafter\renewcommand\csname HundredShallowRigidRadius\endcsname{\ensuremath{1.000}}
\expandafter\providecommand\csname HundredShallowRigidTime\endcsname{}\expandafter\renewcommand\csname HundredShallowRigidTime\endcsname{\ensuremath{48.64}}
\expandafter\providecommand\csname HundredShallowSoftCount\endcsname{}\expandafter\renewcommand\csname HundredShallowSoftCount\endcsname{\ensuremath{3}}
\expandafter\providecommand\csname HundredShallowSoftFirstRecurrence\endcsname{}\expandafter\renewcommand\csname HundredShallowSoftFirstRecurrence\endcsname{\ensuremath{25.15}}
\expandafter\providecommand\csname HundredShallowSoftLastRecurrence\endcsname{}\expandafter\renewcommand\csname HundredShallowSoftLastRecurrence\endcsname{\ensuremath{5.987}}
\expandafter\providecommand\csname HundredShallowSoftRadius\endcsname{}\expandafter\renewcommand\csname HundredShallowSoftRadius\endcsname{\ensuremath{1.045}}
\expandafter\providecommand\csname HundredShallowSoftTime\endcsname{}\expandafter\renewcommand\csname HundredShallowSoftTime\endcsname{\ensuremath{52.68}}
\expandafter\providecommand\csname ManyAngularError\endcsname{}\expandafter\renewcommand\csname ManyAngularError\endcsname{\ensuremath{0.102}}
\expandafter\providecommand\csname ManyCount\endcsname{}\expandafter\renewcommand\csname ManyCount\endcsname{\ensuremath{100}}
\expandafter\providecommand\csname ManyEcc\endcsname{}\expandafter\renewcommand\csname ManyEcc\endcsname{\ensuremath{0.8302}}
\expandafter\providecommand\csname ManyEnergyError\endcsname{}\expandafter\renewcommand\csname ManyEnergyError\endcsname{\ensuremath{0.227}}
\expandafter\providecommand\csname ManyFirstApo\endcsname{}\expandafter\renewcommand\csname ManyFirstApo\endcsname{\ensuremath{382.00}}
\expandafter\providecommand\csname ManyFirstDelay\endcsname{}\expandafter\renewcommand\csname ManyFirstDelay\endcsname{\ensuremath{15831.26}}
\expandafter\providecommand\csname ManyFirstEcc\endcsname{}\expandafter\renewcommand\csname ManyFirstEcc\endcsname{\ensuremath{0.998431}}
\expandafter\providecommand\csname ManyFirstPeri\endcsname{}\expandafter\renewcommand\csname ManyFirstPeri\endcsname{\ensuremath{0.3000}}
\expandafter\providecommand\csname ManyLastApo\endcsname{}\expandafter\renewcommand\csname ManyLastApo\endcsname{\ensuremath{3.2173}}
\expandafter\providecommand\csname ManyLastDelay\endcsname{}\expandafter\renewcommand\csname ManyLastDelay\endcsname{\ensuremath{12.5616}}
\expandafter\providecommand\csname ManyPeri\endcsname{}\expandafter\renewcommand\csname ManyPeri\endcsname{\ensuremath{0.2986}}
\expandafter\providecommand\csname ManyTime\endcsname{}\expandafter\renewcommand\csname ManyTime\endcsname{\ensuremath{37891.84}}
\expandafter\providecommand\csname MiddleInputDepth\endcsname{}\expandafter\renewcommand\csname MiddleInputDepth\endcsname{\ensuremath{0.165551}}
\expandafter\providecommand\csname OldDeepRigidCount\endcsname{}\expandafter\renewcommand\csname OldDeepRigidCount\endcsname{\ensuremath{3}}
\expandafter\providecommand\csname OldDeepRigidFirstRecurrence\endcsname{}\expandafter\renewcommand\csname OldDeepRigidFirstRecurrence\endcsname{\ensuremath{3209}}
\expandafter\providecommand\csname OldDeepRigidLastRecurrence\endcsname{}\expandafter\renewcommand\csname OldDeepRigidLastRecurrence\endcsname{\ensuremath{108.4}}
\expandafter\providecommand\csname OldDeepRigidRadius\endcsname{}\expandafter\renewcommand\csname OldDeepRigidRadius\endcsname{\ensuremath{1.000}}
\expandafter\providecommand\csname OldDeepRigidTime\endcsname{}\expandafter\renewcommand\csname OldDeepRigidTime\endcsname{\ensuremath{3596}}
\expandafter\providecommand\csname OldDeepSoftCount\endcsname{}\expandafter\renewcommand\csname OldDeepSoftCount\endcsname{\ensuremath{3}}
\expandafter\providecommand\csname OldDeepSoftFirstRecurrence\endcsname{}\expandafter\renewcommand\csname OldDeepSoftFirstRecurrence\endcsname{\ensuremath{3208}}
\expandafter\providecommand\csname OldDeepSoftLastRecurrence\endcsname{}\expandafter\renewcommand\csname OldDeepSoftLastRecurrence\endcsname{\ensuremath{145.6}}
\expandafter\providecommand\csname OldDeepSoftRadius\endcsname{}\expandafter\renewcommand\csname OldDeepSoftRadius\endcsname{\ensuremath{1.084}}
\expandafter\providecommand\csname OldDeepSoftTime\endcsname{}\expandafter\renewcommand\csname OldDeepSoftTime\endcsname{\ensuremath{3712}}
\expandafter\providecommand\csname OldMiddleRigidCount\endcsname{}\expandafter\renewcommand\csname OldMiddleRigidCount\endcsname{\ensuremath{12}}
\expandafter\providecommand\csname OldMiddleRigidFirstRecurrenceExp\endcsname{}\expandafter\renewcommand\csname OldMiddleRigidFirstRecurrenceExp\endcsname{\ensuremath{4}}
\expandafter\providecommand\csname OldMiddleRigidFirstRecurrenceMant\endcsname{}\expandafter\renewcommand\csname OldMiddleRigidFirstRecurrenceMant\endcsname{\ensuremath{1.967}}
\expandafter\providecommand\csname OldMiddleRigidLastRecurrence\endcsname{}\expandafter\renewcommand\csname OldMiddleRigidLastRecurrence\endcsname{\ensuremath{27.29}}
\expandafter\providecommand\csname OldMiddleRigidRadius\endcsname{}\expandafter\renewcommand\csname OldMiddleRigidRadius\endcsname{\ensuremath{1.000}}
\expandafter\providecommand\csname OldMiddleRigidTimeExp\endcsname{}\expandafter\renewcommand\csname OldMiddleRigidTimeExp\endcsname{\ensuremath{4}}
\expandafter\providecommand\csname OldMiddleRigidTimeMant\endcsname{}\expandafter\renewcommand\csname OldMiddleRigidTimeMant\endcsname{\ensuremath{2.750}}
\expandafter\providecommand\csname OldMiddleSoftCount\endcsname{}\expandafter\renewcommand\csname OldMiddleSoftCount\endcsname{\ensuremath{10}}
\expandafter\providecommand\csname OldMiddleSoftFirstRecurrenceExp\endcsname{}\expandafter\renewcommand\csname OldMiddleSoftFirstRecurrenceExp\endcsname{\ensuremath{4}}
\expandafter\providecommand\csname OldMiddleSoftFirstRecurrenceMant\endcsname{}\expandafter\renewcommand\csname OldMiddleSoftFirstRecurrenceMant\endcsname{\ensuremath{1.967}}
\expandafter\providecommand\csname OldMiddleSoftLastRecurrence\endcsname{}\expandafter\renewcommand\csname OldMiddleSoftLastRecurrence\endcsname{\ensuremath{104.7}}
\expandafter\providecommand\csname OldMiddleSoftRadius\endcsname{}\expandafter\renewcommand\csname OldMiddleSoftRadius\endcsname{\ensuremath{1.762}}
\expandafter\providecommand\csname OldMiddleSoftTimeExp\endcsname{}\expandafter\renewcommand\csname OldMiddleSoftTimeExp\endcsname{\ensuremath{4}}
\expandafter\providecommand\csname OldMiddleSoftTimeMant\endcsname{}\expandafter\renewcommand\csname OldMiddleSoftTimeMant\endcsname{\ensuremath{2.973}}
\expandafter\providecommand\csname OldShallowRigidCount\endcsname{}\expandafter\renewcommand\csname OldShallowRigidCount\endcsname{\ensuremath{60}}
\expandafter\providecommand\csname OldShallowRigidFirstRecurrenceExp\endcsname{}\expandafter\renewcommand\csname OldShallowRigidFirstRecurrenceExp\endcsname{\ensuremath{5}}
\expandafter\providecommand\csname OldShallowRigidFirstRecurrenceMant\endcsname{}\expandafter\renewcommand\csname OldShallowRigidFirstRecurrenceMant\endcsname{\ensuremath{7.213}}
\expandafter\providecommand\csname OldShallowRigidLastRecurrence\endcsname{}\expandafter\renewcommand\csname OldShallowRigidLastRecurrence\endcsname{\ensuremath{1386}}
\expandafter\providecommand\csname OldShallowRigidRadius\endcsname{}\expandafter\renewcommand\csname OldShallowRigidRadius\endcsname{\ensuremath{1.000}}
\expandafter\providecommand\csname OldShallowRigidTimeExp\endcsname{}\expandafter\renewcommand\csname OldShallowRigidTimeExp\endcsname{\ensuremath{6}}
\expandafter\providecommand\csname OldShallowRigidTimeMant\endcsname{}\expandafter\renewcommand\csname OldShallowRigidTimeMant\endcsname{\ensuremath{1.688}}
\expandafter\providecommand\csname OldShallowSoftCount\endcsname{}\expandafter\renewcommand\csname OldShallowSoftCount\endcsname{\ensuremath{33}}
\expandafter\providecommand\csname OldShallowSoftFirstRecurrenceExp\endcsname{}\expandafter\renewcommand\csname OldShallowSoftFirstRecurrenceExp\endcsname{\ensuremath{5}}
\expandafter\providecommand\csname OldShallowSoftFirstRecurrenceMant\endcsname{}\expandafter\renewcommand\csname OldShallowSoftFirstRecurrenceMant\endcsname{\ensuremath{7.213}}
\expandafter\providecommand\csname OldShallowSoftLastRecurrence\endcsname{}\expandafter\renewcommand\csname OldShallowSoftLastRecurrence\endcsname{\ensuremath{594.5}}
\expandafter\providecommand\csname OldShallowSoftRadius\endcsname{}\expandafter\renewcommand\csname OldShallowSoftRadius\endcsname{\ensuremath{5.201}}
\expandafter\providecommand\csname OldShallowSoftTimeExp\endcsname{}\expandafter\renewcommand\csname OldShallowSoftTimeExp\endcsname{\ensuremath{6}}
\expandafter\providecommand\csname OldShallowSoftTimeMant\endcsname{}\expandafter\renewcommand\csname OldShallowSoftTimeMant\endcsname{\ensuremath{1.397}}
\expandafter\providecommand\csname TightAngular\endcsname{}\expandafter\renewcommand\csname TightAngular\endcsname{\ensuremath{5.557}}
\expandafter\providecommand\csname TightEnergy\endcsname{}\expandafter\renewcommand\csname TightEnergy\endcsname{\ensuremath{80.865}}
\expandafter\providecommand\csname TwentyDeepRigidCount\endcsname{}\expandafter\renewcommand\csname TwentyDeepRigidCount\endcsname{\ensuremath{3}}
\expandafter\providecommand\csname TwentyDeepRigidFirstRecurrence\endcsname{}\expandafter\renewcommand\csname TwentyDeepRigidFirstRecurrence\endcsname{\ensuremath{255.6}}
\expandafter\providecommand\csname TwentyDeepRigidLastRecurrence\endcsname{}\expandafter\renewcommand\csname TwentyDeepRigidLastRecurrence\endcsname{\ensuremath{4.419}}
\expandafter\providecommand\csname TwentyDeepRigidRadius\endcsname{}\expandafter\renewcommand\csname TwentyDeepRigidRadius\endcsname{\ensuremath{1.000}}
\expandafter\providecommand\csname TwentyDeepRigidTime\endcsname{}\expandafter\renewcommand\csname TwentyDeepRigidTime\endcsname{\ensuremath{268.1}}
\expandafter\providecommand\csname TwentyDeepSoftCount\endcsname{}\expandafter\renewcommand\csname TwentyDeepSoftCount\endcsname{\ensuremath{2}}
\expandafter\providecommand\csname TwentyDeepSoftFirstRecurrence\endcsname{}\expandafter\renewcommand\csname TwentyDeepSoftFirstRecurrence\endcsname{\ensuremath{255.5}}
\expandafter\providecommand\csname TwentyDeepSoftLastRecurrence\endcsname{}\expandafter\renewcommand\csname TwentyDeepSoftLastRecurrence\endcsname{\ensuremath{255.5}}
\expandafter\providecommand\csname TwentyDeepSoftRadius\endcsname{}\expandafter\renewcommand\csname TwentyDeepSoftRadius\endcsname{\ensuremath{1.028}}
\expandafter\providecommand\csname TwentyDeepSoftTime\endcsname{}\expandafter\renewcommand\csname TwentyDeepSoftTime\endcsname{\ensuremath{262}}
\expandafter\providecommand\csname TwentyMiddleRigidCount\endcsname{}\expandafter\renewcommand\csname TwentyMiddleRigidCount\endcsname{\ensuremath{2}}
\expandafter\providecommand\csname TwentyMiddleRigidFirstRecurrence\endcsname{}\expandafter\renewcommand\csname TwentyMiddleRigidFirstRecurrence\endcsname{\ensuremath{8.964}}
\expandafter\providecommand\csname TwentyMiddleRigidLastRecurrence\endcsname{}\expandafter\renewcommand\csname TwentyMiddleRigidLastRecurrence\endcsname{\ensuremath{8.964}}
\expandafter\providecommand\csname TwentyMiddleRigidRadius\endcsname{}\expandafter\renewcommand\csname TwentyMiddleRigidRadius\endcsname{\ensuremath{1.000}}
\expandafter\providecommand\csname TwentyMiddleRigidTime\endcsname{}\expandafter\renewcommand\csname TwentyMiddleRigidTime\endcsname{\ensuremath{19.24}}
\expandafter\providecommand\csname TwentyMiddleSoftCount\endcsname{}\expandafter\renewcommand\csname TwentyMiddleSoftCount\endcsname{\ensuremath{2}}
\expandafter\providecommand\csname TwentyMiddleSoftFirstRecurrence\endcsname{}\expandafter\renewcommand\csname TwentyMiddleSoftFirstRecurrence\endcsname{\ensuremath{8.937}}
\expandafter\providecommand\csname TwentyMiddleSoftLastRecurrence\endcsname{}\expandafter\renewcommand\csname TwentyMiddleSoftLastRecurrence\endcsname{\ensuremath{8.937}}
\expandafter\providecommand\csname TwentyMiddleSoftRadius\endcsname{}\expandafter\renewcommand\csname TwentyMiddleSoftRadius\endcsname{\ensuremath{1.019}}
\expandafter\providecommand\csname TwentyMiddleSoftTime\endcsname{}\expandafter\renewcommand\csname TwentyMiddleSoftTime\endcsname{\ensuremath{19.55}}
\expandafter\providecommand\csname TwentyShallowRigidCount\endcsname{}\expandafter\renewcommand\csname TwentyShallowRigidCount\endcsname{\ensuremath{3}}
\expandafter\providecommand\csname TwentyShallowRigidFirstRecurrence\endcsname{}\expandafter\renewcommand\csname TwentyShallowRigidFirstRecurrence\endcsname{\ensuremath{18.14}}
\expandafter\providecommand\csname TwentyShallowRigidLastRecurrence\endcsname{}\expandafter\renewcommand\csname TwentyShallowRigidLastRecurrence\endcsname{\ensuremath{3.732}}
\expandafter\providecommand\csname TwentyShallowRigidRadius\endcsname{}\expandafter\renewcommand\csname TwentyShallowRigidRadius\endcsname{\ensuremath{1.000}}
\expandafter\providecommand\csname TwentyShallowRigidTime\endcsname{}\expandafter\renewcommand\csname TwentyShallowRigidTime\endcsname{\ensuremath{35.09}}
\expandafter\providecommand\csname TwentyShallowSoftCount\endcsname{}\expandafter\renewcommand\csname TwentyShallowSoftCount\endcsname{\ensuremath{3}}
\expandafter\providecommand\csname TwentyShallowSoftFirstRecurrence\endcsname{}\expandafter\renewcommand\csname TwentyShallowSoftFirstRecurrence\endcsname{\ensuremath{18.11}}
\expandafter\providecommand\csname TwentyShallowSoftLastRecurrence\endcsname{}\expandafter\renewcommand\csname TwentyShallowSoftLastRecurrence\endcsname{\ensuremath{4.474}}
\expandafter\providecommand\csname TwentyShallowSoftRadius\endcsname{}\expandafter\renewcommand\csname TwentyShallowSoftRadius\endcsname{\ensuremath{1.049}}
\expandafter\providecommand\csname TwentyShallowSoftTime\endcsname{}\expandafter\renewcommand\csname TwentyShallowSoftTime\endcsname{\ensuremath{37.98}}
\expandafter\providecommand\csname VoneAngularError\endcsname{}\expandafter\renewcommand\csname VoneAngularError\endcsname{\ensuremath{2.474}}
\expandafter\providecommand\csname VoneCases\endcsname{}\expandafter\renewcommand\csname VoneCases\endcsname{\ensuremath{30}}
\expandafter\providecommand\csname VoneDepositionError\endcsname{}\expandafter\renewcommand\csname VoneDepositionError\endcsname{\ensuremath{1.044}}
\expandafter\providecommand\csname VoneEnergyError\endcsname{}\expandafter\renewcommand\csname VoneEnergyError\endcsname{\ensuremath{7.483}}
\expandafter\providecommand\csname VoneEngulfCount\endcsname{}\expandafter\renewcommand\csname VoneEngulfCount\endcsname{\ensuremath{33}}
\expandafter\providecommand\csname VoneEngulfRadius\endcsname{}\expandafter\renewcommand\csname VoneEngulfRadius\endcsname{\ensuremath{5.201}}
\expandafter\providecommand\csname VoneEngulfYears\endcsname{}\expandafter\renewcommand\csname VoneEngulfYears\endcsname{\ensuremath{159.4}}
\expandafter\providecommand\csname VoneHundredMax\endcsname{}\expandafter\renewcommand\csname VoneHundredMax\endcsname{\ensuremath{52.7}}
\expandafter\providecommand\csname VoneHundredMin\endcsname{}\expandafter\renewcommand\csname VoneHundredMin\endcsname{\ensuremath{21.0}}
\expandafter\providecommand\csname VoneReplayAngular\endcsname{}\expandafter\renewcommand\csname VoneReplayAngular\endcsname{\ensuremath{5.5\times10^{-5}}}
\expandafter\providecommand\csname VoneReplayEnergy\endcsname{}\expandafter\renewcommand\csname VoneReplayEnergy\endcsname{\ensuremath{5.9\times10^{-5}}}
\expandafter\providecommand\csname VoneReplayPassages\endcsname{}\expandafter\renewcommand\csname VoneReplayPassages\endcsname{\ensuremath{2}}
\expandafter\providecommand\csname VoneReplayTime\endcsname{}\expandafter\renewcommand\csname VoneReplayTime\endcsname{\ensuremath{0.000135}}
\expandafter\providecommand\csname VoneResponsive\endcsname{}\expandafter\renewcommand\csname VoneResponsive\endcsname{\ensuremath{15}}
\expandafter\providecommand\csname VoneShellCrossing\endcsname{}\expandafter\renewcommand\csname VoneShellCrossing\endcsname{\ensuremath{6}}
\expandafter\providecommand\csname VoneSmoothingMax\endcsname{}\expandafter\renewcommand\csname VoneSmoothingMax\endcsname{\ensuremath{40}}
\expandafter\providecommand\csname VoneStrippedMass\endcsname{}\expandafter\renewcommand\csname VoneStrippedMass\endcsname{\ensuremath{0.0}}
\expandafter\providecommand\csname VoneTidalMax\endcsname{}\expandafter\renewcommand\csname VoneTidalMax\endcsname{\ensuremath{36.5}}
\expandafter\providecommand\csname VoneTidalMin\endcsname{}\expandafter\renewcommand\csname VoneTidalMin\endcsname{\ensuremath{0.0141}}

\providecommand\result[1]{%
  \ifcsname #1\endcsname\csname #1\endcsname\else\textbf{??#1??}\fi}

\expandafter\providecommand\csname BlackHoleMass\endcsname{}\expandafter\renewcommand\csname BlackHoleMass\endcsname{\ensuremath{7}}
\expandafter\providecommand\csname BoundCount\endcsname{}\expandafter\renewcommand\csname BoundCount\endcsname{\ensuremath{92}}
\expandafter\providecommand\csname CoreCount\endcsname{}\expandafter\renewcommand\csname CoreCount\endcsname{\ensuremath{63}}
\expandafter\providecommand\csname CoreMass\endcsname{}\expandafter\renewcommand\csname CoreMass\endcsname{\ensuremath{27.55483}}
\expandafter\providecommand\csname DynTime\endcsname{}\expandafter\renewcommand\csname DynTime\endcsname{\ensuremath{1.899}}
\expandafter\providecommand\csname EnvelopeGravity\endcsname{}\expandafter\renewcommand\csname EnvelopeGravity\endcsname{\ensuremath{1.362\times10^{51}}}
\expandafter\providecommand\csname EnvelopeInternal\endcsname{}\expandafter\renewcommand\csname EnvelopeInternal\endcsname{\ensuremath{9.298\times10^{50}}}
\expandafter\providecommand\csname EnvelopeMass\endcsname{}\expandafter\renewcommand\csname EnvelopeMass\endcsname{\ensuremath{42.44517}}
\expandafter\providecommand\csname EnvelopeNet\endcsname{}\expandafter\renewcommand\csname EnvelopeNet\endcsname{\ensuremath{4.323\times10^{50}}}
\expandafter\providecommand\csname FastInnerDepth\endcsname{}\expandafter\renewcommand\csname FastInnerDepth\endcsname{\ensuremath{0.06433324}}
\expandafter\providecommand\csname FastInnerLimitedHigh\endcsname{}\expandafter\renewcommand\csname FastInnerLimitedHigh\endcsname{\ensuremath{0.832031}}
\expandafter\providecommand\csname FastInnerLimitedLow\endcsname{}\expandafter\renewcommand\csname FastInnerLimitedLow\endcsname{\ensuremath{0.828125}}
\expandafter\providecommand\csname FastInnerRatio\endcsname{}\expandafter\renewcommand\csname FastInnerRatio\endcsname{\ensuremath{2.613}}
\expandafter\providecommand\csname FastInnerRawHigh\endcsname{}\expandafter\renewcommand\csname FastInnerRawHigh\endcsname{\ensuremath{0.832031}}
\expandafter\providecommand\csname FastInnerRawLow\endcsname{}\expandafter\renewcommand\csname FastInnerRawLow\endcsname{\ensuremath{0.828125}}
\expandafter\providecommand\csname FastInnerWork\endcsname{}\expandafter\renewcommand\csname FastInnerWork\endcsname{\ensuremath{1.130\times10^{49}}}
\expandafter\providecommand\csname FastLimitedGamma\endcsname{}\expandafter\renewcommand\csname FastLimitedGamma\endcsname{\ensuremath{0.07219672}}
\expandafter\providecommand\csname FastLimitedNative\endcsname{}\expandafter\renewcommand\csname FastLimitedNative\endcsname{\ensuremath{0.07425148}}
\expandafter\providecommand\csname FastOuterDepth\endcsname{}\expandafter\renewcommand\csname FastOuterDepth\endcsname{\ensuremath{0.08815913}}
\expandafter\providecommand\csname FastOuterLimitedHigh\endcsname{}\expandafter\renewcommand\csname FastOuterLimitedHigh\endcsname{\ensuremath{1.289062}}
\expandafter\providecommand\csname FastOuterLimitedLow\endcsname{}\expandafter\renewcommand\csname FastOuterLimitedLow\endcsname{\ensuremath{1.285156}}
\expandafter\providecommand\csname FastOuterRatio\endcsname{}\expandafter\renewcommand\csname FastOuterRatio\endcsname{\ensuremath{1.663}}
\expandafter\providecommand\csname FastOuterRawHigh\endcsname{}\expandafter\renewcommand\csname FastOuterRawHigh\endcsname{\ensuremath{1.289062}}
\expandafter\providecommand\csname FastOuterRawLow\endcsname{}\expandafter\renewcommand\csname FastOuterRawLow\endcsname{\ensuremath{1.285156}}
\expandafter\providecommand\csname FastOuterWork\endcsname{}\expandafter\renewcommand\csname FastOuterWork\endcsname{\ensuremath{7.191\times10^{48}}}
\expandafter\providecommand\csname FastRawGamma\endcsname{}\expandafter\renewcommand\csname FastRawGamma\endcsname{\ensuremath{0.07220419}}
\expandafter\providecommand\csname FastRawNative\endcsname{}\expandafter\renewcommand\csname FastRawNative\endcsname{\ensuremath{0.07425918}}
\expandafter\providecommand\csname FastShift\endcsname{}\expandafter\renewcommand\csname FastShift\endcsname{\ensuremath{2.77}}
\expandafter\providecommand\csname FastSpeed\endcsname{}\expandafter\renewcommand\csname FastSpeed\endcsname{\ensuremath{1}}
\expandafter\providecommand\csname GammaPressureError\endcsname{}\expandafter\renewcommand\csname GammaPressureError\endcsname{\ensuremath{57.91}}
\expandafter\providecommand\csname GammaSoundError\endcsname{}\expandafter\renewcommand\csname GammaSoundError\endcsname{\ensuremath{8.70}}
\expandafter\providecommand\csname GammaVirial\endcsname{}\expandafter\renewcommand\csname GammaVirial\endcsname{\ensuremath{5.638\times10^{-1}}}
\expandafter\providecommand\csname HydroMedian\endcsname{}\expandafter\renewcommand\csname HydroMedian\endcsname{\ensuremath{5.049\times10^{-4}}}
\expandafter\providecommand\csname HydroTail\endcsname{}\expandafter\renewcommand\csname HydroTail\endcsname{\ensuremath{1.981\times10^{-3}}}
\expandafter\providecommand\csname JointEnergy\endcsname{}\expandafter\renewcommand\csname JointEnergy\endcsname{\ensuremath{2.419\times10^{-9}}}
\expandafter\providecommand\csname JointHigh\endcsname{}\expandafter\renewcommand\csname JointHigh\endcsname{\ensuremath{4.96}}
\expandafter\providecommand\csname JointLow\endcsname{}\expandafter\renewcommand\csname JointLow\endcsname{\ensuremath{2.88}}
\expandafter\providecommand\csname JointRefinement\endcsname{}\expandafter\renewcommand\csname JointRefinement\endcsname{\ensuremath{4.216\times10^{-7}}}
\expandafter\providecommand\csname MachFastInnerLimitedFive\endcsname{}\expandafter\renewcommand\csname MachFastInnerLimitedFive\endcsname{\ensuremath{0.032}}
\expandafter\providecommand\csname MachFastInnerLimitedMedian\endcsname{}\expandafter\renewcommand\csname MachFastInnerLimitedMedian\endcsname{\ensuremath{2.751}}
\expandafter\providecommand\csname MachFastInnerLimitedThree\endcsname{}\expandafter\renewcommand\csname MachFastInnerLimitedThree\endcsname{\ensuremath{17.435}}
\expandafter\providecommand\csname MachFastInnerRawFive\endcsname{}\expandafter\renewcommand\csname MachFastInnerRawFive\endcsname{\ensuremath{0.042}}
\expandafter\providecommand\csname MachFastInnerRawMedian\endcsname{}\expandafter\renewcommand\csname MachFastInnerRawMedian\endcsname{\ensuremath{2.750}}
\expandafter\providecommand\csname MachFastInnerRawThree\endcsname{}\expandafter\renewcommand\csname MachFastInnerRawThree\endcsname{\ensuremath{17.446}}
\expandafter\providecommand\csname MachFastOuterLimitedFive\endcsname{}\expandafter\renewcommand\csname MachFastOuterLimitedFive\endcsname{\ensuremath{0.061}}
\expandafter\providecommand\csname MachFastOuterLimitedMedian\endcsname{}\expandafter\renewcommand\csname MachFastOuterLimitedMedian\endcsname{\ensuremath{2.922}}
\expandafter\providecommand\csname MachFastOuterLimitedThree\endcsname{}\expandafter\renewcommand\csname MachFastOuterLimitedThree\endcsname{\ensuremath{35.401}}
\expandafter\providecommand\csname MachFastOuterRawFive\endcsname{}\expandafter\renewcommand\csname MachFastOuterRawFive\endcsname{\ensuremath{0.077}}
\expandafter\providecommand\csname MachFastOuterRawMedian\endcsname{}\expandafter\renewcommand\csname MachFastOuterRawMedian\endcsname{\ensuremath{2.922}}
\expandafter\providecommand\csname MachFastOuterRawThree\endcsname{}\expandafter\renewcommand\csname MachFastOuterRawThree\endcsname{\ensuremath{35.458}}
\expandafter\providecommand\csname MachFiveMax\endcsname{}\expandafter\renewcommand\csname MachFiveMax\endcsname{\ensuremath{1.92}}
\expandafter\providecommand\csname MachFractionError\endcsname{}\expandafter\renewcommand\csname MachFractionError\endcsname{\ensuremath{1.434\times10^{-4}}}
\expandafter\providecommand\csname MachMedianHigh\endcsname{}\expandafter\renewcommand\csname MachMedianHigh\endcsname{\ensuremath{3.69}}
\expandafter\providecommand\csname MachMedianLow\endcsname{}\expandafter\renewcommand\csname MachMedianLow\endcsname{\ensuremath{2.75}}
\expandafter\providecommand\csname MachReplayError\endcsname{}\expandafter\renewcommand\csname MachReplayError\endcsname{\ensuremath{1.415\times10^{-9}}}
\expandafter\providecommand\csname MachSlowInnerLimitedFive\endcsname{}\expandafter\renewcommand\csname MachSlowInnerLimitedFive\endcsname{\ensuremath{0.195}}
\expandafter\providecommand\csname MachSlowInnerLimitedMedian\endcsname{}\expandafter\renewcommand\csname MachSlowInnerLimitedMedian\endcsname{\ensuremath{3.441}}
\expandafter\providecommand\csname MachSlowInnerLimitedThree\endcsname{}\expandafter\renewcommand\csname MachSlowInnerLimitedThree\endcsname{\ensuremath{99.997}}
\expandafter\providecommand\csname MachSlowInnerRawFive\endcsname{}\expandafter\renewcommand\csname MachSlowInnerRawFive\endcsname{\ensuremath{0.327}}
\expandafter\providecommand\csname MachSlowInnerRawMedian\endcsname{}\expandafter\renewcommand\csname MachSlowInnerRawMedian\endcsname{\ensuremath{3.441}}
\expandafter\providecommand\csname MachSlowInnerRawThree\endcsname{}\expandafter\renewcommand\csname MachSlowInnerRawThree\endcsname{\ensuremath{99.997}}
\expandafter\providecommand\csname MachSlowOuterLimitedFive\endcsname{}\expandafter\renewcommand\csname MachSlowOuterLimitedFive\endcsname{\ensuremath{1.200}}
\expandafter\providecommand\csname MachSlowOuterLimitedMedian\endcsname{}\expandafter\renewcommand\csname MachSlowOuterLimitedMedian\endcsname{\ensuremath{3.690}}
\expandafter\providecommand\csname MachSlowOuterLimitedThree\endcsname{}\expandafter\renewcommand\csname MachSlowOuterLimitedThree\endcsname{\ensuremath{99.986}}
\expandafter\providecommand\csname MachSlowOuterRawFive\endcsname{}\expandafter\renewcommand\csname MachSlowOuterRawFive\endcsname{\ensuremath{1.921}}
\expandafter\providecommand\csname MachSlowOuterRawMedian\endcsname{}\expandafter\renewcommand\csname MachSlowOuterRawMedian\endcsname{\ensuremath{3.691}}
\expandafter\providecommand\csname MachSlowOuterRawThree\endcsname{}\expandafter\renewcommand\csname MachSlowOuterRawThree\endcsname{\ensuremath{99.986}}
\expandafter\providecommand\csname MachWorkError\endcsname{}\expandafter\renewcommand\csname MachWorkError\endcsname{\ensuremath{3.462\times10^{-7}}}
\expandafter\providecommand\csname MassError\endcsname{}\expandafter\renewcommand\csname MassError\endcsname{\ensuremath{7.840\times10^{-5}}}
\expandafter\providecommand\csname MassNormalization\endcsname{}\expandafter\renewcommand\csname MassNormalization\endcsname{\ensuremath{1.000294}}
\expandafter\providecommand\csname MaxGravityRatio\endcsname{}\expandafter\renewcommand\csname MaxGravityRatio\endcsname{\ensuremath{0.829}}
\expandafter\providecommand\csname MaxNetRatio\endcsname{}\expandafter\renewcommand\csname MaxNetRatio\endcsname{\ensuremath{2.613}}
\expandafter\providecommand\csname MaxWork\endcsname{}\expandafter\renewcommand\csname MaxWork\endcsname{\ensuremath{1.130\times10^{49}}}
\expandafter\providecommand\csname MixtureDerivativeError\endcsname{}\expandafter\renewcommand\csname MixtureDerivativeError\endcsname{\ensuremath{3.089\times10^{-11}}}
\expandafter\providecommand\csname MixtureEnergy\endcsname{}\expandafter\renewcommand\csname MixtureEnergy\endcsname{\ensuremath{0.366}}
\expandafter\providecommand\csname MixtureSound\endcsname{}\expandafter\renewcommand\csname MixtureSound\endcsname{\ensuremath{0.010}}
\expandafter\providecommand\csname MixtureWorstEnergy\endcsname{}\expandafter\renewcommand\csname MixtureWorstEnergy\endcsname{\ensuremath{33.000}}
\expandafter\providecommand\csname NativeVirial\endcsname{}\expandafter\renewcommand\csname NativeVirial\endcsname{\ensuremath{2.186\times10^{-5}}}
\expandafter\providecommand\csname NewWorkError\endcsname{}\expandafter\renewcommand\csname NewWorkError\endcsname{\ensuremath{5.963\times10^{-5}}}
\expandafter\providecommand\csname NullResidual\endcsname{}\expandafter\renewcommand\csname NullResidual\endcsname{\ensuremath{4.330\times10^{-7}}}
\expandafter\providecommand\csname OldWorkError\endcsname{}\expandafter\renewcommand\csname OldWorkError\endcsname{\ensuremath{5.72}}
\expandafter\providecommand\csname RadMass\endcsname{}\expandafter\renewcommand\csname RadMass\endcsname{\ensuremath{88.97}}
\expandafter\providecommand\csname RootBalance\endcsname{}\expandafter\renewcommand\csname RootBalance\endcsname{\ensuremath{1.497\times10^{-9}}}
\expandafter\providecommand\csname RootRuns\endcsname{}\expandafter\renewcommand\csname RootRuns\endcsname{\ensuremath{77}}
\expandafter\providecommand\csname RootWidth\endcsname{}\expandafter\renewcommand\csname RootWidth\endcsname{\ensuremath{4.000\times10^{-6}}}
\expandafter\providecommand\csname ScanCount\endcsname{}\expandafter\renewcommand\csname ScanCount\endcsname{\ensuremath{240}}
\expandafter\providecommand\csname SlowInnerDepth\endcsname{}\expandafter\renewcommand\csname SlowInnerDepth\endcsname{\ensuremath{0.16555066}}
\expandafter\providecommand\csname SlowInnerLimitedHigh\endcsname{}\expandafter\renewcommand\csname SlowInnerLimitedHigh\endcsname{\ensuremath{0.593750}}
\expandafter\providecommand\csname SlowInnerLimitedLow\endcsname{}\expandafter\renewcommand\csname SlowInnerLimitedLow\endcsname{\ensuremath{0.589844}}
\expandafter\providecommand\csname SlowInnerRatio\endcsname{}\expandafter\renewcommand\csname SlowInnerRatio\endcsname{\ensuremath{0.331}}
\expandafter\providecommand\csname SlowInnerRawHigh\endcsname{}\expandafter\renewcommand\csname SlowInnerRawHigh\endcsname{\ensuremath{0.589844}}
\expandafter\providecommand\csname SlowInnerRawLow\endcsname{}\expandafter\renewcommand\csname SlowInnerRawLow\endcsname{\ensuremath{0.585938}}
\expandafter\providecommand\csname SlowInnerWork\endcsname{}\expandafter\renewcommand\csname SlowInnerWork\endcsname{\ensuremath{1.431\times10^{48}}}
\expandafter\providecommand\csname SlowLimitedGamma\endcsname{}\expandafter\renewcommand\csname SlowLimitedGamma\endcsname{\ensuremath{0.18585180}}
\expandafter\providecommand\csname SlowLimitedNative\endcsname{}\expandafter\renewcommand\csname SlowLimitedNative\endcsname{\ensuremath{0.18684452}}
\expandafter\providecommand\csname SlowOuterDepth\endcsname{}\expandafter\renewcommand\csname SlowOuterDepth\endcsname{\ensuremath{0.22686250}}
\expandafter\providecommand\csname SlowOuterLimitedHigh\endcsname{}\expandafter\renewcommand\csname SlowOuterLimitedHigh\endcsname{\ensuremath{2.671875}}
\expandafter\providecommand\csname SlowOuterLimitedLow\endcsname{}\expandafter\renewcommand\csname SlowOuterLimitedLow\endcsname{\ensuremath{2.667969}}
\expandafter\providecommand\csname SlowOuterRatio\endcsname{}\expandafter\renewcommand\csname SlowOuterRatio\endcsname{\ensuremath{0.073}}
\expandafter\providecommand\csname SlowOuterRawHigh\endcsname{}\expandafter\renewcommand\csname SlowOuterRawHigh\endcsname{\ensuremath{2.648438}}
\expandafter\providecommand\csname SlowOuterRawLow\endcsname{}\expandafter\renewcommand\csname SlowOuterRawLow\endcsname{\ensuremath{2.644531}}
\expandafter\providecommand\csname SlowOuterWork\endcsname{}\expandafter\renewcommand\csname SlowOuterWork\endcsname{\ensuremath{3.154\times10^{47}}}
\expandafter\providecommand\csname SlowRawGamma\endcsname{}\expandafter\renewcommand\csname SlowRawGamma\endcsname{\ensuremath{0.18595498}}
\expandafter\providecommand\csname SlowRawNative\endcsname{}\expandafter\renewcommand\csname SlowRawNative\endcsname{\ensuremath{0.18695239}}
\expandafter\providecommand\csname SlowShift\endcsname{}\expandafter\renewcommand\csname SlowShift\endcsname{\ensuremath{0.53}}
\expandafter\providecommand\csname SlowSpeed\endcsname{}\expandafter\renewcommand\csname SlowSpeed\endcsname{\ensuremath{0.3}}
\expandafter\providecommand\csname SphereError\endcsname{}\expandafter\renewcommand\csname SphereError\endcsname{\ensuremath{0}}
\expandafter\providecommand\csname StarBinding\endcsname{}\expandafter\renewcommand\csname StarBinding\endcsname{\ensuremath{-3.332\times10^{51}}}
\expandafter\providecommand\csname StarMass\endcsname{}\expandafter\renewcommand\csname StarMass\endcsname{\ensuremath{70}}
\expandafter\providecommand\csname StarRadius\endcsname{}\expandafter\renewcommand\csname StarRadius\endcsname{\ensuremath{90.54738}}
\expandafter\providecommand\csname UnboundCount\endcsname{}\expandafter\renewcommand\csname UnboundCount\endcsname{\ensuremath{85}}

\providecommand\result[1]{%
  \ifcsname #1\endcsname\csname #1\endcsname\else\textbf{??#1??}\fi}

\expandafter\providecommand\csname DeepFastDepth\endcsname{}\expandafter\renewcommand\csname DeepFastDepth\endcsname{\ensuremath{0.06433}}
\expandafter\providecommand\csname DeepFastFraction\endcsname{}\expandafter\renewcommand\csname DeepFastFraction\endcsname{\ensuremath{100.000}}
\expandafter\providecommand\csname DeepFastLoss\endcsname{}\expandafter\renewcommand\csname DeepFastLoss\endcsname{\ensuremath{0.24284}}
\expandafter\providecommand\csname DeepFastMedian\endcsname{}\expandafter\renewcommand\csname DeepFastMedian\endcsname{\ensuremath{5.882}}
\expandafter\providecommand\csname DeepFastSpeed\endcsname{}\expandafter\renewcommand\csname DeepFastSpeed\endcsname{\ensuremath{10}}
\expandafter\providecommand\csname DeepModerateDepth\endcsname{}\expandafter\renewcommand\csname DeepModerateDepth\endcsname{\ensuremath{0.06433}}
\expandafter\providecommand\csname DeepModerateFraction\endcsname{}\expandafter\renewcommand\csname DeepModerateFraction\endcsname{\ensuremath{0.486}}
\expandafter\providecommand\csname DeepModerateLoss\endcsname{}\expandafter\renewcommand\csname DeepModerateLoss\endcsname{\ensuremath{9.96532}}
\expandafter\providecommand\csname DeepModerateMedian\endcsname{}\expandafter\renewcommand\csname DeepModerateMedian\endcsname{\ensuremath{3.134}}
\expandafter\providecommand\csname DeepModerateSpeed\endcsname{}\expandafter\renewcommand\csname DeepModerateSpeed\endcsname{\ensuremath{3}}
\expandafter\providecommand\csname EnergyReplayError\endcsname{}\expandafter\renewcommand\csname EnergyReplayError\endcsname{\ensuremath{1.276}}
\expandafter\providecommand\csname FractionOrderError\endcsname{}\expandafter\renewcommand\csname FractionOrderError\endcsname{\ensuremath{9.254}}
\expandafter\providecommand\csname IntegrationCount\endcsname{}\expandafter\renewcommand\csname IntegrationCount\endcsname{\ensuremath{24}}
\expandafter\providecommand\csname LocalExposureMax\endcsname{}\expandafter\renewcommand\csname LocalExposureMax\endcsname{\ensuremath{9.291}}
\expandafter\providecommand\csname MiddleFastDepth\endcsname{}\expandafter\renewcommand\csname MiddleFastDepth\endcsname{\ensuremath{0.16555}}
\expandafter\providecommand\csname MiddleFastFraction\endcsname{}\expandafter\renewcommand\csname MiddleFastFraction\endcsname{\ensuremath{100.000}}
\expandafter\providecommand\csname MiddleFastLoss\endcsname{}\expandafter\renewcommand\csname MiddleFastLoss\endcsname{\ensuremath{0.01420}}
\expandafter\providecommand\csname MiddleFastMedian\endcsname{}\expandafter\renewcommand\csname MiddleFastMedian\endcsname{\ensuremath{10.263}}
\expandafter\providecommand\csname MiddleFastSpeed\endcsname{}\expandafter\renewcommand\csname MiddleFastSpeed\endcsname{\ensuremath{10}}
\expandafter\providecommand\csname MiddleModerateDepth\endcsname{}\expandafter\renewcommand\csname MiddleModerateDepth\endcsname{\ensuremath{0.16555}}
\expandafter\providecommand\csname MiddleModerateFraction\endcsname{}\expandafter\renewcommand\csname MiddleModerateFraction\endcsname{\ensuremath{8.573}}
\expandafter\providecommand\csname MiddleModerateLoss\endcsname{}\expandafter\renewcommand\csname MiddleModerateLoss\endcsname{\ensuremath{0.88835}}
\expandafter\providecommand\csname MiddleModerateMedian\endcsname{}\expandafter\renewcommand\csname MiddleModerateMedian\endcsname{\ensuremath{4.498}}
\expandafter\providecommand\csname MiddleModerateSpeed\endcsname{}\expandafter\renewcommand\csname MiddleModerateSpeed\endcsname{\ensuremath{3}}
\expandafter\providecommand\csname OuterFastDepth\endcsname{}\expandafter\renewcommand\csname OuterFastDepth\endcsname{\ensuremath{0.30000}}
\expandafter\providecommand\csname OuterFastFraction\endcsname{}\expandafter\renewcommand\csname OuterFastFraction\endcsname{\ensuremath{100.000}}
\expandafter\providecommand\csname OuterFastLoss\endcsname{}\expandafter\renewcommand\csname OuterFastLoss\endcsname{\ensuremath{0.00030}}
\expandafter\providecommand\csname OuterFastMedian\endcsname{}\expandafter\renewcommand\csname OuterFastMedian\endcsname{\ensuremath{17.967}}
\expandafter\providecommand\csname OuterFastSpeed\endcsname{}\expandafter\renewcommand\csname OuterFastSpeed\endcsname{\ensuremath{10}}
\expandafter\providecommand\csname OuterModerateDepth\endcsname{}\expandafter\renewcommand\csname OuterModerateDepth\endcsname{\ensuremath{0.30000}}
\expandafter\providecommand\csname OuterModerateFraction\endcsname{}\expandafter\renewcommand\csname OuterModerateFraction\endcsname{\ensuremath{99.964}}
\expandafter\providecommand\csname OuterModerateLoss\endcsname{}\expandafter\renewcommand\csname OuterModerateLoss\endcsname{\ensuremath{0.02370}}
\expandafter\providecommand\csname OuterModerateMedian\endcsname{}\expandafter\renewcommand\csname OuterModerateMedian\endcsname{\ensuremath{6.988}}
\expandafter\providecommand\csname OuterModerateSpeed\endcsname{}\expandafter\renewcommand\csname OuterModerateSpeed\endcsname{\ensuremath{3}}
\expandafter\providecommand\csname PassingVariants\endcsname{}\expandafter\renewcommand\csname PassingVariants\endcsname{\ensuremath{8}}
\expandafter\providecommand\csname SweptMax\endcsname{}\expandafter\renewcommand\csname SweptMax\endcsname{\ensuremath{0.013}}
\expandafter\providecommand\csname WorkError\endcsname{}\expandafter\renewcommand\csname WorkError\endcsname{\ensuremath{11.643}}

\providecommand\result[1]{%
  \ifcsname #1\endcsname\csname #1\endcsname\else\textbf{??#1??}\fi}

\expandafter\providecommand\csname DragCoulombLog\endcsname{}\expandafter\renewcommand\csname DragCoulombLog\endcsname{\ensuremath{1.4}}
\expandafter\providecommand\csname DragDeepDepth\endcsname{}\expandafter\renewcommand\csname DragDeepDepth\endcsname{\ensuremath{0.064}}
\expandafter\providecommand\csname DragDeepMachBhl\endcsname{}\expandafter\renewcommand\csname DragDeepMachBhl\endcsname{\ensuremath{2.7}}
\expandafter\providecommand\csname DragDeepMachOst\endcsname{}\expandafter\renewcommand\csname DragDeepMachOst\endcsname{\ensuremath{2.1}}
\expandafter\providecommand\csname DragDeepTorqueRatio\endcsname{}\expandafter\renewcommand\csname DragDeepTorqueRatio\endcsname{\ensuremath{23.9}}
\expandafter\providecommand\csname DragDeepWorkRatio\endcsname{}\expandafter\renewcommand\csname DragDeepWorkRatio\endcsname{\ensuremath{19.2}}
\expandafter\providecommand\csname DragMiddleDepth\endcsname{}\expandafter\renewcommand\csname DragMiddleDepth\endcsname{\ensuremath{0.166}}
\expandafter\providecommand\csname DragMiddleMachBhl\endcsname{}\expandafter\renewcommand\csname DragMiddleMachBhl\endcsname{\ensuremath{3.4}}
\expandafter\providecommand\csname DragMiddleMachOst\endcsname{}\expandafter\renewcommand\csname DragMiddleMachOst\endcsname{\ensuremath{3.4}}
\expandafter\providecommand\csname DragMiddleTorqueRatio\endcsname{}\expandafter\renewcommand\csname DragMiddleTorqueRatio\endcsname{\ensuremath{1.7}}
\expandafter\providecommand\csname DragMiddleWorkRatio\endcsname{}\expandafter\renewcommand\csname DragMiddleWorkRatio\endcsname{\ensuremath{1.7}}
\expandafter\providecommand\csname DragOuterDepth\endcsname{}\expandafter\renewcommand\csname DragOuterDepth\endcsname{\ensuremath{0.300}}
\expandafter\providecommand\csname DragOuterMachBhl\endcsname{}\expandafter\renewcommand\csname DragOuterMachBhl\endcsname{\ensuremath{4.7}}
\expandafter\providecommand\csname DragOuterMachOst\endcsname{}\expandafter\renewcommand\csname DragOuterMachOst\endcsname{\ensuremath{4.7}}
\expandafter\providecommand\csname DragOuterTorqueRatio\endcsname{}\expandafter\renewcommand\csname DragOuterTorqueRatio\endcsname{\ensuremath{1.5}}
\expandafter\providecommand\csname DragOuterWorkRatio\endcsname{}\expandafter\renewcommand\csname DragOuterWorkRatio\endcsname{\ensuremath{1.5}}
\expandafter\providecommand\csname DragRatioPeak\endcsname{}\expandafter\renewcommand\csname DragRatioPeak\endcsname{\ensuremath{5.7}}
\expandafter\providecommand\csname DragRatioPeakMach\endcsname{}\expandafter\renewcommand\csname DragRatioPeakMach\endcsname{\ensuremath{1.0}}
\expandafter\providecommand\csname DragSpeed\endcsname{}\expandafter\renewcommand\csname DragSpeed\endcsname{\ensuremath{0.001}}
\expandafter\providecommand\csname TideAnalyticCells\endcsname{}\expandafter\renewcommand\csname TideAnalyticCells\endcsname{\ensuremath{18}}
\expandafter\providecommand\csname TideAnalyticFullHigh\endcsname{}\expandafter\renewcommand\csname TideAnalyticFullHigh\endcsname{\ensuremath{1.41}}
\expandafter\providecommand\csname TideAnalyticFullLow\endcsname{}\expandafter\renewcommand\csname TideAnalyticFullLow\endcsname{\ensuremath{1.27}}
\expandafter\providecommand\csname TideAnalyticRadialHigh\endcsname{}\expandafter\renewcommand\csname TideAnalyticRadialHigh\endcsname{\ensuremath{1.02}}
\expandafter\providecommand\csname TideAnalyticRadialLow\endcsname{}\expandafter\renewcommand\csname TideAnalyticRadialLow\endcsname{\ensuremath{0.94}}
\expandafter\providecommand\csname TideCoefficient\endcsname{}\expandafter\renewcommand\csname TideCoefficient\endcsname{\ensuremath{4.4}}
\expandafter\providecommand\csname TideDeeperMax\endcsname{}\expandafter\renewcommand\csname TideDeeperMax\endcsname{\ensuremath{30}}
\expandafter\providecommand\csname TideDeltaHigh\endcsname{}\expandafter\renewcommand\csname TideDeltaHigh\endcsname{\ensuremath{0.03}}
\expandafter\providecommand\csname TideDeltaLow\endcsname{}\expandafter\renewcommand\csname TideDeltaLow\endcsname{\ensuremath{12.31}}
\expandafter\providecommand\csname TideDeltaMid\endcsname{}\expandafter\renewcommand\csname TideDeltaMid\endcsname{\ensuremath{0.34}}
\expandafter\providecommand\csname TideFirstApocentre\endcsname{}\expandafter\renewcommand\csname TideFirstApocentre\endcsname{\ensuremath{382}}
\expandafter\providecommand\csname TideHillHigh\endcsname{}\expandafter\renewcommand\csname TideHillHigh\endcsname{\ensuremath{1906}}
\expandafter\providecommand\csname TideHillLow\endcsname{}\expandafter\renewcommand\csname TideHillLow\endcsname{\ensuremath{243}}
\expandafter\providecommand\csname TideHillMid\endcsname{}\expandafter\renewcommand\csname TideHillMid\endcsname{\ensuremath{805}}
\expandafter\providecommand\csname TideMissDelta\endcsname{}\expandafter\renewcommand\csname TideMissDelta\endcsname{\ensuremath{1.0}}
\expandafter\providecommand\csname TideOrientations\endcsname{}\expandafter\renewcommand\csname TideOrientations\endcsname{\ensuremath{96}}
\expandafter\providecommand\csname TideSeqCoreCount\endcsname{}\expandafter\renewcommand\csname TideSeqCoreCount\endcsname{\ensuremath{3}}
\expandafter\providecommand\csname TideSeqCoreMax\endcsname{}\expandafter\renewcommand\csname TideSeqCoreMax\endcsname{\ensuremath{27}}
\expandafter\providecommand\csname TideSeqCoreMin\endcsname{}\expandafter\renewcommand\csname TideSeqCoreMin\endcsname{\ensuremath{8}}
\expandafter\providecommand\csname TideSeqHillInner\endcsname{}\expandafter\renewcommand\csname TideSeqHillInner\endcsname{\ensuremath{1000}}
\expandafter\providecommand\csname TideSeqHillMid\endcsname{}\expandafter\renewcommand\csname TideSeqHillMid\endcsname{\ensuremath{3000}}
\expandafter\providecommand\csname TideSeqHillOuter\endcsname{}\expandafter\renewcommand\csname TideSeqHillOuter\endcsname{\ensuremath{10000}}
\expandafter\providecommand\csname TideSeqKeplerCount\endcsname{}\expandafter\renewcommand\csname TideSeqKeplerCount\endcsname{\ensuremath{100}}
\expandafter\providecommand\csname TideSeqKeplerPeri\endcsname{}\expandafter\renewcommand\csname TideSeqKeplerPeri\endcsname{\ensuremath{0.299}}
\expandafter\providecommand\csname TideSeqMidShift\endcsname{}\expandafter\renewcommand\csname TideSeqMidShift\endcsname{\ensuremath{8}}
\expandafter\providecommand\csname TideSeqOrientations\endcsname{}\expandafter\renewcommand\csname TideSeqOrientations\endcsname{\ensuremath{4}}
\expandafter\providecommand\csname TideSeqOuterShift\endcsname{}\expandafter\renewcommand\csname TideSeqOuterShift\endcsname{\ensuremath{0.6}}

\providecommand\result[1]{%
  \ifcsname #1\endcsname\csname #1\endcsname\else\textbf{??#1??}\fi}

\expandafter\providecommand\csname GasAspect\endcsname{}\expandafter\renewcommand\csname GasAspect\endcsname{\ensuremath{0.03}}
\expandafter\providecommand\csname GasCaptureDt\endcsname{}\expandafter\renewcommand\csname GasCaptureDt\endcsname{\ensuremath{0.21}}
\expandafter\providecommand\csname GasCgHigh\endcsname{}\expandafter\renewcommand\csname GasCgHigh\endcsname{\ensuremath{1.7\times10^{-3}}}
\expandafter\providecommand\csname GasCgLow\endcsname{}\expandafter\renewcommand\csname GasCgLow\endcsname{\ensuremath{1.1\times10^{-4}}}
\expandafter\providecommand\csname GasHillOverH\endcsname{}\expandafter\renewcommand\csname GasHillOverH\endcsname{\ensuremath{0.26}}
\expandafter\providecommand\csname GasPairOverReduced\endcsname{}\expandafter\renewcommand\csname GasPairOverReduced\endcsname{\ensuremath{8.1}}
\expandafter\providecommand\csname GasSonicRadius\endcsname{}\expandafter\renewcommand\csname GasSonicRadius\endcsname{\ensuremath{0.39}}
\expandafter\providecommand\csname ResARhBeyondTenth\endcsname{}\expandafter\renewcommand\csname ResARhBeyondTenth\endcsname{\ensuremath{6.2}}
\expandafter\providecommand\csname ResARhFifty\endcsname{}\expandafter\renewcommand\csname ResARhFifty\endcsname{\ensuremath{0.009}}
\expandafter\providecommand\csname ResARhNinety\endcsname{}\expandafter\renewcommand\csname ResARhNinety\endcsname{\ensuremath{0.07}}
\expandafter\providecommand\csname ResCondRatio\endcsname{}\expandafter\renewcommand\csname ResCondRatio\endcsname{\ensuremath{1.016}}
\expandafter\providecommand\csname ResCondStarBinary\endcsname{}\expandafter\renewcommand\csname ResCondStarBinary\endcsname{\ensuremath{0.958}}
\expandafter\providecommand\csname ResCondStarSingle\endcsname{}\expandafter\renewcommand\csname ResCondStarSingle\endcsname{\ensuremath{1.069}}
\expandafter\providecommand\csname ResGridRecords\endcsname{}\expandafter\renewcommand\csname ResGridRecords\endcsname{\ensuremath{1000000}}
\expandafter\providecommand\csname ResOutsideSupport\endcsname{}\expandafter\renewcommand\csname ResOutsideSupport\endcsname{\ensuremath{66}}
\expandafter\providecommand\csname ShearEpsHigh\endcsname{}\expandafter\renewcommand\csname ShearEpsHigh\endcsname{\ensuremath{3.8\times10^{-3}}}
\expandafter\providecommand\csname ShearEpsLow\endcsname{}\expandafter\renewcommand\csname ShearEpsLow\endcsname{\ensuremath{5.0\times10^{-4}}}
\expandafter\providecommand\csname ShearHillColdRatio\endcsname{}\expandafter\renewcommand\csname ShearHillColdRatio\endcsname{\ensuremath{3.5}}
\expandafter\providecommand\csname ShearHillFidRatio\endcsname{}\expandafter\renewcommand\csname ShearHillFidRatio\endcsname{\ensuremath{0.04}}
\expandafter\providecommand\csname ShearHillSonicRatio\endcsname{}\expandafter\renewcommand\csname ShearHillSonicRatio\endcsname{\ensuremath{0.0092}}
\expandafter\providecommand\csname ShearKernelCells\endcsname{}\expandafter\renewcommand\csname ShearKernelCells\endcsname{\ensuremath{66}}
\expandafter\providecommand\csname ShearKernelHotRatio\endcsname{}\expandafter\renewcommand\csname ShearKernelHotRatio\endcsname{\ensuremath{0.0089}}
\expandafter\providecommand\csname ShearKernelMultMax\endcsname{}\expandafter\renewcommand\csname ShearKernelMultMax\endcsname{\ensuremath{1.12}}
\expandafter\providecommand\csname ShearKernelPlanarRatio\endcsname{}\expandafter\renewcommand\csname ShearKernelPlanarRatio\endcsname{\ensuremath{3.5}}
\expandafter\providecommand\csname ShearKernelThinRatio\endcsname{}\expandafter\renewcommand\csname ShearKernelThinRatio\endcsname{\ensuremath{0.8}}
\expandafter\providecommand\csname ShearKernelWarmBound\endcsname{}\expandafter\renewcommand\csname ShearKernelWarmBound\endcsname{\ensuremath{64}}
\expandafter\providecommand\csname ShearKernelWarmRatio\endcsname{}\expandafter\renewcommand\csname ShearKernelWarmRatio\endcsname{\ensuremath{0.058}}

\providecommand\result[1]{%
  \ifcsname #1\endcsname\csname #1\endcsname\else\textbf{??#1??}\fi}

\expandafter\providecommand\csname KinBhMass\endcsname{}\expandafter\renewcommand\csname KinBhMass\endcsname{\ensuremath{8.9}}
\expandafter\providecommand\csname KinDampBhHigh\endcsname{}\expandafter\renewcommand\csname KinDampBhHigh\endcsname{\ensuremath{2.4\times10^{5}}}
\expandafter\providecommand\csname KinDampBhLow\endcsname{}\expandafter\renewcommand\csname KinDampBhLow\endcsname{\ensuremath{7.4\times10^{4}}}
\expandafter\providecommand\csname KinDampScalingYr\endcsname{}\expandafter\renewcommand\csname KinDampScalingYr\endcsname{\ensuremath{1.5\times10^{5}}}
\expandafter\providecommand\csname KinDampStarHigh\endcsname{}\expandafter\renewcommand\csname KinDampStarHigh\endcsname{\ensuremath{1.7\times10^{4}}}
\expandafter\providecommand\csname KinDampStarLow\endcsname{}\expandafter\renewcommand\csname KinDampStarLow\endcsname{\ensuremath{5.4\times10^{3}}}
\expandafter\providecommand\csname KinDampingCoefficient\endcsname{}\expandafter\renewcommand\csname KinDampingCoefficient\endcsname{\ensuremath{0.78}}
\expandafter\providecommand\csname KinGammaFiducial\endcsname{}\expandafter\renewcommand\csname KinGammaFiducial\endcsname{\ensuremath{1\times10^{-3}}}
\expandafter\providecommand\csname KinGammaOneBhHigh\endcsname{}\expandafter\renewcommand\csname KinGammaOneBhHigh\endcsname{\ensuremath{1\times10^{-3}}}
\expandafter\providecommand\csname KinGammaOneBhLow\endcsname{}\expandafter\renewcommand\csname KinGammaOneBhLow\endcsname{\ensuremath{5\times10^{-4}}}
\expandafter\providecommand\csname KinGammaSonicBhHigh\endcsname{}\expandafter\renewcommand\csname KinGammaSonicBhHigh\endcsname{\ensuremath{6\times10^{-3}}}
\expandafter\providecommand\csname KinGammaSonicBhLow\endcsname{}\expandafter\renewcommand\csname KinGammaSonicBhLow\endcsname{\ensuremath{2\times10^{-3}}}
\expandafter\providecommand\csname KinGammaStrong\endcsname{}\expandafter\renewcommand\csname KinGammaStrong\endcsname{\ensuremath{1\times10^{-2}}}
\expandafter\providecommand\csname KinHOverHill\endcsname{}\expandafter\renewcommand\csname KinHOverHill\endcsname{\ensuremath{4.0}}
\expandafter\providecommand\csname KinHcritBhHigh\endcsname{}\expandafter\renewcommand\csname KinHcritBhHigh\endcsname{\ensuremath{0.10}}
\expandafter\providecommand\csname KinHcritBhLow\endcsname{}\expandafter\renewcommand\csname KinHcritBhLow\endcsname{\ensuremath{0.08}}
\expandafter\providecommand\csname KinHcritStarHigh\endcsname{}\expandafter\renewcommand\csname KinHcritStarHigh\endcsname{\ensuremath{0.20}}
\expandafter\providecommand\csname KinHcritStarLow\endcsname{}\expandafter\renewcommand\csname KinHcritStarLow\endcsname{\ensuremath{0.15}}
\expandafter\providecommand\csname KinInclinationRatio\endcsname{}\expandafter\renewcommand\csname KinInclinationRatio\endcsname{\ensuremath{0.5}}
\expandafter\providecommand\csname KinPairFidHigh\endcsname{}\expandafter\renewcommand\csname KinPairFidHigh\endcsname{\ensuremath{3.5}}
\expandafter\providecommand\csname KinPairFidLow\endcsname{}\expandafter\renewcommand\csname KinPairFidLow\endcsname{\ensuremath{2.6}}
\expandafter\providecommand\csname KinPairHotHigh\endcsname{}\expandafter\renewcommand\csname KinPairHotHigh\endcsname{\ensuremath{5.6}}
\expandafter\providecommand\csname KinPairHotLow\endcsname{}\expandafter\renewcommand\csname KinPairHotLow\endcsname{\ensuremath{4.5}}
\expandafter\providecommand\csname KinStarMass\endcsname{}\expandafter\renewcommand\csname KinStarMass\endcsname{\ensuremath{122}}
\expandafter\providecommand\csname KinStirBhHigh\endcsname{}\expandafter\renewcommand\csname KinStirBhHigh\endcsname{\ensuremath{3.1}}
\expandafter\providecommand\csname KinStirBhLow\endcsname{}\expandafter\renewcommand\csname KinStirBhLow\endcsname{\ensuremath{1.7}}
\expandafter\providecommand\csname KinStirStarHigh\endcsname{}\expandafter\renewcommand\csname KinStirStarHigh\endcsname{\ensuremath{1.6}}
\expandafter\providecommand\csname KinStirStarLow\endcsname{}\expandafter\renewcommand\csname KinStirStarLow\endcsname{\ensuremath{0.9}}
\expandafter\providecommand\csname KinTurbCoefficient\endcsname{}\expandafter\renewcommand\csname KinTurbCoefficient\endcsname{\ensuremath{18.2}}
\expandafter\providecommand\csname KinTurbReferenceSigma\endcsname{}\expandafter\renewcommand\csname KinTurbReferenceSigma\endcsname{\ensuremath{2400}}
\expandafter\providecommand\csname KinTurbSourceAmplitude\endcsname{}\expandafter\renewcommand\csname KinTurbSourceAmplitude\endcsname{\ensuremath{0.1}}

\providecommand\result[1]{%
  \ifcsname #1\endcsname\csname #1\endcsname\else\textbf{??#1??}\fi}

\expandafter\providecommand\csname AgnAllEChangeMax\endcsname{}\expandafter\renewcommand\csname AgnAllEChangeMax\endcsname{\ensuremath{2.83}}
\expandafter\providecommand\csname AgnAllEtaMax\endcsname{}\expandafter\renewcommand\csname AgnAllEtaMax\endcsname{\ensuremath{0.014}}
\expandafter\providecommand\csname AgnAllIChangeMax\endcsname{}\expandafter\renewcommand\csname AgnAllIChangeMax\endcsname{\ensuremath{4.49}}
\expandafter\providecommand\csname AgnAlphaDrag\endcsname{}\expandafter\renewcommand\csname AgnAlphaDrag\endcsname{\ensuremath{0.01}}
\expandafter\providecommand\csname AgnAuditSamples\endcsname{}\expandafter\renewcommand\csname AgnAuditSamples\endcsname{\ensuremath{65536}}
\expandafter\providecommand\csname AgnBhCrossBoundMax\endcsname{}\expandafter\renewcommand\csname AgnBhCrossBoundMax\endcsname{\ensuremath{1.81}}
\expandafter\providecommand\csname AgnBhDragFractionMax\endcsname{}\expandafter\renewcommand\csname AgnBhDragFractionMax\endcsname{\ensuremath{0.011}}
\expandafter\providecommand\csname AgnBhDragGravity\endcsname{}\expandafter\renewcommand\csname AgnBhDragGravity\endcsname{\ensuremath{0.014}}
\expandafter\providecommand\csname AgnBhEChangeMax\endcsname{}\expandafter\renewcommand\csname AgnBhEChangeMax\endcsname{\ensuremath{0.004}}
\expandafter\providecommand\csname AgnBhIChangeMax\endcsname{}\expandafter\renewcommand\csname AgnBhIChangeMax\endcsname{\ensuremath{0.008}}
\expandafter\providecommand\csname AgnContactChange\endcsname{}\expandafter\renewcommand\csname AgnContactChange\endcsname{\ensuremath{-0.1608}}
\expandafter\providecommand\csname AgnContactChangeSE\endcsname{}\expandafter\renewcommand\csname AgnContactChangeSE\endcsname{\ensuremath{0.0016}}
\expandafter\providecommand\csname AgnContactRatio\endcsname{}\expandafter\renewcommand\csname AgnContactRatio\endcsname{\ensuremath{0.998392}}
\expandafter\providecommand\csname AgnEpsteinGamma\endcsname{}\expandafter\renewcommand\csname AgnEpsteinGamma\endcsname{\ensuremath{4.78\times10^{-4}}}
\expandafter\providecommand\csname AgnForcePrefactor\endcsname{}\expandafter\renewcommand\csname AgnForcePrefactor\endcsname{\ensuremath{2.19\times10^{4}}}
\expandafter\providecommand\csname AgnQRef\endcsname{}\expandafter\renewcommand\csname AgnQRef\endcsname{\ensuremath{8.71}}
\expandafter\providecommand\csname AgnRadiusPcHigh\endcsname{}\expandafter\renewcommand\csname AgnRadiusPcHigh\endcsname{\ensuremath{0.0957}}
\expandafter\providecommand\csname AgnRadiusPcLow\endcsname{}\expandafter\renewcommand\csname AgnRadiusPcLow\endcsname{\ensuremath{0.0096}}
\expandafter\providecommand\csname AgnStarCrossBoundMax\endcsname{}\expandafter\renewcommand\csname AgnStarCrossBoundMax\endcsname{\ensuremath{11.27}}
\expandafter\providecommand\csname AgnStarDragFractionMax\endcsname{}\expandafter\renewcommand\csname AgnStarDragFractionMax\endcsname{\ensuremath{0.913}}
\expandafter\providecommand\csname AgnStarDragGravity\endcsname{}\expandafter\renewcommand\csname AgnStarDragGravity\endcsname{\ensuremath{2.626}}
\expandafter\providecommand\csname AgnStarEChangeMax\endcsname{}\expandafter\renewcommand\csname AgnStarEChangeMax\endcsname{\ensuremath{0.277}}
\expandafter\providecommand\csname AgnStarIChangeMax\endcsname{}\expandafter\renewcommand\csname AgnStarIChangeMax\endcsname{\ensuremath{0.550}}

\providecommand\result[1]{%
  \ifcsname #1\endcsname\csname #1\endcsname\else\textbf{??#1??}\fi}

\expandafter\providecommand\csname RgBinaryPairDeepest\endcsname{}\expandafter\renewcommand\csname RgBinaryPairDeepest\endcsname{\ensuremath{89}}
\expandafter\providecommand\csname RgClosestBhBh\endcsname{}\expandafter\renewcommand\csname RgClosestBhBh\endcsname{\ensuremath{83}}
\expandafter\providecommand\csname RgContactARhFifty\endcsname{}\expandafter\renewcommand\csname RgContactARhFifty\endcsname{\ensuremath{0.0036}}
\expandafter\providecommand\csname RgContactARhNinety\endcsname{}\expandafter\renewcommand\csname RgContactARhNinety\endcsname{\ensuremath{0.026}}
\expandafter\providecommand\csname RgContactBeyondTenth\endcsname{}\expandafter\renewcommand\csname RgContactBeyondTenth\endcsname{\ensuremath{2.2}}
\expandafter\providecommand\csname RgDetachedFraction\endcsname{}\expandafter\renewcommand\csname RgDetachedFraction\endcsname{\ensuremath{80}}
\expandafter\providecommand\csname RgEncounters\endcsname{}\expandafter\renewcommand\csname RgEncounters\endcsname{\ensuremath{200000}}
\expandafter\providecommand\csname RgExclPrimordial\endcsname{}\expandafter\renewcommand\csname RgExclPrimordial\endcsname{\ensuremath{0.29}}
\expandafter\providecommand\csname RgFixedMemberA\endcsname{}\expandafter\renewcommand\csname RgFixedMemberA\endcsname{\ensuremath{80}}
\expandafter\providecommand\csname RgIsoOverFixed\endcsname{}\expandafter\renewcommand\csname RgIsoOverFixed\endcsname{\ensuremath{0.87}}
\expandafter\providecommand\csname RgIsoOverFixedStarSingle\endcsname{}\expandafter\renewcommand\csname RgIsoOverFixedStarSingle\endcsname{\ensuremath{0.82}}
\expandafter\providecommand\csname RgLowvExchRerun\endcsname{}\expandafter\renewcommand\csname RgLowvExchRerun\endcsname{\ensuremath{33}}
\expandafter\providecommand\csname RgLowvExchVendored\endcsname{}\expandafter\renewcommand\csname RgLowvExchVendored\endcsname{\ensuremath{26}}
\expandafter\providecommand\csname RgLowvIdentical\endcsname{}\expandafter\renewcommand\csname RgLowvIdentical\endcsname{\ensuremath{73}}
\expandafter\providecommand\csname RgNewOverPooled\endcsname{}\expandafter\renewcommand\csname RgNewOverPooled\endcsname{\ensuremath{0.49}}
\expandafter\providecommand\csname RgNewOverPooledBhSingle\endcsname{}\expandafter\renewcommand\csname RgNewOverPooledBhSingle\endcsname{\ensuremath{0.72}}
\expandafter\providecommand\csname RgNewOverPooledStarSingle\endcsname{}\expandafter\renewcommand\csname RgNewOverPooledStarSingle\endcsname{\ensuremath{0.29}}
\expandafter\providecommand\csname RgPrimordialFraction\endcsname{}\expandafter\renewcommand\csname RgPrimordialFraction\endcsname{\ensuremath{63}}
\expandafter\providecommand\csname RgRecords\endcsname{}\expandafter\renewcommand\csname RgRecords\endcsname{\ensuremath{184034}}
\expandafter\providecommand\csname RgStepCap\endcsname{}\expandafter\renewcommand\csname RgStepCap\endcsname{\ensuremath{0.29}}
\expandafter\providecommand\csname RgTideARhMin\endcsname{}\expandafter\renewcommand\csname RgTideARhMin\endcsname{\ensuremath{0.01}}
\expandafter\providecommand\csname RgTideHigh\endcsname{}\expandafter\renewcommand\csname RgTideHigh\endcsname{\ensuremath{0.73}}
\expandafter\providecommand\csname RgTideLow\endcsname{}\expandafter\renewcommand\csname RgTideLow\endcsname{\ensuremath{0.61}}
\expandafter\providecommand\csname RgTidePairs\endcsname{}\expandafter\renewcommand\csname RgTidePairs\endcsname{\ensuremath{50000}}
\expandafter\providecommand\csname RgVendoredExchFixed\endcsname{}\expandafter\renewcommand\csname RgVendoredExchFixed\endcsname{\ensuremath{30}}
\expandafter\providecommand\csname RgVendoredExchIso\endcsname{}\expandafter\renewcommand\csname RgVendoredExchIso\endcsname{\ensuremath{26}}
\expandafter\providecommand\csname RgVendoredOrientation\endcsname{}\expandafter\renewcommand\csname RgVendoredOrientation\endcsname{\ensuremath{0.953}}

\providecommand\result[1]{%
  \ifcsname #1\endcsname\csname #1\endcsname\else\textbf{??#1??}\fi}

\expandafter\providecommand\csname FcCaps\endcsname{}\expandafter\renewcommand\csname FcCaps\endcsname{\ensuremath{20}}
\expandafter\providecommand\csname FcCensorHigh\endcsname{}\expandafter\renewcommand\csname FcCensorHigh\endcsname{\ensuremath{86.6}}
\expandafter\providecommand\csname FcCensorLow\endcsname{}\expandafter\renewcommand\csname FcCensorLow\endcsname{\ensuremath{58.3}}
\expandafter\providecommand\csname FcContactRate\endcsname{}\expandafter\renewcommand\csname FcContactRate\endcsname{\ensuremath{334.7}}
\expandafter\providecommand\csname FcContacts\endcsname{}\expandafter\renewcommand\csname FcContacts\endcsname{\ensuremath{950}}
\expandafter\providecommand\csname FcControlDraws\endcsname{}\expandafter\renewcommand\csname FcControlDraws\endcsname{\ensuremath{768}}
\expandafter\providecommand\csname FcDraws\endcsname{}\expandafter\renewcommand\csname FcDraws\endcsname{\ensuremath{5040}}
\expandafter\providecommand\csname FcEnergyError\endcsname{}\expandafter\renewcommand\csname FcEnergyError\endcsname{\ensuremath{7.0\times10^{-6}}}
\expandafter\providecommand\csname FcHorizons\endcsname{}\expandafter\renewcommand\csname FcHorizons\endcsname{\ensuremath{471}}
\expandafter\providecommand\csname FcNearHigh\endcsname{}\expandafter\renewcommand\csname FcNearHigh\endcsname{\ensuremath{84.7}}
\expandafter\providecommand\csname FcNearLow\endcsname{}\expandafter\renewcommand\csname FcNearLow\endcsname{\ensuremath{77.9}}
\expandafter\providecommand\csname FcNearPercent\endcsname{}\expandafter\renewcommand\csname FcNearPercent\endcsname{\ensuremath{81.3}}
\expandafter\providecommand\csname FcPairedReferencePercent\endcsname{}\expandafter\renewcommand\csname FcPairedReferencePercent\endcsname{\ensuremath{77.8}}
\expandafter\providecommand\csname FcQualityFailures\endcsname{}\expandafter\renewcommand\csname FcQualityFailures\endcsname{\ensuremath{1}}
\expandafter\providecommand\csname FcRmsControlPercent\endcsname{}\expandafter\renewcommand\csname FcRmsControlPercent\endcsname{\ensuremath{82.1}}
\expandafter\providecommand\csname FcSeedOnePercent\endcsname{}\expandafter\renewcommand\csname FcSeedOnePercent\endcsname{\ensuremath{80.9}}
\expandafter\providecommand\csname FcSeedTwoPercent\endcsname{}\expandafter\renewcommand\csname FcSeedTwoPercent\endcsname{\ensuremath{81.7}}
\expandafter\providecommand\csname FcTightDifferenceHigh\endcsname{}\expandafter\renewcommand\csname FcTightDifferenceHigh\endcsname{\ensuremath{12.9}}
\expandafter\providecommand\csname FcTightDifferenceLow\endcsname{}\expandafter\renewcommand\csname FcTightDifferenceLow\endcsname{\ensuremath{-2.4}}
\expandafter\providecommand\csname FcTightPercent\endcsname{}\expandafter\renewcommand\csname FcTightPercent\endcsname{\ensuremath{83.0}}

\def\RIT{Center for Computational Relativity and Gravitation, Rochester Institute of Technology, Rochester, New York   14623, USA}
\def\NTID{National Technical Institute for the Deaf, Rochester Institute of Technology, Rochester, NY 14623, USA}

\providecommand\result[1]{%
  \ifcsname #1\endcsname\csname #1\endcsname\else\textbf{??#1??}\fi}

\expandafter\providecommand\csname MassResponseBsCdfPercent\endcsname{}\expandafter\renewcommand\csname MassResponseBsCdfPercent\endcsname{\ensuremath{0.3}}
\expandafter\providecommand\csname MassResponseBsRatePercent\endcsname{}\expandafter\renewcommand\csname MassResponseBsRatePercent\endcsname{\ensuremath{1.41}}
\expandafter\providecommand\csname MassResponseColdMaxPercent\endcsname{}\expandafter\renewcommand\csname MassResponseColdMaxPercent\endcsname{\ensuremath{0.6}}
\expandafter\providecommand\csname MassResponseSsCdfPercent\endcsname{}\expandafter\renewcommand\csname MassResponseSsCdfPercent\endcsname{\ensuremath{0.6}}
\expandafter\providecommand\csname MassResponseSsRatePercent\endcsname{}\expandafter\renewcommand\csname MassResponseSsRatePercent\endcsname{\ensuremath{0.46}}
\expandafter\providecommand\csname MassResponseStarExponent\endcsname{}\expandafter\renewcommand\csname MassResponseStarExponent\endcsname{\ensuremath{0.753}}

\providecommand\result[1]{%
  \ifcsname #1\endcsname\csname #1\endcsname\else\textbf{??#1??}\fi}

\expandafter\providecommand\csname ColdDomainEpsilon\endcsname{}\expandafter\renewcommand\csname ColdDomainEpsilon\endcsname{\ensuremath{0.00088}}
\expandafter\providecommand\csname ColdDomainFidFe\endcsname{}\expandafter\renewcommand\csname ColdDomainFidFe\endcsname{\ensuremath{2.46}}
\expandafter\providecommand\csname ColdDomainFidFi\endcsname{}\expandafter\renewcommand\csname ColdDomainFidFi\endcsname{\ensuremath{1.23}}
\expandafter\providecommand\csname ColdDomainFidPercent\endcsname{}\expandafter\renewcommand\csname ColdDomainFidPercent\endcsname{\ensuremath{1.0}}
\expandafter\providecommand\csname ColdDomainFidRatio\endcsname{}\expandafter\renewcommand\csname ColdDomainFidRatio\endcsname{\ensuremath{0.0098}}
\expandafter\providecommand\csname ColdDomainMassMaxPercent\endcsname{}\expandafter\renewcommand\csname ColdDomainMassMaxPercent\endcsname{\ensuremath{0.6}}
\expandafter\providecommand\csname ColdDomainThinPercent\endcsname{}\expandafter\renewcommand\csname ColdDomainThinPercent\endcsname{\ensuremath{21}}
\expandafter\providecommand\csname ColdDomainThinRatio\endcsname{}\expandafter\renewcommand\csname ColdDomainThinRatio\endcsname{\ensuremath{0.21}}

\providecommand\result[1]{%
  \ifcsname #1\endcsname\csname #1\endcsname\else\textbf{??#1??}\fi}

\expandafter\providecommand\csname SingleBhDensity\endcsname{}\expandafter\renewcommand\csname SingleBhDensity\endcsname{\ensuremath{0.39}}
\expandafter\providecommand\csname SingleBhDensityRatio\endcsname{}\expandafter\renewcommand\csname SingleBhDensityRatio\endcsname{\ensuremath{124}}
\expandafter\providecommand\csname SingleBhEtaHigh\endcsname{}\expandafter\renewcommand\csname SingleBhEtaHigh\endcsname{\ensuremath{4.0\times10^{-4}}}
\expandafter\providecommand\csname SingleBhEtaLow\endcsname{}\expandafter\renewcommand\csname SingleBhEtaLow\endcsname{\ensuremath{4.0\times10^{-5}}}
\expandafter\providecommand\csname SingleBhFull\endcsname{}\expandafter\renewcommand\csname SingleBhFull\endcsname{\ensuremath{0.60}}
\expandafter\providecommand\csname SingleBhFullHigh\endcsname{}\expandafter\renewcommand\csname SingleBhFullHigh\endcsname{\ensuremath{0.79}}
\expandafter\providecommand\csname SingleBhFullLow\endcsname{}\expandafter\renewcommand\csname SingleBhFullLow\endcsname{\ensuremath{0.60}}
\expandafter\providecommand\csname SingleBhRatio\endcsname{}\expandafter\renewcommand\csname SingleBhRatio\endcsname{\ensuremath{190}}
\expandafter\providecommand\csname SingleBhTrani\endcsname{}\expandafter\renewcommand\csname SingleBhTrani\endcsname{\ensuremath{3.17\times10^{-3}}}
\expandafter\providecommand\csname SingleEtaTe\endcsname{}\expandafter\renewcommand\csname SingleEtaTe\endcsname{\ensuremath{2.14}}
\expandafter\providecommand\csname SinglePairFullCs\endcsname{}\expandafter\renewcommand\csname SinglePairFullCs\endcsname{\ensuremath{0.58}}
\expandafter\providecommand\csname SinglePairFullHill\endcsname{}\expandafter\renewcommand\csname SinglePairFullHill\endcsname{\ensuremath{2.28}}
\expandafter\providecommand\csname SinglePairFullVelocity\endcsname{}\expandafter\renewcommand\csname SinglePairFullVelocity\endcsname{\ensuremath{51.9}}
\expandafter\providecommand\csname SinglePairTraniCs\endcsname{}\expandafter\renewcommand\csname SinglePairTraniCs\endcsname{\ensuremath{0.011}}
\expandafter\providecommand\csname SinglePairTraniHill\endcsname{}\expandafter\renewcommand\csname SinglePairTraniHill\endcsname{\ensuremath{0.044}}
\expandafter\providecommand\csname SinglePairTraniVelocity\endcsname{}\expandafter\renewcommand\csname SinglePairTraniVelocity\endcsname{\ensuremath{1.00}}
\expandafter\providecommand\csname SinglePairVelocityRatio\endcsname{}\expandafter\renewcommand\csname SinglePairVelocityRatio\endcsname{\ensuremath{52}}
\expandafter\providecommand\csname SingleRateCold\endcsname{}\expandafter\renewcommand\csname SingleRateCold\endcsname{\ensuremath{3705.6}}
\expandafter\providecommand\csname SingleRateColdError\endcsname{}\expandafter\renewcommand\csname SingleRateColdError\endcsname{\ensuremath{2.2}}
\expandafter\providecommand\csname SingleRateColdRatio\endcsname{}\expandafter\renewcommand\csname SingleRateColdRatio\endcsname{\ensuremath{29.08}}
\expandafter\providecommand\csname SingleRateDensity\endcsname{}\expandafter\renewcommand\csname SingleRateDensity\endcsname{\ensuremath{556.4}}
\expandafter\providecommand\csname SingleRateDensityError\endcsname{}\expandafter\renewcommand\csname SingleRateDensityError\endcsname{\ensuremath{1.3}}
\expandafter\providecommand\csname SingleRateDensityRatio\endcsname{}\expandafter\renewcommand\csname SingleRateDensityRatio\endcsname{\ensuremath{4.37}}
\expandafter\providecommand\csname SingleRateInclination\endcsname{}\expandafter\renewcommand\csname SingleRateInclination\endcsname{\ensuremath{96.3}}
\expandafter\providecommand\csname SingleRateInclinationError\endcsname{}\expandafter\renewcommand\csname SingleRateInclinationError\endcsname{\ensuremath{0.3}}
\expandafter\providecommand\csname SingleRateInclinationRatio\endcsname{}\expandafter\renewcommand\csname SingleRateInclinationRatio\endcsname{\ensuremath{0.76}}
\expandafter\providecommand\csname SingleRateOurs\endcsname{}\expandafter\renewcommand\csname SingleRateOurs\endcsname{\ensuremath{127.4}}
\expandafter\providecommand\csname SingleRateOursError\endcsname{}\expandafter\renewcommand\csname SingleRateOursError\endcsname{\ensuremath{0.4}}
\expandafter\providecommand\csname SingleRateOursRatio\endcsname{}\expandafter\renewcommand\csname SingleRateOursRatio\endcsname{\ensuremath{1.00}}
\expandafter\providecommand\csname SingleRateOursRms\endcsname{}\expandafter\renewcommand\csname SingleRateOursRms\endcsname{\ensuremath{50.8}}
\expandafter\providecommand\csname SingleRateOursRmsError\endcsname{}\expandafter\renewcommand\csname SingleRateOursRmsError\endcsname{\ensuremath{0.1}}
\expandafter\providecommand\csname SingleRateOursRmsRatio\endcsname{}\expandafter\renewcommand\csname SingleRateOursRmsRatio\endcsname{\ensuremath{0.40}}
\expandafter\providecommand\csname SingleRateRayleigh\endcsname{}\expandafter\renewcommand\csname SingleRateRayleigh\endcsname{\ensuremath{2975.5}}
\expandafter\providecommand\csname SingleRateRayleighError\endcsname{}\expandafter\renewcommand\csname SingleRateRayleighError\endcsname{\ensuremath{2.1}}
\expandafter\providecommand\csname SingleRateRayleighRatio\endcsname{}\expandafter\renewcommand\csname SingleRateRayleighRatio\endcsname{\ensuremath{23.35}}
\expandafter\providecommand\csname SingleRateTrani\endcsname{}\expandafter\renewcommand\csname SingleRateTrani\endcsname{\ensuremath{2867.5}}
\expandafter\providecommand\csname SingleRateTraniError\endcsname{}\expandafter\renewcommand\csname SingleRateTraniError\endcsname{\ensuremath{1.9}}
\expandafter\providecommand\csname SingleRateTraniRatio\endcsname{}\expandafter\renewcommand\csname SingleRateTraniRatio\endcsname{\ensuremath{22.50}}
\expandafter\providecommand\csname SingleStarDensity\endcsname{}\expandafter\renewcommand\csname SingleStarDensity\endcsname{\ensuremath{0.11}}
\expandafter\providecommand\csname SingleStarDensityRatio\endcsname{}\expandafter\renewcommand\csname SingleStarDensityRatio\endcsname{\ensuremath{9}}
\expandafter\providecommand\csname SingleStarFull\endcsname{}\expandafter\renewcommand\csname SingleStarFull\endcsname{\ensuremath{0.28}}
\expandafter\providecommand\csname SingleStarRatio\endcsname{}\expandafter\renewcommand\csname SingleStarRatio\endcsname{\ensuremath{24}}
\expandafter\providecommand\csname SingleStarTrani\endcsname{}\expandafter\renewcommand\csname SingleStarTrani\endcsname{\ensuremath{1.18\times10^{-2}}}
\expandafter\providecommand\csname SingleTotalSamples\endcsname{}\expandafter\renewcommand\csname SingleTotalSamples\endcsname{\ensuremath{400000}}

\providecommand\result[1]{%
  \ifcsname #1\endcsname\csname #1\endcsname\else\textbf{??#1??}\fi}

\expandafter\providecommand\csname HillValColdCoefficient\endcsname{}\expandafter\renewcommand\csname HillValColdCoefficient\endcsname{\ensuremath{0.357}}
\expandafter\providecommand\csname HillValColdEntry\endcsname{}\expandafter\renewcommand\csname HillValColdEntry\endcsname{\ensuremath{4.77}}
\expandafter\providecommand\csname HillValColdPhysicalRatio\endcsname{}\expandafter\renewcommand\csname HillValColdPhysicalRatio\endcsname{\ensuremath{4.61}}
\expandafter\providecommand\csname HillValColdProbability\endcsname{}\expandafter\renewcommand\csname HillValColdProbability\endcsname{\ensuremath{0.075}}
\expandafter\providecommand\csname HillValConvergencePull\endcsname{}\expandafter\renewcommand\csname HillValConvergencePull\endcsname{\ensuremath{2.75}}
\expandafter\providecommand\csname HillValFreeBoundRatio\endcsname{}\expandafter\renewcommand\csname HillValFreeBoundRatio\endcsname{\ensuremath{0.79}}
\expandafter\providecommand\csname HillValFreeDraws\endcsname{}\expandafter\renewcommand\csname HillValFreeDraws\endcsname{\ensuremath{8192}}
\expandafter\providecommand\csname HillValFreeOrbitError\endcsname{}\expandafter\renewcommand\csname HillValFreeOrbitError\endcsname{\ensuremath{0}}
\expandafter\providecommand\csname HillValInnerError\endcsname{}\expandafter\renewcommand\csname HillValInnerError\endcsname{\ensuremath{3.0\times10^{-7}}}
\expandafter\providecommand\csname HillValPlanarSlope\endcsname{}\expandafter\renewcommand\csname HillValPlanarSlope\endcsname{\ensuremath{0.501}}
\expandafter\providecommand\csname HillValPlanarWarmSlope\endcsname{}\expandafter\renewcommand\csname HillValPlanarWarmSlope\endcsname{\ensuremath{0.485}}
\expandafter\providecommand\csname HillValSpatialSlope\endcsname{}\expandafter\renewcommand\csname HillValSpatialSlope\endcsname{\ensuremath{1.002}}

\newcommand{\rtA}{191}
\newcommand{\rtEcc}{0.9984}
\newcommand{\rtApo}{382}
\newcommand{\rtApoAU}{161}
\newcommand{\rtYears}{82}
\newcommand{\rtRadius}{90.5}

\providecommand\result[1]{%
  \ifcsname #1\endcsname\csname #1\endcsname\else\textbf{??#1??}\fi}

\expandafter\providecommand\csname SepLogBbRate\endcsname{}\expandafter\renewcommand\csname SepLogBbRate\endcsname{\ensuremath{13.1}}
\expandafter\providecommand\csname SepLogBeta\endcsname{}\expandafter\renewcommand\csname SepLogBeta\endcsname{\ensuremath{-1.0}}
\expandafter\providecommand\csname SepLogBinaryPercent\endcsname{}\expandafter\renewcommand\csname SepLogBinaryPercent\endcsname{\ensuremath{67.8}}
\expandafter\providecommand\csname SepLogBsRate\endcsname{}\expandafter\renewcommand\csname SepLogBsRate\endcsname{\ensuremath{255.2}}
\expandafter\providecommand\csname SepLogContactA\endcsname{}\expandafter\renewcommand\csname SepLogContactA\endcsname{\ensuremath{0.182}}
\expandafter\providecommand\csname SepLogDetachedPercent\endcsname{}\expandafter\renewcommand\csname SepLogDetachedPercent\endcsname{\ensuremath{0.803}}
\expandafter\providecommand\csname SepLogRatio\endcsname{}\expandafter\renewcommand\csname SepLogRatio\endcsname{\ensuremath{1.00}}
\expandafter\providecommand\csname SepLogTotalRate\endcsname{}\expandafter\renewcommand\csname SepLogTotalRate\endcsname{\ensuremath{395.4}}
\expandafter\providecommand\csname SepTightBbRate\endcsname{}\expandafter\renewcommand\csname SepTightBbRate\endcsname{\ensuremath{30.9}}
\expandafter\providecommand\csname SepTightBeta\endcsname{}\expandafter\renewcommand\csname SepTightBeta\endcsname{\ensuremath{-2.0}}
\expandafter\providecommand\csname SepTightBinaryPercent\endcsname{}\expandafter\renewcommand\csname SepTightBinaryPercent\endcsname{\ensuremath{82.6}}
\expandafter\providecommand\csname SepTightBsRate\endcsname{}\expandafter\renewcommand\csname SepTightBsRate\endcsname{\ensuremath{572.2}}
\expandafter\providecommand\csname SepTightContactA\endcsname{}\expandafter\renewcommand\csname SepTightContactA\endcsname{\ensuremath{0.098}}
\expandafter\providecommand\csname SepTightDetachedPercent\endcsname{}\expandafter\renewcommand\csname SepTightDetachedPercent\endcsname{\ensuremath{0.398}}
\expandafter\providecommand\csname SepTightRatio\endcsname{}\expandafter\renewcommand\csname SepTightRatio\endcsname{\ensuremath{1.85}}
\expandafter\providecommand\csname SepTightTotalRate\endcsname{}\expandafter\renewcommand\csname SepTightTotalRate\endcsname{\ensuremath{730.2}}
\expandafter\providecommand\csname SepWideBbRate\endcsname{}\expandafter\renewcommand\csname SepWideBbRate\endcsname{\ensuremath{1.7}}
\expandafter\providecommand\csname SepWideBeta\endcsname{}\expandafter\renewcommand\csname SepWideBeta\endcsname{\ensuremath{0.5}}
\expandafter\providecommand\csname SepWideBinaryPercent\endcsname{}\expandafter\renewcommand\csname SepWideBinaryPercent\endcsname{\ensuremath{20.7}}
\expandafter\providecommand\csname SepWideBsRate\endcsname{}\expandafter\renewcommand\csname SepWideBsRate\endcsname{\ensuremath{31.6}}
\expandafter\providecommand\csname SepWideContactA\endcsname{}\expandafter\renewcommand\csname SepWideContactA\endcsname{\ensuremath{3.008}}
\expandafter\providecommand\csname SepWideDetachedPercent\endcsname{}\expandafter\renewcommand\csname SepWideDetachedPercent\endcsname{\ensuremath{0.990}}
\expandafter\providecommand\csname SepWideRatio\endcsname{}\expandafter\renewcommand\csname SepWideRatio\endcsname{\ensuremath{0.41}}
\expandafter\providecommand\csname SepWideTotalRate\endcsname{}\expandafter\renewcommand\csname SepWideTotalRate\endcsname{\ensuremath{160.4}}

\providecommand\result[1]{%
  \ifcsname #1\endcsname\csname #1\endcsname\else\textbf{??#1??}\fi}

\expandafter\providecommand\csname FracA\endcsname{}\expandafter\renewcommand\csname FracA\endcsname{\ensuremath{8945.7}}
\expandafter\providecommand\csname FracBbReference\endcsname{}\expandafter\renewcommand\csname FracBbReference\endcsname{\ensuremath{13.18}}
\expandafter\providecommand\csname FracBsReference\endcsname{}\expandafter\renewcommand\csname FracBsReference\endcsname{\ensuremath{255.2}}
\expandafter\providecommand\csname FracCommonThreshold\endcsname{}\expandafter\renewcommand\csname FracCommonThreshold\endcsname{\ensuremath{0.486}}
\expandafter\providecommand\csname FracCss\endcsname{}\expandafter\renewcommand\csname FracCss\endcsname{\ensuremath{127.11}}
\expandafter\providecommand\csname FracD\endcsname{}\expandafter\renewcommand\csname FracD\endcsname{\ensuremath{18811.6}}
\expandafter\providecommand\csname FracInterpError\endcsname{}\expandafter\renewcommand\csname FracInterpError\endcsname{\ensuremath{4.4}}
\expandafter\providecommand\csname FracNodeAB\endcsname{}\expandafter\renewcommand\csname FracNodeAB\endcsname{\ensuremath{17148.2}}
\expandafter\providecommand\csname FracNodeABeta\endcsname{}\expandafter\renewcommand\csname FracNodeABeta\endcsname{\ensuremath{-2.0}}
\expandafter\providecommand\csname FracNodeAContactA\endcsname{}\expandafter\renewcommand\csname FracNodeAContactA\endcsname{\ensuremath{4044.3}}
\expandafter\providecommand\csname FracNodeAContactB\endcsname{}\expandafter\renewcommand\csname FracNodeAContactB\endcsname{\ensuremath{3355.8}}
\expandafter\providecommand\csname FracNodeBB\endcsname{}\expandafter\renewcommand\csname FracNodeBB\endcsname{\ensuremath{17928.1}}
\expandafter\providecommand\csname FracNodeBBeta\endcsname{}\expandafter\renewcommand\csname FracNodeBBeta\endcsname{\ensuremath{-1.5}}
\expandafter\providecommand\csname FracNodeBContactA\endcsname{}\expandafter\renewcommand\csname FracNodeBContactA\endcsname{\ensuremath{2993.7}}
\expandafter\providecommand\csname FracNodeBContactB\endcsname{}\expandafter\renewcommand\csname FracNodeBContactB\endcsname{\ensuremath{2722.4}}
\expandafter\providecommand\csname FracNodeCB\endcsname{}\expandafter\renewcommand\csname FracNodeCB\endcsname{\ensuremath{18539.0}}
\expandafter\providecommand\csname FracNodeCBeta\endcsname{}\expandafter\renewcommand\csname FracNodeCBeta\endcsname{\ensuremath{-1.0}}
\expandafter\providecommand\csname FracNodeCContactA\endcsname{}\expandafter\renewcommand\csname FracNodeCContactA\endcsname{\ensuremath{1651.4}}
\expandafter\providecommand\csname FracNodeCContactB\endcsname{}\expandafter\renewcommand\csname FracNodeCContactB\endcsname{\ensuremath{1800.7}}
\expandafter\providecommand\csname FracNodeDB\endcsname{}\expandafter\renewcommand\csname FracNodeDB\endcsname{\ensuremath{18514.1}}
\expandafter\providecommand\csname FracNodeDBeta\endcsname{}\expandafter\renewcommand\csname FracNodeDBeta\endcsname{\ensuremath{-0.5}}
\expandafter\providecommand\csname FracNodeDContactA\endcsname{}\expandafter\renewcommand\csname FracNodeDContactA\endcsname{\ensuremath{673.7}}
\expandafter\providecommand\csname FracNodeDContactB\endcsname{}\expandafter\renewcommand\csname FracNodeDContactB\endcsname{\ensuremath{928.5}}
\expandafter\providecommand\csname FracNodeEB\endcsname{}\expandafter\renewcommand\csname FracNodeEB\endcsname{\ensuremath{18233.0}}
\expandafter\providecommand\csname FracNodeEBeta\endcsname{}\expandafter\renewcommand\csname FracNodeEBeta\endcsname{\ensuremath{0.0}}
\expandafter\providecommand\csname FracNodeEContactA\endcsname{}\expandafter\renewcommand\csname FracNodeEContactA\endcsname{\ensuremath{281.5}}
\expandafter\providecommand\csname FracNodeEContactB\endcsname{}\expandafter\renewcommand\csname FracNodeEContactB\endcsname{\ensuremath{496.6}}
\expandafter\providecommand\csname FracNodeFB\endcsname{}\expandafter\renewcommand\csname FracNodeFB\endcsname{\ensuremath{18055.4}}
\expandafter\providecommand\csname FracNodeFBeta\endcsname{}\expandafter\renewcommand\csname FracNodeFBeta\endcsname{\ensuremath{0.5}}
\expandafter\providecommand\csname FracNodeFContactA\endcsname{}\expandafter\renewcommand\csname FracNodeFContactA\endcsname{\ensuremath{150.4}}
\expandafter\providecommand\csname FracNodeFContactB\endcsname{}\expandafter\renewcommand\csname FracNodeFContactB\endcsname{\ensuremath{331.6}}
\expandafter\providecommand\csname FracRss\endcsname{}\expandafter\renewcommand\csname FracRss\endcsname{\ensuremath{37870.9}}
\expandafter\providecommand\csname FracSamples\endcsname{}\expandafter\renewcommand\csname FracSamples\endcsname{\ensuremath{400000}}

\providecommand\result[1]{%
  \ifcsname #1\endcsname\csname #1\endcsname\else\textbf{??#1??}\fi}

\expandafter\providecommand\csname ApproxBbRatio\endcsname{}\expandafter\renewcommand\csname ApproxBbRatio\endcsname{\ensuremath{0.0516}}
\expandafter\providecommand\csname ApproxBsBh\endcsname{}\expandafter\renewcommand\csname ApproxBsBh\endcsname{\ensuremath{0.04}}
\expandafter\providecommand\csname ApproxBsCdfError\endcsname{}\expandafter\renewcommand\csname ApproxBsCdfError\endcsname{\ensuremath{1.2}}
\expandafter\providecommand\csname ApproxBsCellError\endcsname{}\expandafter\renewcommand\csname ApproxBsCellError\endcsname{\ensuremath{7}}
\expandafter\providecommand\csname ApproxBsNorm\endcsname{}\expandafter\renewcommand\csname ApproxBsNorm\endcsname{\ensuremath{10.20}}
\expandafter\providecommand\csname ApproxBsRateError\endcsname{}\expandafter\renewcommand\csname ApproxBsRateError\endcsname{\ensuremath{0.1}}
\expandafter\providecommand\csname ApproxBsStar\endcsname{}\expandafter\renewcommand\csname ApproxBsStar\endcsname{\ensuremath{0.76}}
\expandafter\providecommand\csname ApproxSamplingScale\endcsname{}\expandafter\renewcommand\csname ApproxSamplingScale\endcsname{\ensuremath{10^{-2}}}
\expandafter\providecommand\csname ApproxSsBh\endcsname{}\expandafter\renewcommand\csname ApproxSsBh\endcsname{\ensuremath{0.54}}
\expandafter\providecommand\csname ApproxSsCdfError\endcsname{}\expandafter\renewcommand\csname ApproxSsCdfError\endcsname{\ensuremath{1.7}}
\expandafter\providecommand\csname ApproxSsCellError\endcsname{}\expandafter\renewcommand\csname ApproxSsCellError\endcsname{\ensuremath{26}}
\expandafter\providecommand\csname ApproxSsNorm\endcsname{}\expandafter\renewcommand\csname ApproxSsNorm\endcsname{\ensuremath{4.61}}
\expandafter\providecommand\csname ApproxSsRateError\endcsname{}\expandafter\renewcommand\csname ApproxSsRateError\endcsname{\ensuremath{2.0}}
\expandafter\providecommand\csname ApproxSsStar\endcsname{}\expandafter\renewcommand\csname ApproxSsStar\endcsname{\ensuremath{1.28}}
\expandafter\providecommand\csname ApproxSystematicScale\endcsname{}\expandafter\renewcommand\csname ApproxSystematicScale\endcsname{\ensuremath{10^{-1}}}

\providecommand\result[1]{%
  \ifcsname #1\endcsname\csname #1\endcsname\else\textbf{??#1??}\fi}

\expandafter\providecommand\csname AvgWBhSlope\endcsname{}\expandafter\renewcommand\csname AvgWBhSlope\endcsname{\ensuremath{0.576}}
\expandafter\providecommand\csname AvgWColdPercent\endcsname{}\expandafter\renewcommand\csname AvgWColdPercent\endcsname{\ensuremath{2.53}}
\expandafter\providecommand\csname AvgWFiniteBh\endcsname{}\expandafter\renewcommand\csname AvgWFiniteBh\endcsname{\ensuremath{0.57590}}
\expandafter\providecommand\csname AvgWFiniteRadius\endcsname{}\expandafter\renewcommand\csname AvgWFiniteRadius\endcsname{\ensuremath{-0.25279}}
\expandafter\providecommand\csname AvgWFiniteStar\endcsname{}\expandafter\renewcommand\csname AvgWFiniteStar\endcsname{\ensuremath{1.49467}}
\expandafter\providecommand\csname AvgWFitBh\endcsname{}\expandafter\renewcommand\csname AvgWFitBh\endcsname{\ensuremath{0.577}}
\expandafter\providecommand\csname AvgWFitMaxPercent\endcsname{}\expandafter\renewcommand\csname AvgWFitMaxPercent\endcsname{\ensuremath{12.3}}
\expandafter\providecommand\csname AvgWFitRmsPercent\endcsname{}\expandafter\renewcommand\csname AvgWFitRmsPercent\endcsname{\ensuremath{4.63}}
\expandafter\providecommand\csname AvgWFitStar\endcsname{}\expandafter\renewcommand\csname AvgWFitStar\endcsname{\ensuremath{1.48}}
\expandafter\providecommand\csname AvgWFixedQSlope\endcsname{}\expandafter\renewcommand\csname AvgWFixedQSlope\endcsname{\ensuremath{2.07}}
\expandafter\providecommand\csname AvgWGeometryError\endcsname{}\expandafter\renewcommand\csname AvgWGeometryError\endcsname{\ensuremath{1.11\times10^{-16}}}
\expandafter\providecommand\csname AvgWNu\endcsname{}\expandafter\renewcommand\csname AvgWNu\endcsname{\ensuremath{0.973}}
\expandafter\providecommand\csname AvgWPopulationMissingPercent\endcsname{}\expandafter\renewcommand\csname AvgWPopulationMissingPercent\endcsname{\ensuremath{0.255}}
\expandafter\providecommand\csname AvgWRadiusSlope\endcsname{}\expandafter\renewcommand\csname AvgWRadiusSlope\endcsname{\ensuremath{-0.253}}
\expandafter\providecommand\csname AvgWReferenceRatio\endcsname{}\expandafter\renewcommand\csname AvgWReferenceRatio\endcsname{\ensuremath{2.59}}
\expandafter\providecommand\csname AvgWRefinementChangePercent\endcsname{}\expandafter\renewcommand\csname AvgWRefinementChangePercent\endcsname{\ensuremath{0.473}}
\expandafter\providecommand\csname AvgWStarSlope\endcsname{}\expandafter\renewcommand\csname AvgWStarSlope\endcsname{\ensuremath{1.49}}
\expandafter\providecommand\csname AvgWTableChainBh\endcsname{}\expandafter\renewcommand\csname AvgWTableChainBh\endcsname{\ensuremath{0.57590}}
\expandafter\providecommand\csname AvgWTableChainRadius\endcsname{}\expandafter\renewcommand\csname AvgWTableChainRadius\endcsname{\ensuremath{-0.25279}}
\expandafter\providecommand\csname AvgWTableChainStar\endcsname{}\expandafter\renewcommand\csname AvgWTableChainStar\endcsname{\ensuremath{1.49467}}
\expandafter\providecommand\csname AvgWTailBenchmarkZ\endcsname{}\expandafter\renewcommand\csname AvgWTailBenchmarkZ\endcsname{\ensuremath{2.6}}
\expandafter\providecommand\csname AvgWTailPercent\endcsname{}\expandafter\renewcommand\csname AvgWTailPercent\endcsname{\ensuremath{0.0499}}

\providecommand\result[1]{%
  \ifcsname #1\endcsname\csname #1\endcsname\else\textbf{??#1??}\fi}

\expandafter\providecommand\csname BatIncident\endcsname{}\expandafter\renewcommand\csname BatIncident\endcsname{\ensuremath{1873.97}}
\expandafter\providecommand\csname BatIsolated\endcsname{}\expandafter\renewcommand\csname BatIsolated\endcsname{\ensuremath{194.70}}
\expandafter\providecommand\csname BatLowLMax\endcsname{}\expandafter\renewcommand\csname BatLowLMax\endcsname{\ensuremath{4.39}}
\expandafter\providecommand\csname BatMaxShift\endcsname{}\expandafter\renewcommand\csname BatMaxShift\endcsname{\ensuremath{0.00964}}
\expandafter\providecommand\csname BatQ\endcsname{}\expandafter\renewcommand\csname BatQ\endcsname{\ensuremath{0.84164}}
\expandafter\providecommand\csname BatQHigh\endcsname{}\expandafter\renewcommand\csname BatQHigh\endcsname{\ensuremath{0.8523}}
\expandafter\providecommand\csname BatQLow\endcsname{}\expandafter\renewcommand\csname BatQLow\endcsname{\ensuremath{0.8306}}
\expandafter\providecommand\csname BatQuadError\endcsname{}\expandafter\renewcommand\csname BatQuadError\endcsname{\ensuremath{1.7\times10^{-5}}}
\expandafter\providecommand\csname BatRmsQ\endcsname{}\expandafter\renewcommand\csname BatRmsQ\endcsname{\ensuremath{0.84321}}
\expandafter\providecommand\csname BatSeedOneQ\endcsname{}\expandafter\renewcommand\csname BatSeedOneQ\endcsname{\ensuremath{0.84135}}
\expandafter\providecommand\csname BatSeedTwoQ\endcsname{}\expandafter\renewcommand\csname BatSeedTwoQ\endcsname{\ensuremath{0.84193}}
\expandafter\providecommand\csname BatShiftHigh\endcsname{}\expandafter\renewcommand\csname BatShiftHigh\endcsname{\ensuremath{0.000632}}
\expandafter\providecommand\csname BatShiftLow\endcsname{}\expandafter\renewcommand\csname BatShiftLow\endcsname{\ensuremath{-0.00362}}
\expandafter\providecommand\csname BatStateEvaluations\endcsname{}\expandafter\renewcommand\csname BatStateEvaluations\endcsname{\ensuremath{8118}}
\expandafter\providecommand\csname BatTidal\endcsname{}\expandafter\renewcommand\csname BatTidal\endcsname{\ensuremath{163.87}}

\begin{document}

\title{What kinds of strong star--black-hole encounters in AGN disks occur, and how often?}
\author{R.~O'Shaughnessy}
\affiliation{\RIT}
\author{Z.~Rosenberg}
\affiliation{\RIT}
\author{J.~Nordhaus}
\affiliation{\RIT}
\affiliation{\NTID}

\begin{abstract}
Disks surrounding  active galactic nuclei contain many stars and black holes, which will inevitably encounter one
another.  The outcome of these encounters depends on how 
closely and how rapidly the bodies meet.
Extending previous studies, we account for initial relative velocities seeded by stochastic effects in the disk;
incorporate a strong-encounter cross-section accounting for single- and binary-mediated interactions computed via direct two-body (Hill) or three-body (isolated) interactions; and present a repeated-encounter
model accounting both for energy loss due to stellar encounters as well as the SMBH tidal field.
We convolve our encounter model with a parametric family of  natal stellar and binary populations, focused on varying binary
fraction and period distribution.
In our fiducial disk model,
even though single-single interactions supply most encounters, binary-mediated interactions supply
$\result{ChaoticFractionCollisions}$\% of collisions; most encounters are well-approximated as parabolic;
and stochastically-driven  relative velocities are comparatively substantial and driven by gravitational perturbations from disk
fluctuations and neighboring bodies. 
Our tools are ready to use in downstream applications, including several easily-expressed semianalytic approximations to
key interaction phenomena. 
\end{abstract}
\maketitle

\section{Introduction}
\label{sec:intro}

Encounters between stars and black holes (BHs) in active galactic nucleus
(AGN) disks connect stellar evolution, the assembly of gravitational-wave
(GW) sources, and the observable activity of galactic nuclei
~\cite{2021ApJ...910...94C,2025ApJ...990..217M,2026arXiv260823677C,2025ApJ...993L..25A}.
These encounters can reshape the compact binary population, as gas capture, accretion, and migration bring stars and
compact remnants into close proximity
~\cite{2020MNRAS.499.2608F,2020ApJ...898...25T,2024ApJ...972..193D,2026arXiv260922074R}.
For the BH population, accretion of stellar debris can change BH masses and spins, providing a route
to massive, rapidly spinning GW progenitors
~\cite{2019ApJ...877...56L,2025ApJ...983L...9K,2025ApJ...994L..37K}.
For the AGN disk itself, these stars exchange
mass with the gas, modify its chemical composition, and contribute to its
heating
~\cite{2021ApJ...910...94C,2021ApJ...916...48D,2022ApJ...929..133J,2024ApJ...967...88C,2024ApJ...974..106C};
collisions and stellar transformations can dramatically punctuate this process
~\cite{2022ApJ...929..133J,2024ApJ...967...88C,2026arXiv260823677C}.
Given the substantial energy liberated over short timescales and later through BH accretion, these collisions should
produce strong, observationally-accessible transient electromagnetic emission 
~\cite{2016ApJ...823..113P,2021ApJ...911..104K,2022ApJ...932...84M,2022ApJ...933L..28Y}.
Finally, these encounters may contribute to the early  formation and rapid growth of massive BHs, with potential
connections to the 
``little red dots'' discovered by JWST
~\cite{2026arXiv260827596S,2025A&A...702A..57H}.
Studying the frequency and nature of these encounters therefore provides an essential ingredient to a wide range of phenomena in
AGN disks.

Semianalytic and hydrodynamic studies of star--BH encounters without an
ambient AGN gas disk have examined tidal capture, stripping, and disruption
of stars by more massive BHs
~\cite{2016ApJ...823..113P,2019ApJ...877...56L,2022ApJ...933..203K,2023ApJ...948...89K,2024A&A...685A..45V}.
Disruptions of lower-mass stars by stellar-mass BHs (micro-tidal disruptions)
can produce accretion-powered flares analogous to tidal disruption flares
from supermassive BHs
~\cite{2016ApJ...823..113P,2021ApJ...911..104K,2022ApJ...933..203K}.
Hydrodynamic calculations have mapped partial and complete disruption,
asymmetric mass loss, and the capture or ejection of surviving stellar cores
~\cite{2019ApJ...877...56L,2022ApJ...933..203K,2023ApJ...948...89K,2024A&A...685A..45V}.
Much of this work adopts initially parabolic approaches motivated by dense
star clusters; calculations of stars on bound orbits also follow repeated
stripping~\cite{2024ApJ...961..149X,2025ApJ...979...40L}.
Together these studies establish how sensitively the surviving stellar core
and the supply of accreting gas depend on the mass ratio, encounter depth,
and stellar structure.

Recent studies have begun to investigate  a different regime: a stellar-mass BH
encountering a star more massive than itself, as may occur after runaway stellar
growth in an AGN disk~\cite{2021ApJ...910...94C,2024ApJ...967...88C}.
In this regime a BH can penetrate and be captured within a surviving stellar
envelope. A study by Shi et al.\ combines semianalytic trajectories, three-dimensional
hydrodynamics, and stellar relaxation calculations to demonstrate this route
to a BH embedded in a massive star~\cite{2026arXiv260827596S}.
Hu et al.\ explore how a subsequent compact object can form an eccentric
binary within that stellar envelope~\cite{2026arXiv260827595H}.
Ma et al.\ provide an independent grid-hydrodynamic calculation of a BH
encountering a massive star that already contains a central BH, following
whether the incoming BH is captured or escapes, together with stellar
heating and eccentric orbital evolution~\cite{2026arXiv260916290M}.
The survival and observable consequences of accreting AGN stars also depend
on feedback, as studied by Cantiello et al.~\cite{2026arXiv260823677C}.
These studies identify possible outcomes for selected stellar models and
encounter conditions. To weight those outcomes in a disk population, we need
the rates of encounters at each mass ratio, penetration depth, and incident
speed.

Our objective is to calculate the tight-interaction rate and incident
conditions supplied by specified stellar and BH populations, including the
abundances and separations of binaries. These interactions provide the
conditions for stellar contact and capture.
 For a star--BH pair that remains
bound, we also model how dissipation during stellar passages and the SMBH
tide between passages change subsequent approaches.

Encounter rates depend both on how objects reach the same disk region and
on what happens when they meet. Orbital-alignment and population-evolution
studies describe gas-driven delivery~\cite{2020MNRAS.499.2608F,2020ApJ...898...25T,2025ApJ...990..217M},
while migration-driven resonance blocking can prevent close approaches
within that region~\cite{2026MNRAS.545f2217M}. Given an incoming population,
N-body integrations of single and binary BHs quantify the dependence of
scattering on random motions~\cite{2026arXiv260922074R}; micro-tidal-disruption
estimates and cluster calculations address stellar encounters under their
respective population and kinematic assumptions
~\cite{2022ApJ...933L..28Y,2019ApJ...881...75K,2026A&A...707A.217R}.
For the local dynamics, the equal-mass Hill treatment of Dodici and
Tremaine~\cite{2024ApJ...972..193D} separates friction-driven binary formation
from the environmental scales. Statistical three-body treatments describe
binary--single outcomes conditional on incoming conditions
~\cite{2019Natur.576..406S,2021PhRvX..11c1020G}, and calculations retaining
the SMBH tide show that the disk's gravitational field can modify those
interactions~\cite{2024MNRAS.535.3630F,2024A&A...683A.135T}.
Combining local encounter descriptions with population weights is therefore
needed to predict which star--BH passages a specified disk population supplies.

The challenge is to combine these dynamical descriptions into a common
measure of passages through a finite-size star, retaining the population
weights needed to compare their astrophysical importance. In this paper,
we construct that measure in BH-to-stellar mass ratio, pericenter in stellar
radii, and relative (asymptotic) velocity in stellar dynamical units. We combine Hill-regime
single--single approaches with binary--single encounters and a separately
identified binary--binary prescription. In the fiducial disk population,
we find that binary--single interactions supply most stellar contacts even though
single--single approaches dominate the total encounter count.
We organize the precomputed encounter outcomes into large tables and use
suitable weights to describe how often the different outcomes occur for
specified incoming conditions. We abbreviate such an outcome table, used
with these weights, as a ``kernel.'' Where an analytic prescription replaces
a table, we use the same term for that prescription. Keeping these outcomes
separate from the assumed population abundances lets us reuse the tables
for other disk populations within the range of conditions already studied.
Our response model accounts for the initial capture and repeated encounters, to  provide a framework for future
investigations coupling orbital evolution to the stellar response.

We also critically reassess the initial conditions of these encounters.
Recent prescriptions for gas-supported motions, turbulent density forcing,
and stochastic gas drag predict different residual orbital motions and
encounter speeds~\cite{2026PhRvD.113f3036T,2025arXiv251012895E,2025A&A...703A...6T}.
We combine density forcing, stochastic drag, and gravitational stirring
in a common balance with gas damping~\cite{2008ApJ...686.1292I,2004ApJ...602..388T},
then propagate the resulting relative motions into encounter rates.
Appendices~\ref{app:gamma} and~\ref{app:single-comparison} examine the forcing
assumptions and compare prescriptions at fixed gas and population parameters.

This paper is organized as follows. Section~\ref{sec:disk} defines the
encounter coordinates, Hill scales, and illustrative embedded populations.
Section~\ref{sec:channels} describes the mechanisms that bring binaries together.  Section~\ref{sec:return} discusses
how the ambient environment (SMBH) and collisions regulate how many and what kind of repeated encounters the star-BH system can undergo.   Section~\ref{sec:discuss}
examines the model's range of applicability and limitations. Section~\ref{sec:conclude}
summarizes the results.

\section{Encounter coordinates and Hill kinematics}
\label{sec:disk}

We define the incident encounter coordinates and the local Hill scales,
then specify the assumed stellar, BH, and binary populations used to
calculate encounter rates.

\subsection{Reduced encounter coordinates}
\label{sec:disk:coords}

We report incident passages using three dimensionless coordinates:
\begin{equation}\label{eq:coords}
  q = \frac{\Mbh}{\Mstar},\qquad
  x = \frac{\rp}{\Rstar},\qquad
  u = \frac{\vinf}{\sqrt{G\Mstar/\Rstar}} .
\end{equation}
Here $\Mbh$ and $\Mstar$ are the BH and stellar masses, $\Rstar$ is the
stellar radius, and $\rp$ is the pericenter of the hypothetical point-mass
star--BH orbit. The speed scale $\sqrt{G\Mstar/\Rstar}$ makes $u$
dimensionless. The assignment of $\vinf$ depends on the encounter channel,
as described below.
For a single--single encounter, $\vinf$ is the relative speed the BH and star
would have at infinite separation if only their mutual gravity acted.
We construct this hypothetical isolated encounter from their separation
and relative velocity at closest approach in the point-mass calculation.
The velocity is measured in a freely falling, nonrotating frame, rather than
by differentiating positions in corotating disk axes. For positive pair
energy, $\vinf^2=v_{\rm rel}^2-2G(\Mstar+\Mbh)/\rp$.
The corresponding point-mass pericenter sets $x$. This construction uses
the small size and short duration of the inner encounter compared with the
SMBH tidal scales; Appendix~\ref{app:velocity-handoff} states the assumptions
and gives the frame conversion.

We use these quantities to calculate collision rates for specified
encounter conditions and to represent the encounter population in histograms.
A subtlety arises for systems that become bound and undergo repeated
encounters: their orbital energy is negative, so a real speed at infinity,
and hence $u$, does not exist. For bookkeeping in these distributions, we
group such encounters in the $u=0$ bin. When evolving repeated encounters
with the return map, however, we retain the system's actual orbital energy
and angular momentum. As discussed later, the speed during a close stellar
passage can nevertheless be accurately approximated by the parabolic value,
even though the complete orbit is bound.

Contact means $x\le1$; tidal disruption approximately occurs when 
$x\le q^{1/3}$ and is generally more restrictive when the BH is less massive than the star ($q<1$) as is usually adopted
in this work.   For illustrations and many calculations in this work, we assume zero-age stellar radii follow a scaling relation
$\Rstar\propto\Mstar^{\gamma_R}$;
for calculations involving passage through a star, however, we will perform more detailed stellar modeling
for which this scaling is only an approximation.

The label $\rp$ belongs to the hypothetical orbit in which both objects
are point masses. It is the physical closest separation for an isolated
passage that stays outside a spherical star. If that orbit would enter the
star, the extended stellar potential changes the interior turning point,
and drag can change it further. We therefore match position and velocity
outside the stellar surface, preserving orbital energy and angular momentum,
rather than treating $\rp$ as the depth reached inside the star.
Appendix~\ref{app:velocity-handoff} gives this conversion and explains how
bound returns are matched to the stellar model.

\subsection{Hill's problem and the encounter rate}
\label{sec:disk:hill}

For a star and BH of total mass $\Mstar+\Mbh$ on a circular orbit of radius $r$ around a SMBH
of mass $\Msmbh$, let
$\Omega=(G\Msmbh/r^3)^{1/2}$. In the corotating frame, let
$\hat{\bm x}$ point radially outward, $\hat{\bm y}$ along the orbit, and
$\hat{\bm z}$ normal to the disk. The relative separation
$\bm \rho=(X,Y,Z)$ obeys Hill's equations
~\cite{1986CeMec..38...67H,2024ApJ...972..193D}
\begin{equation}\label{eq:hill-eom}
 \ddot{\bm \rho}+2\Omega\hat{\bm z}\times\dot{\bm \rho}
 =-\frac{G(\Mstar+\Mbh)\,\bm \rho}{|\bm \rho|^3}
   +\Omega^2\left(3X\,\hat{\bm x}-Z\,\hat{\bm z}\right).
\end{equation}
The mutual Hill radius and shear speed are
\begin{equation}\label{eq:hill}
 \begin{aligned}
  \rH &= r\left(\frac{\Mstar+\Mbh}{3\Msmbh}\right)^{1/3},\\
  \vH &= \Omega\,\rH = \left(\frac{G(\Mstar+\Mbh)}{3 \rH}\right)^{1/2}.
 \end{aligned}
\end{equation}
For initially circular neighboring orbits separated radially by $b$, the
shear speed is $\tfrac32\Omega b$. At the reference speed $\vH$ the
two-body focusing radius $2G(\Mstar+\Mbh)/\vH^2=6\rH$ exceeds the Hill radius.
To zeroth approximation, the encounter rate can be approximated using the Hill radius, orbital speed, and perturber
surface density $\Sigma_n$  via
~\cite{2004ARA&A..42..549G,2020ApJ...898...25T}
\begin{equation}\label{eq:rate}
  \Gamma_{\rm H} = 2\,\Sigma_n\,\Omega\, \rH^2 ,
\end{equation}
This is a nominal rate scale. The numerical coefficient $F$ uses the
explicit incoming shear flux; its normalization is specified in
Appendix~\ref{app:hill-kernel-method}.
For the fiducial SMBH mass $\Msmbh=10^8\,\Msun$, if we define
$\hat r=r/(10^4r_g)$, $\hat m=(\Mstar+\Mbh)/(115\,\Msun)$, and
$\hat\Sigma_n=\Sigma_n/\Sigma_{n,0}$, where
$\Sigma_{n,0}=N_{\rm BH}/[\pi(r_{\rm out}^2-r_{\rm in}^2)]$
is the assumed uniform BH surface density for $N_{\rm BH}=200$ between
inner disk radius $r_{\rm in}=300r_g$ and outer disk radius $r_{\rm out}=5\times10^4r_g$, then
\begin{align}
 \rH &= \result{StarMidHillRadius}\,\unit{AU}\,
       \hat r\,\hat m^{1/3}, \label{eq:hill-scaled-radius}\\
 \vH &= \result{StarMidHillVelocity}\,\unit{km\,s^{-1}}\,
       \hat r^{-1/2}\hat m^{1/3}, \label{eq:hill-scaled-speed}\\
 \Gamma_{\rm H} &=
 \frac{\hat\Sigma_n\hat r^{1/2}\hat m^{2/3}}
 {\result{StarMidEncounterTime}\times10^4\,\unit{yr}},
 \label{eq:hill-scaled-rate}\\
 \tau_{\rm H}\equiv\Gamma_{\rm H}^{-1}
 &=\result{StarMidEncounterTime}\times10^4\,\unit{yr}\,
 \hat\Sigma_n^{-1}\hat r^{-1/2}\hat m^{-2/3}.
 \label{eq:hill-scaled-time}
\end{align}

\subsection{Embedded populations}
\label{sec:disk:model}

We use a simple toy-model population, motivated by more detailed calculations~\cite{2003MNRAS.341..501S,2025ApJ...990..217M}.  Distances
are in $r_g=G\Msmbh/c^2$. For stars and BHs the assumed mass and
radial distributions factor as
\begin{equation}\label{eq:population-measure}
 \frac{\mathrm{d}^{2}N_i}{\mathrm{d}m\,\mathrm{d}A}
 = A_i m^{\alpha_i}r^{\beta_i}
   \mathbf{1}_{[m_{i,-},m_{i,+}]}(m)
   \mathbf{1}_{[r_{i,-},r_{i,+}]}(r),
\end{equation}
where $A_i$ is set by the chosen population size. Table~\ref{tab:population}
states both the single-object and binary inputs.

We choose simple order-of-magnitude normalizations of $N_\star=1000$
stars and $N_{\rm BH}=200$ black holes (Table~\ref{tab:population}), using
Nathaniel's population model as a benchmark~\cite{2025PhDT........74N}.
That model begins with approximately $1000$ low-mass stars, sets the initial
embedded BH population from the nuclear-cluster population and disk geometry,
and follows subsequent stellar capture and growth. Here the stellar count
specifies the assumed massive population in our selected radial band;
its relation to the initial seed population depends on growth, migration,
and losses. These counts provide illustrative encounter-rate normalizations;
we do not derive the census self-consistently for our adopted mass and
radial distributions.

For each adopted binary composition, we draw the two component masses
independently from the corresponding stellar or BH mass distributions in
Eq.~\eqref{eq:population-measure}, neglecting any additional physical effects that introduce 
correlations  by formation,
exchange, accretion, and survival. The adopted binary abundances and orbital
distributions are likewise prescribed inputs.

The stellar distribution describes the massive population after growth in
the disk. Captured stars can accrete rapidly, becoming much more massive
than their initial nuclear-cluster masses. Where accretion and mixing
replenish core hydrogen faster than nuclear burning consumes it, stars can
remain in an ``immortal'' hydrogen-burning state while the disk supplies
gas~\cite{2021ApJ...910...94C,2022ApJ...929..133J}.
Because stars will be captured from the nuclear star cluster, the population of stars should be steadily replenished and
offer many targets for star-BH encounters ~\cite{2025PhDT........74N}.

The resulting stellar mass distribution depends on the capture history,
gas accretion, winds, mergers, and the local disk conditions; stellar
feedback also changes the population that the disk can
support~\cite{2021ApJ...910...94C,2024ApJ...967...88C}.
For simplicity, we approximate this uncertain evolved distribution with the power law
$\mathrm{d}N_\star/\mathrm{d}m\propto m^{-1}$ over the mass range in
Table~\ref{tab:population}, ignoring any correlations with position or time-dependent accretion.
Self-consistent evolution of the stellar population in the disk can predict much more complex
structure~\cite{2025PhDT........74N}; our calculations here use an illustrative model, keeping in mind downstream users
can apply our tools immediately within their own applications.
Rather than self-consistently incorporate capture as in that work, for simplicity we use a 
finite-reservoir estimate in Sec.~\ref{sec:channels:rates}, 
assuming no replenishment and that each destructive event consumes a fresh
star.

We compare two schematic stellar-radius prescriptions. For stars evolving
approximately in isolation, each post-main-sequence stage is represented by
its lifetime fraction and associated mean radius from the adopted stellar-evolution tracks. We
interpolate between the $20$ and $100\,\Msun$ tracks and hold the latter
prescription above $100\,\Msun$. The alternative accreting-star
prescription keeps stars at main-sequence radii, motivated by models in
which sustained disk accretion delays ordinary post-main-sequence expansion
~\cite{2021ApJ...910...94C}.

The implications of evolving populations and correlated mass--position
distributions are discussed in Sec.~\ref{sec:discuss:populations}.

\begin{table*}[t]
\caption{\textbf{Fiducial population.} Inputs to
Eq.~(\ref{eq:population-measure}). Masses are in $\Msun$, radii are in
$r_g$, and $\Sigma_i\propto r^{\beta_i}$. The stars are restricted to
the band occupied by the accreting population, whereas black holes can occur
throughout the adopted disk radii.\llmfig}

\label{tab:population}
\begin{ruledtabular}
\begin{tabular}{lrrrrr}
population & $N_i$ & mass range & $\mathrm{d}N/\mathrm{d}m$ & radii & $\Sigma_i(r)$ \\
\hline
SMBH / disk & -- & $M_{\rm SMBH}=10^{8}\,\Msun$ & -- & $300$--$5\times10^4$ & Sirko--Goodman-like \\
massive stars & 1000 & $50$--$300$ & $m^{-1}$ & $2\times10^3$--$2\times10^4$ & $r^{-2}$ \\
black holes & 200 & $5$--$40$ & $m^{-2}$ & full disk & constant \\
binaries & $f_{\bullet\bullet}=0.10$; $f_{\star\bullet}=0.05$ & -- & -- & -- & $p(a)\propto a^{-1}$, $0.05\,\mathrm{AU}<a<r_H/3$ \\

\end{tabular}
\end{ruledtabular}
\end{table*}

\subsubsection{Binary separations and residence times}
\label{sec:disk:binary-separations}

At fixed disk radius $r$ and binary mass $m_B=m_a+m_b$, we assume
$p_a(a\mid r,m_B,\beta_a)\propto a^{\beta_a}$ for semi-major axes
$a\in[a_-,a_+]$, normalized so that
$\int_{a_-}^{a_+}p_a(a\mid r,m_B,\beta_a)\,\mathrm da=1$.
We impose a common minimum imposed semi-major axis  $a_-$, but allow  tidal truncation to set the maximum
semi-major axis,
\begin{equation}\label{eq:binary-a-hill-cutoff}
 a_+=f_H r\left(\frac{m_B}{3\Msmbh}\right)^{1/3},
 \qquad f_H=\frac{1}{3}.
\end{equation}
The upper cutoff uses the binary's mass, while the binary--single
encounter Hill scale uses the mass of all three bodies.
 Star-bearing binaries must additionally satisfy the
periastron-detachment selection of Appendix~\ref{sec:channels:bs}.

We adopt the logarithmically uniform prescription, $\beta_a=-1$, as
our reference and compare it with $\beta_a=-2$ and $1/2$ over the same
separation bounds, component mass distributions, and prescribed binary
abundances. These alternatives isolate sensitivity to the separation
distribution; the $\beta_a=1/2$ case weights binaries toward wider
separations. Their physical interpretation follows from the residence
time rather than an assumed evolved AGN binary census.

Their connection to binary evolution follows from number conservation.
Let $n_a=\mathrm dN_B/\mathrm da$ be the instantaneous binary number density
in separation at fixed masses and location. If binaries enter at the
outer boundary and shrink monotonically through a stationary interval
without sources or losses inside it, the inward number flux
$\mathcal J=|\dot a|n_a$ is constant. Thus
\begin{equation}\label{eq:binary-a-residence}
 \frac{\mathrm dN_B}{\mathrm d\ln a}
      =\mathcal J\frac{a}{|\dot a|}
      =\mathcal J t_a,
 \qquad t_a\equiv\frac{a}{|\dot a|}.
\end{equation}
A law $|\dot a|\propto a^\gamma$ gives
$p_a\propto a^{-\gamma}$ under these assumptions. A constant hardening
rate $\mathrm d(1/a)/\mathrm dt$ yields $\beta_a=-2$, as in the
fixed-background stellar-scattering limit~\cite{2017MNRAS.464..946S};
a constant fractional
shrinkage rate $|\dot a|/a$ yields $\beta_a=-1$. If the shrinkage time is
proportional to the internal binary period, $t_a\propto a^{3/2}$ at fixed
mass, then $\beta_a=1/2$. This last example is an orbital-timescale ansatz,
used as a wide-weighted sensitivity case. It requires slower evolution
at large separation, but does not require a zero of $\dot a$ or a stall.
Increasing a hardening rate at fixed
injection flux also decreases the standing binary abundance; our
fixed-abundance slope comparison isolates the distribution's shape.

For the gas-processed reference, contracting hydrodynamic solutions give
fractional orbital decay rates of order the specific mass-accretion
rate~\cite{2023MNRAS.522.1881L}. We represent this scaling locally by
$\dot a/a=-C_a\dot m_B/m_B$, with positive dimensionless $C_a$.
If $C_a$ and $\dot m_B/m_B$ vary slowly across the separation interval,
$t_a$ is approximately independent of $a$, motivating $\beta_a=-1$.
This is our preferred simple gas-processing prescription; the
$\beta_a=-2$ case represents stellar-scattering hardening and the
$\beta_a=1/2$ case tests a stronger preference for wide binaries.
The source calculations also find dependence on thermodynamics and
binary separation, so this argument motivates a reference model rather
than fixing the exponent for every AGN disk.

Gas-driven evolution can couple separation, eccentricity, mass ratio,
and orientation~\cite{2024ApJ...970..107C,2025ApJ...990..137D}.
Distributed formation, removal through contact or disruption, migration,
and finite disk lifetimes can invalidate the stationary reduction in
Eq.~\eqref{eq:binary-a-residence}. We therefore keep the thermal,
detachment-limited eccentricity model and imposed abundances explicit
when interpreting this comparison. We retain binaries on both sides of
the incident hard--soft boundary; ionization remains an outcome of the
conditional scattering kernel rather than an additional cut on $a$.

All separation variants apply the adopted Hill-scale encounter rate to
close-encounter outcomes drawn from the scattering simulations.
For single--single interactions, we evaluate the incoming flux and
close-approach probabilities with Hill trajectories. For binary--single
interactions, we adopt the Hill-scale normalization as an ansatz rather
than separately weighting the fraction of incoming trajectories that
reach the binary. This fraction can depend on binary separation,
incoming vertical motion, and the orientation and phase of the binary
relative to the incoming trajectory. The separation-law comparison
retains this prescription and the imposed binary census.

\subsection{From the embedded population to encounter velocities}
\label{sec:disk:kinematics}

The masses and spatial distributions in Sec.~\ref{sec:disk:model} do not
alone determine the incident coordinates of Sec.~\ref{sec:disk:coords}.
The drawn masses and stellar radius fix $q$ and the conversion from physical
pericenter to $x$, but the encounter kernel also needs the bodies' relative
motion. Random velocities change both the frequency of close approaches and
their incident energies. This choice controls the channel ranking: changing to circular,
coplanar orbits reverses the fiducial ordering of the contact-producing
channels. We therefore derive the adopted
velocity amplitudes before constructing the channel measure.

\subsubsection{Conditional stochastic orbital eccentricity}
Tomar, Hopkins, and Kremer~\cite{2026PhRvD.113f3036T} argue that embedded
objects inherit or relax toward the gas velocity dispersion, which can
suppress gravitationally focused encounters in thick disks. Our question is
whether the bodies retain that motion in the particular thin disk of
Table~\ref{tab:population}. Our closure balances linear gas damping against excitation by turbulent
density fluctuations, stochastic gaseous drag, and gravitational encounters. We estimate
their balance separately for each body mass and disk radius.

Gravitational encounters produce small kicks $\Delta v$ and thus small eccentricity changes.
To order of magnitude,  a radial excursion of order $er$ occurs on a timescale
$\Omega^{-1}$, giving a random velocity $v_{\rm rand}\sim er\Omega=e v_{\rm K}$,
where $v_{\rm K}=r\Omega$ is the circular speed around the SMBH.
A small velocity kick therefore changes the eccentricity vector by order
$\Delta e\sim\Delta v/v_{\rm K}$, with order-unity factors depending on the
kick direction and orbital phase. Damping this random motion on a timescale
$\te$ gives $\dot e=-e/\te$; the variance will decay with a rate $2/t_e$.
For independent kicks with zero mean, the eccentricity variance 
grows at a rate of order $\Gamma(\Delta v/v_{\rm K})^2$, where $\Gamma$ is
the kick frequency. In a strong Hill encounter the relative velocity changes
by order $\Omega r_{{\rm H},j}$; the test body receives the fraction
$m_j/(\mt+m_j)$ of this change. Its eccentricity kick is thus of order
$(r_{{\rm H},j}/r)m_j/(\mt+m_j)$.

Turbulent gas motion also produces small kicks through the fluctuating
gravity of its density perturbations. A patch of horizontal size $\ell$
and excess surface density $\delta\Sigma_g$ has excess mass
$\delta M\sim\delta\Sigma_g\ell^2$, giving an acceleration of order
$G\delta\Sigma_g$ at a distance $\ell$. If this acceleration remains
coherent for a time $\tau_c$, it produces a velocity kick
$\Delta v\sim G\delta\Sigma_g\tau_c$ and hence an eccentricity kick
$\Delta e\sim\Delta v/v_{\rm K}
\sim(\delta\Sigma_g/\Sigmag)(\Sigmag r^2/\Msmbh)\Omega\tau_c$.
Independent kicks then give
$\mathrm d\langle e^2\rangle/\mathrm dt\sim(\Delta e)^2/\tau_c$.
For $\tau_c\sim\Omega^{-1}$ and fractional surface-density fluctuations
proportional to the turbulence amplitude $\gamma$, this yields the scaling
$\mathrm d\langle e^2\rangle/\mathrm dt
\propto\gamma^2(\Sigmag r^2/\Msmbh)^2\Omega$.

Combining these two arguments together, we adopt a simple stochastic model for eccentricity excitation: if
 $e$ denotes a body's ensemble RMS orbital eccentricity, $\te$ its
amplitude-damping time, and $\Sturb$ and $\Sstir$ the rates at
which fluctuating-density gravity and encounters increase $e^2$, then we use
\begin{equation}\label{eq:kin-evolution}
 \frac{\mathrm d e^2}{\mathrm dt}
 =-\frac{2 e^2 }{\te}+\Sturb+\Sstir(e)+\Sdrag.
\end{equation}
The factor two follows because $\te$ damps $e$, not $e^2$. A stationary
solution therefore has $e^2=\te(\Sturb+\Sstir+\Sdrag)/2$, providing our fiducial eccentricity estimate within the adopted disk closure.
For a test mass $\mt$ at disk radius $r$, we use
\begin{align}
 \te&=\frac{1}{\result{KinDampingCoefficient}\,\Omega}\,
      \frac{\Msmbh}{\mt}\,
      \frac{\Msmbh}{\Sigmag r^2}\left(\frac{h}{r}\right)^4,
      \label{eq:kin-damping}\\
 \Sturb&=\left(\Cturb\gamma\frac{64\Sigmag r^2}{\pi\Msmbh}\right)^2
      \frac{\Omega}{2\pi},\label{eq:kin-turbulence}\\
 \Sstir&=\sum_{i\in\{\star,\bullet\}}
      \int_{m_{i,-}}^{m_{i,+}}\!\mathrm dm\,n_i(m,r)\,
      2\Omega r_{\rm H}^2(m)\nonumber\\
      &\quad\times\left(\frac{r_{\rm H}(m)}{r}
      \frac{m}{\mt+m}\right)^2
      \min\!\left[1,f^{-2}(m)\right].
      \label{eq:kin-stirring}
\end{align}
and $\Sdrag$ is a negligibly small contribution from stochastic drag that
we address in Appendix~\ref{app:gamma} and include only for clarity.
Here $\Sigmag$ and $h$ are the gas surface density and scale height,
$c_s=h\Omega$ is an effective-thermal assumption. In the source turbulence
model $\gamma\simeq\delta\rho/(3\rho)$; the AGN forcing also depends
on the spatial and temporal structure of the density fluctuations.
The damping law of Eq.~\eqref{eq:kin-damping} is the small-eccentricity
result of Ref.~\cite{2004ApJ...602..388T}; see their Eqs.~(45) and (49).
The turbulent forcing of Eq.~\eqref{eq:kin-turbulence} follows
Ref.~\cite{2008ApJ...686.1292I}; see their Eq.~(13) and Eqs.~(27)--(29).
The coefficient $\Cturb=\result{KinTurbCoefficient}$ converts their
solar-nebula normalization to the dimensionless disk surface density used
here. Appendix~\ref{app:kinematics} gives the conversion.

We use $\gamma=\result{KinGammaFiducial}$ as the reference forcing
parameter.
Appendix~\ref{app:gamma} traces this normalization to
Ref.~\cite{2008ApJ...686.1292I}, examines its transfer to AGN support
regimes, and adds stochastic wake forcing following
Ref.~\cite{2025A&A...703A...6T}.

The encounter term  Eq.~\eqref{eq:kin-stirring} requires additional explanation.  First, we must sum over all types of
object our fiducial star can encounter.  Second, the expression has an added factor
that prevents additional collisions from exciting random motion beyond the natural characteristic Hill velocity.
In this expression, the mutual Hill radius is
$r_{{\rm H},j}=r[(\mt+m_j)/(3\Msmbh)]^{1/3}$, and
$f_j=e_{{\rm rel},j}r/r_{{\rm H},j}$ with
$e_{{\rm rel},j}=\sqrt{2}e$. In the calculations we use each population's local surface
number density $\Sigma_i(r)=\int n_i(m,r)\,\mathrm dm$ and
number-weighted median mass $\widetilde m_i$, rather than evaluating
the mass integral: $n_i(m,r)\simeq\Sigma_i(r)\delta(m-\widetilde m_i)$.

We use the stationary root below $e=h/r$ and set $e=h/r$ above it.
The damping time can be written as
\begin{equation}\label{eq:kin-damping-scaled}
 \begin{aligned}
 \te\simeq{}&\result{KinDampScalingYr}\,\unit{yr}\,
 \left(\frac{\Msmbh}{10^8\,\Msun}\right)
 \left(\frac{\mt}{10\,\Msun}\right)^{-1}\\
 &\times\left(\frac{\Sigmag}{10^4\,\unit{g\,cm^{-2}}}\right)^{-1}
 \left(\frac{r}{10^4r_g}\right)^{-1/2}
 \left(\frac{h/r}{0.03}\right)^4.
 \end{aligned}
\end{equation}
Here $r_g=G\Msmbh/c^2$, so the explicit SMBH-mass dependence is at
fixed $r/r_g$ and local $\Sigmag$. In the fiducial profile,
$\Sigmag=10^4\,\unit{g\,cm^{-2}}[r/(10^4r_g)]^{-1}$ and $h/r=0.03$,
so $\te\propto r^{1/2}$ at fixed masses. Across the stellar band, the
representative star and BH therefore have damping times shorter than the
assumed $10^{\result{TAgn}}\,\unit{yr}$ disk lifetime. Thus the adopted linear damping can erase inherited small-amplitude motion
within this effective-thermal model, while the forcing sustains nonzero
orbital eccentricities.
In Appendix \ref{app:single-comparison} we compare our prescription to other work estimating stochastically-driven
eccentricity and inclination.

\subsubsection{From single-body orbits to relative motion and close approach }
At every radius, rather than be in a planar circular orbit, individual objects orbiting the SMBH will have
stochastically excited orbital inclinations and eccentricites.

The next step converts individual eccentricities to relative motion.
The eccentricity (Runge--Lenz) vector $\bm e$ points toward periapsis
and has magnitude $e$. It is a three-dimensional vector constrained to
lie perpendicular to its own orbital angular momentum.
For nearly coplanar prograde orbits, its leading components lie in the disk
plane. The orbit's orientation is specified by the unit angular-momentum
vector $\hat{\bm l}$. Near the disk normal its two transverse components
form a two-dimensional tilt vector of magnitude $i$ to first order, with
orientation set by the ascending node.    For pairs of objects orbiting in the disk, we can characterize their proximity
by $|\mathbf{e}_1-\mathbf{e}_2|^2$ and similarly.  Because $e$ and $i$ are isotropically oriented in the plane, this
  difference reduces to a sum in quadrature.
With RMS eccentricities $e_1,e_2$ and
small inclinations $i_1,i_2$, we find
\begin{equation}\label{eq:kin-relative}
 \begin{aligned}
 e_{\rm rel}^2&=e_1^2+e_2^2,& i_{\rm rel}^2&=i_1^2+i_2^2,\\
 f_e&=\frac{r}{\rH}e_{\rm rel},& f_I&=\frac{r}{\rH}i_{\rm rel}.
 \end{aligned}
\end{equation}
Here
$\rH$ uses the total pair
mass.
We call the relative motion dynamically cold when $f_e\ll1$ and $f_I\ll1$:
the radial and vertical epicyclic excursions are small compared with $\rH$,
and the random velocities are small compared with $\vH=\Omega\rH$.
This describes orbital motion, not the gas temperature. The circular,
coplanar limit has $f_e=f_I=0$; differential orbital shear still supplies
relative motion. A cold population can have finite vertical thickness,
which matters when the stellar contact scale is much smaller than $\rH$
(Sec.~\ref{sec:channels:mass-response}).

This approximation and its downstream application assume small eccentricity and inclination.
The linear gas-damping law additionally requires motions small compared
with $h/r$.
Appendix~\ref{app:kinematics} states the averaging assumptions and range of validity.

The Hill integrations average the orbital phases at fixed amplitudes.
For small eccentricities and inclinations, each body's random orbital
state is described by two two-component vectors,
\begin{equation}
 \begin{aligned}
 \boldsymbol e_k&=\epsilon_k(\cos\varpi_k,\sin\varpi_k),\\
 \boldsymbol i_k&=\iota_k(\cos\Omega_{{\rm node},k},
                         \sin\Omega_{{\rm node},k}),
 \end{aligned}
\end{equation}
Here $\epsilon_k$ and $\iota_k$ are instantaneous magnitudes, whose RMS
values are the $e_k$ and $i_k$ used above; $k$ labels the body, $\varpi_k$ is its longitude of periapsis, and
$\Omega_{{\rm node},k}$ is its longitude of ascending node.
We model each vector as an isotropic, zero-mean bivariate Gaussian,
with independent eccentricity and tilt vectors and independent bodies.
If $a$ denotes either vector's magnitude and $a_{\rm rms}$ its RMS
value, each component has variance $a_{\rm rms}^2/2$, giving
\begin{equation}
 p(a)=\frac{2a}{a_{\rm rms}^2}
       \exp\!\left(-\frac{a^2}{a_{\rm rms}^2}\right),\qquad a\geq0.
\end{equation}
Thus eccentricity and inclination have Rayleigh magnitude distributions.
Their phases are uniform; the conditional Hill kernels have already
averaged over those phases, so the population calculation integrates
only over the relative magnitudes. Differences of independent Gaussian
vectors are Gaussian, with component variances that add. Their RMS
magnitudes therefore obey Eq.~\eqref{eq:kin-relative}, and rescaling by
$r/r_H$ gives the independent Rayleigh amplitudes $f_e$ and $f_I$.
These distributions describe two components for eccentricity and two
for tilt in the linearized orbital model; they are not the magnitude
distribution of a three-component isotropic Gaussian velocity.
Binary incoming velocities also retain their orbital-phase distribution,
as detailed in Appendix~\ref{app:kinematics}.

Trani and Di Cintio provide approximate stationary eccentricity and
inclination distributions for subsonic, quasi-embedded
orbits~\cite{2025A&A...703A...6T}. Appendix~\ref{app:kinematics} discusses
using these distributions to average encounter rates;
Appendix~\ref{app:single-comparison} compares their single-body moments
and single--single contact rates with ours at fixed gas parameters.

For the non-drag sources we prescribe
$i_i/e_i=\result{KinInclinationRatio}$; the vertical drag source of
Appendix~\ref{app:gamma} gives a small correction.
With $\gamma=\result{KinGammaFiducial}$, the representative star--BH pair
has $f_e=\result{KinPairFidLow}$--$\result{KinPairFidHigh}$ across the
stellar band. The population integration evaluates the balance for the
masses and radius of each draw.
The circular, coplanar bracket sets $f_e=f_I=0$. The gas-speed
bracket sets each body's $e_i=h/r$, giving $f_e=\sqrt{2}h/\rH$ for
independent epicycles with the same prescribed inclination ratio.
Increasing $\gamma$ raises the forcing while retaining the mass-dependent
balance; the gas-speed bracket sets both bodies at the cap.

\subsubsection{Encounter-rate coefficient and passage depth}
For relative epicyclic amplitudes $(f_e,f_I)$, let
$F(<\epsilon;f_e,f_I)$ be the cumulative dimensionless encounter-rate
coefficient, with $\epsilon=\rp/\rH$. It is computed by flux-weighted
numerical integrations of Hill's equations. The rate per target and the
conditional depth probability for an encounter that enters the Hill sphere
are, respectively,
\begin{equation}\label{eq:hill-rate-and-probability}
 \begin{aligned}
 \Gamma(<\epsilon)&=\Sigma_n\Omega\rH^2F(<\epsilon;f_e,f_I),\\
 P_H(<\epsilon)&=\frac{F(<\epsilon;f_e,f_I)}{F(<1;f_e,f_I)}.
 \end{aligned}
\end{equation}
For population rates we first form the orbital average
\begin{equation}\label{eq:hill-population-average}
 \overline F(<\epsilon)=\int df_e\,df_I\,
 p(f_e,f_I)F(<\epsilon;f_e,f_I).
\end{equation}
The corresponding entry rate uses $\overline F(<1)$, and the depth
distribution is $\overline F(<\epsilon)/\overline F(<1)$.
Equivalently, amplitudes drawn from the population carry their own
entry-rate weight before conditional encounters are sampled. We retain
the conditional kernels for the scaling arguments below.

Appendix~\ref{app:hill-kernel-method} gives the numerical estimator,
absolute normalization, and physical limiting cases; the supplemental
material gives implementation details and numerical checks.
Specifically, if we have some pair energy and threshold, then we determine $x=r/R_*=\epsilon r_H/R_*$ and $u$ in 
Eq.~\eqref{eq:coords}
using $\varepsilon_{\rm pair}>0$ at closest approach:
\begin{equation}\label{eq:kin-to-coordinates}
 u=\frac{\sqrt{2\varepsilon_{\rm pair}}}{\sqrt{G\Mstar/\Rstar}}.
\end{equation}
Bound samples receive $u=0$ under the convention of
Sec.~\ref{sec:disk:coords}; a finite Hill-scale random motion therefore does
not require a large positive stellar-scale excess speed.
For binary--single encounters we apply the same eccentricity closure to the
single body and the binary center of mass, treating the latter as a body of
the binary's total mass. Their relative epicycle has an RMS speed
$v_{\rm disp}=\vH(5f_e^2/8+f_I^2/2)^{1/2}$. We add each sampled orbital velocity to the local shear before
selecting the resonance kernel.
Appendix~\ref{app:kinematics} derives this conversion and states the numerical
implementation. The binary--binary channel retains its scaled-kernel
prescription of Sec.~\ref{sec:channels:kernels}.

This chain connects the assumed population to the delivered encounter
measure: damping and excitation set $e_i$; independent motions set
$(f_e,f_I)$; the channel kernels set rates, pericenters, and energies.
The damping times and encounter velocities depend on the adopted disk
thickness and gas surface density.

\section{Encounter channels and the stacked distribution}
\label{sec:channels}

All encounters in this section are supplied by orbital shear and the
prescribed random motions within an already embedded population.
We do not include radial migration as a source of encounters: there is no
migration-driven convergence, delivery to traps, or replenishment of the
interacting population. The calculation holds the adopted surface densities
and orbital distributions fixed and evaluates their instantaneous encounter
flux. It does not evolve an initial population through an AGN episode.
This deliberately restricted calculation omits a potentially important
additional supply mechanism.

The physical clock comes from relative orbital motion, not from the number
of Monte Carlo draws. For example, the single--single contact rate is
\begin{equation}\label{eq:instantaneous-contact-flux}
 \begin{aligned}
 \Gamma_{\rm ss,contact}&=\int\mathrm dA\,
 \mathcal N_\star(r)\mathcal N_{\rm BH}(r)\\
 &\quad\times\left\langle\Omega r_H^2
 F(<R_\star/r_H;f_e,f_I)\right\rangle .
 \end{aligned}
\end{equation}
Here the $\mathcal N_i$ are surface number densities and the average uses the
adopted local mass and orbital distributions. The coefficient
$\Omega r_H^2F$ has units of area per time and includes the incoming shear
flux (Appendix~\ref{app:hill-kernel-method}). Binary-mediated channels also
receive dimensional encounter-flux weights before their conditional
scattering outcomes are sampled.

The three channels deposit encounters differently: single--single approaches
dominate the total count, but binary-mediated interactions supply most of the
stellar contacts (Fig.~\ref{fig:marginals} and Table~\ref{tab:channels}).
This distinction concerns close approaches. The rate of distinct
destructive events also depends on the supply of fresh stars.

In practical terms, we look up encounter outcomes in the precomputed tables
and weight their contributions by the encounter rates of the adopted disk
population. The tables specify which outcomes occur for given incoming
conditions; the population weights specify how often those conditions are
supplied. The ``kernel'' below is the weighted outcome table introduced in
the Introduction, or its analytic replacement.

For the single--single and binary--single channels, a tight-interaction rate
counts each incoming interaction once if its minimum separation crosses the
chosen threshold. Several close passages within that interaction can affect
whether it reaches the stellar surface, but they do not contribute separate
counts to this rate. The scaled binary--binary channel additionally includes
the prescribed multiplicity of Table~\ref{tab:population}. The minimum-separation statistic already includes repeated approaches
resolved within the point-mass scattering histories. The return map of
Sec.~\ref{sec:return} serves a subsequent purpose: it follows a surviving
pair through stellar passages, including dissipation, to help identify the
encounters that merit detailed hydrodynamic treatment. We demonstrate that
evolution for specified systems without propagating the population samples
through it or adding a count for each later stellar passage.

To state the calculation before its channel details, let $c$ label
single--single, binary--single, or the scaled binary--binary channel, and let
$\lambda_c$ collect its disk radius, component masses, orbital elements, and
encounter geometry. The population parameters $\Theta_{\rm pop}$ are specified
in Table~\ref{tab:population}. Each channel supplies a rate measure
$\mathrm{d}\Gamma_c/\mathrm{d}\lambda_c$ and a conditional kernel
$K_c(q,x,u\mid\lambda_c;\mathcal E_c)$ for the encounter coordinates of
Eq.~\eqref{eq:coords}, where $\mathcal E_c$ denotes its calibration or analytic
prescription. Their population-weighted differential rate is
\begin{equation}\label{eq:stack-measure}
 \frac{\mathrm{d}^{3}\Gamma}{\mathrm{d}q\,\mathrm{d}x\,\mathrm{d}u}
 =\sum_c\int \mathrm{d}\lambda_c\,
 \frac{\mathrm{d}\Gamma_c}{\mathrm{d}\lambda_c}
 K_c(q,x,u\mid\lambda_c;\mathcal E_c).
\end{equation}
Writing $\Gamma_{\rm tot}=\sum_c\Gamma_c$ and
$\mathcal E=\{\mathcal E_c\}$, the normalized encounter distribution is
$p_{\rm eff}(q,x,u\mid\Theta_{\rm pop},\mathcal E)
=\Gamma_{\rm tot}^{-1}\mathrm{d}^3\Gamma/
(\mathrm{d}q\,\mathrm{d}x\,\mathrm{d}u)$.
Thus the channel fractions are rate-weighted marginals:
changing the mass functions, binary population,
or disk kinematics changes $p_{\rm eff}$.

The Python implementation returns rate-weighted encounter samples and a
normalized measure in $(q,x,u)$, using the analytic and tabulated kernels.
It evaluates alternative population assumptions without new scattering
integrations within their calibration domain. Appendix~\ref{app:reuse}
describes the interface and an abundance-reweighting example.
The resonance records are
averaged over the adopted component-mass distributions rather than conditioned
on every component mass. A change to those mass distributions therefore reuses
that averaged kernel; a population with substantially different mass ratios
requires reassessing the calibration.

\subsection{Encounter kernels}
\label{sec:channels:kernels}

For single--single approaches, the analytic circular, coplanar shear approximation gives a
cumulative encounter-rate coefficient proportional to
$(r_p/r_H)^{1/2}$. It is a useful scale argument, but the actual kernel
comes from integrations of Hill's equation: horseshoe trajectories exclude
the smallest guiding-center offsets, while a chaotic band supplies the
close approaches. We evaluate that kernel at
the finite radial and vertical amplitudes of
Sec.~\ref{sec:disk:kinematics}; it also records the osculating energy used
to assign $u$. Appendix~\ref{app:hill-kernel-method} gives the numerical
construction and its focusing limits.

For binary--single encounters we use a large table of precomputed outcomes
from isolated Newtonian star--BH--BH scattering, following the standard
binary--single three-body setup~\cite{1983ApJ...268..319H}.
Only the mutual gravitational forces of the three objects are included:
the SMBH potential and disk-gas forces are absent during these integrations.
The AGN model supplies the incident populations and rate weights. The
table is organized by incident speed relative to the binary binding energy,
composition, and detached initial binary state.
Each record identifies the pair responsible for the closest star--BH
approach and records how the interaction ends: exchange of a binary member,
a flyby, a merger, or disruption of the binary (ionization). Recording only
the smallest separation among all three pairs would not establish whether
a star--BH contact occurred. The calibration ensemble and its limitations are given
in Appendix~\ref{app:channels}. Binary--binary encounters have a rate
computed from their population abundance, but their $(q,x,u)$ shape is
borrowed from the binary--single kernel, not calibrated with four-body
integrations. We report the binary--binary contribution separately in
Table~\ref{tab:channels}.

The conditional resonant kernel also assumes that exterior gas does not
reshape the encounter while it unfolds. Hydrodynamic BH binary--single
calculations find gas-dependent chaotic branches and contraction
~\cite{2025MNRAS.539.1501R}. An independent series finds changes in both
the number of temporary binary states and the close-periapsis probability
per state~\cite{2025ApJ...983..114W,2025ApJ...993..139W}.
A gas-modified resonance model would therefore need both the number of
temporary binary states and the close-periapsis probability per state.

\subsection{Rate-weighted encounters}
\label{sec:channels:stack}

\begin{figure*}[t]
\includegraphics[width=\textwidth]{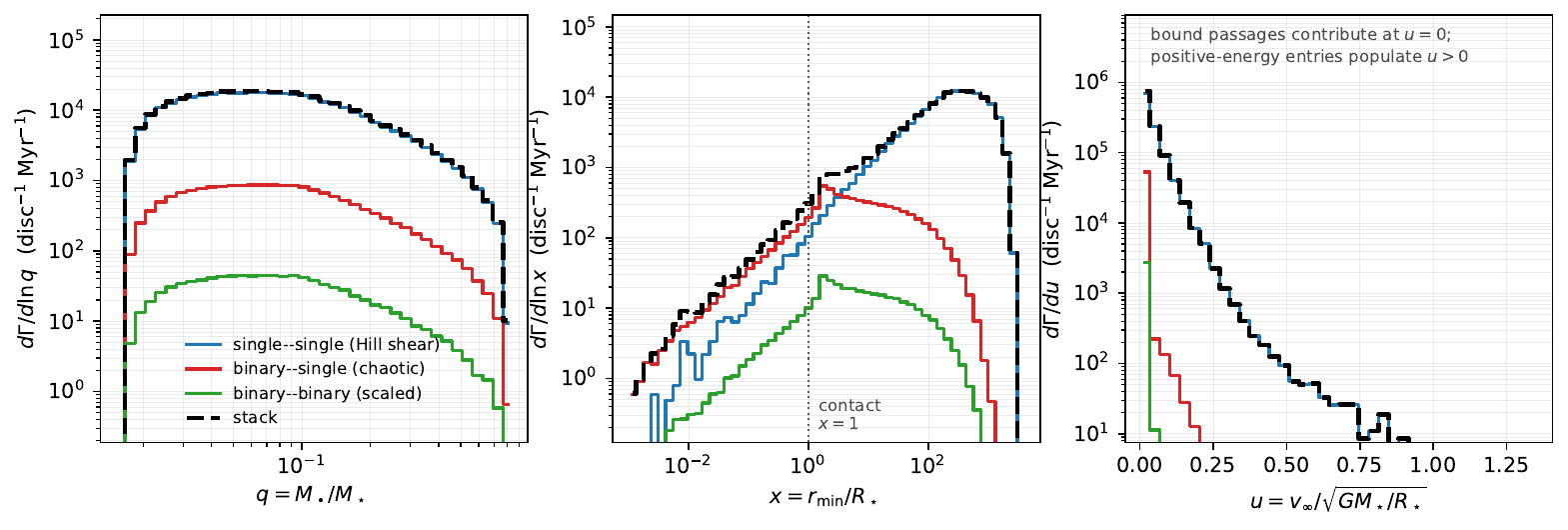}
\caption{\textbf{The three channels deposit encounters in different places.}
Rate-weighted marginals of the stacked differential rate in the mass ratio $q$
(left), the scaled pericenter $x$ (center) and the scaled speed $u$ (right), per
channel and summed. The mass-ratio marginals reflect the adopted stellar and BH mass functions
together with each channel's rate weighting. The pericenter marginal separates the channels
completely: the shear channel peaks two decades outside the star while the
resonant channels peak inside it, so only
$\result{SingleSingleCollisionFraction}$ per cent of the former and
$\result{BinarySingleCollisionFraction}$ per cent of the latter reach $x\le1$.
The velocity marginal is confined to $u\le\result{StackUpFive}$ at the $95$th
percentile. The spike at $u=0$ is the bookkeeping bin for bound single--single samples
and non-ionizing binary-mediated contributions (Sec.~\ref{sec:disk:coords}
and Appendix~\ref{app:velocity-handoff}).
It does not imply zero speed during a stellar passage or establish that each
orbit is exactly parabolic. The right panel uses an approximation described in
Appendix~\ref{app:hill-kernel-method}.\llmfig}

\label{fig:marginals}
\end{figure*}

Figure~\ref{fig:marginals} and Table~\ref{tab:channels} show the
population-weighted result of Eq.~\eqref{eq:stack-measure}. The
single--single channel peaks well outside the stellar surface; resonant
encounters begin on binary scales and more often reach contact.
Selecting $x\leq1$ therefore reverses the ranking of the channels:
binary-mediated interactions account for
$\result{ChaoticFractionCollisions}$ per cent of contacts in the
fiducial population. Supplemental Material, Sec.~\ref{sup-sup:population},
shows the depth-cut and radial dependence: the rate grows toward larger disk
radii while the contact fraction decreases. The population integral therefore
includes both the local rate and the probability of reaching the star.

The stack row in Table~\ref{tab:channels} sums rates over channels.
Contacts with $q\le0.2$ occur at a total rate of
$\result{StackQsmallXsmallRate}\,\unit{Myr^{-1}}$ per disk; their channel
rates are listed directly in the table.

We convert a per-disk rate $\Gamma$ into a volumetric rate through
$\mathcal R=n_{\rm AGN}\Gamma$. At the assumed number density
$n_{\rm AGN}=\result{NAgnRuleOfThumb}\times10^{-4}\,\unit{Mpc^{-3}}$,
a rate of $1\,\unit{Myr^{-1}}$ per disk corresponds to
$1\,\unit{Gpc^{-3}\,yr^{-1}}$, so the numerical rate entries in
Table~\ref{tab:channels} also give volumetric rates at this density.

The velocity marginal in Fig.~\ref{fig:marginals} lies near the parabolic
boundary. Deep passages in bound resonances are represented at $u=0$;
this convention omits their negative pair energies rather than measuring
exactly parabolic orbits. For unbound approaches,
the single--single channel spans a narrow range set by speeds of order
$v_H$, small relative to the stellar dynamical speed in
Eq.~\eqref{eq:coords}. In this population, $95$ per cent of contacts
have $u\leq\result{ContactUNinetyFive}$. This is a statement about the
adopted incident measure, not a prediction of post-contact flow speed.

\begin{figure}[t]
\includegraphics[width=\columnwidth]{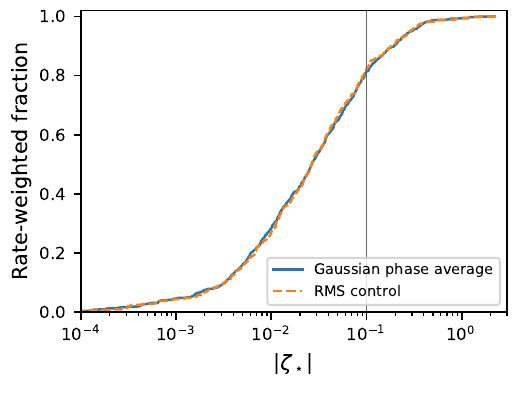}
\caption{\textbf{Calibrated binary--single surface speeds.}
The cumulative rate-weighted distribution of the absolute signed-energy
diagnostic $\zeta_\star$ at first contact. The vertical line marks
$|\zeta_\star|=0.1$; $\result{FcNearPercent}$\% of contacts lie below it.
The Gaussian orbital average and RMS control use the same original
population draws. Intervals describe sampling of resolved first contacts.
See Appendix~\ref{app:passage-energy} for more information.
Produced by \texttt{binary\_analysis.py}.\llmfig}
\label{fig:first-contact-calibration}
\end{figure}

We independently calibrate the binary--single first-contact energy
distribution with isolated Newtonian three-body integrations of the joint
fiducial population, with two independent orbital-phase draws
(Appendix~\ref{app:passage-energy}). At the first inward crossing
of the stellar surface, $\result{FcNearPercent}$\% of resolved rate-weighted contacts
satisfy $|\zeta_\star|<0.1$, where
$\zeta_\star=v_{\rm rel}^2/v_{\rm esc}^2-1$ and
$v_{\rm esc}^2=2G(\Mstar+\Mbh)/\Rstar$.
This corresponds to approximately $5$\% accuracy in the parabolic surface
speed. The $95$\% Monte Carlo interval is
$\result{FcNearLow}$--$\result{FcNearHigh}$\%
(Fig.~\ref{fig:first-contact-calibration}). This calibration measures the
relative speed of the actual star--BH pair before penetration; the incident
coordinate convention and its $u=0$ compression remain as defined above.

The contact ordering depends on the embedded bodies' random motion:
circular, coplanar orbits increase single--single contacts enough to
reverse the fiducial ordering. The binary-mediated share is therefore
conditional on the kinematic closure of Sec.~\ref{sec:disk:kinematics}.
In particular, the
binary--single kernel omits exterior gas evolution and has no SMBH tide: integrations of wide resonances
show suppression, whereas the contact-producing binaries are typically
tighter. We do not extrapolate that wide-binary correction into a
validated contact-rate adjustment (Appendix~\ref{app:channels}).

\begin{table*}[t]
\caption{\textbf{Channel rates.} Rates are per disk per
Myr. ``Contact'' is $x\le1$; ``micro-TDE'' is the stricter $x\le q^{1/3}$
criterion of Ref.~\cite{2022ApJ...933L..28Y}. The $q\le0.2\,\&\,x\le1$
column gives the rate of contacts with low mass ratio. The stack sums the
unrounded channel rates; rounding can affect sums of the displayed values.
Medians are rate-weighted within each channel. The binary--binary row is a
scaled channel (Sec.~\ref{sec:channels:bs}).\llmfig}

\label{tab:channels}
\begin{ruledtabular}
\begin{tabular}{lrrrrrrr}
channel & rate & contact & micro-TDE & $q\le0.2$ & median & median & median \\
 & (Myr$^{-1}$) & (Myr$^{-1}$) & (Myr$^{-1}$) & $\&\,x\le1$ (Myr$^{-1}$) & $q$ & $x$ & $u$ \\
\hline
single--single (Hill shear) & \result{SingleSingleRate} & \result{SingleSingleCollisionRate} & \result{SingleSingleMicroTdeRate} & \result{SingleSingleQsmallXsmallRate} & \result{SingleSingleMedianQ} & \result{SingleSingleMedianX} & \result{SingleSingleMedianU} \\
binary--single (chaotic) & \result{BinarySingleRate} & \result{BinarySingleCollisionRate} & \result{BinarySingleMicroTdeRate} & \result{BinarySingleQsmallXsmallRate} & \result{BinarySingleMedianQ} & \result{BinarySingleMedianX} & \result{BinarySingleMedianU} \\
binary--binary (scaled) & \result{BinaryBinaryRate} & \result{BinaryBinaryCollisionRate} & \result{BinaryBinaryMicroTdeRate} & \result{BinaryBinaryQsmallXsmallRate} & \result{BinaryBinaryMedianQ} & \result{BinaryBinaryMedianX} & \result{BinaryBinaryMedianU} \\
\hline
\textbf{stack} & \result{StackRate} & \result{StackCollisionRate} & \result{StackMicroTdeRate} & \result{StackQsmallXsmallRate} & -- & -- & -- \\

\end{tabular}
\end{ruledtabular}
\end{table*}

\subsection{Mass functions and the contact response}
\label{sec:channels:mass-response}

A change in the mass functions changes both the pairs available to meet and
the rate at which each pair reaches the stellar surface. To separate these
effects, let $\phi_i(m)$ be the normalized number distribution of population
$i$, and define the population-averaged contact response
$\overline{\mathcal W}_c(\Mstar,\Mbh)$ by
\begin{equation}\label{eq:mass-response-definition}
 \frac{\mathrm d^2\Gamma_{c,\mathrm{contact}}}
 {\mathrm d\Mstar\,\mathrm d\Mbh}
 =N_\star N_\bullet\,\phi_\star(\Mstar)\phi_\bullet(\Mbh)
 \overline{\mathcal W}_c(\Mstar,\Mbh).
\end{equation}
The response has units of inverse time and includes the spatial overlap,
incident motions, and probability of contact. For a binary channel it also
includes the binary abundance and the conditional average over the other
components and orbital elements. It is therefore conditional on those
population choices. Transforming from BH mass to $q$ gives
\begin{equation}\label{eq:mass-response-convolution}
 \begin{aligned}
 \frac{\mathrm d\Gamma_{\mathrm{contact}}}{\mathrm dq}
 &=N_\star N_\bullet\sum_c\int\mathrm d\Mstar\,\Mstar\,\phi_\star(\Mstar)\\
 &\quad\times\phi_\bullet(q\Mstar)\overline{\mathcal W}_c(\Mstar,q\Mstar).
 \end{aligned}
\end{equation}
The factor $\Mstar$ is the Jacobian. The integral includes only masses
within both adopted mass ranges. Heavier BHs shift the available pairs toward
larger $q$; heavier stars shift them toward smaller $q$. The response weights
these demographic changes before the contact distribution is normalized.

For single--single encounters the local response at fixed orbital
amplitudes has a particularly simple form. Write $s_i(r)=\Sigma_i(r)/N_i$, with $\int s_i\,\mathrm dA=1$, for the
normalized surface profile. Under the factorization of
Eq.~\eqref{eq:population-measure},
\begin{align}
 \mathcal W_{\mathrm{ss}}(\Mstar,\Mbh;r)
 &=\Omega r_H^2 F(<R_\star/r_H;f_e,f_I),
 \label{eq:mass-response-local}\\
 \overline{\mathcal W}_{\mathrm{ss}}
 &=\int\mathrm dA\,s_\star s_\bullet
   \left\langle\mathcal W_{\mathrm{ss}}\right\rangle_{R_\star,\mathrm{orb}}.
 \label{eq:mass-response-spatial}
\end{align}
Here $\mathcal W_{\mathrm{ss}}$ has units of area per time, $F$ is the cumulative
Hill rate kernel. The orbital average is taken at each mass, disk radius,
and stellar radius using Eq.~\eqref{eq:hill-population-average}; the
remaining average uses the specified stellar-radius history.
The Hill entry rate and the contact probability combine into the single
cumulative factor $F$; no separate entry normalization is required.

For nearly parabolic encounters, gravitational focusing gives a transverse
width proportional to $r_p^{1/2}$ at fixed incident speed
~\cite{2004ARA&A..42..549G}. In the circular, coplanar limit, Hill encounters have entry speeds of order
$\vH$, giving the focused width
$b_{\rm graze}\sim[2G(\Mstar+\Mbh)r_p]^{1/2}/\vH$.
In a disk thinner than this width, the intercepted flux scales as a
linear width, $b_{\rm graze}/\rH\propto(r_p/\rH)^{1/2}$,
rather than an impact-parameter area. The Hill entry normalization is
independent of the small target size, giving
$F(<\epsilon)\propto\epsilon^{1/2}$ for $\epsilon=r_p/\rH$.
Finite thickness or random motion can change this scaling; the focused
width here is measured within the Hill encounter, not along an unperturbed
far-field shear trajectory.
Thus the square-root scaling applies in the circular, coplanar limit and
requires a vertically unresolved target. For a population with finite
thickness, $f_I\ll1$ alone does not ensure this: the vertical excursion
$f_I\rH$ must also be small compared with the focused width $b_{\rm graze}$.
At fixed disk radius and SMBH mass, the circular, coplanar limit yields
\begin{align}
 \mathcal W_{\mathrm{ss}}&\propto
 r_H^{3/2}R_\star^{1/2}
 \propto(\Mstar+\Mbh)^{1/2}R_\star^{1/2},
 \label{eq:mass-response-cold}\\
 \mathcal W_{\mathrm{ss}}&\propto
 \Mstar^{(1+\gamma_R)/2}(1+q)^{1/2},\qquad
 R_\star\propto \Mstar^{\gamma_R}.
 \label{eq:mass-response-cold-q}
\end{align}
Increasing stellar mass raises both the encounter scale and the target
size; increasing BH mass raises only the encounter scale. For $q<1$,
the explicit $(1+q)^{1/2}$ dependence is weak. The adopted zero-age
radius law gives a stellar-mass exponent of
$\result{MassResponseStarExponent}$. At $r=10^4r_g$, after normalization
at $(\Mstar,\Mbh)=(100,15)\,\Msun$, Eqs.~\eqref{eq:mass-response-cold}
and~\eqref{eq:mass-response-cold-q} reproduce the circular, coplanar kernel
across $50\leq\Mstar/\Msun\leq300$ and $5\leq\Mbh/\Msun\leq40$ with a
maximum fractional discrepancy of approximately
$\result{ColdDomainMassMaxPercent}$\%. This checks the mass scaling at
$f_e=f_I=0$; it does not establish the same accuracy for finite random motions.

\begin{figure}[t]
\includegraphics[width=\columnwidth]{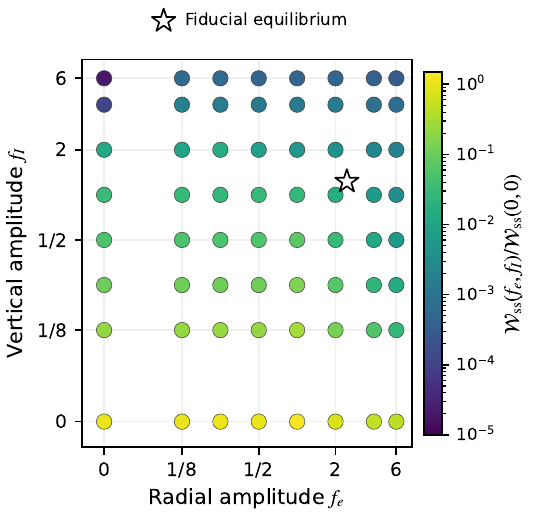}
\caption{\textbf{Vertical random motion strongly reduces the contact response.}
The color gives the single--single response at one fixed target size,
divided by its circular,
coplanar value for a zero-age $100\,\Msun$ star and a $15\,\Msun$ BH at
$r=10^4r_g$, where $R_\star/\rH=\result{ColdDomainEpsilon}$.
Circles mark the evaluated orbital amplitudes; the white star marks the
fiducial RMS input, evaluated by interpolation. Its conditional response is
approximately $\result{ColdDomainFidPercent}$\% of the circular, coplanar value.
The response is evaluated at fixed $R_\star/r_H$;
Eq.~\eqref{eq:hill-geometric-scaling} describes its target-size dependence.\llmfig}
\label{fig:mass-response-cold}
\end{figure}

Figure~\ref{fig:mass-response-cold} shows how the contact response changes
as radial and vertical random motions increase, holding both masses and
the disk radius fixed. At $f_e=0$, a vertical amplitude
$f_I=1/8$ reduces the response to
$\result{ColdDomainThinPercent}$\% of its circular, coplanar value. The fiducial
RMS amplitudes for the reference pair are
$(f_e,f_I)=(\result{ColdDomainFidFe},\result{ColdDomainFidFi})$, where the
conditional response at those amplitudes is approximately
$\result{ColdDomainFidPercent}$\% of that value. Orbital averaging gives
$\result{AvgWColdPercent}$\% of the cold response, a factor
$\result{AvgWReferenceRatio}$ above evaluation at the RMS point.

Random motions change both the contact rate and its dependence on
stellar size. At fixed incoming speed $v_{\rm in}$, gravitational focusing
gives $b_{\rm foc}\simeq[2G(\Mstar+\Mbh)R_\star]^{1/2}/v_{\rm in}$.
The relative vertical thickness is
$H_{\rm rel}=f_Ir_H=r i_{\rm rel}$.
Using $\epsilon=R_\star/r_H$, $s=v_{\rm in}/(\Omega r_H)$,
and $\Omega^2r_H^3=G(\Mstar+\Mbh)/3$ in the geometric ansatz removes
the explicit Hill scales from the response:
\begin{equation}\label{eq:hill-geometric-scaling}
 \begin{aligned}
 \mathcal W_{\mathrm{ss}}&\simeq
 A(f_e)\sqrt{\frac{G(\Mstar+\Mbh)R_\star}{3}}\\
 &\quad\times\min\!\left(1,\frac{b_{\rm foc}}{H_{\rm rel}}\right).
 \end{aligned}
\end{equation}
Here $A$ describes the planar incoming flux and access to close approaches;
the second factor is one for coplanar motion. The intercepted flux scales
as a width when $H_{\rm rel}\ll b_{\rm foc}$ and as a focused area divided
by thickness when $H_{\rm rel}\gg b_{\rm foc}$.
Direct Hill integrations validate these limiting scalings and the turnover
with vertical thickness. We use the numerical kernel for quantitative rates.

Writing $M_{\rm tot}=\Mstar+\Mbh$, the limiting mass and radius dependence is
\begin{equation}\label{eq:mass-response-general}
 \frac{\mathcal W_{\mathrm{ss}}}{A(f_e)}\propto
 \begin{cases}
 M_{\rm tot}^{1/2}R_\star^{1/2}, & H_{\rm rel}\ll b_{\rm foc},\\
 \dfrac{M_{\rm tot}R_\star}{v_{\rm in}H_{\rm rel}}, & H_{\rm rel}\gg b_{\rm foc}.
 \end{cases}
\end{equation}
At fixed $A$, physical incoming speed, vertical thickness, and mass ratio
$q$, a radius law $R_\star\propto\Mstar^{\gamma_R}$ gives
$\mathcal W_{\rm ss}\propto\Mstar^{(1+\gamma_R)/2}$ in the thin limit
$H_{\rm rel}\ll b_{\rm foc}$ and
$\mathcal W_{\rm ss}\propto\Mstar^{1+\gamma_R}$ in the thick limit
$H_{\rm rel}\gg b_{\rm foc}$.
At fixed stellar mass, the explicit dependence on mass ratio is
$(1+q)^{1/2}$ in the thin limit and $(1+q)$ in the thick limit.
Orbital conditions still enter through $A(f_e)$, $v_{\rm in}$, and
$H_{\rm rel}$, which can themselves depend on mass.
These are conditional scalings. We retain their full kernels inside the
orbital average for each mass pair, rather than insert an RMS amplitude
into a conditional scaling law.

The preceding scalings describe the response at a specified orbital state.
The population calculation averages that response over the orbital
distribution before averaging stellar radii and spatial overlap. Let
$\boldsymbol{f}=(f_e,f_I)$ and let $p(\boldsymbol{f}\mid\lambda)$ denote its density at
fixed physical pair properties $\lambda$. The orbital average is
\begin{equation}\label{eq:W-orbital-average}
 \mathcal W_{\rm ss}^{\rm av}(\lambda)
 =\int d\boldsymbol{f}\,p(\boldsymbol{f}\mid\lambda)\mathcal W_{\rm ss}(\lambda;\boldsymbol{f}).
\end{equation}
For a stellar-radius change at fixed mass, disk properties, and orbital
distribution, we define the conditional slope
$\nu_R=\partial\ln\mathcal W_{\rm ss}/\partial\ln R_\star$.
Differentiating the rate integral gives
\begin{equation}\label{eq:W-average-radius-slope}
 \nu_R^{\rm av}=
 \frac{\int d\boldsymbol{f}\,p\mathcal W_{\rm ss}\nu_R}
      {\int d\boldsymbol{f}\,p\mathcal W_{\rm ss}}.
\end{equation}
To illustrate the magnitude average, consider the vertical orbital amplitude
$H=f_Ir_H$, which is Rayleigh distributed under the two-component Gaussian
tilt model. Here $H$ is the oscillation amplitude, rather than the
instantaneous vertical position, and $H_0$ is its RMS value. Define
$y=H/H_0$ and $x=b_{\rm foc}/H_0$, where $b_{\rm foc}$ is the maximum
impact parameter for stellar contact at the specified incoming speed,
including gravitational focusing. The density of $y$ is $2y e^{-y^2}$,
so the mean geometric factor is
\begin{equation}\label{eq:W-Rayleigh-geometry}
 \begin{aligned}
 g(x)&\equiv\int_0^\infty 2y e^{-y^2}\min(1,x/y)\,\mathrm dy\\
 &=1-e^{-x^2}+\sqrt\pi x\,\operatorname{erfc}(x).
 \end{aligned}
\end{equation}
Here $\operatorname{erfc}$ is the complementary error function.
For $x\ll1$, $g(x)\simeq\sqrt\pi x$; for $x\gg1$ it approaches unity.
In the gravitational-focusing limit, $b_{\rm foc}\propto R_\star^{1/2}$
and the contact rate scales as $b_{\rm foc}g(b_{\rm foc}/H_0)$.
At fixed incoming speed and $H_0$, the interpolating radius exponent
in this example can be approximated as
\begin{equation}\label{eq:W-Rayleigh-radius-slope}
 \nu_R^{\rm av}=\frac12+
 \frac{\sqrt\pi x\operatorname{erfc}(x)}{2g(x)}.
\end{equation}
The contact rate therefore approaches a linear dependence on stellar radius
when vertical excursions are large compared with $b_{\rm foc}$, and a
square-root dependence when they are small. For the population results,
we average the numerical Hill kernel over both eccentricity and inclination;
stellar evolution is included by averaging the rate over the radius history.

At the reference mass pair and disk radius of
Fig.~\ref{fig:mass-response-cold}, the numerical orbital average gives a
target-radius exponent of $\result{AvgWNu}$.
Mass derivatives also include the mass dependence of the orbital
distribution; Appendix~\ref{app:mass-response} gives that derivative.

Binary-mediated contacts require a separate calculation because their
rates also depend on the companion masses, binary separation, and whether
the initial binary is detached.

The response separates two questions: how many pairs of each mass are
available, and how often those pairs make contact. One can change the
assumed mass distributions and reuse the response in
Eq.~\eqref{eq:mass-response-convolution} if the orbital motions, radial
profiles, stellar radii, and binary properties at fixed masses stay the
same. Changing those properties requires recalculating the response.
For example, changing the amount of stirring changes $f_e$ and $f_I$ and
therefore the contact rate even at fixed masses. Appendix~\ref{app:mass-response}
gives the weighted-sum response used in our calculations, followed by a fitted
separable approximation and its semianalytic mass-ratio distribution,
including the scaled binary--binary contribution.
Our binary calculation uses the mass-averaged resonance prescription of
Sec.~\ref{sec:channels:kernels}; it does not calibrate scattering separately
for every combination of component masses.

For evolved stars, the same calculation averages over the radii supplied
by the population's ages and compositions. The simple scaling in
Eq.~\eqref{eq:mass-response-cold} applies only when its orbital and geometric
conditions hold. Its weak explicit dependence on $q$ does not imply weak
mass selection for the population as a whole: migration, embedding,
capture, and binary evolution can change which pairs reach an encounter
and the conditions under which they meet.

\subsection{Sensitivity to the binary separation distribution}
\label{sec:channels:binary-separations}

We apply the same encounter calculation to the three illustrative separation
laws of Sec.~\ref{sec:disk:binary-separations}, holding the mass distributions,
spatial profiles, stellar radii, and standing binary abundances fixed.
Table~\ref{tab:binary-separations} gives the resulting contact rates.
The $\beta_a=-1$ case remains the reference population for the other figures
and tables. The slope family tests how strongly the predicted contact
population depends on the assumed binary separations, without requiring a
separately evolved population model.

Weighting the binaries toward smaller separations ($\beta_a=-2$) raises
the total contact rate to $\result{SepTightRatio}$ times the logarithmically
uniform reference; weighting them toward larger separations
($\beta_a=1/2$) lowers it to $\result{SepWideRatio}$ times that reference.
The binary-mediated contact share changes from
$\result{SepTightBinaryPercent}$\% to $\result{SepWideBinaryPercent}$\%
across these examples (Table~\ref{tab:binary-separations}). Thus the channel
ranking depends on the separation distribution even at fixed orbital
excitation. The single--single samples and their contact rate are identical
in all three calculations.

The comparison holds the number of detached star--BH binaries fixed:
initial states that admit no detached eccentricity are removed and the
remaining population is renormalized to that number. It therefore does
not predict depletion of the binary population by rapid hardening or
stellar contact. All rows also retain the adopted incoming binary flux,
the mass-averaged isolated scattering records, and the scaled
binary--binary channel. These rates retain the binary encounter-rate ansatz of
Sec.~\ref{sec:disk:binary-separations}. The wide-weighted example additionally places more
contact-producing binaries at separations where the SMBH tide during
scattering may matter (Appendix~\ref{app:population-integration}).

\begin{table*}[t]
\caption{\textbf{Illustrative binary separation laws at fixed binary abundance.}
The exponent is defined by $p_a(a)\,\mathrm da\propto a^{\beta_a}\mathrm da$,
with the same lower separation and binary-mass Hill cutoff in every row.
Rates are contacts per disk per Myr under the adopted incoming-flux
prescription and scale with the assumed numbers of stars and BHs in the
disk relative to our fiducial choices (Table~\ref{tab:population}). The ratio
compares the total contact rate with the
$\beta_a=-1$ reference. The last column is the rate-weighted median initial
binary separation among binary--single encounters reaching stellar contact.
The single--single contribution is held fixed across these separation models; binary--binary rates retain
the scaled shape prescription. Numerical entries describe the prescribed
models, without marginalizing over their physical uncertainties.
Produced by \texttt{ansatz-prototype/binary\_separation\_sweep.py}.\llmfig}
\label{tab:binary-separations}
\begin{ruledtabular}
\begin{tabular}{rrrrrr}
$\beta_a$ & binary--single & binary--binary & total & ratio & median $a$ (AU) \\
\hline
\result{SepTightBeta} & \result{SepTightBsRate} & \result{SepTightBbRate} & \result{SepTightTotalRate} & \result{SepTightRatio} & \result{SepTightContactA} \\
\result{SepLogBeta} & \result{SepLogBsRate} & \result{SepLogBbRate} & \result{SepLogTotalRate} & \result{SepLogRatio} & \result{SepLogContactA} \\
\result{SepWideBeta} & \result{SepWideBsRate} & \result{SepWideBbRate} & \result{SepWideTotalRate} & \result{SepWideRatio} & \result{SepWideContactA} \\

\end{tabular}
\end{ruledtabular}
\end{table*}

\subsubsection{Dependence on the two binary abundances}
\label{sec:channels:binary-fractions}

Let $f_{\bullet\bullet}$ be the fraction of BHs in BH--BH binaries and
$f_{\star\bullet}$ the fraction of stars in star--BH binaries, as in
Table~\ref{tab:population}.
We vary these at fixed ambient single densities, kinematics, and kernels,
not at fixed total object counts. Let $k_1$ and $k_2$ denote rates per
unit abundance for star+(BH--BH) and BH+(star--BH) encounters, with
contact-selected coefficients $k_{1,C}$ and $k_{2,C}$. Then
\begin{equation}\label{eq:binary-fraction-bs}
 \begin{aligned}
 \Gamma_{\rm bs}&=f_{\bullet\bullet}k_1+f_{\star\bullet}k_2,\\
 \Gamma_{{\rm bs},C}&=f_{\bullet\bullet}k_{1,C}+f_{\star\bullet}k_{2,C},\\
 \Gamma_{\rm bb}&=f_{\bullet\bullet}f_{\star\bullet}k_{\rm bb}.
 \end{aligned}
\end{equation}
Rate subscripts bs and bb denote encounter channels; fraction
subscripts denote binary compositions. The BH--BH system-count factor
$1/2$ is included in $k_1,k_{1,C}$, and passage multiplicity in $k_{\rm bb}$.
The borrowed binary--binary shape follows the abundance-dependent
binary--single mixture, giving
\begin{equation}\label{eq:binary-fraction-contact}
 \begin{aligned}
 \Gamma_{{\rm bin},C}
 &=\Gamma_{{\rm bs},C}\left(1+\frac{\Gamma_{\rm bb}}{\Gamma_{\rm bs}}\right),\\
 F_{{\rm bin}\mid C}&=
 \frac{\Gamma_{{\rm bin},C}}{\Gamma_{{\rm ss},C}+\Gamma_{{\rm bin},C}}.
 \end{aligned}
\end{equation}
The bb contribution is its resampling expectation;
$\Gamma_{{\rm bin},C}=0$ when both abundances vanish. The unrestricted binary share is
$(\Gamma_{\rm bs}+\Gamma_{\rm bb})/
(\Gamma_{\rm ss}+\Gamma_{\rm bs}+\Gamma_{\rm bb})$.

We define $s$ as a common abundance multiplier relative to the fiducial
fractions: $(f_{\bullet\bullet},f_{\star\bullet})=s(0.10,0.05)$.
Thus $s=1$ is the reference population, while $s=2$ doubles both binary
fractions at fixed ambient single densities. This variation preserves the
ratio of the two binary fractions; it is a controlled abundance change,
rather than a binary-formation or depletion model. Since both fractions
must lie between zero and one, $0\leq s\leq10$.
Along this sequence,
\begin{equation}\label{eq:binary-fraction-common-scale}
 \Gamma_{{\rm bin},C}(s)
 =s\Gamma_{{\rm bs},C}(1)+s^2\Gamma_{{\rm bb},C}(1).
\end{equation}
At $\beta_a=-1$, the coefficients in
Table~\ref{tab:binary-fraction-coefficients} give
\begin{equation}\label{eq:binary-fraction-numerical}
 \Gamma_{{\rm bin},C}(s)\simeq
 \left(\result{FracBsReference}s+\result{FracBbReference}s^2\right)
 \,\mathrm{Myr}^{-1},
\end{equation}
exceeding $\Gamma_{{\rm ss},C}=\result{FracCss}\,\mathrm{Myr}^{-1}$
for $s>\result{FracCommonThreshold}$ within the adopted rate prescription.

For intermediate $\beta_a$, interpolate $\ln k$ linearly in $\beta_a$
between the entries in Table~\ref{tab:binary-fraction-coefficients}.

\begin{table}[t]
\caption{\textbf{Binary-abundance coefficients.}
Rates per disk per Myr per unit abundance in
Eq.~\eqref{eq:binary-fraction-bs}, at fixed ambient singles.
The $\beta_a$-independent quantities are $k_1=\result{FracA}$,
$k_{\rm bb}=\result{FracD}$ (per abundance product), and
$\Gamma_{\rm ss}=\result{FracRss}\,\mathrm{Myr}^{-1}$.
The coefficient $k_2$ includes detached-population selection and
renormalization at fixed star--BH abundance. Each composition uses
$\result{FracSamples}$ draws per coefficient evaluation.
Produced by \texttt{ansatz-prototype/binary\_fraction\_scaling.py}.\llmfig}
\label{tab:binary-fraction-coefficients}
\begin{ruledtabular}
\begin{tabular}{rrrr}
$\beta_a$ & $k_2$ & $k_{1,C}$ & $k_{2,C}$ \\
\hline
\result{FracNodeABeta} & \result{FracNodeAB} & \result{FracNodeAContactA} & \result{FracNodeAContactB} \\
\result{FracNodeBBeta} & \result{FracNodeBB} & \result{FracNodeBContactA} & \result{FracNodeBContactB} \\
\result{FracNodeCBeta} & \result{FracNodeCB} & \result{FracNodeCContactA} & \result{FracNodeCContactB} \\
\result{FracNodeDBeta} & \result{FracNodeDB} & \result{FracNodeDContactA} & \result{FracNodeDContactB} \\
\result{FracNodeEBeta} & \result{FracNodeEB} & \result{FracNodeEContactA} & \result{FracNodeEContactB} \\
\result{FracNodeFBeta} & \result{FracNodeFB} & \result{FracNodeFContactA} & \result{FracNodeFContactB} \\

\end{tabular}
\end{ruledtabular}
\end{table}

\subsection{Rate normalization and stellar supply}
\label{sec:channels:rates}

Equation~\eqref{eq:stack-measure} gives an instantaneous encounter
rate per disk for the prescribed population. Sustaining it requires that
population to remain available; migration and replenishment are not
calculated here. An assumed AGN number density converts these into volumetric rates;
that conversion does not improve the encounter model's physical calibration. A star cannot supply repeated destructive
first contacts, so the maximum destructive-event yield, assuming each contact destroys a fresh star
and there is no replenishment, is
\begin{equation}\label{eq:supply-limit}
 \Gamma_{\rm event}
 =\min\!\left(\Gamma_{x\leq1},\frac{N_\star}{t_{\rm AGN}}\right),
 \qquad
 \mathcal{R}_{\rm event}=n_{\rm AGN}\Gamma_{\rm event}.
\end{equation}
The contact rate in Table~\ref{tab:channels} can exceed the supply of fresh
stars under this assumption. Survival and replenishment then control the
rate of destructive events. Supplemental Material, Sec.~\ref{sup-sup:rates},
collects the rate conversions and the comparison with the different
population of Ref.~\cite{2022ApJ...933L..28Y}.

\section{The orbital return map}
\label{sec:return}

The encounter measure identifies the rate and incident conditions of
systems that reach a specified separation. Its point-mass scattering
histories already include repeated approaches through their minimum
separation. Once the star is affected by a passage, hydrodynamic modeling
also requires identifying which encounter determines the outcome. The
first contact need not be the most consequential one.
A surviving pair can return to a star altered by earlier passages, while
dissipation and external torques change its orbit. Following that sequence
is necessary to identify the final encounter relevant to the outcome of
interest.

The return map provides an economical way to connect those encounters.
Given an initial orbital state and a stellar model, it propagates the pair
through a stellar transit and the exterior return, retaining the orbital
state and timing needed to initialize the next passage.
Coupled to a prescription for the changing star, this evolution can help
target hydrodynamic resources on the final, physically consequential
encounters. Which encounter is most impactful depends on the stellar
response and the chosen outcome; it need not be the passage with the
smallest separation. The map organizes the orbital sequence needed for
that selection, while the response prescription determines how the star
survives or changes between passages.

Here we develop the map and demonstrate its dynamics for specified initial
states. It produces the isolated-return, tidal-return, and recurrence
sequences in Supplemental Material, Figs.~\ref{sup-fig:returnmap},
\ref{sup-fig:tidal-scaling-test}, and~\ref{sup-fig:clock}. We do not propagate the population samples
through this map or compute a population distribution of the most impactful
passages. The population
rates and distributions in Table~\ref{tab:channels} retain the counting
convention of Sec.~\ref{sec:channels}. The map supplies the orbital tool
needed to turn those incident conditions into a sequence of simulation
targets.
For simplicity, our illustrative sequences use prescribed stellar-response
closures, without a calibrated treatment of mass loss or structural
evolution between passages; see, however, idealized treatments of mass
ejection following impulsive heating~\cite{2024ApJ...967...33C} and of the
surviving star's response to prescribed mass loss~\cite{2025ApJ...987...16B}.

The map describes a surviving bound star--BH pair after any strong
interaction with additional encounter participants has ended. We evolve it
with the composition of two maps: a dissipative transit through the stellar
interior and a return outside the star. For an isolated binary the return is
an exact Kepler orbit; inside an AGN disk it is an orbit of Hill's problem,
Eq.~(\ref{eq:hill-eom}), of which the Kepler return is the $\Omega\to0$ limit.
In the illustrative figures, lengths and times are reported in units of the
initial stellar radius $R_{\star,0}$ and dynamical time
$[R_{\star,0}^3/(G M_{\star,0})]^{1/2}$; these plotting scales do not change the
dimensional equations below. We first develop the transit equations and the
role of angular-momentum loss, then restore the SMBH tide. Supplemental
Material gives the isolated-return examples, recurrence clocks, and
numerical and physical closure tests.

\subsection{Energy and angular momentum through a transit}
\label{sec:return:map}

We retain physical units and define the reduced mass
$\mu=\Mstar\Mbh/(\Mstar+\Mbh)$ and $q=\Mbh/\Mstar$.
Let $\psi(r)$ be the positive gravitational potential depth of the spherical
star, with $\psi(r)=G\Mstar/r$ outside its surface. The orbital energy and
angular momentum of the relative motion are
$E=\mu[v^2/2-(1+q)\psi]$ and
$L=\mu|\bm r\times\bm v|$; the specific angular momentum is $L/\mu$.
For gas at rest relative to the stellar center, the interior trajectory and its
two bookkeeping variables obey
\begin{equation}\label{eq:transit-ode}
 \begin{aligned}
  \dot{\bm r}&=\bm v,&
  \dot{\bm v}&=(1+q)\bm\nabla\psi-\frac{F_D}{\mu v}\bm v,\\
  \dot{\mathcal W}&=F_Dv,&
  \dot{L}_D&=\frac{F_D L}{\mu v}.
 \end{aligned}
\end{equation}
Thus the solver advances the orbit and the accumulated drag work and torque in
the same integration.  We use the standard gravitational-focusing form from
common-envelope wind-tunnel calculations
\cite{2015ApJ...803...41M,2017ApJ...838...56M},
\begin{equation}\label{eq:drag}
  F_D = C_D\,\pi R_{\rm eff}^2\,\rho v^2,\qquad
  R_a = \frac{2G\Mbh}{v^2+c_s^2},
\end{equation}
with either $R_{\rm eff}=R_a$ or the locally limited
$R_{\rm eff}=\min(R_a,H_\rho,r)$, where $H_\rho$ is the density scale height and
$C_D$ is an untuned scalar.  The alternative is gaseous dynamical
friction~\cite{1999ApJ...513..252O},
\begin{equation}\label{eq:ostriker}
  \begin{aligned}
    F_D &= \frac{4\pi G^2\Mbh^2\rho}{v^2}\,I(\Mach),\\
    I(\Mach) &= \begin{cases}
      \min\!\left[\ln\Lambda,\ \tfrac12\ln\tfrac{1+\Mach}{1-\Mach}-\Mach\right],
        & \Mach<1,\\
      \ln\Lambda, & \Mach\geq1,
    \end{cases}
  \end{aligned}
\end{equation}
with $\Mach=v/c_s$ and the finite-time singularity of Ostriker's $I$ at
$\Mach=1$ removed by the regularization of Ref.~\cite{2023MNRAS.522.1763G}, as adopted by Ref.~\cite{2026arXiv260827596S}.
Written the same way, Eq.~(\ref{eq:drag}) with $R_{\rm eff}=R_a$ has
$I_{\rm BHL}=C_D\Mach^4/(1+\Mach^2)^2$, so their high-Mach normalizations differ by $\ln\Lambda/C_D$,
with an additional dependence on Mach number in the transonic regime
(Supplemental Material, Sec.~\ref{sup-sec:return:mach}). Hydrodynamic
comparisons are needed to establish their applicability. Except in
the drag-law comparison in Supplemental Material, Sec.~\ref{sup-sec:return:mach},
the closure is fixed to the raw focusing radius, so
the calculation tests the orbital map rather than calibrating drag.

Integrating Eq.~(\ref{eq:transit-ode}) from the inward to the outward surface
crossing defines the transit operator
\begin{equation}\label{eq:map}
 \mathcal P_R:(E^-,L^-)\mapsto(E^+,L^+),\qquad
 \begin{cases}
 E^+=E^- -\mathcal W,\\
 L^+=L^- -L_D.
 \end{cases}
\end{equation}
The superscripts denote the two crossings and
$\mathcal W=\int F_Dv\,\mathrm dt$ and
$L_D=\int F_D L/(\mu v)\,\mathrm dt$.  At a surface $R$, the state is
admissible when
\begin{equation}\label{eq:surface-state}
 v_r^2=2\left(E^-/\mu+G(\Mstar+\Mbh)/R\right)-[L^-/(\mu R)]^2>0.
\end{equation}
For a nearly parabolic first passage, $E^-\simeq0$ and $e\simeq1$.
A simple estimate follows directly from the work and torque integrals.
If the drag acts over a chord of length $\ell$ at characteristic speed $v_D$,
define $\epsilon_D=F_D\ell/(\mu v_D^2)$. For weak drag,
\begin{equation}\label{eq:passage-estimate}
 \begin{aligned}
 \Delta E&\sim-\mu v_D^2\epsilon_D,&
 \frac{\Delta L}{L}&\sim-\epsilon_D,\\
 a^+&\simeq\frac{G\Mstar\Mbh}{2\mathcal W}.
 \end{aligned}
\end{equation}
Thus even a small fractional loss of angular momentum can bind an initially
parabolic orbit: $E^+\simeq-\mathcal W$, while $L$ changes little.
For Eq.~(\ref{eq:drag}), the same estimate is
$\epsilon_D\sim C_D\pi R_{\rm eff}^2\Sigma_{\rm chord}/\mu$, with
$\Sigma_{\rm chord}=\int\rho\,ds$ and characteristic values of the speed
and effective radius along the chord. The resulting orbit can have a large
apocenter even though its binding energy is set by the passage.
To estimate the pericenter change, expand the osculating Kepler relation
$r_p=L^2/[\mu^2G(\Mstar+\Mbh)(1+e)]$ about $e=1$:
\begin{equation}\label{eq:peri-differential}
 \Delta\ln r_p\simeq-2\frac{L_D}{L}
       +\frac{r_p\mathcal W}{\mu G(\Mstar+\Mbh)}.
\end{equation}
Both terms are small for a weak passage. Drag torque lowers $r_p$, while
energy loss at fixed $L$ raises it.

\subsection{Repeated encounters}
\label{sec:return:kepler}
\label{sec:return:tide}
After capture,  to a first approximation the orbits between close encounters  are nearly Keplerian.  Given the stellar radius $R_n$ expected for any encounter, the
time between encounters can be computed in terms of the appropriate stellar radii as $T(R_n)+T(R_{n+1})$, where
\begin{equation}\label{eq:return}
  T(R)=\sqrt{\frac{a^3}{G(\Mstar+\Mbh)}}[\pi-\eta_R+e\sin\eta_R],
\end{equation}
with $\eta_R=\arccos[(1-R/a)/e]$.
This process halts when the orbit is entirely within the stellar surface.
In the Supplemental Material (Sec.~\ref{sup-sup:isolated}), we illustrate this simplified process the evolved massive-star structure
of Ref.~\cite{2026ApJ...999..123M}\footnote{B.~Mukhija and A.~Kashi, dataset for ``Envelope Inflation and Mass Loss Driven by Energy Deposition in Massive Stars'' (2025), Zenodo, \url{https://doi.org/10.5281/zenodo.15577892}.}.
We find that while stellar encounters can efficiently reduce energy and therefore the apocenter, they do not always
provide sufficient torque to guarantee the pericenter also decreases. Drag alone can nevertheless drive rapid inspiral in massive AGN stars,
as demonstrated by the hydrodynamic simulations of Ma et al.
~\cite{2026arXiv260916290M}.

The SMBH supplies gravitational torque to the binary's orbit;  encounters with large apocenters experience generally
experience greater torque. To account for this torque, instead of using Keplerian orbits, we instead evolve orbits
numerically in the local tidal potential (and corotating frame): the familiar Hill problem of Eq.~(\ref{eq:hill-eom}).
At re-entry the inertial
relative velocity $\dot{\bm r}+\Omega\hat{\bm z}\times\bm r$ defines the
osculating $(E^-,L^-)$ handed to the next transit and the plane in which it
occurs. The environment enters through the SMBH-centric orbital frequency
$\Omega$ multiplied by the characteristic stellar timescale:
\begin{equation}\label{eq:omega-hat}
 \hat\Omega=\Omega\left(\frac{\Rstar^3}{G\Mstar}\right)^{1/2}.
\end{equation}
Here $\Omega$ is the SMBH-centric orbital frequency. The exterior calculation retains the orientation and
phase of the orbit; it omits ambient-gas drag, the stellar quadrupole, and
eccentricity of the binary's center-of-mass orbit about the SMBH.

For an eccentric binary, a weak tidal acceleration can nevertheless change
its small angular momentum substantially: the torque acts at large separation
throughout the long exterior return. Its magnitude is of order
$\mu\Omega^2a^2$ over a time $[a^3/G(\Mstar+\Mbh)]^{1/2}$. With
$L\simeq\mu[2G(\Mstar+\Mbh)r_p]^{1/2}$ and
$\Omega^2=G(\Mstar+\Mbh)/(3\rH^3)$, this gives
\begin{equation}\label{eq:tide-delta}
 |\Delta\ln r_p|\simeq C_\delta\,\delta,\qquad
 \delta\equiv\left(\frac{a}{\rH}\right)^3
             \left(\frac{a}{r_p}\right)^{1/2}.
\end{equation}
Here $r_p$ denotes the osculating Kepler pericenter, which need not equal the
closest approach inside an extended star. The sign depends on the orbital
plane, apsidal direction, and phase relative to the SMBH; prograde or
retrograde motion alone does not determine it. In the small-kick limit,
orientations with opposite projected torques raise and lower $r_p$ with
comparable frequency. This symmetry does not imply that half of a captured
population reaches the core. The isotropic-orientation test in Supplemental
Material, Sec.~\ref{sup-sup:torque-tests}, gives
$C_\delta\simeq\result{TideCoefficient}$ over its tested small-kick range.

The nearly parabolic estimate, Eq.~(\ref{eq:peri-differential}), explains
why drag can bind the binary while barely changing its pericenter.
For the first bound transit of the shallow-capture example, its torque and
energy terms are approximately $-8.2\times10^{-4}$ and
$+7.9\times10^{-4}$; the integrated map gives
$\Delta\ln r_p=-3.77\times10^{-5}$.
By comparison, at $\rH=3000\,\Rstar$ the initial exterior orbit has
$\delta=0.0065$, so Eq.~(\ref{eq:tide-delta}) estimates a typical tidal
change $|\Delta\ln r_p|\simeq0.03$, nearly three orders of magnitude larger
than the drag change. At $\rH=1000\,\Rstar$, $\delta=0.18$ indicates that
large kicks are possible, requiring the full Hill integration rather than
a linear estimate. The tide dominates the initial pericenter evolution
because it acts on the small angular momentum of a highly eccentric orbit
throughout the long excursion, whereas the work and torque of a shallow
stellar transit nearly cancel in the pericenter change.
Numerical tidal-return and fixed-star sequence tests are given in
Supplemental Material, Sec.~\ref{sup-sup:torque-tests}.

These initial returns are very wide. The example has $q=0.1$,
$a=\rtA\,\Rstar$, $e=\rtEcc$, and
$r_a=\rtApo\,\Rstar$. For its $70\,\Msun$,
$\Rstar=\rtRadius\,\Rsun$ reference star, this is an apocenter of
$\rtApoAU$ AU and a first Kepler return time of $\rtYears$ yr.
Environmental effects from other bodies or the ambient medium can perturb the orbit during these excursions.
We do not incorporate these additional effects in our proof-of-concept code;  see
Appendix \ref{app:triple-truncation-example} for a preliminary estimate of the modest impacts of tidal truncation on our results.

\section{Discussion}
\label{sec:discuss}

\subsection{Existing hydrodynamics simulations may miss expected events}
\label{sec:discuss:weight}

Existing hydrodynamic simulations of star-BH merger are modestly mismatched to our predictions in one key respect:
 our contact distribution favors shallow passages near the stellar surface (Fig.~\ref{fig:marginals} and
Table~\ref{tab:channels}), whereas the parabolic star--BH simulations of
Ref.~\cite{2026arXiv260827596S} have exterior pericenters $x=0$ or $0.125$.

Existing parabolic calculations already probe roughly the incident-speed
regime favored by our AGN population~\cite{2026arXiv260827596S,2026arXiv260827595H}.
Our fiducial population also favors small mass ratios, but this conclusion
is highly sensitive to the assumed BH population. The embedded BH and
stellar mass functions are population ansatzes, and their convolution with
the contact response sets the predicted $q$ distribution
[Eq.~\eqref{eq:mass-response-convolution}]. Changing the BH mass spectrum or
its mass limits can therefore shift the preferred mass ratios substantially;
the small-$q$ preference is conditional on these inputs.

\subsection{Population dependence and future applications}
\label{sec:discuss:populations}

The separability of mass
and disk position is an assumption, rather than a consequence of mass segregation.
Continuous-mass nuclear-cusp models correlate mass with orbital binding
energy~\cite{2022ApJ...940..101L,2024ApJ...977....7R,2025ApJ...991..146R}.
Delfavero et al.\ explore segregation-motivated BH mass functions in
McFACTS~\cite{2025arXiv250813412D}, but a spherical source cusp does not fix
the embedded disk census. Disk capture and alignment select that source,
and migration subsequently changes its radial distribution
~\cite{2024MNRAS.528.4958W}. A source-to-disk calculation therefore generally
requires a joint mass--position measure in place of
Eq.~\eqref{eq:population-measure}.

More detailed stellar models offer a direct replacement for these radius
prescriptions. The semianalytic \texttt{starsam} model evolves bulk
composition and supplies stellar mass, radius, and luminosity, within a
chemically homogeneous core-burning approximation
~\cite{2025ApJ...979..245D}. Its accretion-feedback extension changes both
runaway growth and equilibrium stellar properties
~\cite{2026arXiv260422913D}. Radiation-hydrodynamic and stratified-disk
calculations also distinguish thermal feedback from geometric supply limits
~\cite{2024ApJ...974..106C,2025ApJ...987..188C}. Mixing, nuclear evolution,
and the requirement that stars contribute to disk thermal balance link
stellar age, luminosity, abundance, and remnant production
~\cite{2025ApJ...981...16F,2025MNRAS.537.1556F,2026ApJ...997..206X}.

Population models supply complementary ways to relax the fixed census.
McFACTS follow-ups vary hierarchical growth, encounter and disk prescriptions,
while migration and N-body calculations follow pair-up locations and
resonantly interacting populations
~\cite{2025ApJ...993..163C,2025ApJ...989...67D,2026A&A...708A.171V,2026arXiv260610823V,2026arXiv260813641G}.
Disk-structure implementations and thermal-torque calculations show how
migration selection depends on the background disk and on gap opening
~\cite{2024MNRAS.530.3689G,2024MNRAS.530.2114G,2024MNRAS.535.1693G,2025ApJ...982L..13G}.
The thermal-torque prescription must include its published gap-domain
correction~\cite{2024MNRAS.535.1693G}. Resonance criteria additionally depend
on precession, turbulent forcing, and stellar scattering
~\cite{2025arXiv251012895E}. Such models evolve the delivery of pairs;
their pair-up or merger criteria do not supply a finite-star contact outcome.

Some alternatives also require different orbital distributions. Strongly
magnetized turbulent quasar-disk calculations produce stellar wandering,
radially varying mass functions, and altered multiplicity
~\cite{2024OJAp....7E..71H}. Feedback-coupled disks evolve gas and embedded
populations together~\cite{2025MNRAS.537.3396E}. These are alternatives to
the fixed thin-disk population, and cannot in general be represented by
reweighting masses within the present low-amplitude encounter tables.

Future investigations can apply the encounter measure to stellar-response
and emission models, propagating alternative populations into predictions
for BH growth, mergers, and transients. The return map provides the orbital
connection between successive passages.

\subsection{Gas torques and the supply of stellar contacts}
\label{sec:discuss:validity}

Gas-driven binary evolution can change the contact rate in two competing
ways. Hardening can increase the probability of stellar contact in a close
interaction, but it can also drive a binary into contact before the next
interloper arrives, removing it from the detached population. The
binary-mediated contribution therefore depends on the residence time at each
separation and eccentricity as well as the contact probability of an
individual resonance. Our imposed binary census and vacuum resonance
histories specify one limit of this competition. A self-consistent census
would require formation, gas-driven evolution, encounters, and removal to be
balanced together; increasing the binary abundance alone cannot describe
that evolution.

A useful measure of proximity to stellar contact is the exterior
osculating pericenter in stellar-radius units. For a bound pair, write $\chi=a_b(1-e_b)/\Rstar$. Its evolution obeys
\begin{equation}\label{eq:gas-pericenter-timescale}
 \frac{\mathrm d\ln\chi}{\mathrm dt}
 =\frac{\dot a_b}{a_b}-\frac{\dot e_b}{1-e_b}
  -\frac{\dot R_\star}{\Rstar}.
\end{equation}
Here $a_b$ and $e_b$ describe the relative orbit. The criterion $\chi\leq1$
indicates contact only while the exterior Kepler approximation applies. Shrinking $a_b$ brings contact closer, but
circularization raises the pericenter at fixed $a_b$ and can oppose it.
Eccentricity growth reinforces hardening, with particular sensitivity near
$e_b=1$. Stellar expansion can instead bring the surface to an orbit whose
pericenter changes little. Thus a hardening timescale by itself determines
neither the time to contact nor the depth of the next transit.

Embedded BH-binary calculations illustrate why the eccentricity term cannot
be assigned a universal sign. Retrograde binaries can undergo eccentricity
growth that accelerates as the orbit becomes more eccentric
~\cite{2024ApJ...970..107C}. For sufficiently unequal masses, a retained
primary minidisk can instead reverse the pumping and produce an eccentricity
attractor~\cite{2025ApJ...990..137D}. Inclined systems also undergo alignment
and secular excitation~\cite{2024ApJ...964...61D}. These results identify
possible torque mechanisms, but do not fix their strength for a star--BH
binary: an extended stellar envelope replaces the compact primary and its
minidisk, and the tested mass ratios do not reach our main small-$q$
hydrodynamic domain. The required physical input is the joint evolution of
$a_b$, $e_b$, and the stellar surface for the gas flow around that pair.

The comparison must also use the appropriate interval. To preserve an
assumed binary distribution, gas-driven changes must be small over the
waiting time to the next encounter, or be included in its evolution.
To preserve an individual resonant trajectory, they must be small over the
interaction itself. A binary can therefore be strongly processed by gas
between encounters even when its next short interaction is well approximated
by vacuum dynamics. Extrapolating a local torque over either interval also
requires the global disk and migration to evolve slowly enough for the
local flow to remain representative~\cite{2026ApJ...999...91K}.

For newly captured pairs, the same competition is concentrated in the
first wide excursions. At fixed pericenter, the tidal perturbation scales
as $a^{7/2}$ [Eq.~\eqref{eq:tide-delta}], while the Kepler return time scales
as $a^{3/2}$. A weakly bound orbit is therefore most susceptible to tidal
changes precisely when gas, other bodies, and the star have the longest time
to evolve before re-entry. A tidal kick toward a deeper passage can trigger
stronger drag and contraction, suppressing subsequent tidal kicks; a kick
that lifts the pericenter can interrupt the sequence. A surviving third
body adds an angular-momentum reservoir~\cite{2016ARA&A..54..441N}, and a
strong three-body interaction requires following all three bodies through
the passage. Predicting recurrence thus requires comparing the first return
time with the times for environmental perturbations and stellar change,
and retaining the captured orbit's orientation and phase.

\section{Conclusions}
\label{sec:conclude}

We constructed a differential model for the tight star--BH interactions
supplied by an AGN disk. We focused on the tight-interaction rate: the rate
at which incoming systems reach a specified separation, counting each
single--single or binary--single interaction once. This supplies conditions
for stellar contact and capture, rather than a count of every collision in
a repeated-passage sequence. Repeated approaches resolved within the point-mass histories already
contribute through their minimum separation. We have not propagated these
population samples through the stellar return map or counted each later
stellar passage as a separate event.
A population rate for all
collision passages would additionally require the number of passages
produced by each interaction and the survival and evolution of the star. Selecting stellar contacts changes which dynamics matter:
binary--single interactions dominate contacts in the fiducial population,
even though single--single approaches dominate the total encounter count
(Fig.~\ref{fig:marginals} and Table~\ref{tab:channels}). Binaries bring the
bodies together on smaller scales and therefore reach the star more readily.
Our direct three-body calibration shows that most binary--single first
contacts reach the surface close to the local parabolic speed
(Fig.~\ref{fig:first-contact-calibration}), identifying an incident regime
that deserves priority in hydrodynamic studies. The contact rate can also exceed the supply of fresh stars,
making survival and replenishment central to any forecast of destructive
events.

The encounter population is a dynamically selected sample of the objects
embedded in the disk. The ambient BH and stellar mass distributions, their
spatial distributions, and their participation in binaries enter through the
channel rates and contact selection. Our framework connects these inputs to
the masses, mass ratios, penetration depths, and speeds of the delivered
encounters. Within its calibration domain, it provides a way to compare
alternative ambient populations and propagate their assumptions into
population-weighted outcomes. Predicting BH growth or a transient population
therefore requires these encounter weights together with the response to
individual passages.

The return map connects the population-supplied incident conditions to
the later encounters of a surviving system. Its purpose is to help follow
that system toward its final relevant encounter, concentrating detailed
hydrodynamic calculations on the passages that determine the stellar
outcome. At fixed angular momentum, energy loss tightens a Keplerian orbit
while raising its pericenter. In a disk, SMBH torques can instead deepen
or lift the next passage. A first contact therefore does not specify the
conditions of the most impactful later encounter. Identifying that
encounter requires the orbital sequence together with the changing
stellar structure and a criterion for the outcome of interest. We do not include disruption of
intermediate, long-lived binaries by subsequent encounters with ambient
objects. This mechanism can be added straightforwardly by assigning a
disruption probability over each return interval, given an interaction
model and the local density and velocity distribution of perturbers.

Our population model omits radial migration and uses a simplistic binary
prescription, including independent component-mass draws from the ambient
star and BH populations. Binary formation and evolution can select
correlated component masses and orbital properties, changing the encounter
weights and contact rates. The framework can readily be applied to any
chosen binary population by supplying its joint mass and orbital
distributions and abundances, with conditional encounter kernels covering
those states.

We have provided reusable tools for incorporating these encounter dynamics
into calculations of stellar populations in dense AGN disks. The supplied
kernel builders, interpolation routines, and portable encounter measures
allow rates and correlated encounter parameters to be recomputed, saved,
and sampled for a specified disk population. These tools give population
codes such as McFACTS a concrete route to include star--BH encounters:
the population calculation supplies the evolving disk and embedded objects,
while the encounter model supplies their encounter rates and incident
conditions within its calibration domain. Coupling this dynamical input to
stellar-response and orbital-evolution prescriptions will allow such codes
to follow how encounters transform or destroy stars, feed embedded BHs,
and change the disk's sources of mass and energy. This connects the growth
of BHs, their GW progenitors, and stellar transients to the evolution of
the same population hidden inside AGN disks.

\begin{acknowledgments}
R.~O'Shaughnessy acknowledges support from National Science Foundation
awards PHY-2012057, PHY-2309172, and PHY-2206321, and assistance from
Claude and Codex during the implementation, operation, and drafting of the
work described in this manuscript.  In addition to performing the underlying simulations and creating the figures, these
tools produced the first draft of all text presented in this work, as well as supported repeated edits during its preparation. The use of these tools is described in Appendix~\ref{app:llm}.

J.~Nordhaus acknowledges support from U.S. National Science Foundation
awards AST-2009713, AST-2319326, and AST-2511139.
\end{acknowledgments}

\appendix
\section{Velocity frames and matching to a stellar passage}
\label{app:velocity-handoff}

\subsection{Separating the encounter from the SMBH tidal field}

Here the pair is a star and a BH, whether supplied by a single--single
encounter or left bound after a binary-mediated encounter. An additional
encounter participant must be dynamically negligible during the modeled
stellar passage; the SMBH tide is treated separately below. We call the
short segment near and through the star the inner passage, and the orbit
between stellar passages the exterior return. A transit is the passage
from an inward to an outward crossing of the stellar surface.

We distinguish the pair's mutual gravity from the slowly varying SMBH
field, following the scale-separation approach of Thorne and
Hartle~\cite{1985PhRvD..31.1815T}. We use its Newtonian form and assume
nonrelativistic speeds. Let $r_d$ be the disk radius of the pair's CM and
$s$ the BH--star separation. The external gravitational field is that of
the SMBH, with $\Mstar+\Mbh\ll\Msmbh$ and circular orbital angular speed
$\Omega=(G\Msmbh/r_d^3)^{1/2}$. We join the inner and exterior descriptions
at a sphere of separation $R_0$ outside the star, requiring
\begin{equation}\label{eq:velocity-scales}
 \Rstar\lesssim R_0\ll\rH\ll r_d,\qquad
 \rH=\left(\frac{G(\Mstar+\Mbh)}{3\Omega^2}\right)^{1/3}.
\end{equation}
Here $\rH$ is the mutual Hill radius of Eq.~\eqref{eq:hill}. We assume
speeds small compared with $c$ and $G\Msmbh/(r_dc^2)\ll1$.
The SMBH field varies on spatial scale $r_d$ and orbital time
$\Omega^{-1}$; it is neglected only during the short inner passage and
retained during the exterior return.

Choose an origin following the pair's CM on a reference circular free-fall
orbit, with axes that do not rotate relative to local inertial directions.
Let $\bm r_{\rm rel}=\bm r_\bullet-\bm r_\star$ be the displacement from
the star to the BH, so that $s=|\bm r_{\rm rel}|$. The SMBH accelerates
both bodies almost equally. The origin's acceleration
removes that common force, leaving the difference of the SMBH forces across
the pair. Outside the stellar surface, expanding the external potential
about the origin gives the leading relative equation
\begin{equation}\label{eq:velocity-tidal-expansion}
 \begin{aligned}
 \ddot{\bm r}_{\rm rel}
 &=-\frac{G(\Mstar+\Mbh)\bm r_{\rm rel}}{s^3}\\
 &\quad-\left.(\bm r_{\rm rel}\cdot\bm\nabla)\bm\nabla
       \Phi_{\rm SMBH}\right|_{\bm R(t)}
  +O\!\left(\frac{\Omega^2s^2}{r_d}\right),
 \end{aligned}
\end{equation}
where $\Phi_{\rm SMBH}(\bm R)=-G\Msmbh/|\bm R|$, and the potential
 derivatives are evaluated along the reference orbit $\bm R(t)$.
Let $a_{\rm tide}$ be the magnitude of the differential SMBH acceleration
across the pair and $a_{\rm pair}$ that of its mutual gravitational
acceleration. At the matching separation $R_0$, their ratio is
\begin{equation}\label{eq:velocity-scale-ratios}
 \frac{a_{\rm tide}}{a_{\rm pair}}
 \sim\frac{\Omega^2R_0^3}{G(\Mstar+\Mbh)}
   =\frac13\left(\frac{R_0}{\rH}\right)^3\ll1.
\end{equation}
Thus mutual gravity dominates the local approach and the external tide
changes little during a direct inner passage. We neglect the tide on that
short segment and retain it on the exterior return. An additional
encounter participant, such as the other BH in a binary--single encounter,
is distinct from the SMBH. If its mass is $m_3$ and its distance from the
pair's CM is $D_3$, we also require
$[m_3/(\Mstar+\Mbh)](R_0/D_3)^3\ll1$ throughout the inner passage.
An ongoing strong three-body resonance need not satisfy this condition and
must be modeled with the additional body retained.
Drag through the stellar gas is included in the transit, while the exterior
return model neglects ambient-gas drag.

\subsection{Pair energy, velocity frames, and recoil}

An exterior state specifies the pair's relative position and velocity at
a sampled location outside the star, for example at the matching sphere
$s=R_0$. To define its isolated two-body energy, we keep that state and
consider the hypothetical continuation under mutual gravity alone, with the
external tide removed. The energy per unit reduced mass of this isolated
orbit is
\begin{equation}\label{eq:velocity-energy}
 \varepsilon_K=\frac{v_{\rm rel}^2}{2}-\frac{G(\Mstar+\Mbh)}{s}.
\end{equation}
The reduced mass is $\mu=\Mstar\Mbh/(\Mstar+\Mbh)$, the orbital angular
momentum is $\bm L=\mu\bm r_{\rm rel}\times\bm v_{\rm rel}$, and the orbital
energy is $E_{\rm orb}=\mu\varepsilon_K$. If the energy is positive,
conservation of energy gives
\begin{equation}\label{eq:velocity-infinity}
 v_{\infty,K}^2=2\varepsilon_K
 =v_{\rm rel}^2-\frac{2G(\Mstar+\Mbh)}{s}.
\end{equation}
This is the relative speed at infinity of the hypothetical isolated
star--BH orbit, measured in nonrotating axes. The actual star--BH trajectory
inside the AGN disk need not reach a region where both mutual gravity and
the SMBH tide are negligible. In particular, at small $v_{\infty,K}$ the
focusing length $G(\Mstar+\Mbh)/v_{\infty,K}^2$ can exceed $\rH$.
Infinity therefore defines the continuation of the isolated two-body
problem, rather than a location the disk trajectory must visit.

The tidal potential has scale $\Omega^2R_0^2$. Although small relative to
the pair potential, it can change a nearly zero $\varepsilon_K$ appreciably.
The Hill tide changes
pair energy and angular momentum even when the Jacobi integral is conserved.
Negative $\varepsilon_K$ describes a bound isolated orbit with no real
speed at infinity; the return map retains that signed energy.

The Hill solver stores relative positions $\bm\rho$ in corotating axes
and their derivatives $\dot{\bm\rho}$. If $\mathsf Q(t)$ rotates these
axes into the nonrotating axes above, the physical relative position and
velocity are
\begin{equation}\label{eq:velocity-frame}
 \bm r_{\rm rel}=\mathsf Q\bm\rho,\qquad
 \bm v_{\rm rel}=\mathsf Q\bm w.
\end{equation}
Here $\bm w=\dot{\bm\rho}+\bm\Omega\times\bm\rho$, so
$v_{\rm rel}^2=w^2$. This coordinate transformation
retains the tidal force; neglecting it on the short inner passage is the
separate scale approximation of Eq.~\eqref{eq:velocity-scale-ratios}.

This use of relative velocity does not require $q\ll1$. In the pair CM frame
the individual velocities obey
\begin{equation}\label{eq:velocity-recoil}
 \begin{aligned}
 \bm v_\bullet-\bm V_{\rm CM}&=\frac{\bm v_{\rm rel}}{1+q},\\
 \bm v_\star-\bm V_{\rm CM}&=-\frac{q\bm v_{\rm rel}}{1+q}.
 \end{aligned}
\end{equation}
An instantaneous change to the stellar rest frame gives the BH velocity
$\bm v_{\rm rel}$ for any mass ratio. The pair's CM frame approximates
stellar rest only when $q\ll1$. The star also accelerates under the BH's
gravity; its recoil is included through the total mass in the relative
acceleration and the reduced mass in orbital energy and angular momentum.

Let $\bm v_{g,\star}$ be the local stellar-gas velocity measured relative
to the stellar center. Drag depends on the BH--gas speed
$|\bm v_{\rm rel}-\bm v_{g,\star}|$; for the prescribed gas at rest,
$\bm v_{g,\star}=0$ and this reduces to $v_{\rm rel}$. The corresponding
Mach number is this local speed divided by the local sound speed. It must
be evaluated along the passage through the star, rather than replaced by
$v_{\infty,K}$.

\subsection{How each channel assigns its velocity coordinate}

The single--single Hill integrations record the energy of Eq.~\eqref{eq:velocity-energy}
at closest approach in the point-mass calculation using Eq.~\eqref{eq:velocity-frame}.
These scattering integrations use point masses, including at separations
later labeled as contact. The condition $x\le1$ therefore marks a candidate
stellar penetration in the adopted incident measure. It does not resolve
the trajectory inside the extended star or determine the hydrodynamic
outcome. A stellar-passage calculation must instead use the exterior state
and match it to the interior potential as described below. The population sampler draws a closest separation, a bound flag, and, for
unbound samples, a positive energy from the tabulation. It uses
$v_{\infty,K}$ in $u$. Bound samples are assigned $u=0$ rather than
retaining their negative energies. This is a compression of the incident
measure; an exactly parabolic orbit and a bound orbit with finite binding
energy are not thereby physically identical.

For a binary--single encounter, $V_{\rm bs}$ instead refers to the incoming
single's motion relative to the initial binary CM. Its kinetic energy is
$\mu_s V_{\rm bs}^2/2$ with the reduced mass $\mu_s$ of
Eq.~\eqref{eq:critical-speed}. The phase-resolved shear-plus-epicycle prescription of
Appendix~\ref{app:kinematics} supplies this speed to the isolated-scattering
kernel through $\eta=V_{\rm bs}/v_c$. It is not the speed of the close
star--BH pair inside the ensuing three-body resonance.
The retained records supply minimum separations for each pair and the final encounter
outcome, but not the signed pair-energy history needed to reconstruct each
inner orbit. The current population prescription assigns $u=0$ to a
non-ionizing branch and $u=V_{\rm bs}/\sqrt{G\Mstar/\Rstar}$ to an ionizing
branch. This assignment reuses the incoming binary--single speed; it does
not measure the close pair's energy or an escaping pair's relative speed.
The scaled binary--binary shape inherits these assignments.

\subsection{Passage energy and the parabolic approximation}
\label{app:passage-energy}

For a star--BH pair, the signed osculating specific energy is
$\epsilon_{\rm pair}=v_{\rm pair}^2/2-G(\Mstar+\Mbh)/r$, where
$\Mstar$ and $\Mbh$ are the stellar and BH masses. A useful measure of the error in treating its
pericenter speed as parabolic is
\begin{equation}\label{eq:parabolic-energy}
 \zeta=\frac{\epsilon_{\rm pair}\rp}{G(\Mstar+\Mbh)},\qquad
 \frac{v_p^2}{2G(\Mstar+\Mbh)/\rp}-1=\zeta.
\end{equation}
The second relation applies to the osculating two-body orbit at pericenter;
it makes $|\zeta|\ll1$ the criterion for a nearly parabolic passage.
For an unbound passage in our coordinates,
$\zeta=u^2x/[2(1+q)]$. For a bound binary with instantaneous semi-major axis
$a_{\rm pair}$, $\zeta=-\rp/(2a_{\rm pair})$; the initial binary's
semi-major axis need not equal $a_{\rm pair}$ during an exchange.
The adopted $u=0$ representation of bound passages discards this signed
energy. The archived resonance kernel retains minimum separations and
final outcomes, so its $u=0$ contribution does not independently measure
the distribution of $\zeta$ at those passages.

We measure first-contact energies in a separate set of isolated Newtonian
three-body integrations~\cite{1983ApJ...268..319H}, retaining the fiducial
population weights and Roche-detachment condition. For each encounter we
record the chronologically first inward stellar-surface crossing by either
BH, before any stellar penetration, and define
\begin{equation}\label{eq:first-contact-energy}
 \zeta_\star=\frac{\epsilon_{\rm pair}\Rstar}{G(\Mstar+\Mbh)}
 =\frac{v_{\rm rel}^2}{2G(\Mstar+\Mbh)/\Rstar}-1.
\end{equation}
This identity measures the surface speed even if third-body perturbations
are still important there; $\zeta$ in Eq.~\eqref{eq:parabolic-energy}
instead refers to the osculating pericenter.

The REBOUND IAS15 integrations preserve the specified incoming energy and
impact parameter. Escape requires outward motion, positive outer energy,
and separation beyond both the initial binary scale and the instantaneous
inner apocenter; unfinished encounters remain censored. The rate-weighted
distribution and bootstrap interval appear in
Fig.~\ref{fig:first-contact-calibration}. These runs calibrate the
conditional first-contact velocity distribution under the adopted isolated
three-body model. They do not replace the channel rates, whose yields
depend on the return horizon and escape prescription. The interval measures
Monte Carlo uncertainty of resolved contacts, excluding population
uncertainty and censored late returns. The two phase samples give
$\result{FcSeedOnePercent}$\% and $\result{FcSeedTwoPercent}$\%.
A paired $\result{FcControlDraws}$-state control increases the starting
separation and horizon and tightens the integration tolerance; its
near-parabolic fraction changes from
$\result{FcPairedReferencePercent}$\% to $\result{FcTightPercent}$\%,
with a paired interval for the difference spanning zero.
The primary ensemble contains $\result{FcHorizons}$ horizon stops,
$\result{FcCaps}$ step caps, and $\result{FcQualityFailures}$
energy-screen failure. Allowing unresolved incident weight to yield
later contacts gives the finite-model bounds
$\result{FcCensorLow}$--$\result{FcCensorHigh}$\%, distinct from the
Monte Carlo interval.

\subsection{Surface matching, depth labels, and bound returns}

For fixed component masses, the changes of the exterior orbital elements
obey
\begin{equation}\label{eq:decc}
 \mathrm d(e^2)=
 \frac{2[L^2\,\mathrm dE+2EL\,\mathrm dL]}
 {\mu^3[G(\Mstar+\Mbh)]^2}.
\end{equation}
In particular, for $E<0$, decreasing $E$ at fixed $L$ decreases $e$ and gives
\begin{equation}\label{eq:dperi}
 \left.\mathrm dr_p\right|_{\mathrm dL=0}
 =-\frac{L^2\,\mathrm de}
 {\mu^2G(\Mstar+\Mbh)(1+e)^2}>0.
\end{equation}
The sign applies to the fixed-angular-momentum contribution only. Drag
also reduces $L$, so the net change in pericenter must retain both terms
in Eq.~\eqref{eq:decc}.

Once an additional encounter participant no longer strongly perturbs the
surviving star--BH pair, we initialize its stellar passage using
$(\varepsilon_K,\bm L)$ and the orbital plane normal to $\bm L$.
An inward crossing of the matching sphere means the separation decreases
through $s=R_0\ge\Rstar$, with radial speed $v_r<0$. Under the isolated
pair's gravity, this speed is
\begin{equation}\label{eq:velocity-surface}
 v_r=-\left[2\left(\varepsilon_K+\frac{G(\Mstar+\Mbh)}{R_0}\right)
                    -\frac{L^2}{\mu^2 R_0^2}\right]^{1/2}.
\end{equation}
The tangential speed is $v_t=L/(\mu R_0)$, and the quantity inside the
square root must be positive. Choosing the radius direction and
orbital plane completes the vector state. The scale assumptions in Eq.~\eqref{eq:velocity-scales} permit this
matching to the spherical-star model; a strong three-body resonance must
instead be evolved with its third body.

Care is also needed with the pericenter label. The incident encounter
measure uses $x_K=r_{p,K}/\Rstar$, the point-mass pericenter. In the
dimensional notation of Sec.~\ref{sec:return:map}, it corresponds to
\begin{equation}\label{eq:velocity-point-depth}
 L_K^2=\mu^2G\Mstar\Rstar[u^2x_K^2+2(1+q)x_K]
 \quad(\varepsilon_K\ge0).
\end{equation}
The illustrated interior passages and return sequences choose
$x_{\rm int}=r_{\rm turn}/\Rstar$, where $r_{\rm turn}$ is the turning
radius in the extended stellar potential before drag is applied. The
positive potential depth $\psi(r)$ is defined in
Sec.~\ref{sec:return:map}. This choice sets
\begin{equation}\label{eq:velocity-interior-depth}
 \begin{aligned}
 L_{\rm int}^2={}&\mu^2(x_{\rm int}\Rstar)^2\\
 &\times\left[\frac{G\Mstar}{\Rstar}u^2
                  +2(1+q)\psi(x_{\rm int}\Rstar)\right].
 \end{aligned}
\end{equation}
Their input depth is an interior label before drag; dissipation can move
the actual closest approach. Inside the star $\psi(r)$ generally differs
from $G\Mstar/r$, so supplying the same numerical $x$ to the two formulas changes
the angular momentum. This distinction persists for a small BH: reducing
$q$ does not remove the difference between the potentials.
A population-to-transit conversion must preserve energy and angular
momentum, using Eq.~\eqref{eq:velocity-point-depth} followed by
Eq.~\eqref{eq:velocity-surface}, or solving
Eq.~\eqref{eq:velocity-interior-depth} with $L_{\rm int}=L_K$ for the corresponding
interior turning radius. The illustrated return sequences specify their own initial orbital states;
they do not propagate the full encounter population. A chosen parabolic
start has zero orbital energy. A bound population sample can occupy the
same $u=0$ histogram bin while having negative orbital energy. To initialize
its bound return, we therefore need the signed energy and angular momentum
in addition to the histogram label. Once initialized, the return map retains
and evolves that orbital state.

After a transit, we express its outgoing position and velocity in the
three-dimensional orbital basis retained by the exterior solver. We convert
that velocity to the Hill coordinate derivative by subtracting
$\bm\Omega\times\bm\rho$. The Hill solver advances that full state to the
next inward crossing, where Eq.~\eqref{eq:velocity-frame} restores the
velocity used for $(E,\bm L)$. The new radius and tangential velocity set
the transit plane. The interior solver receives the signed $E$ and the
magnitude of $L$, while the exterior calculation retains the orientation
basis. Subsequent entries are never reset to an incident $(x,u)$ sample.
Thus the return map retains a captured binary's negative energy even though
the incident population representation compresses bound approaches to
$u=0$.

\section{Channel-kernel construction}
\label{app:channels}

This appendix specifies the conditional encounter kernels and their
population integration in Eq.~\eqref{eq:stack-measure}.
Table~\ref{tab:fewbody-calibration} summarizes the binary--single sampling distribution.

\subsection{Single--single encounters}
\label{sec:channels:ss}

We integrate Hill's equation, Eq.~\eqref{eq:hill-eom}, at the relative
radial and vertical amplitudes $(f_e,f_I)$ supplied by
Sec.~\ref{sec:disk:kinematics}. Each trajectory records the minimum
separation over all passages and the pair energy there. The flux-weighted
coefficient $F(<\epsilon;f_e,f_I)$ gives the rate below $\epsilon\rH$
in units of $\Sigma_n\Omega\rH^2$; stellar contact requires
$\epsilon\leq\Rstar/\rH$. Appendix~\ref{app:hill-kernel-method} specifies
the incoming measure and normalization, and
Appendix~\ref{app:velocity-handoff} specifies the velocity assignment.

\subsection{Binary-mediated encounters}
\label{sec:channels:bs}

The second channel is a chaotic resonance: a star meets a BH--BH binary, or a BH
meets a star--BH binary. The interaction cycles through non-hierarchical
intermediate states before one body escapes, the regime described by ergodic
and flux-based statistical theories~\cite{2019Natur.576..406S,2021PhRvX..11c1020G,2021CeMDA.133...17K}.
We denote the initial binary masses by $m_a$ and $m_b$, the incoming single
mass by $m_s$, and the initial binary semi-major axis and eccentricity by $a$
and $e_0$. With $m_B=m_a+m_b$, $M=m_B+m_s$, and
$\mu_s=m_s m_B/M$, the hard--soft critical speed is defined by zero total
three-body energy, $\mu_s v_c^2/2=Gm_a m_b/(2a)$:
\begin{equation}\label{eq:critical-speed}
 v_c=\left(\frac{Gm_a m_b M}{m_s m_B a}\right)^{1/2}.
\end{equation}
Here $V_{\rm bs}$ is the incoming single body's speed relative to the
initial binary CM, rather than the speed of an inner star--BH pair.
Thus $\eta\equiv V_{\rm bs}/v_c$ measures the incident speed relative to the
binary binding energy: $\eta<1$ is energetically hard and $\eta>1$ permits
ionization~\cite{1983ApJ...268..319H}.

We also report the initial total angular momentum in units of the
circular value at the same binding energy. The initial three-body energy is
\begin{equation}\label{eq:threebody-energy}
 E_0=-\frac{Gm_a m_b}{2a}+\frac12\mu_s V_{\rm bs}^2.
\end{equation}
For $E_0<0$, define $\epsilon_0=-E_0$ and $\mu_B=m_a m_b/m_B$.
The binary and incoming orbital angular momenta have magnitudes
$L_B=\mu_B\sqrt{Gm_Ba(1-e_0^2)}$ and $L_{\rm orb}=\mu_s V_{\rm bs}b$,
where $b$ is the binary--single impact parameter. Their relative angle
is sampled isotropically, and
$L_0=|\boldsymbol L_B+\boldsymbol L_{\rm orb}|$.
With $L_c=G\mu_B\sqrt{m_a m_b m_B/(2\epsilon_0)}$, the normalized coordinate is
\begin{equation}\label{eq:angular-momentum-coordinate}
 \widetilde L_0=L_0/L_c.
\end{equation}
For $E_0\geq0$, we assign $\widetilde L_0=\infty$.
This is the normalization of the Stone--Leigh
 distributions~\cite{2019Natur.576..406S}; it characterizes the initial
conditions, while the outcomes come from the scattering ensemble below.

We construct $\result{RgEncounters}$ realizations of binary--single
encounters from a specified sampling distribution, using isolated Newtonian
three-body scattering experiments~\cite{1983ApJ...268..319H}.
For both compositions, star $+$ (BH BH) and BH $+$ (star BH), we draw
masses from the AGN mass functions of Table~\ref{tab:population}, initial
eccentricities from $p(e_0)=2e_0$, $\eta$ log-uniformly over three decades
across the hard--soft boundary, isotropic binary orientations, and impact
parameters from $p(b)\propto b$ up to the focused strong-interaction radius
(Table~\ref{tab:fewbody-calibration}). We evolve each realization with
REBOUND under the bodies' mutual gravity, omitting the SMBH potential and
gas forces. Each
encounter records, for every pair separately, the minimum separation reached
at any time and the terminal outcome. For the present calculation the required
outputs are the deepest \emph{star--BH} separation, which BH made it, and the
final outcome,
\begin{equation}\label{eq:resonance-map}
 (x_{\rm res},\,j,\,o)\sim P\!\left(\frac{r_{\min}^{\star\bullet}}{a},\,j,\,o
       \mathrel{\Big|}\eta,\,c,\,e_0\leq e_{\max}\right).
\end{equation}
where $c$ is the composition, $j$ labels the BH, and
$o\in\{\text{exchange},\text{flyby},\text{merger},\text{ionization}\}$. We construct this conditional distribution from the numerical ensemble.
For each population sample, we draw the three outputs jointly from one
record and set $r_{\min}=a x_{\rm res}$, preserving their correlations. The
cap $e_{\max}$ keeps star--BH binaries physical: such a binary is only one the
encounter can meet if it is detached at periastron,
$\Rstar\leq a(1-e_0)r_L$ with $r_L$ the star's Roche-lobe fraction, and we remove binaries that admit no detached eccentricity at fixed binary
number. This excludes stellar contacts that predate the encounter.

We retain pair identities throughout: a BH--BH closest approach is not a
stellar contact, and a surviving star--BH pair can contain either the
original companion or the incoming BH. Binary orientations are isotropic,
and terminal outcomes are recorded jointly with the deepest star--BH
approach. Table~\ref{tab:fewbody-calibration} specifies the ensemble.

\begin{table*}[t]
\caption{\textbf{Binary--single scattering ensemble.} The kernel of
Eq.~\eqref{eq:resonance-map} is resampled from these records, conditioned on
composition, $\eta$ and the detachment cap; it is not a pooled table.\llmfig}

\label{tab:fewbody-calibration}
\begin{tabular}{@{}lp{0.35\textwidth}p{0.42\textwidth}@{}}
\toprule
quantity & scattering ensemble & application here \\
\midrule
size & $\result{RgEncounters}$ REBOUND encounters, $\result{RgRecords}$ energy-clean star-bearing records & one record drawn per stack sample \\
bodies & a star and two BHs from the AGN mass functions (Table~\ref{tab:population}), both compositions & the same laws; the composition sets which bodies are drawn \\
binary scale & fixed $a$; minimum separations stored per pair in units of $a$ & $x_{\rm res}=r_{\min}/a$ scaled to each sample's $a$ \\
binary state & thermal $p(e_0)=2e_0$ & truncated at $e_{\max}$ so star--BH binaries are detached at periastron \\
incident speed & $\eta$ log-uniform over three decades across $\eta=1$ & $\eta$ from Eq.~\eqref{eq:critical-speed} with the shear and random speeds of Sec.~\ref{sec:disk:kinematics} \\
geometry & $p(b)\propto b$ to the focused strong-interaction radius; isotropic binary orientation & as sampled \\
\bottomrule
\end{tabular}
\end{table*}

Binary semi-major axes follow the power law of Sec.~\ref{sec:disk:binary-separations}, with
the binary-mass Hill cutoff of Eq.~\eqref{eq:binary-a-hill-cutoff}.
This upper separation limit is comparable to the binary-formation scale used by
Ref.~\cite{2026arXiv260922074R}, rather than a general stability boundary.
The binary--single incident speed selects the scattering kernel;
Appendix~\ref{app:velocity-handoff} specifies the close-pair velocities.

The third channel, binary--binary, is computed for its rate and assumed for its
shape. A star-bearing binary meets a BH--BH binary at Eq.~(\ref{eq:rate})
evaluated with the four-body mass. We then reuse the binary--single distribution
in $(q,x,u)$ with the multiplicity specified in Table~\ref{tab:population}.
Because our calibration does not describe four-body resonances, every such
sample is flagged as scaled. Its contribution appears separately in Table~\ref{tab:channels}.

\subsection{Population integration}
\label{app:population-integration}

For a binary--single draw we choose either
$\star+(\bullet\bullet)$ or $\bullet+(\star\bullet)$. We draw the star from
$p_\star(m)\propto m^{-1}$, each BH independently from
$p_\bullet(m)\propto m^{-2}$, the disk radius from the stated surface-density
profiles, $a$ from the power law of Sec.~\ref{sec:disk:binary-separations} (the reference
case has $\beta_a=-1$),
$e_0$ from $p_e(e_0)=2e_0$ subject to the detachment cap, and the orientation
isotropically. For
$\lambda_c=(r,\Mstar,M_{\bullet,i},a,e_0,b,\cos\theta,\ldots)$, the
population measure $\mathrm d\Pi_c=\pi_c\,\mathrm d\lambda_c$ has the
schematic factorization
\begin{equation}\label{eq:channel-population-measure}
 \begin{aligned}
 \pi_c&\propto p_{r,c}(r)\,p_\star(\Mstar)
   \prod_i p_\bullet(M_{\bullet,i})\\[-0.2ex]
 &\quad\times p_a(a\mid r,\boldsymbol m)\,p_e(e_0)
   \,p_b(b\mid r,\boldsymbol m)\,p_\Omega(\cos\theta).
 \end{aligned}
\end{equation}
with $p_\Omega=1/2$ on $-1\leq\cos\theta\leq1$.
We evaluate Eq.~\eqref{eq:stack-measure} by importance sampling in $\ln r$:
each sample $j$ contributes its rate weight $w_{cj}$ at
$(q_{cj},x_{cj},u_{cj})$, with $\sum_jw_{cj}=\Gamma_c$.
Unit-conservation tests check the reported marginals; the calibration
sample counts, input hashes, and source commit are recorded with the
generated results.

The isolated binary--single records omit the SMBH tide. Matched Hill-frame
tests suppress deep approaches in wide resonances, while contact-producing
binaries are predominantly tight relative to $\rH$. We retain the isolated
kernel for those contacts without extrapolating the wide-binary correction
to their rate.

\label{app:reuse}

The implementation in \texttt{encounter\_outcomes.agn\_encounters} separates
population inputs from the stored Hill and resonance calibrations.
\texttt{stack\_channels} returns the channel samples and their rate weights;
\texttt{build\_population\_weight} converts their sum into a normalized
$(q,x,u)$ measure. The population inputs specify the disk, stellar and BH
mass distributions, binary abundances and separations, and random-motion
prescription. Input-table provenance is recorded in the calibration manifest.

An abundance-only example holds the incident kinematics, coordinate
mapping, and single-object surface densities fixed. Multiplying both adopted
binary fractions by a factor $s$ rescales the channel rates as
\begin{equation}\label{eq:abundance-reweight}
 (\Gamma_{\rm ss}',\Gamma_{\rm bs}',\Gamma_{\rm bb}')
 =(\Gamma_{\rm ss},s\Gamma_{\rm bs},s^2\Gamma_{\rm bb}).
\end{equation}
The same factors multiply the corresponding channel sample weights.
Section~\ref{sec:channels:binary-fractions} gives the separate dependence
on the two binary fractions, including the change in the borrowed
binary--binary contact probability when their ratio changes.
The total rate changes, and the encounter distribution is normalized by
that new total. Given an outcome model marginalized over a specified stellar-state
distribution, let $P_A(q,x,u)$ denote its conditional probability of outcome
$A$. This probability can be integrated against the updated measure to
obtain the encounter rate:
\begin{equation}\label{eq:weighted-outcome}
 \Gamma_A=\int P_A(q,x,u)\,
 \frac{\mathrm d^3\Gamma}{\mathrm dq\,\mathrm dx\,\mathrm du}
 \,\mathrm dq\,\mathrm dx\,\mathrm du.
\end{equation}
Changing stellar radii or incident kinematics requires reevaluating the
coordinate mapping and kernel selection as well as the weights. Changes
outside the calibrated range require new dynamical calibration; a
mass-function change reuses the mass-averaged resonance kernel described in
Sec.~\ref{sec:channels}.

\subsection{Measuring and approximating the contact rate~$\mathcal W$}
\label{app:mass-response}

We compute the contact response from rate-weighted samples. For mass bins
$I$ and $J$, let $P_{\star,I}=\int_I\phi_\star\,\mathrm dm$ and
$P_{\bullet,J}=\int_J\phi_\bullet\,\mathrm dm$. The input-weighted bin average is
\begin{equation}\label{eq:mass-response-estimator}
 \widehat{\overline{\mathcal W}}_{c,IJ}
 =\frac{\sum_{k\in(I,J)}w_{c,k}\,\mathbf1_{x_k\leq1}}
 {N_\star N_\bullet P_{\star,I}P_{\bullet,J}},
\end{equation}
where $w_{c,k}$ is a sample's rate contribution. For single--single
encounters we integrate the conditional contact probability in
Eq.~\eqref{eq:mass-response-local} at the sampled orbital amplitudes,
including their distribution in the population integral. This avoids noise
from rare contact flags while retaining the amplitude dependence.

For the fiducial populations, we approximate this measured response by
\begin{equation}\label{eq:response-separable}
 \overline{\mathcal W}_c\simeq A_c
 \left(\frac{\Mstar}{100\Msun}\right)^{a_c}
 \left(\frac{\Mbh}{15\Msun}\right)^{b_c}.
\end{equation}
The coefficients in Table~\ref{tab:response-approximation} are fitted to
population-averaged contact responses over the adopted mass ranges.
We fit the bin-averaged power laws in logarithmic response,
including the input mass distributions within each bin.

\begin{table}[b]
\caption{\textbf{Fiducial contact-response approximation.} Coefficients of
Eq.~\eqref{eq:response-separable}; $A_c$ is in
$10^{-4}\,\mathrm{Myr}^{-1}$ per available star--BH pair.\llmfig}
\label{tab:response-approximation}
\begin{ruledtabular}
\begin{tabular}{lrrr}
channel & $A_c/(10^{-4}\,\mathrm{Myr}^{-1})$ & $a_c$ & $b_c$\\
\hline
single--single & \result{ApproxSsNorm} & \result{ApproxSsStar} & \result{ApproxSsBh}\\
binary--single & \result{ApproxBsNorm} & \result{ApproxBsStar} & \result{ApproxBsBh}\\
\end{tabular}
\end{ruledtabular}
\end{table}

With the adopted $\phi_\star\propto\Mstar^{-1}$ and
$\phi_\bullet\propto\Mbh^{-2}$, this gives the semianalytic mass-ratio
 distribution
\begin{equation}\label{eq:response-q-approximation}
 \frac{\mathrm d\Gamma_c}{\mathrm dq}\propto
 q^{b_c-2}\,
 \frac{m_{\rm hi}^{a_c+b_c-1}-m_{\rm lo}^{a_c+b_c-1}}{a_c+b_c-1},
\end{equation}
where $m=\Mstar/\Msun$, $m_{\rm lo}=\max(50,5/q)$,
$m_{\rm hi}=\min(300,40/q)$, and the rate is zero when
$m_{\rm hi}\leq m_{\rm lo}$. We estimate a relative sampling error
$O(\result{ApproxSamplingScale})$ on integrated rates and a systematic
error $O(\result{ApproxSystematicScale})$ in the approximation of
Eq.~\eqref{eq:response-separable}.

The binary--binary prescription borrows the rate-weighted binary--single
outcomes, so in expectation
\begin{equation}\label{eq:response-bb-scaling}
 \overline{\mathcal W}_{\rm bb}
 =\frac{\Gamma_{\rm bb}}{\Gamma_{\rm bs}}\,
  \overline{\mathcal W}_{\rm bs}.
\end{equation}
Here $\Gamma_{\rm bb}$ includes the prescribed passage multiplicity. The
fiducial ratio is approximately $\result{ApproxBbRatio}$; the normalized
binary--binary and binary--single mass-ratio distributions coincide.
This is the assumed four-body shape, rather than an independent
four-body scattering result.

Mass scalings also depend on changes in the orbital distribution. For a
smooth density on a fixed domain and a parameter $y$,
\begin{equation}\label{eq:W-average-general-slope}
 \frac{\partial\ln\mathcal W^{\rm av}}{\partial\ln y}
 =\left\langle
 \left.\frac{\partial\ln\mathcal W}{\partial\ln y}\right|_{\boldsymbol{f}}
 +\frac{\partial\ln p}{\partial\ln y}
 \right\rangle_{p\mathcal W}.
\end{equation}
The normalized weight in this expectation is
$p\mathcal W/\mathcal W^{\rm av}$. For independent Rayleigh amplitudes
with RMS values $s_e$ and $s_I$, the distribution term is
\begin{equation}\label{eq:W-Rayleigh-score}
 \frac{\partial\ln p}{\partial\ln y}
 =2\sum_{j=e,I}
 \left(\frac{f_j^2}{s_j^2}-1\right)
 \frac{\partial\ln s_j}{\partial\ln y}.
\end{equation}
Consequently the conditional mass powers are retained, while the
population powers include the mass dependence of damping, stirring,
thickness, and incident velocities. The mass-function convolution uses
the resulting averaged response without changing its Jacobian or input
mass distributions.

These coefficients summarize the adopted spatial profiles, stellar-radius
history, velocity closure, binary abundances and separations, and companion
spectrum. The single--single exponents include the mass dependence of the
orbital motions.

\section{The single--single Hill rate coefficient}
\label{app:hill-kernel-method}

\subsection{Incoming measure and normalization}

The quantity $F(<\epsilon;f_e,f_I)$ is the cumulative dimensionless
encounter-rate coefficient for an incoming pair with relative epicyclic
amplitudes $(f_e,f_I)$ to attain separation below $\epsilon\rH$.
An incoming encounter is counted once, using its smallest separation over
all resolved passages. The rate per target is $\Gamma(<\epsilon)=\Sigma_n K(<\epsilon)$,
with dimensional cumulative rate kernel
\begin{equation}\label{eq:hill-kernel-units}
 K(<\epsilon)=\Omega\rH^2 F(<\epsilon).
\end{equation}
Thus $K$ has dimensions of area per time, and $F$ is not restricted to
the interval $[0,1]$. For distinct target and perturber species, multiplying
by the target surface density gives the rate per disk area. A further
factor $1/2$ is needed only when counting unordered pairs of identical
species.

Conditioning on an encounter that enters the mutual Hill sphere gives
the normalized depth distribution used by the sampler:
\begin{equation}\label{eq:hill-kernel-conditional}
 P_H(<\epsilon;f_e,f_I)
 =\frac{F(<\epsilon;f_e,f_I)}{F(<1;f_e,f_I)},
 \qquad 0\le\epsilon\le1.
\end{equation}
The denominator is the Hill-entry coefficient $F(<1)$, rather than the
nominal rate prefactor two in Eq.~\eqref{eq:rate}.

We integrate the relative point-mass equations~\eqref{eq:hill-eom},
following the standard Hill encounter setup
\cite{1986CeMec..38...67H}. These trajectories include the SMBH tide and
Coriolis force, but no gas drag or extended stellar potential. The stellar
surface is imposed afterward as the threshold $\epsilon=\Rstar/\rH$.
To keep the dimensional scales explicit in this section, write
\begin{equation}\label{eq:hill-kernel-k}
 k\equiv G(\Mstar+\Mbh),\qquad
 \rH=\left(\frac{k}{3\Omega^2}\right)^{1/3}.
\end{equation}
Thus $k$ has dimensions of length cubed per time squared.
Let $a=b\rH$ be the signed dimensional guiding-center offset,
$A_e=f_e\rH$ and $A_I=f_I\rH$ the epicyclic amplitudes, and
$\theta_j=\Omega t+\phi_j$. Without mutual gravity the exact solution is
\begin{equation}\label{eq:hill-kernel-epicycle}
 \begin{aligned}
 x&=a-A_e\cos\theta_e,& \dot x&=\Omega A_e\sin\theta_e,\\
 y&=y_{{\rm gc},0}-\tfrac32\Omega at+2A_e\sin\theta_e,\\
 \dot y&=-\tfrac32\Omega a+2\Omega A_e\cos\theta_e,\\
 z&=A_I\sin\theta_I,& \dot z&=\Omega A_I\cos\theta_I.
 \end{aligned}
\end{equation}
Dots here denote dimensional time derivatives. Both phases are uniform
and independent. The guiding centers drift past one another at
$3\Omega|a|/2$, independent of epicycle phase. The differential incoming
rate coefficient, with dimensions of area per time, is therefore
\begin{equation}\label{eq:hill-kernel-measure}
 \mathrm d\mathcal K=\frac32\Omega|a|\,\mathrm da\,
 \frac{\mathrm d\phi_e\mathrm d\phi_I}{(2\pi)^2}.
\end{equation}
Integration over signed $a$ already includes approaches from both sides;
there is no additional factor two. If $r_{\min}$ is the smallest separation
attained along a trajectory, the cumulative dimensional coefficient is
$K(<\epsilon)=\int\mathrm d\mathcal K\,\mathbf1[r_{\min}<\epsilon\rH]$.
Dividing by the scale $\Omega\rH^2$ and changing variables to $b=a/\rH$
gives
\begin{equation}\label{eq:hill-kernel-integral}
 F(<\epsilon;f_e,f_I)
 =\int_{-B}^{B}\frac32|b|\,\mathrm db\,
 \left\langle\mathbf1[r_{\min}/\rH<\epsilon]\right\rangle_{\phi_e,\phi_I}.
\end{equation}
The finite offset window is $B=3+4\max(f_e,f_I)$ and carries total
coefficient $K_{\rm window}=3\Omega\rH^2B^2/2$.
Thus $F/(3B^2/2)$ conditions on the computational incoming window,
whereas Eq.~\eqref{eq:hill-kernel-conditional} conditions on encounters
that enter the Hill sphere. The supplement gives the importance-sampling
estimator and its numerical implementation.

For each incoming offset and phase pair, we integrate the relative Hill
trajectory and retain its minimum separation over all resolved passages.
The stored table is cumulative in this minimum, so repeated pericenter
passages do not count as independent encounters. The population calculation
interpolates this table at the relative amplitudes $(f_e,f_I)$ and samples
passage depths from the conditional distribution $P_H$.
The supplemental material gives the finite-start prescription, integration
and event tolerances, treatment of finite horizons, table construction,
and interpolation rules, together with numerical control tests.

The population calculation starts from the nominal rate scale
$2\Sigma_n\Omega\rH^2$ of Eq.~\eqref{eq:rate}, multiplies it by
$F(<1)/2$, and samples from $P_H$. Consequently the expected rate below
$\epsilon$ is
\begin{equation}\label{eq:hill-kernel-rate-normalization}
 2\Sigma_n\Omega\rH^2\,
 \frac{F(<1)}{2}\,
 \frac{F(<\epsilon)}{F(<1)}
 =\Sigma_n\Omega\rH^2F(<\epsilon).
\end{equation}
The division by two cancels the nominal encounter-window convention;
it is unrelated to identical-species pair counting. No further
normalization factor is needed.

\subsection{Semianalytic limits}

\begin{figure}[t]
\includegraphics[width=\columnwidth]{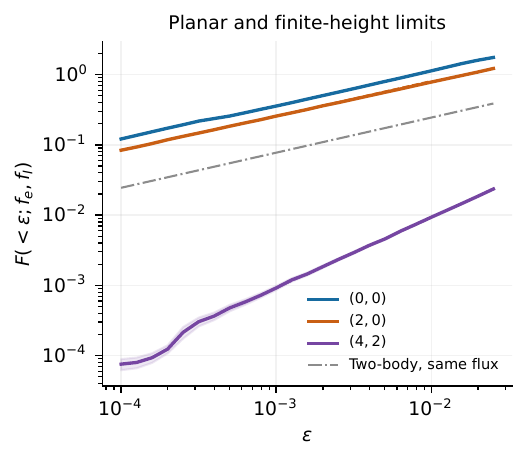}
\caption{\textbf{Planar and finite-height limits of the encounter-rate coefficient.}
The full-Hill coefficients recover square-root behavior in a plane and
approximately linear behavior at finite vertical amplitude. Labels give
$(f_e,f_I)$; shading gives Monte Carlo standard errors. Dashed power-law
guides are anchored at $\epsilon=10^{-3}$ and test exponents, not amplitudes.
The gray curve is the isolated two-body approximation with the same incoming
flux, $\sqrt{6\epsilon}$. Numerical normalization, inner-passage, and
starting-distance checks appear in the supplemental material.\llmfig}
\label{fig:hill-kernel-validation}
\end{figure}

\paragraph{Planar and finite-height focusing.}
In a deep, approximately Keplerian passage, angular momentum satisfies
$\ell^2\simeq2G(\Mstar+\Mbh)r_p$. For a smooth incoming angular-momentum
flux near zero, an exactly planar target selects an interval of angular
momentum, giving $F\propto\epsilon^{1/2}$. A target with resolved vertical
extent selects an area in the two transverse components, giving
$F\propto\epsilon$. This distinction underlies the usual thin- and
thick-target collision scalings~\cite{2004ARA&A..42..549G}.
The limits need not commute: an arbitrarily thin but nonzero vertical
distribution is resolved if the selected encounter depth is sufficiently
small. Figure~\ref{fig:hill-kernel-validation} compares these limiting exponents
with the Hill integrations.

The isolated two-body shear approximation also gives a useful reference.
With the dimensional scale $k$ of Eq.~\eqref{eq:hill-kernel-k},
$v_\infty=3\Omega|a|/2$ and strong focusing give
$r_p\simeq a^2v_\infty^2/(2k)=9\Omega^2a^4/(8k)$.
Substituting $a=b\rH$ and $k=3\Omega^2\rH^3$ yields
$\epsilon\simeq3b^4/8$. With the same incoming flux as
Eq.~\eqref{eq:hill-kernel-integral}, its coefficient is
\begin{equation}\label{eq:hill-kernel-two-body-limit}
 F_{\rm 2b}^{\rm flux}(<\epsilon)
 \simeq\frac32\sqrt{\frac{8\epsilon}{3}}
 =\sqrt{6\epsilon}.
\end{equation}
The archived \texttt{two\_body} comparison instead uses
$2\sqrt{8\epsilon/3}$, retaining the nominal prefactor in
Eq.~\eqref{eq:rate}; it is $4/3$ times the equal-flux reference.
Neither expression supplies the absolute normalization of the full Hill
problem, whose chaotic incoming band differs from isolated focusing
\cite{2023MNRAS.518.5653B}. The comparison in Fig.~\ref{fig:hill-kernel-validation} recovers the
square-root exponent, while the absolute rate remains set by the Hill
trajectories.

\subsection{Completing the amplitude integral}

The amplitude integral uses the measured conditional table, augmented
by integrations at $(f_e,f_I)=(1,0.03),(1,0.06),(2,0.03),(2,0.06)$
within its first vertical interval. Beyond the measured rectangle we
use an explicit focused kinetic continuation. In Hill units, a
conjunction phase $\phi$ has speed
$v^2=f_e^2(\sin^2\phi+\cos^2\phi/4)+f_I^2$ and focused width
$\rho^2=\epsilon^2+6\epsilon/v^2$. The harmonic vertical average is
\begin{align}
 F_{\rm kin}(<\epsilon)&=\langle2v\rho\,h(f_I/\rho)\rangle_\phi,\\
 h(a)&=\frac2\pi\int_0^{\pi/2}
 \sqrt{\max(1-a^2\sin^2\vartheta,0)}\,d\vartheta.
\end{align}
If $\bm a_0$ is the nearest boundary amplitude pair, the continuation is
$F(\bm a)=F(\bm a_0)F_{\rm kin}(\bm a)/F_{\rm kin}(\bm a_0)$
at each depth. This retains the measured boundary access factor while
integrating the full population tail. Independent cells at $(8,4)$,
$(12,6)$ and $(8,1)$ test the continuation, with largest audited
standardized residual $\result{AvgWTailBenchmarkZ}$; low effective
counts in several deep bins limit that test. At the reference pair
the continuation supplies $\result{AvgWTailPercent}$\% of contacts.
This contribution is modeled, not numerical coverage to infinite amplitude.

For cumulative bound fraction $B(<\epsilon)$ the shell bound rate is
$d(FB)$, rather than $B\,dF$. We restrict shell bound rates to
$0\leq d(FB)\leq dF$ and reconstruct the cumulative law used in
both sampling and direct bound-contact integrals. Energy histograms
are differenced between their stored cumulative thresholds; their
within-shell shape remains a finite-resolution approximation. For entries
above the largest stored energy threshold, we reuse the outer available
shell shape. This is an additional approximation in the all-entry velocity
marginal and does not change the cumulative contact-rate integral. Empty
positive-energy shells use a flagged kinetic continuation. The
high-amplitude tail likewise uses a positive-energy kinetic model.
These prescriptions define the joint depth--energy approximation
without conflating bound Hill passages with permanent capture.

\section{Random motions: closure, forcing, and comparison}
\label{app:random-motions}

This appendix collects the moment closure, its forcing sources, and
comparisons at matched gas parameters. We first connect single-body
moments to the channel kernels, then specify turbulent excitation and
compare the resulting contact rates.

\subsection{From single-body moments to encounter velocities}
\label{app:kinematics}

The closure of Sec.~\ref{sec:disk:kinematics} is evaluated at each body's
mass and disk radius. The stellar and BH stirrers are represented by their
population median masses and local surface number densities. In the stirring
estimate $e_{{\rm rel},j}=\sqrt{2}e$ is used separately for each test body;
each test body is therefore heated by the prescribed median-mass
populations. The stationary balance is solved below $h/r$ and capped there
when its root lies above the small-eccentricity domain. The implementation
interpolates $\ln e$ on a grid in $\ln m$ and $\ln r$ before forming the
relative amplitudes of Eq.~\eqref{eq:kin-relative}.

For small-amplitude eccentricity vectors $\boldsymbol e_1$ and
$\boldsymbol e_2$, the ensemble identity is
\begin{equation}\label{eq:kin-vector-covariance}
 \begin{aligned}
 \left\langle|\boldsymbol e_1-\boldsymbol e_2|^2\right\rangle
 &=\left\langle e_1^2\right\rangle+\left\langle e_2^2\right\rangle\\
 &\quad-2\left\langle\boldsymbol e_1\cdot\boldsymbol e_2\right\rangle .
 \end{aligned}
\end{equation}
An analogous relation holds for small-angle inclination vectors. Independent
uniform apsidal and nodal orientations remove the cross terms, giving
Eq.~\eqref{eq:kin-relative}; individual relative amplitudes still vary. The
population calculation integrates a specified amplitude distribution
through the conditional Hill kernel. In general,
\begin{equation}\label{eq:kin-nonlinear-average}
 \left\langle F(<\epsilon;f_e,f_I)\right\rangle
 \ne F(<\epsilon;f_{e,{\rm rms}},f_{I,{\rm rms}}).
\end{equation}
The sign and size of the resulting rate bias depend on the distribution,
correlations, and selected depth. Correlated phases or large eccentricities
and inclinations require an orbit-resolved incident measure; increasing an
RMS amplitude alone does not supply it.

We adopt independent zero-mean isotropic Gaussian eccentricity and
small-angle tilt vectors, whose relative amplitudes have density
\begin{equation}\label{eq:kin-Rayleigh-population}
 p(f_j\mid s_j)=\frac{2f_j}{s_j^2}
 \exp\!\left(-\frac{f_j^2}{s_j^2}\right),\qquad j=e,I,
\end{equation}
where $s_j$ is the corresponding relative RMS amplitude. A zero RMS is
a point mass at zero. Eccentricity and tilt amplitudes are independently
drawn, so their RMS ratio is not imposed on individual realizations.
Linear damping with additive Gaussian noise gives this law; adopting it
at the moments supplied by our nonlinear stirring balance is a
distributional closure, rather than a derivation of that stationary law.
The cap on an RMS does not truncate the Gaussian-vector distribution.
Its tails beyond the small-amplitude damping domain remain a physical
extrapolation of that closure.

Trani and Di Cintio derive stochastic eccentricity- and inclination-vector
evolution with approximate stationary amplitude distributions in the
subsonic, quasi-embedded regime~\cite{2025A&A...703A...6T}. These supply a
way to average over amplitudes while retaining finite forcing correlation
times and checking the phase-average assumptions.
Hydrodynamic calculations of outer inclination damping distinguish
disk-crossing from partially embedded motion
~\cite{2025MNRAS.543..132R,2025MNRAS.543.3768W}.
Imai et al.\ vary the outer eccentricity and epicyclic phase of initially
unbound BH encounters: moderate excitation can widen the gas-capture
window, whereas stronger excitation reduces the gas reservoir carried
into the encounter~\cite{2026arXiv260919281I}. These results motivate
retaining the joint orbital and gas states when averaging over an
encounter population.

At fixed gas surface density, radius, and body mass,
Eq.~\eqref{eq:kin-damping} gives $\te\propto(h/r)^4$. The damping time
must therefore be compared with the disk lifetime before replacing an
initial gas-scale dispersion by the stationary closure.

The speed used for binary--single kernel selection follows from the
small-amplitude epicycle. Relative to the circular shear at its instantaneous
radial position, the unforced epicyclic velocity components can be written
\begin{equation}\label{eq:kin-epicycle}
 \begin{aligned}
 \delta v_X&=f_e\vH\sin\phi_e,\\
 \delta v_Y&=\tfrac12 f_e\vH\cos\phi_e,\\
 \delta v_Z&=f_I\vH\cos\phi_I.
 \end{aligned}
\end{equation}
Averaging over independent uniform phases gives
\begin{equation}\label{eq:kin-rms}
 \left\langle |\delta\bm v|^2\right\rangle
 =\vH^2\left(\frac58f_e^2+\frac12f_I^2\right).
\end{equation}
The $5/8$ is the sum of the radial contribution $1/2$ and the azimuthal
contribution $1/8$. Its contribution to the binary--single incident-speed second moment is
\begin{equation}\label{eq:kin-binary-speed}
 v_{\infty,{\rm bs,rms}}^2=v_{\rm shear}^2+
 \vH^2\left(\frac58f_{e,{\rm bs}}^2+\frac12f_{I,{\rm bs}}^2\right),
\end{equation}
where $\rH$ and $\vH$ refer to the total mass of the single and binary.
The sampled incoming speed selects the binary kernel through
Eq.~\eqref{eq:critical-speed}. The binary center
of mass is assigned the same damping/excitation closure as a point body of
its total mass.

For binary encounters we use the corresponding local Gaussian
epicyclic velocity marginal. At a physical radial offset $B r_H$,
\begin{equation}\label{eq:binary-Gaussian-velocity}
 \frac{\bm V_{\rm bs}}{v_H}
 =\left(E_x,-\frac32B+\frac12 E_y,I_x\right),
\end{equation}
with independent $E_x,E_y\sim\mathcal N(0,s_e^2/2)$ and
$I_x\sim\mathcal N(0,s_I^2/2)$. It recovers the second moment above.
This is a local orbital-distribution closure, not reconstructed phases
of the Hill-entry trajectories. The nominal binary Hill-flux
normalization is retained as the conditional encounter prescription;
phase-dependent entry selection and vertical overlap are not separately
calibrated. The same realized speed is used for kernel selection and
the focused incoming states of the direct and statistical calculations.

Single--single encounters use the conditional Hill kernel of
Appendix~\ref{app:hill-kernel-method}, integrated with
Eq.~\eqref{eq:kin-Rayleigh-population}. We interpolate $\log F$ in
the measured domain and include targeted low-inclination integrations.
Beyond that domain a boundary-matched focused kinetic continuation
supplies the amplitude tail. Its population and contact-rate
contributions are tracked separately; it is an explicit conditional
model, rather than clipping amplitudes to the boundary.

\subsection{Turbulent forcing and stochastic drag}
\label{app:gamma}

\subsubsection{What the planet--disk analogy supplies}

Keplerian dynamics determines the orbital response to an imposed
fluctuating gravitational field. The gas physics determines the field
statistics. The random-walk prescription of
Ida, Guillot, and Morbidelli~\cite{2008ApJ...686.1292I} supplies the
normalization in Eq.~\eqref{eq:kin-turbulence}. Their discussion following
Eq.~(13) adopts $\gamma=10^{-3}$ and suggests $10^{-3}$--$10^{-2}$ for
the MRI model considered there. We adopt these values as our reference
and strong-forcing choices.
Okuzumi and Ormel~\cite{2013ApJ...771...43O}, especially their
Secs.~2 and 5.2, subsequently relate stirring to force correlations and
show that density-wave geometry and vertical structure affect the
conversion from density fluctuations to orbital diffusion. Their
Eq.~(54), $\gamma\simeq5\times10^{-4}(\alpha/0.01)^{1/2}$, applies to
fully active protoplanetary MRI disks; their Eq.~(55) already modifies it
in the presence of a dead zone.

Writing the implemented source as
$\Sturb=A_\gamma\gamma^2(\Sigmag r^2/\Msmbh)^2\Omega$ gives
$A_\gamma=(64\Cturb/\pi)^2/(2\pi)$.
The thin-disk stochastic-force estimate in Sec.~IV.A of
Epstein-Martin, Stone, and Becker~\cite{2025arXiv251012895E} instead assumes
fractional column-density fluctuations proportional to $\sqrt{\alpha}$
and coherence time $\Omega^{-1}$. With their convention
$D_e=\langle(\Delta e)^2\rangle/(2t)$, their Eq.~(48) gives
$\Sturb\sim(\alpha/2)(\Sigma_{\min}r^2/\Msmbh)^2\Omega$.
Here $\Sigma_{\min}$ allows local depletion. Comparing this AGN
prescription with ours requires matching the source rates and diffusion
conventions.

The normalization in Eq.~\eqref{eq:kin-turbulence} can be recovered directly
from Ref.~\cite{2008ApJ...686.1292I}.
Their Eq.~(13) gives a random walk in $e$ per orbital period,
$T_{\rm K}=2\pi/\Omega$.
Their Eqs.~(8) and (27)--(29) fix its reference surface density and
dimensionless gravitational forcing. With
$\Gamma_g=64\Sigmag r^2/(\pi\Msmbh)$ and
$\Sigma_{g,1}=\result{KinTurbReferenceSigma}\,\unit{g\,cm^{-2}}$, the coefficient is
\begin{equation}\label{eq:kin-turb-normalization}
 \Cturb=\frac{\result{KinTurbSourceAmplitude}\,\pi\Msun}
 {64\Sigma_{g,1}(1\,\unit{AU})^2}
 \simeq\result{KinTurbCoefficient}.
\end{equation}
Squaring their random-walk amplitude gives
$\mathrm d\langle e^2\rangle/\mathrm dt=(\Cturb\gamma\Gamma_g)^2/T_{\rm K}$,
which is Eq.~\eqref{eq:kin-turbulence}.
We extend their solar-mass calculation to the SMBH disk using
$\Gamma_g\propto\Sigmag r^2/\Msmbh$ and the local orbital period.

\subsubsection{Support, compressibility, and the inner AGN disk}

Our prescribed $h/r$, gas surface density, and identification $c_s=h\Omega$
define an effective-thermal disk. In a radiation- or magnetically supported
disk, the support speed, thermal response, and turbulent velocity need not
coincide. Radiation diffusion changes compressibility and density contrasts
~\cite{2003ApJ...593..992T}, while opacity and vertical transport can alter
the turbulent support~\cite{2016ApJ...827...10J,2020ApJ...900...25J}.
Magnetic-disk calculations likewise find different support and fluctuation
regimes under different thermal and boundary conditions
~\cite{2024OJAp....7E..19H,2025OJAp....8E..48H,2025OJAp....8E..39S,2026OJAp....963966G}.

Consequently the forcing amplitude $\gamma$ is a model input, rather than
a universal conversion from disk thickness or stress. Applying the closure
to another support regime
requires consistent forcing, damping, and thermodynamics.

\subsubsection{The force statistic that must be calibrated}

For a small perturbing acceleration with radial and azimuthal components
$(a_r,a_\phi)$ on a circular orbit, the eccentricity-vector derivative is
\begin{equation}\label{eq:gamma-vector-forcing}
 \bm b_e(t)\equiv\dot{\bm e}\big|_{\rm force}
 =\frac{2a_\phi\hat{\bm r}-a_r\hat{\bm\phi}}{r\Omega}.
\end{equation}
In a stationary weak-forcing diffusion limit its variance source is
\begin{equation}\label{eq:gamma-correlation}
 S_e=2\int_0^\infty\!\mathrm d\tau\,
       \left\langle\bm b_e(t)\cdot\bm b_e(t+\tau)\right\rangle.
\end{equation}
The orbital basis rotates within this correlation. For exponentially
correlated acceleration components, orbital filtering introduces a
factor involving $[1+(\Omega\tau_c)^2]^{-1}$, suppressing diffusion
for long coherence times
~\cite{2013ApJ...771...43O}.
An AGN calibration therefore requires the spatial and temporal density
correlations, their gravitational field along an orbit, and separate
vertical forcing. Gas velocity noise provides an additional wake-force
source. Bodies sampling the same gas structures can receive correlated
kicks; our quadrature prescription assumes independent kicks.

\subsubsection{Adding stochastic wake forcing without duplicating damping}

Trani and Di Cintio~\cite{2025A&A...703A...6T} prescribe a stochastic gas
velocity field acting through gaseous dynamical friction. They neglect
the disk gravitational field; their mechanism is complementary to our
fluctuating-density gravity source. For the Gaussian midplane density
and local coefficient $\eta_*$ of Eq.~\eqref{eq:single-drag-match}, take
$\sigma_u^2=\alpha_{\rm drag}c_s^2$, $\tau_c=\Omega^{-1}$, and
$\alpha_{\rm drag}=\result{AgnAlphaDrag}$.
Multiplying their published slow-damping stationary moments,
Eq.~\eqref{eq:single-trani-moments}, by their variance-decay rate
$2\eta_*$ gives the additional source prescription
\begin{equation}\label{eq:gamma-drag-sources}
 \begin{pmatrix}S_{\rm drag,e}\\S_{\rm drag,i}\end{pmatrix}
 =\begin{pmatrix}10\\4\end{pmatrix}
   \alpha_{\rm drag}\left(\frac{h}{r}\right)^2
   \frac{\eta_*^2}{\Omega},\qquad \eta_*\ll\Omega.
\end{equation}
The density-force and drag parameters are independent model inputs.
We assume uncorrelated noise when summing the sources. For correlated
forcing in the same gas, a cross source
$S_{\rm cross}$ would obey
$|S_{\rm cross}|\leq2\sqrt{\Sturb S_{\rm drag,e}}$ for the
corresponding diffusion covariance. This term requires a joint gas model
for the density and velocity fluctuations.

We retain $\te$ as the total adopted amplitude-damping time. The local
$\eta_*$ and disk-wave $1/\te$ estimate the same wake response with
different approximations, so adding $\eta_*$ as a second deterministic
damping rate would double count it. Here $\eta_*$ sets the stochastic
source normalization; replacing it there by $1/\te$ gives a normalization
sensitivity. For inclination we adopt the
same amplitude-damping time as an explicit moment closure:
\begin{equation}\label{eq:gamma-inclination-balance}
 \frac{\mathrm d i^2}{\mathrm dt}
 =-\frac{2i^2}{\te}+k^2(\Sturb+\Sstir)+S_{\rm drag,i},
 \qquad k=\result{KinInclinationRatio}.
\end{equation}
This preserves our prescribed $i/e=k$ for the non-drag sources and adds
the distinct vertical drag excitation.
The circular, coplanar and capped brackets retain their prescribed amplitudes.

For the fiducial population, the additional stochastic-drag source is small
relative to density forcing and stirring. A paired population calculation
changes the single--single contact rate by
$(\result{AgnContactChange}\pm\result{AgnContactChangeSE})\%$, where the
uncertainty is Monte Carlo sampling error on the paired change
(Table~\ref{tab:single-contact-comparison}).

\subsection{Comparison with other single-body prescriptions}
\label{app:single-comparison}

We compare the orbital motions of single embedded bodies before an encounter,
and the resulting single--single contact rate. The central question is how
much the predicted eccentricities, inclinations, and relative speeds differ
when the gas and embedded populations are held fixed. Our adopted closure
gives substantially larger amplitudes than the stochastic-drag prediction of
Trani and Di Cintio~\cite{2025A&A...703A...6T}, while retaining relative
speeds comparable to, but below, the gas-support speed used by
Tomar, Hopkins, and Kremer~\cite{2026PhRvD.113f3036T}.
Figure~\ref{fig:single-comparison} compares the amplitudes directly; Table~\ref{tab:single-contact-comparison}
propagates them through the same single--single encounter kernel.

\subsubsection{A common disk and directly comparable moments}

\begin{figure*}[t]
\includegraphics[width=\textwidth]{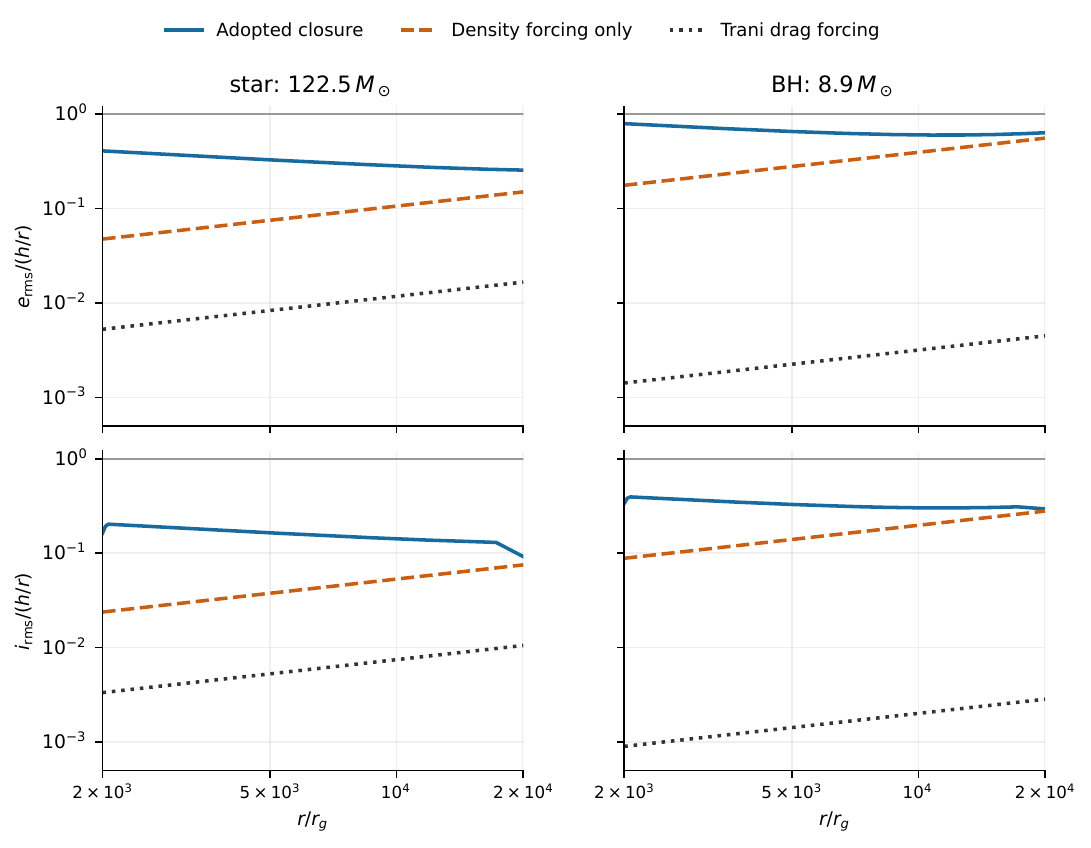}
\caption{\textbf{Predicted single-body orbital amplitudes in the same gas.}
Columns show the representative star and BH; rows show RMS eccentricity
and inclination divided by $h/r$. The solid curves use our full closure,
the dashed curves retain only our density-fluctuation forcing, and the
dotted curves use the Trani slow-damping moments at $\alpha=0.01$.
The density-only curve imposes $i/e=1/2$; the full closure adds a small
vertical stochastic-drag correction. The Trani curves predict
$i/e=\sqrt{2/5}$. The horizontal line marks unity, and none of these
fiducial eccentricity curves reaches the imposed cap.
\llmfig}
\label{fig:single-comparison}
\end{figure*}

All evaluations here use our fiducial gas profile and embedded populations:
$\Msmbh=10^8\,\Msun$, $h/r=0.03$, and
$\Sigmag=10^4\,\unit{g\,cm^{-2}}[r/(10^4r_g)]^{-1}$, with
$r_g=G\Msmbh/c^2$. We retain $\gamma=10^{-3}$ for density-fluctuation
forcing and adopt $\alpha=0.01$ for the stochastic velocity field in the
Trani prescription. These are independent inputs; no relation between $\gamma$ and $\alpha$
is assumed. The full closure also includes the additive stochastic-drag
sources of Appendix~\ref{app:gamma}, while the Trani curves isolate that
mechanism with its original damping normalization. Figure~\ref{fig:single-comparison} shows the resulting amplitudes for a
representative star and BH. Applying the BH drag formula to the stellar
single is our point-mass extension.

We use $e_{\rm rms}^2=\langle e^2\rangle$ and
$i_{\rm rms}^2=\langle i^2\rangle$, rather than the variance of the scalar
amplitude about its nonzero mean. The vector-moment convention is explicit
in their derivation preceding Eq.~(55). For a Rayleigh amplitude with scale
$s$, $\langle e^2\rangle=2s^2$ and
$\langle e\rangle=s\sqrt{\pi/2}$. Consequently a reported mean must be
multiplied by $2/\sqrt{\pi}$ to obtain an RMS amplitude; their published
Eq.~(60) is used here directly as a vector second moment, without another
factor of two.

\subsubsection{Comparison with the gas-support velocity scale}

Tomar et al.\ set $v_{\rm rel}\sim v_{\rm eff}\sim h\Omega$, allowing
thermal, turbulent, and magnetic support to contribute to $v_{\rm eff}$
in their Eqs.~(1)--(2)~\cite{2026PhRvD.113f3036T}. Our effective-thermal identification
is $c_s=h\Omega$. The phase-averaged relative random speed of two
independent singles is
\begin{equation}\label{eq:single-relative-speed}
 v_{{\rm rel},{\rm rms}}^2
 =v_{\rm K}^2\left[\frac58(e_1^2+e_2^2)
                      +\frac12(i_1^2+i_2^2)\right].
\end{equation}
This excludes the local circular shear and mutual gravitational
acceleration during approach. For the adopted amplitudes, the relative random motion exceeds the mutual
Hill speed but remains comparable to the gas-support speed. Damping
therefore does not restore the circular, shear-only limit.

Their velocity scaling therefore remains a useful order-of-magnitude
description of our fiducial result, although our individual amplitudes
depend on mass. This comparison does not test their magnetically
supported disk profiles: fixing $\Sigmag$ while changing $h/r$ is not a
self-consistent change of disk support, and $h\Omega$ need not then be the
thermal sound speed. Nor are the event definitions identical. Their
gravitational scale $b\sim Gm/v_{\rm rel}^2$ yields a per-target rate
$\Gamma\propto n v_{\rm rel}^{-3}$, whereas a fixed-radius, strongly
focused stellar contact has
$b_{\rm contact}^2\simeq2G(\Mstar+\Mbh)\Rstar/v_{\rm rel}^2$ and hence
$\Gamma_{\rm contact}\propto n v_{\rm rel}^{-1}$ at fixed number density.
The single--single rates below instead use the same Hill kernel and
population weights for every velocity prescription, avoiding an
identification of these different rates.

\subsubsection{Comparison with stochastic drag predictions}

Trani and Di Cintio~\cite{2025A&A...703A...6T} couple a gas-velocity
process to drag, with $\sigma_{\rm turb}^2=\alpha c_s^2$ and correlation
time $\Omega^{-1}$ [their Eqs.~(24)--(26)]. To evaluate their damping rate
at our local gas parameters, use a Gaussian vertical profile with
$\rho_0=\Sigmag/(\sqrt{2\pi}h)$ and
$\eta_*=4\pi G^2\rho_0\mt/(3c_s^3)$. The damping-time ratio is
\begin{equation}\label{eq:single-drag-match}
 \eta_*\te=\frac{4\pi}{3\sqrt{2\pi}\,0.78}
             \simeq\result{SingleEtaTe}.
\end{equation}
The normalization differs from our disk-wave damping by only a factor
of approximately two. Their published Eq.~(60) gives
\begin{equation}\label{eq:single-trani-moments}
 \frac{(\langle e^2\rangle,\langle i^2\rangle)}{(h/r)^2}
 \simeq\begin{cases}
  (10\alpha,\,2\alpha/3),&\eta_*\gg\Omega,\\
  (5\alpha\eta_*/\Omega,\,2\alpha\eta_*/\Omega),&\eta_*\ll\Omega.
 \end{cases}
\end{equation}
Our fiducial disk selects the slow-damping branch. Figure~\ref{fig:single-comparison}
shows that removing gravitational stirring still leaves density-forced
amplitudes above the matched-gas stochastic-drag prediction. The latter
gives shear-dominated singles, whereas the adopted closure gives random
motions above the Hill speed.

The opposite mass trends help explain these results. At fixed gas
properties, our density-force source is independent of $\mt$, giving
$e_{\rm density}^2=\te\Sturb/2\propto\mt^{-1}$.
The slow-damping drag prediction instead gives
$\langle e^2\rangle\propto\eta_*\propto\mt$ at fixed $\alpha$.
These are different forcing mechanisms, not competing calibrations of
the same turbulent source. The small difference between the damping coefficients cannot account for
the much larger amplitude difference. We therefore find qualitative
agreement that stochastic forcing prevents complete circularization,
but not quantitative agreement on the eccentricity floor at matched gas
parameters and the stated forcing amplitudes.

The slow-damping drag prediction has $i/e=\sqrt{2/5}$, compared with our
imposed non-drag ratio $1/2$ and its small vertical drag correction.
The moment closure does not specify the full amplitude distribution;
Equation~\eqref{eq:hill-population-average} performs the corresponding
orbital average before constructing contact rates.

\subsubsection{Consequences for single--single stellar contacts}

We substitute these results into the same cumulative Hill kernel,
\begin{equation}\label{eq:single-contact-integral}
 \begin{aligned}
 \Gamma_{\rm contact}^{\rm ss}
 ={}&\int 2\pi r\,\mathrm dr\,\mathrm dm_\star\,\mathrm dm_\bullet\,
 n_\star(m_\star,r)n_\bullet(m_\bullet,r)\\
 &\times\Omega\rH^2\,
 \left\langle F\!\left(<\frac{\Rstar}{\rH};f_e,f_I\right)
 \right\rangle_{\Rstar,\mathrm{orb}|m_\star},
 \end{aligned}
\end{equation}
where $n_i=\mathrm d^2N_i/(\mathrm dm_i\mathrm dA)$ and the average
is over the adopted orbital and stellar-radius distributions at fixed
stellar mass.
Masses, radial populations, and stellar-radius draws are identical between
prescriptions. We evaluate cumulative contact probabilities directly on common population
draws. Gaussian amplitude tails use the same explicit conditional-kernel
continuation as the main population calculation.

\begin{table*}[t]
\caption{\textbf{Effect of the single-body predictions on single--single
contacts.} Rates are raw contacts per disk per Myr before imposing a
stellar-supply limit; no binary channel is included. Ratios use the
adopted closure as reference. Rate errors are Monte Carlo standard errors
only, excluding uncertainty in the kernel, forcing prescription, and
stellar response. The Trani rows extend the point-mass moments to stellar
singles. The adopted and forcing-only predictions average independent
Gaussian orbital vectors; RMS rows retain point controls at the same
second moments.\llmfig}
\label{tab:single-contact-comparison}
\begin{ruledtabular}
\begin{tabular}{lrr}
single-body prescription & contacts/Myr & ratio\\
\hline
Adopted moments, Rayleigh average & $127.4\pm0.4$ & $1.00$ \\
Adopted $e$, $i/e=\sqrt{2/5}$, Rayleigh average & $96.3\pm0.3$ & $0.76$ \\
Density forcing only, Rayleigh average & $556.4\pm1.3$ & $4.37$ \\
Trani moments, RMS evaluation & $2867.5\pm1.9$ & $22.50$ \\
Trani moments, Rayleigh amplitudes & $2975.5\pm2.1$ & $23.35$ \\
Circular coplanar, $e=i=0$ & $3705.6\pm2.2$ & $29.08$ \\
Adopted moments, RMS point control & $50.8\pm0.1$ & $0.40$ \\

\end{tabular}
\end{ruledtabular}
\end{table*}

Table~\ref{tab:single-contact-comparison} compares the resulting rates,
including a Rayleigh-amplitude average and a change of inclination ratio
at fixed eccentricity. The adopted, hotter closure supplies fewer
single--single contacts than the matched-gas stochastic-drag prescription.
The comparison isolates sensitivity to eccentricity, vertical motion,
and amplitude averaging; it does not select the correct forcing model.

\section{Tidal truncation of resonant encounters}
\label{app:triple-truncation-example}

\begin{figure}[t]
\includegraphics[width=\columnwidth]{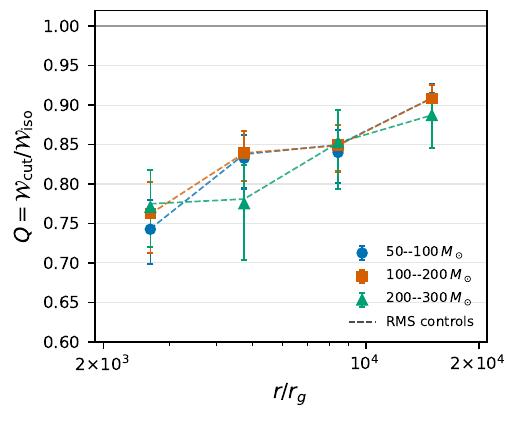}
\caption{\label{fig:triple-truncation-example}
\textbf{Tidal truncation modifies the population-averaged contact coefficient.}
Points give the ratio of contact-rate integrals in radial and stellar-mass
intervals, combining both binary--single compositions. Bars show paired
95\% bootstrap sampling intervals.
The calculation uses the population and incoming scattering measure
of Appendix~\ref{app:channels}, including Roche detachment and the
stellar-radius distribution. Incoming speeds are drawn from the Gaussian epicyclic distribution;
the dashed curves retain the RMS controls. The isolated reference uses the same statistical
kernel and incident supply as the truncated calculation. The statistical
branch counts final-binary contacts at formation.\llmfig}
\end{figure}

The SMBH tidal field can interrupt a resonant encounter by removing
its most extended intermediate triples. We calculate the contact
response using the intermediate-state distributions of
\citet{2019Natur.576..406S} and \citet{2026arXiv260611294M}, terminating a resonance
when an intermediate outer apocenter exceeds the total-triple Hill radius.
An inner star--BH pair already in contact takes precedence.
We integrate both the isolated and truncated models over the adopted
encounter distribution, including the masses, stellar radii, binary sizes,
incoming amplitudes and phases, eccentricities and orientations, with
identical incident weights. We integrate the isolated and truncated
contact coefficients first, then form their ratio:
\begin{equation}\label{eq:triple-truncation-example}
 \mathcal W_C^{\rm tidal}(r)=Q(r)\,\mathcal W_C^{\rm iso}(r).
\end{equation}
Figure~\ref{fig:triple-truncation-example} shows the resulting correction
in stellar-mass intervals. The Hill boundary introduces the additional
scale $r_{H,\mathrm{tot}}(r)/a$.
The alternative statistical kernel supplies both reference and truncated resonance
probabilities. Energetically unbound encounters retain their isolated
contact contribution. We adopt the immediate apocenter veto as the
environmental model.

\section{Use of large language models}
\label{app:llm}

ROS used large language model (LLM) agents, including Claude and OpenAI Codex,
to assist with software development, calculations, analysis, literature
searches, and drafting and revising this manuscript. The agents carried out
assigned calculations, returned preliminary findings, and proposed text and
code changes under the authors' scientific direction. They operated the
simulation, analysis, and plotting tools used to generate the figures in
this paper, including its appendices, and its Supplemental Material, and
produced the initial drafts of the text.

The authors reviewed proposed changes and determined what to include.
Numerical tests and comparisons used to check the calculations are described
in the validation appendices and Supplemental Material. The workflow and its
documentation and review conventions are described in
Ref.~\cite{whitepaper-transparent-agentic-science}. The authors take full
responsibility for the content of this paper.

\bibliography{dynamics}

\end{document}


\title{Supplemental Material: Star--black-hole encounters in AGN disks}
\author{R.~O'Shaughnessy}
\affiliation{\RIT}
\author{Z.~Rosenberg}
\affiliation{\RIT}
\author{J.~Nordhaus}
\affiliation{\RIT}
\maketitle
\renewcommand\thesection{S\Roman{section}}
\renewcommand\thefigure{S\arabic{figure}}
\renewcommand\thetable{S\Roman{table}}
\renewcommand\theequation{S\arabic{equation}}

This supplement contains additional population projections, illustrative return sequences, and numerical and physical
closure diagnostics supporting the encounter measure and return map.
The random-motion closure and its implementation remain in the main paper
and its attached Appendix~\ref{main-app:kinematics}.
Unprefixed equation, figure, table, and section numbers refer to the main paper.

\section{Depth and radial dependence of the encounter measure}\label{sup:population}
The following projections supplement the population-weighted marginals
in main-text Fig.~\ref{main-fig:marginals}. They resolve the crossover as a
function of depth cut and disk radius.
\begin{figure*}[t]
\includegraphics[width=\textwidth]{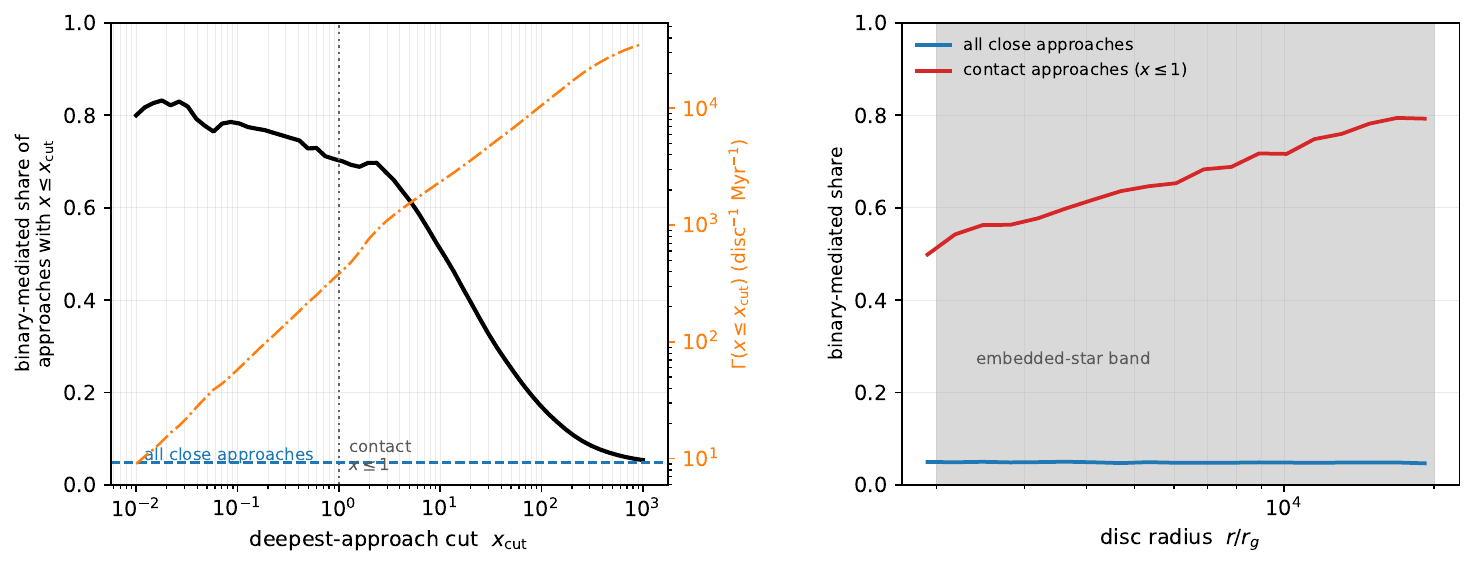}
\caption{\textbf{Chaotic resonances are a minority of encounters and a large
share of contacts.} Left: the binary-mediated share of close approaches with
$x\le x_{\rm cut}$ (black, left axis) against the cumulative rate (orange, right
axis). The share rises from $\result{ChaoticFractionAll}$ per cent with no cut
to $\result{ChaoticFractionCollisions}$ per cent at contact. At smaller depth cuts the relative contributions
depend on the orbital average and conditional binary pericenter law. Right: the same two shares as a function of disk radius, across
the embedded-star band (shaded), showing that the inversion is a property of the
channels and not of a particular annulus.\llmfig}

\label{fig:chaotic}
\end{figure*}
\begin{figure}[t]
\includegraphics[width=\columnwidth]{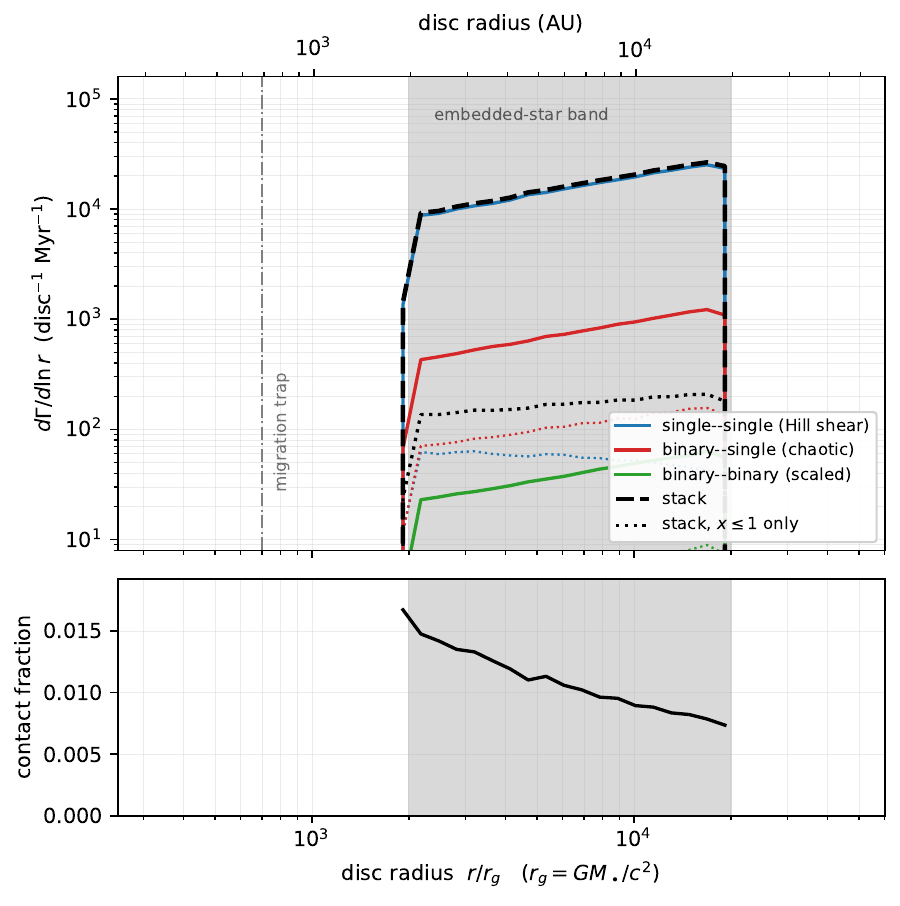}
\caption{\textbf{The stacked rate across the disk.} Top: rate per unit $\ln r$
per channel (solid) and restricted to contact approaches (dotted), with the
stack in black; the shaded band is where the embedded stars are placed and the
dot-dashed line marks the McFACTS migration trap. Bottom: the fraction of close
approaches reaching contact, which falls outward as the Hill radius grows
against a fixed stellar radius. The two trends nearly cancel, so the contact
rate is close to flat in $\ln r$ across the star band.\llmfig}

\label{fig:radius}
\end{figure}
\section{Rate conversion and stellar supply}\label{sup:rates}
The encounter rates and the no-replenishment supply cap of main-text
Eq.~\eqref{main-eq:supply-limit} are placed on the same volumetric normalization
below. The external micro-tidal-disruption estimate uses a different
population and SMBH mass~\cite{2022ApJ...933L..28Y}; numerical proximity is
not a calibration of either model.
\begin{table}[t]
\caption{\textbf{Encounters and supply-limited events.} Per-disk rates
and their volumetric conversion at the explicitly stated rule-of-thumb AGN
number density $n_{\rm AGN}=\result{NAgnRuleOfThumb}\times10^{-4}\,
\unit{Mpc^{-3}}$. The first three rows are encounter rates from
Eq.~(\ref{main-eq:stack-measure}); the fourth applies Eq.~(\ref{main-eq:supply-limit}).
The final row is an external comparison on the same volumetric units.\llmfig}

\label{tab:rate-normalization}
\begin{ruledtabular}
\begin{tabular}{lrr}
quantity & disk$^{-1}$ Myr$^{-1}$ & Gpc$^{-3}$ yr$^{-1}$ \\
\hline
all close approaches & \result{StackRate} & \result{VolAllRuleOfThumb} \\
stellar contact, $x\leq1$ & \result{StackCollisionRate} & \result{VolCollisionsRuleOfThumb} \\
micro-TDE, $x\leq q^{1/3}$ & \result{StackMicroTdeRate} & \result{VolMicroTdeRuleOfThumb} \\
supply-limited events & \result{SupplyLimitedRate} & \result{VolSupplyRuleOfThumb} \\
Ref.~\cite{2022ApJ...933L..28Y} micro-TDE & -- & \result{YangMicroTdeRate} \\

\end{tabular}
\end{ruledtabular}
\end{table}
\section{Stellar response and repeated encounters}
\label{sec:return:implications}

The return map determines the interval and incident state of each subsequent
passage. It does not reset the star. Heating, expansion, stripping, and thermal
relaxation can change both the next chord and the time of re-entry.
The recurrence sequences below use prescribed stellar responses; their
endpoints test those closures rather than predict stellar outcomes.

The drag law introduces a separate physical uncertainty. Near-parabolic
entry does not imply high Mach number where the orbital work is deposited:
the AGN-like reference chords do most of their work in the transonic regime.
There the focusing interpolation and gaseous dynamical-friction prescription
can give different work, torque, and even first-passage outcomes
(Sec.~\ref{sec:return:mach}, Table~\ref{tab:drag-law}). A high-speed drag test
alone therefore cannot calibrate capture or the recurrence clock. These
sequences require a stellar-response model and a drag law appropriate to
the local flow.

\subsection{Isolated-return example}\label{sup:isolated}
\begin{figure}[t]
\includegraphics[width=\columnwidth]{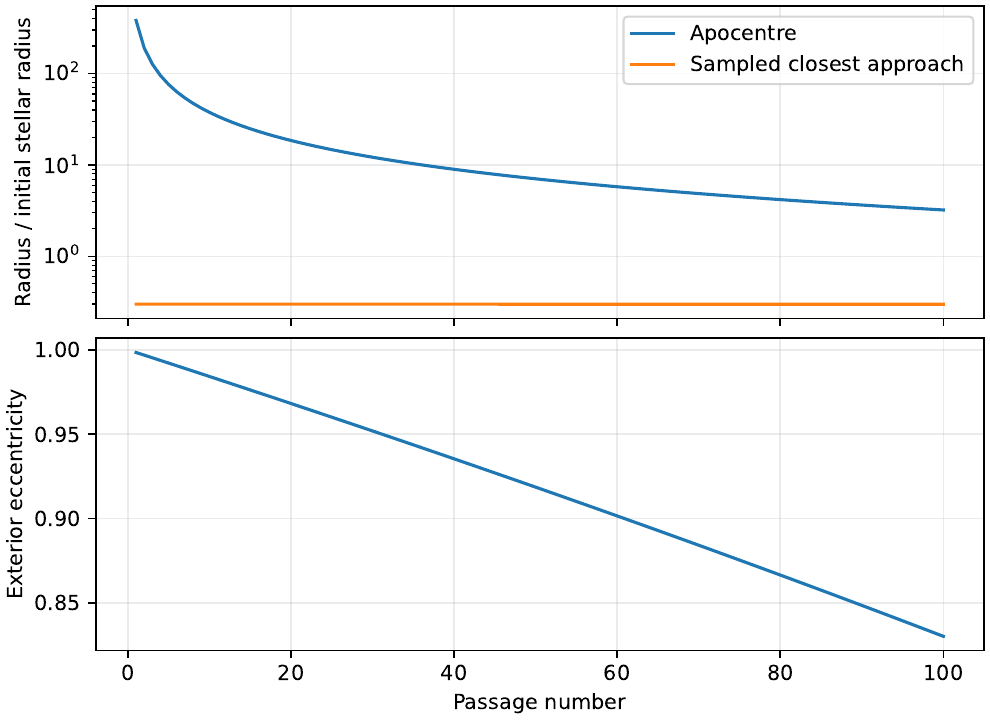}
\caption{\textbf{Repeated drag can leave a shallow pericenter nearly unchanged.}
Top: exterior apocenter and the sampled interior closest approach, in initial
stellar-radius units, over the $\result{ManyCount}$ bound transits of the
shallow fixed-structure sequence. The apocenter falls from
$\result{ManyFirstApo}$ to $\result{ManyLastApo}$ while the closest approach
moves from $\result{ManyFirstPeri}$ to $\result{ManyPeri}$. Bottom: the exterior
eccentricity, falling from $\result{ManyFirstEcc}$ to $\result{ManyEcc}$; this
is Eq.~(\ref{main-eq:decc}) with the energy term dominant. The star is held fixed,
which for the very long first return is a strong physical assumption.\llmfig}

\label{fig:returnmap}
\end{figure}

Figure~\ref{fig:returnmap} distinguishes orbital binding from deeper
penetration for a shallow,
near-parabolic orbit through the evolved massive-star structure of
Ref.~\cite{2026ApJ...999..123M}\footnote{B.~Mukhija and A.~Kashi, dataset for ``Envelope Inflation and Mass Loss Driven by Energy Deposition in Massive Stars'' (2025), Zenodo, \url{https://doi.org/10.5281/zenodo.15577892}.}.  The apocenter contracts strongly
while the sampled closest approach changes weakly, and the eccentricity falls
consistent with energy loss dominating the angular-momentum balance in Eqs.~(\ref{main-eq:decc})--(\ref{main-eq:dperi}).  The contrast with the
shorter, deeper sequences is reported in Table~\ref{tab:return-clock}; it is a
consequence of the two-component map, not an inference from a scalar energy
budget.

\subsection{Tidal-return tests}\label{sup:torque-tests}
\begin{figure*}[t]
\includegraphics[width=\textwidth]{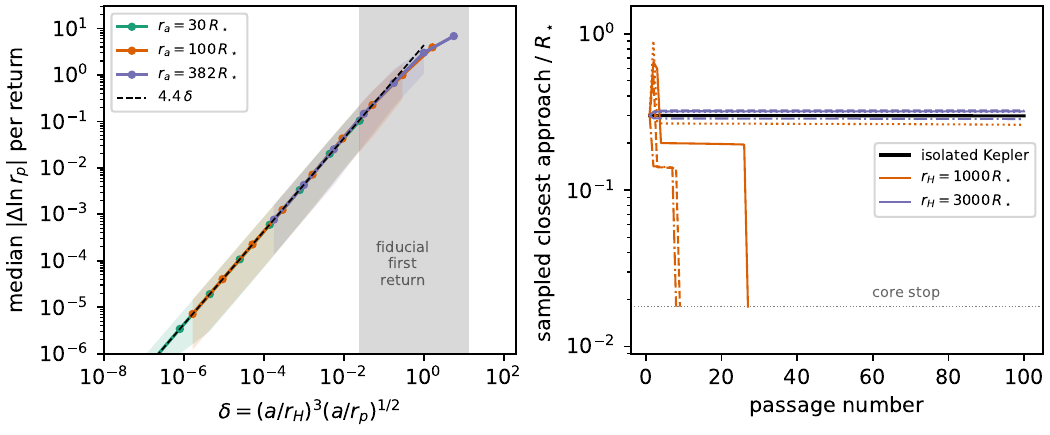}
\caption{Numerical tidal-return tests. Left: median $|\Delta\ln r_p|$ versus
$\delta$ for 96 isotropic orientations at three apocenters; bands show the
16th--84th percentiles and the dashed line is $4.4\delta$. The gray band
marks the fiducial first-return range over the 5th--95th percentile Hill radii.
Right: illustrative fixed-star sequences of closest approach versus passage
number, with isolated Kepler returns (black) and Hill returns at two Hill
radii (colors) and four orientations (line styles).\llmfig}
\label{fig:tidal-scaling-test}
\end{figure*}

The torque parameter and the Hill-return comparison are defined in
main-text Eq.~\eqref{main-eq:tide-delta} and Fig.~\ref{fig:tidal-scaling-test}.
Figure~\ref{fig:tidal-scaling-test} retains the numerical scaling test.
The coefficient has an analytic counterpart.
While the orbital period is short compared with $1/\Omega$, the first-order
kick is $\Delta\bm L=3\mu\Omega^2(\mathsf T\hat{\bm x})\times\hat{\bm x}$ with
$\mathsf T=\int\bm r\bm r^{\rm T}\,dt$ over the unperturbed exterior arc, the
frozen-axis tidal torque of phase-space reductions of the binary encounter
problem (O'Shaughnessy, in preparation). Evaluated on the same orientations in
the $\result{TideAnalyticCells}$ cells where it applies, it reproduces the
measured median to within a factor $\result{TideAnalyticRadialLow}$--$\result{TideAnalyticRadialHigh}$;
adding the vertical tide gives $\result{TideAnalyticFullLow}$--$\result{TideAnalyticFullHigh}$,
an estimate that omits the rotation of the tidal axis during the orbit. Returns deepen and lift the pericenter with
comparable frequency, with up to $\result{TideDeeperMax}$ per cent deepening
it by more than ten per cent. Beyond $\delta\simeq\result{TideMissDelta}$ at
least half of the returns miss the star or leave the Hill sphere, so the map is
no longer a sequence of transits but a temporary capture of the kind described
by Refs.~\cite{2023MNRAS.518.5653B,2024ApJ...972..193D}.

The shallow capture overlaps a tide-sensitive regime of the AGN measure. Over
contact approaches in the stack of Sec.~\ref{main-sec:channels}, the rate-weighted
$\rH/\Rstar$ runs from $\result{TideHillLow}$ to $\result{TideHillHigh}$ (5th to
95th percentile), with median $\result{TideHillMid}$. The shallow
near-parabolic capture of Fig.~\ref{fig:returnmap} leaves the binary with
apocenter $\result{TideFirstApocentre}\,\Rstar$, which gives
$\delta=\result{TideDeltaHigh}$--$\result{TideDeltaLow}$ over that range and
$\delta=\result{TideDeltaMid}$ at the median. By contrast, the drag of a shallow
transit changes $r_{p,K}$ by a fraction of a per cent per passage
(Fig.~\ref{fig:returnmap}), so the tide dominates the pericenter evolution
during the wide initial returns of this example.

The right panel of Fig.~\ref{fig:tidal-scaling-test} carries the same shallow sequence
through the Hill-frame map for a fixed star. The isolated map holds the closest
approach near $\result{TideSeqKeplerPeri}\,\Rstar$ for all
$\result{TideSeqKeplerCount}$ transits; this is a property of the integrated drag sequence, not a consequence of holding $L$ fixed. At
$\rH=\result{TideSeqHillInner}\,\Rstar$, inside the delivered range,
$\result{TideSeqCoreCount}$ of $\result{TideSeqOrientations}$ orientations
instead reach the core within $\result{TideSeqCoreMin}$--$\result{TideSeqCoreMax}$
transits: the tide lowers the pericenter during the first returns, the deeper
transits then do enough work and torque to shrink the orbit, and the sequence
finishes as an ordinary drag-driven inspiral. At
$\rH=\result{TideSeqHillMid}\,\Rstar$ the pericenter moves by at most
$\result{TideSeqMidShift}$ per cent, and at $\result{TideSeqHillOuter}\,\Rstar$
by $\result{TideSeqOuterShift}$ per cent. Because $\delta\propto a^{7/2}$, the
tide acts while the orbit is wide and switches off as drag contracts it, which
is why each sequence becomes Keplerian after its first few returns.

\subsection{Recurrence sequences and numerical accuracy}
\label{sec:return:clock}

The interval between stellar entries is the observable clock.  For passage
$n$, main-text Eq.~(\ref{main-eq:return}) gives
\begin{equation}\label{eq:recurrence}
 \begin{aligned}
  \Delta t_n={}&T(R_n;E_n^+,L_n^+)\\
                &+T(R_{n+1};E_n^+,L_n^+),\\
  t_{n+1}={}&t_n+\Delta t_n.
 \end{aligned}
\end{equation}
The recurrence sequence therefore follows from the same two state variables as
the penetration depth; it is not an independently fitted timescale.

\begin{figure}[t]
\includegraphics[width=\columnwidth]{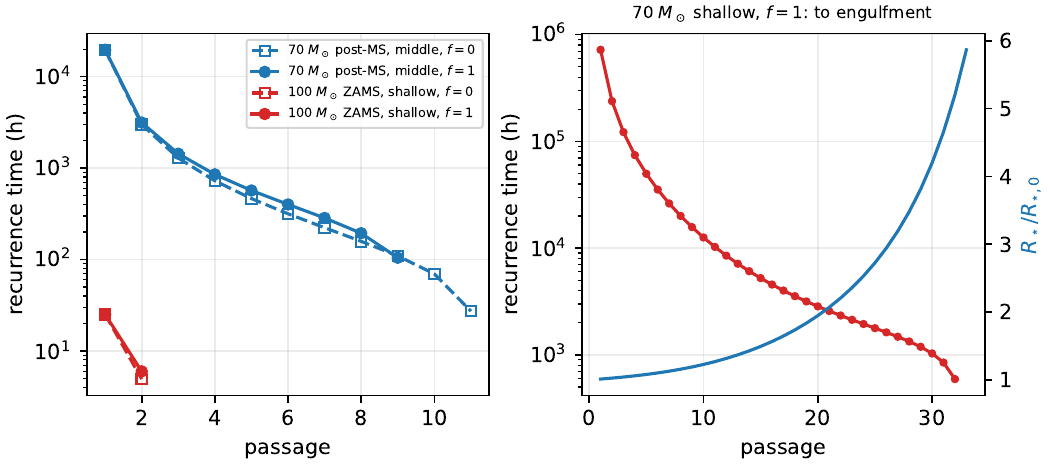}
\caption{\textbf{The recurrence clock runs down within a single capture.}
Left: recurrence time against passage number for a rigid star (dashed, open
markers) and a responsive one (solid), for the evolved $70\,\Msun$ structure at
middle depth and the $100\,\Msun$ zero-age main-sequence structure at shallow
depth; the latter reaches the core stop after three transits, which is why its
sequence is short. Right: the shallow responsive $70\,\Msun$ sequence carried to
engulfment, with the cumulative stellar expansion on the right axis; the surface
reaches apocenter after $\result{VoneEngulfCount}$ transits and
$\result{VoneEngulfYears}\,\unit{yr}$. The responsive branch is an illustration
of the clock's sensitivity to the star, not a calibrated stellar-response
calculation.\llmfig}

\label{fig:clock}
\end{figure}

Figure~\ref{fig:clock} compares a rigid star with a deliberately simple response
closure.  The transit work is binned by enclosed stellar mass $m_{\rm enc}$, and a retained
specific energy $\epsilon(m_{\rm enc})$ displaces each shell according to
\begin{equation}\label{eq:shell-response}
 \begin{aligned}
 \frac{Gm_{\rm enc}}{r}-\frac{Gm_{\rm enc}}{r'}&=f\epsilon(m_{\rm enc}),\\
 r'&=\frac{r}{1-f\epsilon(m_{\rm enc})r/(Gm_{\rm enc})}.
 \end{aligned}
\end{equation}
The deposition is smoothed by the minimum mass-coordinate width that prevents
shell crossing, after which pressure is reconstructed from hydrostatic
equilibrium.  Equation~(\ref{eq:shell-response}) conserves the prescribed shell
energy, but it is not stellar evolution: it omits transport, radiation, spin,
and a thermal timescale.  The responsive curves therefore measure the
sensitivity of Eq.~(\ref{eq:recurrence}) to the stellar state rather than predict
that state.

The plotted clocks and their endpoints are collected in
Table~\ref{tab:return-clock}.  Every sequence shortens its recurrence interval,
but response can change both the number of passages and the terminal condition.
The compact main-sequence-star sequences occupy the same time range as the
embedded damping calculation of Ref.~\cite{2026arXiv260827596S}, although the
physical problems differ: that calculation follows one collision to embedding,
whereas the map alternates interior transits and exterior returns.  The shallow
evolved-star sequence instead ends when the expanded surface reaches apocenter,
which is the explicit boundary of the map.

Table~\ref{tab:return-audit} separates numerical closure from physical closure.
Energy and angular momentum are conserved far more tightly than the model
systematics. The tabulated residuals use
$\widehat E=E R_{\star,0}/(G M_{\star,0}^{2})$ and
$\widehat L=L/[\mu(G M_{\star,0}R_{\star,0})^{1/2}]$,
so the accuracy measures retain their original dimensionless normalization.  No run in this set removes mass, while the separately reported
tidal proxy becomes non-negligible for the shallow evolved-star case.  A
mode-overlap calculation is therefore required before interpreting that clock
as a prediction, and chaotic stellar modes can make the passage-to-passage
exchange irregular~\cite{2026OJAp....961765H}; neither effect is contained in
Eqs.~(\ref{main-eq:transit-ode})--(\ref{eq:shell-response}).

\begin{table*}[t]
\caption{\textbf{Return-clock endpoints for the sequences displayed in
Fig.~\ref{fig:clock}.}  Times are in hours.  ``Responsive'' denotes the
$f=1$ limit of Eq.~(\ref{eq:shell-response}); it is a closure test, not a
calibrated stellar model.  The last row is the shallow evolved-star endpoint
shown in the right panel.\llmfig}

\label{tab:return-clock}
\begin{ruledtabular}
\begin{tabular}{lllrlrrr}
structure & geometry & response & $N_{\rm pass}$ & stop &
$\Delta t_1$ & $\Delta t_{\rm last}$ & $t_{\rm elapsed}$ \\
\hline
70 $M_\odot$ post-MS & middle & rigid & $\result{OldMiddleRigidCount}$ & core entry & $\result{OldMiddleRigidFirstRecurrenceMant}\times10^{\result{OldMiddleRigidFirstRecurrenceExp}}$ & $\result{OldMiddleRigidLastRecurrence}$ & $\result{OldMiddleRigidTimeMant}\times10^{\result{OldMiddleRigidTimeExp}}$ \\
70 $M_\odot$ post-MS & middle & responsive & $\result{OldMiddleSoftCount}$ & core entry & $\result{OldMiddleSoftFirstRecurrenceMant}\times10^{\result{OldMiddleSoftFirstRecurrenceExp}}$ & $\result{OldMiddleSoftLastRecurrence}$ & $\result{OldMiddleSoftTimeMant}\times10^{\result{OldMiddleSoftTimeExp}}$ \\
100 $M_\odot$ ZAMS & shallow & rigid & $\result{HundredShallowRigidCount}$ & core entry & $\result{HundredShallowRigidFirstRecurrence}$ & $\result{HundredShallowRigidLastRecurrence}$ & $\result{HundredShallowRigidTime}$ \\
100 $M_\odot$ ZAMS & shallow & responsive & $\result{HundredShallowSoftCount}$ & core entry & $\result{HundredShallowSoftFirstRecurrence}$ & $\result{HundredShallowSoftLastRecurrence}$ & $\result{HundredShallowSoftTime}$ \\
70 $M_\odot$ post-MS & shallow & responsive & $\result{OldShallowSoftCount}$ & engulfment & $\result{OldShallowSoftFirstRecurrenceMant}\times10^{\result{OldShallowSoftFirstRecurrenceExp}}$ & $\result{OldShallowSoftLastRecurrence}$ & $\result{OldShallowSoftTimeMant}\times10^{\result{OldShallowSoftTimeExp}}$ \\

\end{tabular}
\end{ruledtabular}
\end{table*}

\begin{table*}[t]
\caption{\textbf{Numerical and closure audit of the return map.}  The energy
and angular-momentum columns give the maximum absolute balance residual in
fixed reference units.  The last column is the range of the uncalibrated tidal
proxy as a percentage of cumulative drag work; it is never added to the drag
work.\llmfig}

\label{tab:return-audit}
\begin{ruledtabular}
\begin{tabular}{lrrrr}
map & cases & $|\delta\widehat E|_{\max}$ & $|\delta\widehat L|_{\max}$ & tidal/drag (\%) \\
\hline
fixed structure (v0) & -- & $\result{EnergyError}\times10^{-8}$ & $\result{AngularError}\times10^{-8}$ & -- \\
layer response (v1) & $\result{VoneCases}$ & $\result{VoneEnergyError}\times10^{-9}$ & $\result{VoneAngularError}\times10^{-10}$ & $\result{VoneTidalMin}$--$\result{VoneTidalMax}$ \\

\end{tabular}
\end{ruledtabular}
\end{table*}

\subsection{Where the drag work is produced, and which drag law applies}
\label{sec:return:mach}

A reduced encounter model is only as good as the flow regime in which its drag
closure is evaluated, and the relevant question is not how fast the BH is moving
but how fast it is moving where the work is done.  For any trajectory we define
the work-weighted cumulative distribution
\begin{equation}\label{eq:mach-weight}
 \begin{aligned}
  P_W(<\Mach_0)&=
  \frac{\int \Theta(\Mach_0-v/c_s)F_Dv\,\mathrm dt}
       {\int F_Dv\,\mathrm dt},\\
  f_{W,>5}&=1-P_W(<5).
 \end{aligned}
\end{equation}
The median of Eq.~(\ref{eq:mach-weight}), its high-Mach tail, and the fractional
orbital-energy loss are reported together in Table~\ref{tab:mach-audit}.  The
table shows why neither $u$ nor the maximum Mach number classifies the drag
regime: at fixed $u$, moving the chord inward increases the energy loss while
moving most of the work to lower Mach number.

The AGN measure of Sec.~\ref{main-sec:channels:stack} has no support in the rows that
are dominated by high-Mach work.  Its first passages are near parabolic and do
their work at moderate Mach number.  High-Mach encounters therefore test a
force law under controlled local conditions, whereas capture tests whether the
integral in Eq.~(\ref{main-eq:map}) changes the sign of $E$.  Success in one test does
not validate the other.

For the AGN measure, moderate means transonic. Table~\ref{tab:drag-law}
repeats the three reference chords at $u=\result{DragSpeed}$, inside the
delivered velocity support. The work-weighted median Mach number lies between
$\result{DragDeepMachOst}$ and $\result{DragOuterMachBhl}$, which is where
Eqs.~(\ref{main-eq:drag}) and (\ref{main-eq:ostriker}) differ most: for
$\ln\Lambda=\result{DragCoulombLog}$, the value adopted in
Ref.~\cite{2026arXiv260827596S}, the ratio $I/I_{\rm BHL}$ peaks at
$\result{DragRatioPeak}$ near $\Mach\simeq\result{DragRatioPeakMach}$ and
approaches $\ln\Lambda/C_D$ at high Mach. That reference likewise finds that a
BH entering a massive star moves transonically and treats the drag as gaseous
dynamical friction. On the shallow and middle chords Ostriker's law does
$\result{DragOuterWorkRatio}$ and $\result{DragMiddleWorkRatio}$ times the
work, with the same factors in the torque. On the deepest chord the stronger
drag slows the BH into the transonic regime within the first transit, where
the two laws diverge further; the work grows by a factor
$\result{DragDeepWorkRatio}$ and the BH reaches the core instead of exiting
bound. The drag law is therefore not a detail for AGN encounters. It moves the
capture threshold by order-unity factors at shallow depth and decides the
outcome of a deep passage. Hydrodynamic calibration is needed to establish
its applicability; here it bounds how far the clocks of
Table~\ref{tab:return-clock} can move.

\begin{table}[t]
\caption{\textbf{Drag law on AGN-like chords.} First transits at
$u=\result{DragSpeed}$ and $q=0.1$ through the evolved-star structure, under
the Bondi--Hoyle interpolation of Eq.~(\ref{main-eq:drag}) (BHL) and gaseous
dynamical friction, Eq.~(\ref{main-eq:ostriker}), with
$\ln\Lambda=\result{DragCoulombLog}$ (DF). $\Mach_{50,W}$ is the work-weighted
median Mach number of Eq.~(\ref{eq:mach-weight}); the ratios are DF over BHL
for the drag work $\mathcal W$ and torque $\mathcal J_D$ of
Eq.~(\ref{main-eq:map}).\llmfig}

\label{tab:drag-law}
\begingroup
\hyphenpenalty=10000
\begin{ruledtabular}
\begin{tabular}{@{}rrrrrp{0.51in}p{0.51in}@{}}
$x$ & $\Mach_{50,W}^{\rm BHL}$ & $\Mach_{50,W}^{\rm DF}$ &
$\mathcal W$ ratio & $\mathcal J_D$ ratio & BHL fate & DF fate \\
\hline
\result{DragDeepDepth} & \result{DragDeepMachBhl} & \result{DragDeepMachOst} & \result{DragDeepWorkRatio} & \result{DragDeepTorqueRatio} & bound exit & core entry \\
\result{DragMiddleDepth} & \result{DragMiddleMachBhl} & \result{DragMiddleMachOst} & \result{DragMiddleWorkRatio} & \result{DragMiddleTorqueRatio} & bound exit & bound exit \\
\result{DragOuterDepth} & \result{DragOuterMachBhl} & \result{DragOuterMachOst} & \result{DragOuterWorkRatio} & \result{DragOuterTorqueRatio} & bound exit & bound exit \\

\end{tabular}
\end{ruledtabular}
\endgroup
\end{table}

\begin{table}[t]
\caption{\textbf{Work-weighted flow regime.}  The median and high-Mach fraction
use Eq.~(\ref{eq:mach-weight}); $\Delta E/E_\infty$ is the fractional incoming
orbital-energy loss.  Each row is a distinct $(u,x)$ trajectory through the
same evolved-star structure.\llmfig}

\label{tab:mach-audit}
\begin{ruledtabular}
\begin{tabular}{rrrrr}
$u$ & $x$ & $\Mach_{50,W}$ & $f_{W,>5}$ (\%) & $\Delta E/E_\infty$ (\%) \\
\hline
\result{DeepModerateSpeed} & \result{DeepModerateDepth} & \result{DeepModerateMedian} & \result{DeepModerateFraction} & \result{DeepModerateLoss} \\
\result{DeepFastSpeed} & \result{DeepFastDepth} & \result{DeepFastMedian} & \result{DeepFastFraction} & \result{DeepFastLoss} \\
\result{MiddleModerateSpeed} & \result{MiddleModerateDepth} & \result{MiddleModerateMedian} & \result{MiddleModerateFraction} & \result{MiddleModerateLoss} \\
\result{MiddleFastSpeed} & \result{MiddleFastDepth} & \result{MiddleFastMedian} & \result{MiddleFastFraction} & \result{MiddleFastLoss} \\
\result{OuterModerateSpeed} & \result{OuterModerateDepth} & \result{OuterModerateMedian} & \result{OuterModerateFraction} & \result{OuterModerateLoss} \\
\result{OuterFastSpeed} & \result{OuterFastDepth} & \result{OuterFastMedian} & \result{OuterFastFraction} & \result{OuterFastLoss} \\

\end{tabular}
\end{ruledtabular}
\end{table}

\section{Prescription accounting}\label{sup:validity}
These fractions describe where the adopted prescriptions enter the stacked
rate; they are not probabilities that a physical model is accurate.
The shear kernel depends on total pair mass, and resonance records use the
adopted AGN mass functions. Pair identification and detached initial
star--BH binaries reduce contacts relative to the pooled calibration of
main-text Sec.~\ref{main-sec:channels:stack}; terminal branching includes
ionization. The binary abundances and separations are imposed. The
binary--binary shape is borrowed from binary--single encounters; resonance
records omit the SMBH tide and sample initial phase uniformly in true anomaly.
The tidal test constrains wide binaries, while its correction for the tighter
binaries supplying contacts remains unmeasured. These choices affect the
conditional contact kernel as well as its channel weight.

The return sequences omit ambient-gas dissipation and a surviving third body.
Their stellar-response prescriptions include neither thermal transport and
relaxation nor spin evolution. Core entry, shell crossing, and engulfment
therefore delimit the illustrative sequences rather than establish physical
endpoints.
\begin{table}[t]
\caption{\textbf{Validity accounting.} Rate-weighted percentage of the stacked
rate relying on each extrapolation or convention. The mass-extrapolation
row is empty because the resonance records are drawn at the adopted AGN masses.\llmfig}

\label{tab:validity}
\begin{ruledtabular}
\begin{tabular}{lr}
extrapolation or convention & \% \\
\hline
few-body calibration outside its masses & \result{ValidityExtrapMass} \\
outcome from N-body records or table & \result{ValidityFromTable} \\
$\tilde L_0$ clamped into range & \result{ValidityClamped} \\
democratic-escaper fallback & \result{ValidityFallback} \\
ionization analytic only & \result{ValidityAnalyticIon} \\
pair identity assumed, not measured & \result{ValidityPairIdentity} \\
contact, not GW capture & \result{ValidityFiniteRadius} \\
calibration phase uniform in true anomaly & \result{ValidityPhaseTrue} \\
deep passage parabolic ($u=0$) & \result{ValidityParabolic} \\
$(q,x,u)$ shape assumed & \result{ValidityScaled} \\
shear-dominated regime & \result{ValidityShear} \\

\end{tabular}
\end{ruledtabular}
\end{table}

\section{Single--single trajectory integration and numerical validation}
\label{sec:hill-numerics}
\begin{figure*}[t]
\includegraphics[width=\textwidth]{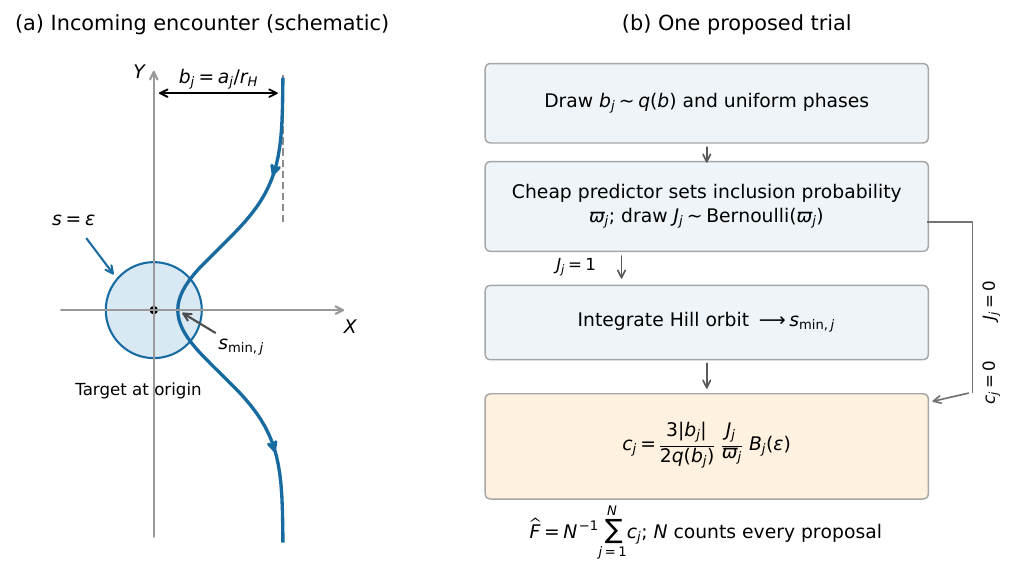}
\caption{\textbf{Geometry and sampling weights in Eq.~\eqref{eq:hill-kernel-estimator}.}
(a) Schematic planar projection: $b_j$ is the asymptotic guiding-center offset in Hill radii, not the pericenter. The threshold is $s=\epsilon$ and the smallest separation is $s_{\min,j}$; an integrated orbit has $B_j(\epsilon)=1$ if it crosses the threshold, and zero otherwise. Incoming epicycle phases are independently uniform. The drawn curve is illustrative, not an integrated trajectory.
(b) The proposal density $q$ determines how often offsets are drawn; the weight $3|b_j|/[2q(b_j)]$ restores the physical incoming shear flux. The separate inclusion probability $\varpi_j$ determines how often an expensive integration is performed; $J_j/\varpi_j$ corrects that selection. Skipped trials contribute zero, and $N$ includes every proposed trial. A trial that is integrated can still have $B_j=0$.\llmfig}
\label{fig:hill-estimator-diagram}
\end{figure*}

Write $s=r/\rH$ for dimensionless separation and $b=a/\rH$ for the incoming offset.
The implementation uses a normalized proposal density $q(b)$ and
uniform phases. A pilot of $200{,}000$ draws per cell uses
$q(b)=|b|/B^2$. Production combines this broad proposal, with fraction
$0.3$, with a proposal concentrated on the pilot's close-approach flux.
To reduce cost, a trial is integrated with probability $\varpi_j$ based
on the closest separation of its gravity-free continuation; the floor is
$\varpi_j\ge0.02$. For trial $j$, let $J_j$ indicate whether we perform the integration,
and define the encounter-outcome indicator
$B_j(\epsilon)=\mathbf1[s_{\min,j}<\epsilon]$ for an integrated trial.
Figure~\ref{fig:hill-estimator-diagram} illustrates the two distinct selections.
The estimator is
\begin{equation}\label{eq:hill-kernel-estimator}
 \widehat F(<\epsilon)=\frac1N\sum_{j=1}^{N}
 \frac{3|b_j|}{2q(b_j)}\,
 \frac{J_j}{\varpi_j}\,
 B_j(\epsilon).
\end{equation}
The denominator $N$ counts all proposed trials, including those not
integrated. A skipped trial has contribution zero; its minimum separation need not be computed. Proposal and inclusion probabilities are frozen from the
pilot before production. Inverse inclusion weighting preserves the flux
integral for this prescribed incoming ensemble; the gravity-free predictor
controls computational efficiency, not encounter selection. Standard
errors are obtained from the trial contributions in
Eq.~\eqref{eq:hill-kernel-estimator}, including their zero contributions.
Conditioned on a proposed initial state, $\mathbb E[J_j/\varpi_j]=1$;
together with the proposal correction this recovers the flux integral
of Appendix~\ref{main-app:hill-kernel-method}.

\subsection{Relation to established Hill-encounter calculations}
Conservative Hill integrations and phase-averaged collision and temporary-capture
rates have a direct precedent in planetesimal calculations, including
Iwasaki and Ohtsuki~\cite{2007MNRAS.377.1763I}, whose length and time units
are the mutual Hill radius and $\Omega^{-1}$.
The flux weighting follows the same shear-encounter construction as
Dodici and Tremaine~\cite{2024ApJ...972..193D}, whose
Eqs.~(20)--(23) estimate a binary-formation rate by summing Boolean
outcomes over incoming offsets and orbital phases. We use their notation
$B_j$ for the outcome, replacing binary formation by the event
$s_{\min,j}<\epsilon$. Our proposal density $q$ and inclusion indicator
$J_j$ are computational sampling devices; they do not change this
physical encounter criterion.

Their incoming radial-offset parameter $f_\xi$ corresponds to our
asymptotic $b$, and their relative epicyclic amplitudes $f_e,f_I$
correspond to ours. Their phase conventions translate as
$\delta_e=\phi_e$ and $\delta_I=\phi_I-\pi/2$; uniform phase averages are
unchanged. Their Cartesian length unit is $3^{1/3}\rH$, whereas ours is
$\rH$, so $(X,Y,Z)=3^{1/3}(\xi,\eta,\zeta)$ and both use
$\tau=\Omega t$. Our locally defined gravitational parameter $k$ is not their friction coefficient $k$.
These are asymptotic translations: our Jacobi-matched
finite-start offset $b_s$, described below, is not a new physical
impact parameter.

The rate convention also matters. For the same symmetric outcome and
incoming ensemble, their positive-offset dimensionless rate is
\begin{equation}\label{eq:hill-kernel-literature-rate}
 \widetilde{\mathcal R}
 =\frac32\int_0^B b\,\langle B_j\rangle\,\mathrm db
 =\frac12F.
\end{equation}
This is also the convention made explicit by Qian et al.~\cite{2024ApJ...962..143Q}
in their Eqs.~(24)--(26), where the signed-offset rate carries an
identical-pair factor $1/2$. It multiplies $n^2\Omega\rH^2$ to give an unordered identical-body pair
rate per disk area. Our signed-offset coefficient multiplies
$n\Omega\rH^2$ to give a rate per target; multiplying by $n/2$ gives
the same area rate. For distinct BH and stellar populations the area
rate is instead $\Sigma_{\rm BH}\Sigma_\star\Omega\rH^2F$ without the
identical-pair factor. The translation concerns normalization and
sampling, not equality of binary-formation and close-passage outcomes.

\begin{figure*}[t]
\includegraphics[width=\textwidth]{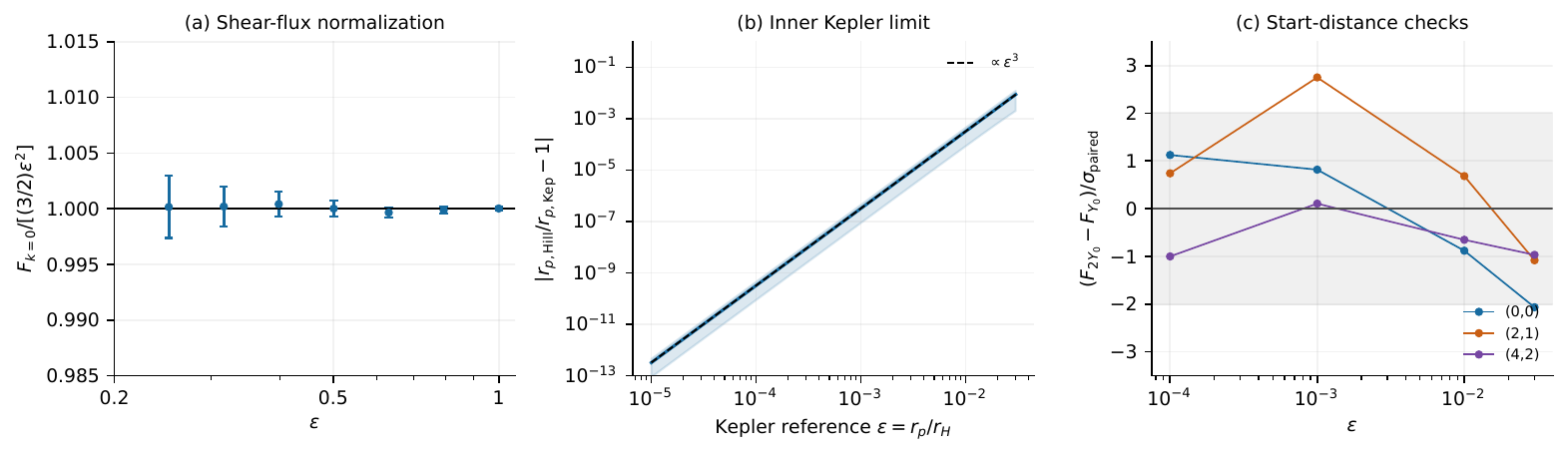}
\caption{\textbf{Numerical controls for the single--single rate coefficient.}
(a) Gravity-free planar shear recovers $F=3\epsilon^2/2$; error bars bound the finite-stratum quadrature error.
(b) Full-Hill pericenters approach isolated Kepler values at fixed $R_0/r_p=10$. The line gives the median over sixteen orientations and shading the 16th--84th percentiles; the cubic guide is anchored at the largest radius.
(c) Changes on doubling the upstream distance, divided by paired Monte Carlo standard errors; the gray band spans two such errors. These checks do not recalibrate the production kernel.\llmfig}
\label{fig:hill-numerical-controls}
\end{figure*}

\subsection{Trajectory integration and the stored table}
Positions in this section are measured in $\rH$, time in $\Omega^{-1}$, and velocities in $\Omega\rH$. Dots denote derivatives with respect to $\tau=\Omega t$. The dimensional gravitational parameter is $k=G(\Mstar+\Mbh)$, as defined in Appendix~\ref{main-app:hill-kernel-method}; its dimensionless coefficient is $k/(\Omega^2\rH^3)=3$.

Initial guiding centers are placed at
$Y_{{\rm gc},0}=\operatorname{sgn}(b)Y_0$, with
$Y_0=40+2f_e+f_I$. The finite-start correction preserves the incoming
Jacobi constant
\begin{equation}\label{eq:hill-kernel-jacobi}
 C=3X^2-Z^2+\frac6s-|\dot{\bm\rho}|^2
   =\frac34b^2-f_e^2-f_I^2.
\end{equation}
The code iterates the starting guiding-center offset $b_s$ so that
$b_s^2=b^2-8/s_0$, retaining the asymptotic $b$ in the flux weight.
Draws without a real $b_s$ are recorded as unreachable under this
finite-start prescription and contribute zero. This prescription is
checked by changing the starting distance below.

The generator \texttt{hill\_shear\_grid.py} calls the adaptive
eighth-order DOP853 integrator in \texttt{hill\_shear\_grid.c}.
Its nominal relative and absolute tolerances are $10^{-12}$ and
$10^{-14}$; inside $s=0.1$ they tighten in proportion to $s/0.1$,
with a relative-tolerance floor of $10^{-15}$.
Pericenters are located by root finding at negative-to-positive zeros
of $\bm\rho\cdot\dot{\bm\rho}$ within a step. The code retains the
minimum of these event radii and the accepted-step radii, and records
the osculating energy there using the inertial relative velocity
$\dot{\bm\rho}+\hat{\bm z}\times\bm\rho$.
The Jacobi drift, repeated-passage counts, and integration fates are
stored separately.

Integrations end after outward separation or a finite horizon based on
the upstream drift time plus $4000$ units of $\Omega^{-1}$, capped at
$60000$ units. Time-limited trajectories are retained and flagged;
their known passages contribute, but a future closer passage remains
censored. They are not assigned permanent escape. Step underflow below
the depth grid is flagged as a subgrid plunge; such an encounter already
qualifies at all tabulated depths. Failed integrations are excluded from
the close-approach sum and counted separately. The archived production
run reports no integration failures.

The regular amplitude axis is
$(0,1/8,1/4,1/2,1,2,4,6)$, giving $64$ interpolation cells;
$(6,3)$ and $(3,1.5)$ are additional diagnostic cells.
The $61$ depth thresholds are logarithmically spaced from $10^{-6}$ to
$1$. The stored coefficient is cumulative in the minimum separation,
not in the number of pericenter passages. The runtime implementation,
\texttt{hill\_shear\_kernel.py}, interpolates $\log F$ bilinearly in
the linear amplitude coordinates and interpolates in $\log\epsilon$
when selecting a depth. The population calculation averages amplitudes and uses the explicit
boundary-matched continuation outside the measured domain; it extends
below the smallest depth by a local power law with slope at least $1/2$.
Those numerical rules do not validate encounters outside the table.

In the population code, \texttt{channel\_single\_single} first constructs
a nominal weight proportional to $2\Sigma_n\Omega\rH^2$, multiplies it
by $F(<1)/2$, and draws depth from $P_H$. The expected weight below
$\epsilon$ is therefore
\begin{equation}\label{eq:hill-kernel-cancellation}
 \underbrace{2\Sigma_n\Omega\rH^2}_{\text{nominal rate}}
 \frac{F(<1)}{2}\,
 \frac{F(<\epsilon)}{F(<1)}
 =\Sigma_n\Omega\rH^2F(<\epsilon).
\end{equation}
This division by two cancels the nominal encounter-window convention;
it is unrelated to identical-species pair counting. The numerical kernel
needs no further normalization factor.

\subsection{Numerical controls}
\paragraph{Unforced planar shear.}
With mutual gravity switched off and $f_e=f_I=0$, the exact trajectory
has $X=b$, $Y=Y_{{\rm gc},0}-3b\tau/2$, and
$s_{\min}=|b|$. Equation~\eqref{main-eq:hill-kernel-integral} then gives
\begin{equation}\label{eq:hill-kernel-free-limit}
 F_{k=0}(<\epsilon)=\frac32\epsilon^2,
 \qquad \epsilon<B.
\end{equation}
This fixes the absolute flux normalization independently of gravitational
focusing. Figure~\ref{fig:hill-numerical-controls}(a) recovers it with
the production integrator and estimator on
$\result{HillValFreeDraws}$ equal-flux strata in $b^2$.
There is one jittered draw per stratum. The cumulative quadrature error
is bounded by the flux of one stratum, shown as error bars.

\paragraph{Inner Kepler limit.}
The SMBH tidal acceleration relative to mutual gravity is of order
$(r/\rH)^3$. We initialize parabolic Kepler passages with specified
$r_p$, at matching radius $R_0=10r_p$, convert the inertial velocity to
the rotating frame, and integrate the full Hill equations. Shrinking
$r_p/\rH$ at fixed $R_0/r_p$ should therefore recover the isolated
pericenter with a cubic tidal correction.
Figure~\ref{fig:hill-numerical-controls}(b) shows this behavior over
sixteen planar orientations; near $r_p/\rH=10^{-3}$ the median
fractional difference is $\result{HillValInnerError}$.
This checks the inner passage and velocity-frame conversion, not the
flux normalization of the outer chaotic band.

\paragraph{Numerical sensitivity.}
The archived checks repeat selected incoming trials at twice the starting
distance and at relative tolerance $10^{-13}$. Panel (c) shows paired
differences for three cells. Across all six checked cells and four depths,
the largest change is $\result{HillValConvergencePull}$ paired standard
errors. Tighter-tolerance changes are no larger than one paired standard
error. The rare deep-event start-distance checks can have substantial
fractional uncertainty; they do not establish uniform percent-level
accuracy throughout the grid. These checks and the semianalytic limits
support the stated normalization and limiting behavior while retaining
finite-start, finite-window, and sampling limitations.

\clearpage
\section{Use of large language models}
\label{app:llm}

Large language model (LLM) agents, including Claude and OpenAI Codex,
assisted with implementation, calculations, numerical validation, plotting,
and text preparation for this supplement. The figures were generated using
simulation, analysis, and plotting tools operated with LLM assistance.
The authors directed the work, reviewed proposed changes, and take full
responsibility for the content. The full disclosure and workflow reference
are given in Appendix~\ref{main-app:llm} of the main paper.

\bibliography{dynamics}